\documentclass[twocolumn,prb,preprintnumbers,amsmath,amssymb,floatfix]{revtex4}

\usepackage{graphicx}
\usepackage{dcolumn}
\usepackage{lscape}
\usepackage{float}

\newcommand{\Ry}{\ensuremath{\mathrm{Ry^*}}}
\newcommand{\aB}{\ensuremath{a^*_{\mathrm{B}}}}
\newcommand{\tOuter}{\ensuremath{\tilde{t}}}
\newcommand{\tInner}{\ensuremath{t}}
\newcommand{\Nsites}{\ensuremath{N_s}}
\newcommand{\Nelec}{\ensuremath{N_e}}
\newcommand{\Ntelec}{\ensuremath{N_e^{tot}}}
\newcommand{\Nsystem}{\ensuremath{N_{sys}}}
\newcommand{\vecr}{\ensuremath{\vec{r}}}
\newcommand{\rholoc}{\ensuremath{\rho_{\mbox{\scriptsize loc}}}}
\newcommand{\emass}{\ensuremath{m^*}}

\begin{document}

\title{Search for Ferromagnetism in doped semiconductors in the absence of transition metal ions}

\author{Erik Nielsen$^{1,2}$}
\author{R. N. Bhatt$^{1,3}$}
\affiliation{$^1$Department of Electrical Engineering, Princeton University, Princeton, NJ 08544-5263}
\affiliation{$^2$Sandia National Laboratories, P.O. Box 5800, Albuquerque, NM 87185 (current)} 
\affiliation{$^3$Princeton Center for Theoretical Science, Jadwin Hall, Princeton, NJ 08544}

\date{\today}

\begin{abstract}
In contrast to semiconductors doped with transition metal magnetic elements (\emph{e.g.}~Ga$_{1-x}$Mn$_x$As), which become ferromagnetic at temperatures below $\sim 10^2\mathrm{K}$, semiconductors doped with non-magnetic ions (\emph{e.g.} silicon doped with phosphorous) have not shown evidence of ferromagnetism down to millikelvin temperatures.  This is despite the fact that for low densities the system is expected to be well modeled by the Hubbard model, which is predicted to have a ferromagnetic ground state at $T=0$ on 2- or 3-dimensional bipartite lattices in the limit of strong correlation near half-filling.  We examine the impurity band formed by hydrogenic centers in semiconductors at low densities, and show that it is described by a generalized Hubbard model which has, in addition to strong electron-electron interaction and disorder, an intrinsic electron-hole asymmetry.  With the help of mean field methods as well as exact diagonalization of clusters around half filling, we can establish the existence of a ferromagnetic ground state, at least on the nanoscale, which is more robust than that found in the standard Hubbard model.  This ferromagnetism is most clearly seen in a regime inaccessible to bulk systems, but attainable in quantum dots and 2D heterostructures.  We present extensive numerical results for small systems that demonstrate the occurrence of high-spin ground states in both periodic and positionally disordered 2D systems.  We consider how properties of real doped semiconductors, such as positional disorder and electron-hole asymmetry, affect the ground state spin of  small 2D systems.  We also discuss the relationship between this work and diluted magnetic semiconductors, such as Ga$_{1-x}$Mn$_x$As, which though disordered, show ferromagnetism at relatively high temperatures.
\end{abstract}

\maketitle

\section{Introduction}
Originally proposed in the early 1960s\cite{Hubbard_1963,Gutzwiller_1963,Kanamori_1963,Anderson_HubModel_1963}, the Hubbard model combines tight binding hopping between nearest neighbors on a lattice with an on-site Coulomb repulsion between electrons in the same orbital state. Though it is one of the simplest interacting models, its on-site intra-orbital correlations are believed to be the most important source of correlations in solids.  Indeed, the Hubbard model displays great diversity of transport and magnetic properties, giving rise to insulating, metallic, and superconducting phases as well as ferromagnetic (FM), antiferromagnetic (AF) and paramagnetic spin order.  It has been used to study a wide range of correlated systems, including Mott-insulator oxides,\cite{MottBook} high-$\mathrm{T}_c$ superconductors,\cite{AndersonCuprates_1987,LeeCupratesRMP_2006,IzyumovTJmodel_1997,MacridinCuprates_2005} organic materials,\cite{Pyo_2005,Wu_2004,Sing_2003} $\sqrt{3}$-adlayer structures,\cite{Weitering_1997} vanadium oxides,\cite{McWhan_1973,Carter_1991} nickel sulphide-selenide alloys,\cite{Ogawa_1979,ThioBennett_1994,ThioBennett_1995} hydrogenic centers in doped semiconductors\cite{FerreiraSpecificHeat_1981,RefolioKSiInterface_1996}, and quantum dots.\cite{Massimo_hubInQDots_1999}  Such great interest and applications have resulted in analyses of the model on different lattices,\cite{Hanisch_diffLatt_1997,Wegner_diffLatt_1998} with multiple\cite{Penc_multiBand} and degenerate\cite{Fresard_degenBand,Kuei_degenBand} bands, and with binary alloy disorder.\cite{Byczuk_alloyDisorder}   Many studies restrict themselves to the infinite $U/t$ limit,\cite{Becca_largeU,Obermeier_largeU} which can be realized most effectively in optical lattices,\cite{JakschZoller_2005} but can be approached in semiconductor systems as well.  We will be concerned with the case of semiconductors doped with shallow hydrogenic impurities.  Here the model is particularly appropriate at low densities (\emph{i.e.} in the insulating phase, where carriers are bound to a few sites and the Coulomb interaction is large compared to the kinetic energy).  In this low density limit each site is treated as an effective hydrogen atom with a corresponding effective Rydberg and Bohr radius:
\begin{equation}
\Ry = \frac{\emass e^4}{2 \epsilon^2 \hbar^2} \qquad \qquad \aB = \epsilon \hbar^2 / \emass e^2 \label{eqIntro_effRyAndBohrRad}
\end{equation}
where $\emass$ is the effective mass in the appropriate band and $\epsilon$ is the dielectric constant of the host material. 
  In doped semiconductors, typically $\epsilon \sim 10 - 20$ and $\emass$ is 0.05 to 0.5 times the free electron mass, so that $\aB \sim 10 - 500$ \AA~and $\Ry \approx 1-50 \mbox{meV}$.  Since the $\Ry$ is usually much smaller than the bandgap of the host semiconductor, the lattice lacks low-energy electronic excitations on the energy scale of the impurity electrons and essentially plays the role of an inert vacuum.  Realistic effects like valley degeneracy and mass anisotropy must be included for quantitative calculations but are unnecessary for the qualitative phenomena of interest to us.\cite{Thomas_1981,AndresBhatt_1981}  We will assume that all relevant energy scales are much smaller than the gap between the lowest and higher orbital states on an isolated dopant, so that we need only care about the $1s$ orbital of each dopant, which consists of two electronic spin-degenerate states at energy denoted $E_0$.  A hydrogenic center, like a hydrogen atom, is known to bind up to two electrons.\cite{Pekeris_1962}  With a single electron the problem is that of atomic hydrogen ($H$), and the electron is bound with 1 \Ry.  The two electron case corresponds to the $H^-$ ion, which has a spin singlet ground state bound by 0.0555 \Ry.\cite{MottBook,BS_QMbook_1977,BhattRice_1981} 

We begin with a review of the Hubbard model and its properties on a lattice in section \ref{secHubbardBackground}.  The absence of certain magnetic properties, namely ferromagnetism, in real materials leads to a discussion of disorder and reveals the need to incorporate it into the model.  This is done in section \ref{secHubbardForHydrogenic} where we motivate and define a model appropriate for doped semiconductors.  The parameter ranges of interest for this model are also given in section \ref{secHubbardForHydrogenic}, along with details of the model's solution.  Results on finite lattices, selected symmetric clusters, and small random clusters are presented in section \ref{secHubbardResults}.  Large systems of random impurities are treated in section \ref{secVaryDensityClusters} by dividing them into smaller clusters which can be solved exactly.  Section \ref{secConclusion} highlights our major conclusions and discusses topics for continued work.

\section{Background: the Hubbard model \label{secHubbardBackground}}

\subsection{Definition and general properties}

The Hamiltonian of the Hubbard model on a \emph{lattice} with $\Nsites$ sites is given by:
\begin{equation}
\mathcal{H} = - t\sum_{\langle i,j\rangle\sigma} \left( c^\dag_{i\sigma} c_{j\sigma} + \mbox{h.c.} \right) + U\sum_{i=1}^{N_s} n_{i\uparrow}n_{i\downarrow} \label{eqnHubHamOriginal}
\end{equation}
where $i$ and $j$ range from $1$ to $\Nsites$, and the first sum is over all distinct nearest neighbor pairs.  Operators $c^\dag_{i\sigma}$ and $ c_{i\sigma}$ create and annihilate, respectively, an electron of spin $\sigma \in \{\uparrow,\downarrow\}$ on site $i$, and satisfy canonical fermion anticommutation relations
\begin{eqnarray}
\left\{ c^\dag_{i\sigma}, c_{i'\sigma'} \right\} &=& \delta_{ii'} \delta_{\sigma\sigma'} \\
\left\{ c^\dag_{i\sigma}, c^\dag_{i'\sigma'} \right\} &=& 0 \\
\Big\{ c_{i\sigma}, c_{i'\sigma'} \Big\} &=& 0
\end{eqnarray}
for any $i = 1\ldots \Nsites$ and $\sigma \in \{\uparrow,\downarrow\}$.  The number operator $n_{i\sigma} = c^\dag_{i\sigma}c_{i\sigma}$, and has eigenvalues 0 and 1.  The Hilbert space of each site has as a basis the four states where the site is occupied by an up-spin, a down-spin, neither, or both.  The total Hilbert space is the direct product of site Hilbert spaces, and therefore has dimension $4^{\Nsites}$.  The parameter $t$ is the quantum mechanical hopping amplitude between (nearest-neighbor) sites, and $U$ is the strength of the on-site Coulomb repulsion.  We include a minus sign in front of the kinetic term, so that for the familiar example of the tight-binding model with hydrogenic wavefunctions,\cite{Bhatt_1981} $t(r)=2(1+r/a_{\mathrm{B}})\exp(-r/a_{\mathrm{B}})$ is positive, and restrict ourselves to $U>0$ (repulsive interaction).  Note that the eigenstates of each term independently are trivial: they are states of definite momentum when $U=0$ and states of definite position when $t=0$.  Thus, inherent in the Hubbard model is a competition between the extended (wave-like) and localized (particle-like) nature of the electrons, and there is no clear classical analogue.

\begin{figure}
\begin{center}
\begin{tabular}{ll}
\includegraphics[height=1.5in]{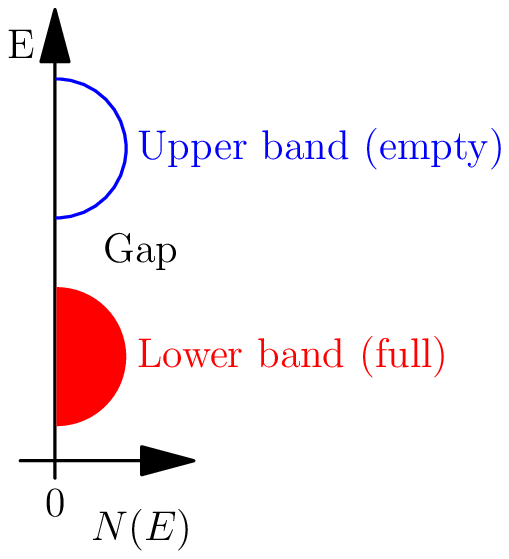} 
& \includegraphics[height=1.5in]{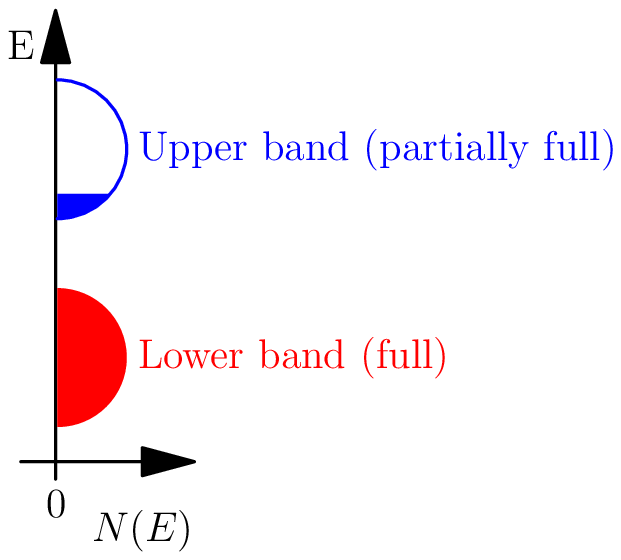} \\
\hspace{1cm} a) &  \hspace{1cm} b)
\end{tabular}
\caption{(Color online) Schematic figure showing a system at half-filling (a), and slightly above half-filling (b).  At half-filling the lower impurity band is completely full and there is a gap to charge excitations. Above half-filling there are electrons present in the upper (unfilled) band that can act as carriers if they occupy extended states (as they do in a lattice).  Note also that each band's density of states $N(E)$ is not actually semicircular, but drawn this way for convenience.\label{figBandFillings}  }
\end{center}
\end{figure}

We will be primarily concerned with the strong correlation limit $U \gg t$, so that at half-filling (\emph{i.e.}~one electron per site) the single particle (charge) spectrum has a gap, and the system is insulating (Fig.~\ref{figBandFillings}(a)).  Nonetheless, the system can have low lying spin excitations.  At large $U/t$ and half-filling, the Hubbard model at low energies is effectively a Heisenberg model,\cite{AndersonEffHeisenberg_1963} in which fermionic electron operators are represented by spin operators.  On a bipartite lattice, the Heisenberg model, given by Hamiltonian  $\mathcal{H}_{Heis} = J \sum_{<ij>} \vec{S}_i\cdot \vec{S}_j$, has an AF ground state.\cite{ManousakisRMP_1991}  Away from half-filling, where there are carriers (see Fig.~\ref{figBandFillings}(b)), one must use a more general low-energy theory that includes a kinetic term, the $t-J$ model:\cite{ChaoSpalekOles_1977,ChaoSpalekOles_1978}

\parbox{3in}{
\begin{eqnarray}
\mathcal{H}_{tJ} &=& - t \sum_{<ij>\sigma} \left( (1-n_{i\bar{\sigma}})c^\dag_{i\sigma}c_{j\sigma}(1-n_{j\bar{\sigma}}) + \mbox{h.c.}\right) \nonumber \\
 & & +\, J \sum_{<ij>} \left(\vec{S}_i\cdot \vec{S}_j - \frac{1}{4}n_i n_j \right)\,. \nonumber 
\end{eqnarray}} \hfill \parbox{0.1cm}{\begin{eqnarray} \label{eqtJModel} \end{eqnarray}}

\begin{figure}
\begin{center}
\includegraphics[width=1.3in]{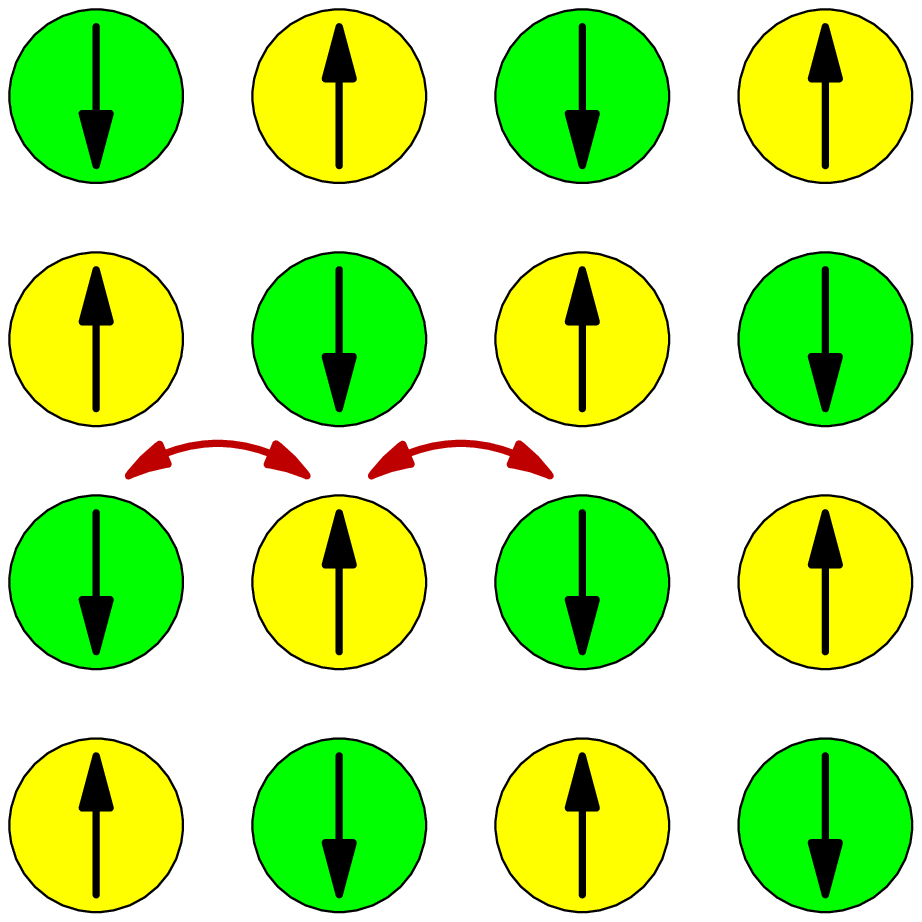} \hspace{1cm}
\includegraphics[width=1.3in]{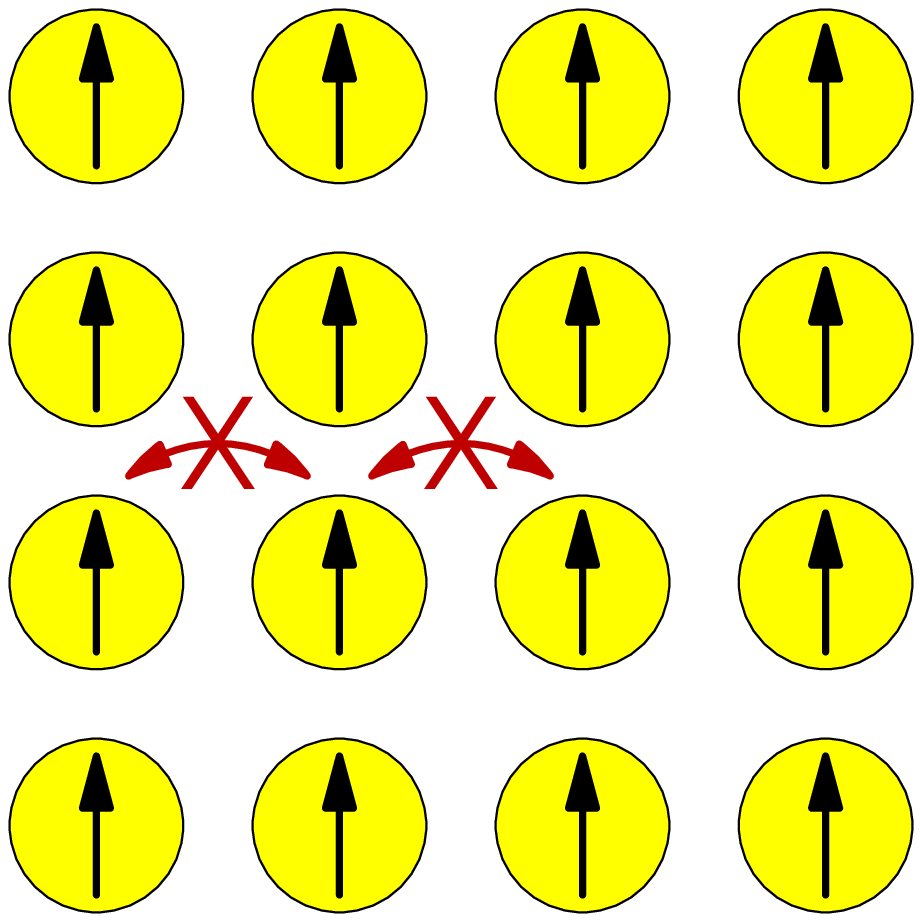}
\caption{(Color online) Diagrams which graphically tell the origin of the AF exchange interaction term of the $t-J$ model: in an AF configuration (left) electrons and virtually hop to a neighboring site and back (shown by the arrows), resulting in a net lowering of the energy by second order perturbation theory.  In a FM configuration (right), however, Pauli exclusion forbids such virtual processes, and the system cannot lower it's energy in this way.\label{figExchangeTermOrigin}}
\end{center}
\end{figure}

\noindent Note that the Hamiltonian operates on the restricted Hilbert space which excludes doubly-occupied sites.  From the Heisenberg and $t-J$ models we see that the inclusion of electron-electron interactions results in an AF exchange interaction $\sim \vec{S}_i \cdot \vec{S}_j$, where $\vec{S}_i = \sum_{\alpha\beta} c^\dag_{i\alpha} \sigma_{\alpha\beta} c_{i\beta}$.  The exchange term is due to the fact that virtual hopping of electrons between neighboring sites is allowed when their spins are oppositely oriented but not when their spins are parallel (as in a FM configuration), as shown in Fig.~\ref{figExchangeTermOrigin}.\cite{AndersonExchange_1959}  

The Hubbard Hamiltonian can also be written in terms of spin operators using the identity $\sum_i \left(\vec{S}_i \right)^2 = \sum_i \left(\frac{1}{4}n_i - \frac{3}{2}n_{i\uparrow}n_{i\downarrow}\right)$, where $n_i=n_{i\uparrow}+n_{i\downarrow}$, casting Eq.~\eqref{eqnHubHamOriginal} into the form:
\begin{equation}
\mathcal{H} = -t\sum_{\langle i,j\rangle\sigma} \left( c^\dag_{i\sigma} c_{j\sigma} + \mbox{h.c.} \right) - \frac{2U}{3}\sum_{i=1}^{N_s} \left(\vec{S}_i\right)^2 + \frac{\Nelec U}{6} \label{eqnHubHamOriginal_spinForm}
\end{equation}
where $\Nelec$ is the total number of electrons.  This form clearly shows the total spin SU(2) invariance of the Hubbard model, and also that when $U>0$ the interaction energy is lowest when the total spin on each site is maximized, suggesting the existence of ground states with high values of total spin at large $U$.  On a bipartite lattice with disjoint sublattices $A$ and $B$, the sign of $t$ can be changed via the transform:

\parbox{2in}{
\begin{eqnarray} 
c_{i\sigma} &\rightarrow& +c_{i\sigma} \qquad \mbox{if} \quad i \in A \nonumber \\
c_{i\sigma} &\rightarrow& -c_{i\sigma} \qquad \mbox{if} \quad i \in B \nonumber
\end{eqnarray}} \hfill
\parbox{1cm}{\begin{eqnarray}\label{eqBipartiteTransform}\end{eqnarray}}

\noindent which does not change the canonical commutation relations and thus leaves the spectrum invariant.  The Hubbard model also possesses particle-hole symmetry on a bipartite lattice, where $U$ maps to $-U$ and total charge is interchanged with total spin (for a detailed explanation of symmetries in the Hubbard model, see Ref.~\onlinecite{FradkinBook_1991}).

Even when it is applied to simple systems (\emph{e.g.}~1-, 2-, and 3-dimensional lattices), the Hubbard model yields interesting and non-trivial properties, seen through the nature of its excitations, density of states, spectral weight, transport, and optical and magnetic behavior.\cite{GeorgesKotliar_1996,UlmkeJanisVollhart_1995,ChandraKollarVollhart_1999,Eckstein_2007}  Here we will concentrate on the nature of magnetic correlations in the ground state, which are then used to construct the ground state (\emph{i.e.}~$\mathrm{T} = 0$) phase diagram.

\subsection{Magnetic Properties}
\begin{figure}
\begin{center}
\begin{tabular}{cc}
\parbox{1.5in}{\includegraphics[width=1.2in]{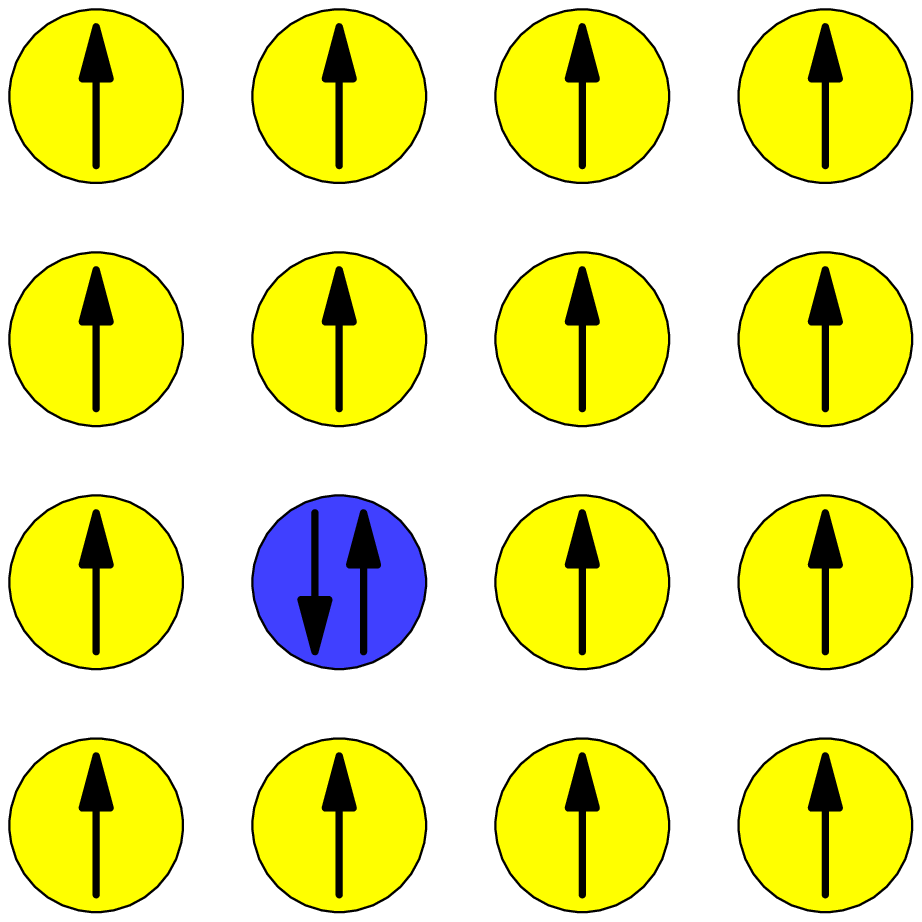}}
& \parbox{1.5in}{\includegraphics[width=1.2in]{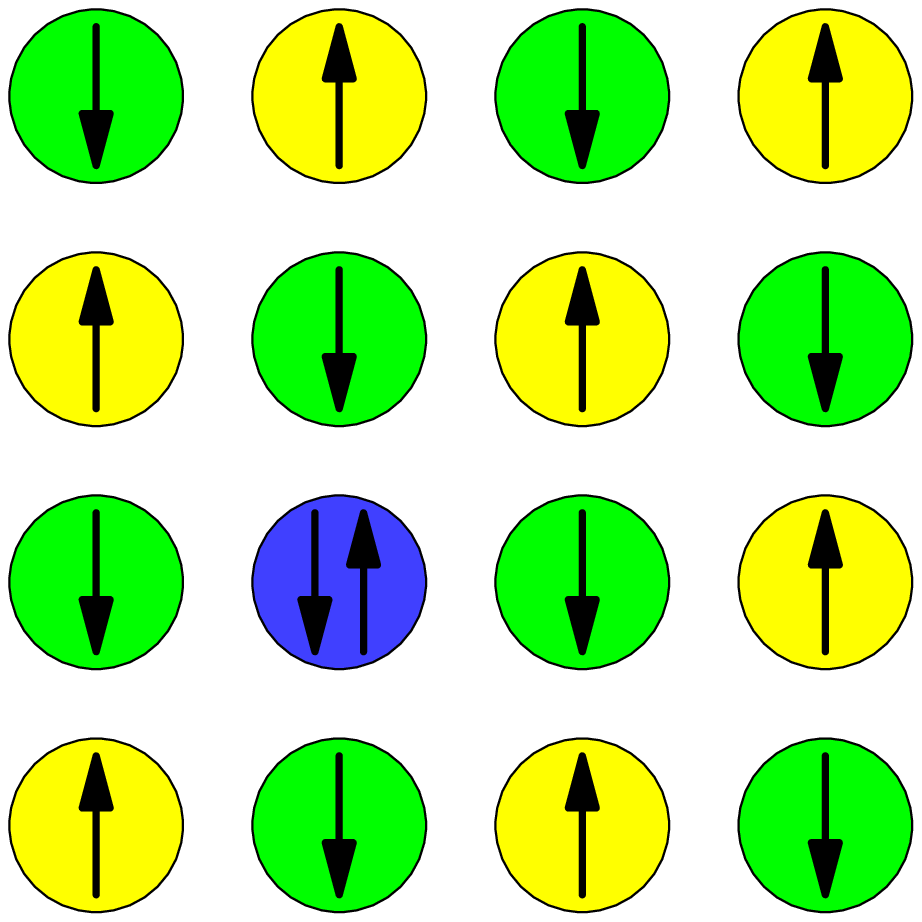}} \\ 
 a) &  b) \\
 & \\
 \multicolumn{2}{c}{ \parbox{1.5in}{\includegraphics[width=1.2in]{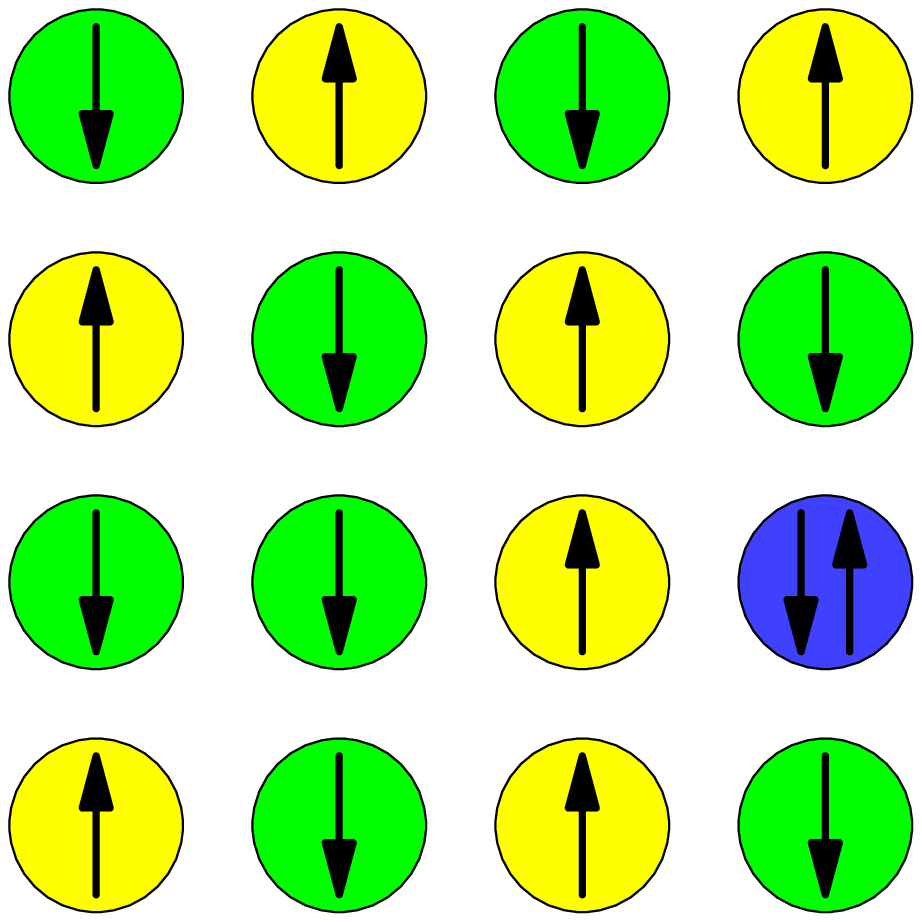}}} \\
 \multicolumn{2}{c} { c) }
\end{tabular}
\caption{(Color online) Diagrams showing why the kinetic term favors a ferromagnetic state: in a) the down spin on the single doubly-occupied site can move freely without disturbing the underlying FM background.  However, if the background is AF as in b), motion of electrons on doubly-occupied sites scramble the Neel order.  Diagram c) shows the result of the doubly occupied site in b) moving two sites to the right.\label{figAFscramble}  }
\end{center}
\end{figure}

The magnetic properties of Hubbard systems can be very rich due to competition between two or more magnetic phases.  Consider the Hubbard model at half-filling on a bipartite lattice, where there is no classical magnetic frustration, and let $U/t$ be large.   The model's quantum ground state is a superposition of ``Neel antiferromagnet states'' where spins on each sublattice are aligned and oppositely oriented to those of the other sublattice as well as ``spin-flip states'' which differ from the Neel AF states by exchanging one or more pairs of spins (states with a greater number of flips occur with lower weight).  In other words, the ground state is a superposition of states with long-range Neel order.  Since $U$ is large, the $t-J$ approximation (Eq.~\eqref{eqtJModel}) is valid, introducing an exchange energy $J\sim t^2/U$  between neighboring spins.  The kinetic term of the $t-J$ Hamiltonian does not play a role since at half-filling there are no mobile carriers.  Thus, at half-filling the exchange directly gives rise to an antiferromagnet.  When the system is above or below half-filling, however, the kinetic term plays a competing role by favoring a ferromagnetic spin configuration.  This is so because as carriers hop from site to site they do not disturb an underlying FM spin configuration, whereas they necessarily scramble an AF one (see Fig.~\ref{figAFscramble}).  This scrambling leads to an unfavorable increase in energy, and thus the preference for ferromagnetism.\cite{BrinkmanRice_1970, ShraimonSiggia_1988}  Relative to an AF state, a FM system with carrier (electron or hole) density $\delta$ gains kinetic energy of order $t\delta$ due to carrier delocalization and loses magnetic energy of order is $J=4t^2/U$.  Thus, at a fixed small $\delta$, when $U$ is large enough, $t \delta \gg J$, and the system prefers a FM configuration over the AF one because it allows carriers to be less confined.  Understanding the applicability and validity of this argument, and more generally the factors that govern the magnetic competition found in the Hubbard model, has been the topic of much work.  Indeed, it has led to most (if not all) of the rigorous results that are known concerning the Hubbard model.

Despite the apparent simplicity of the Hubbard model and voluminous literature surrounding it, few rigorous theoretical results have been proven about it.  Most striking among them is the result of Nagaoka,\cite{Nagaoka_1966} which states that in the infinite correlation limit $U/t\rightarrow\infty$, the Hubbard model on certain finite lattices of dimension $d\ge 2$ with periodic boundary conditions, $t<0$, and a single hole (away from half-filling), has a FM ground state (\emph{i.e.}~the total spin $S^2$, where $\vec{S}=\sum_i \vec{S}_i$, attains it's maximal value).  This result, dubbed the Nagaoka Theorem, applies to most standard lattices, including the square, simple cubic, triangular, kagom\'{e}, bcc, and fcc (hcp).\cite{Nagaoka_1966,Tasaki_2003}  In the case of bipartite lattices, such as the square, simple cubic, and bcc, $t$ can be taken positive (the physical sign in the tight binding model) by the transform of Eq.~\eqref{eqBipartiteTransform}.  This can be understood from the preceding discussion of a bipartite system, where, upon setting $U=\infty$, the criterion $t \delta \gg J$ is satisfied for any $\delta > 0$ and thus only a single hole is needed to produce a FM ground state.  Even though this criterion also predicts ferromagnetism for a finite density of carriers at large $U$, a rigorous result even for the case of a few holes has proved difficult.\cite{TianFewHoles_1991,Trugman_1990}  Along with the rigorous proofs in Nagaoka's and Thouless' work, simpler and more modern mathematical proofs are given by Tian\cite{TianNagaokaProof_1990} and Tasaki.\cite{TasakiNagaokaProof_1998}.  Another rigorous theorem regarding magnetism in the Hubbard model by Lieb\cite{LiebFerrimagnetism_1989} states that a half-filled bipartite system whose sublattices have different numbers of sites will have an unsaturated FM ground state.  It later became clear that the tendency toward ferromagnetism was due to the single particle density of states being dispersionless, or flat, at the center of the band (the Fermi level at half-filling).  Later, results of Mielke\cite{MielkeFlatBands_1991} and Tasaki\cite{TasakiFlatBands_1992} generalized this idea to characterize a broader class of half-filled systems with dispersionless single-particle spectra and saturated FM ground states, said to exhibit ``flat-band ferromagnetism.''  We hasten to point out, however, that half-filled Hubbard systems are generally antiferromagnetic (when on a bipartite lattice) or paramagnetic, and that completely or nearly flat bands should be viewed at least as a non-generic case. 
 This fact underscores the surprising result of Nagaoka, which describes the transition from an antiferromagnet to a ferromagnet upon the addition of a single hole or electron.

The bulk of this section investigates the possibility of saturated ``Nagaoka ferromagnetism'', and it is worthwhile at this point to consider the progress of past work toward understanding the phenomenon.  The topic has generated sizable interest, since the Nagaoka Theorem is at the same time striking and of only limited use, saying nothing about the thermodynamic limit where there is relevance to experiment.  Many theoretical studies\cite{Becca_largeU,Obermeier_largeU,Denteneer_1996,LongZotos_1993,Chiappe_1993} work in the large or infinite $U$ limit, which is where saturated FM is most likely to occur.  In the $U=\infty$ limit doubly-occupied states are eliminated from the Hilbert space, which then has a dimension that scales as $3^{\Nsites}$ -- substantially less than $4^{\Nsites}$ and thereby a great relief for numericists!  Indeed, much computational work has been done setting $U=\infty$, including one which relates the system far below half-filling to one of hard-core bosons.\cite{LongZotos_1993}

Investigating the existence, extent, and stability of the Nagaoka state has established several conditions that are known to favor a stable, saturated FM ground state.  By considering the stability of the fully polarized state to a single spin flip, it is shown\cite{PastorHirschMuhlschlegel_1994,BarbieriRieraYoung_1990,Hanisch_diffLatt_1997} that an asymmetric density of states with a peak at the appropriate band edge (lower edge of the upper Hubbard band if doping above half-filling; upper edge of lower Hubbard band if doping below) is one such condition.  This makes intuitive sense, since having large density of states at the Fermi level diminishes the kinetic cost of filling additional single-particle electronic states and causes the large $U$, which favors spin alignment, to prevail.  This generalizes the condition of a flat band discussed earlier, in which the density of states is infinite at the band edge.  From the geometrical optimization of finite Hubbard clusters, Pastor \emph{et al.}\cite{PastorHirschMuhlschlegel_1994,PastorHirschMuhlschlegel_1996}~find that saturated ferromagnetism coincides with clusters which are \emph{non-bipartite} and have a large number of frustrated ``tight'' triangular loops. They also find that doping clusters above rather than below half-filling yields a density of states with higher weight at the band edge, and leads to FM ground states.  The asymmetry with respect to doping and correspondence between magnetism and triangular loops is corroborated by our results, and appears to be a quite general feature with important experimental ramifications, which we discuss in more detail in section \ref{secGeomDistorted}.  It is also known that adding back into the single-band Hubbard model physical interactions that it neglects, particularly a (direct) ferromagnetic Heisenberg exchange interaction, can be important for stabilizing ferromagnetism near half-filling for finite $U$.\cite{StrackVollhardt_1994,KollarStrackVollhardt_1996,Wahle_1998}  Additionally, the next-nearest-neighbor (NNN) hopping amplitude $t'$ is believed to play an important role: decreasing $t'/t$ (especially below zero) stabilizes saturated FM to higher hole-doping in the $U=\infty$ Hubbard model on a square lattice.\cite{Park_2008}  In one-dimension, where the Lieb-Mattis theorem\cite{LiebMattis_1962} forbids FM in the standard Hubbard model, the addition of (NNN) hopping $t'$ such that $t'/t < 0$ ($t > 0$) results in a widespread FM phase.\cite{DaulNoack_1998}  Alternatively, the generalization to a multi-band model (appropriate for many transition metals) with a ferromagnetic exchange interaction between electrons in different orbitals (``Hund's rule couplings'') also abets the stability of a FM state.\cite{Vollhardt_1999}  (The inclusion of  multiple bands, however, is not crucial to FM stability in 2 and 3 dimensions.\cite{BarbieriYoung_1991})  We do not consider either of these routes, and restrict our study to a nearest-neighbor model with one orbital and on-site Coulomb interaction.



It is important to remember, however, that saturated ferromagnetism in the Hubbard-like models is not ubiquitous.  Other work has shown it to be a subtle effect, depending on dimension and lattice geometry.  For instance, another rigorous result, due to Lieb and Mattis,\cite{LiebMattis_1962} proves that in finite one-dimensional systems with zero-wavefunction or zero-derivative boundary conditions, the ground state must be a singlet (no spin polarization).  More recently Haerter and Shastry\cite{HaerterShastryAFTriangle_2005} have shown that on the frustrated triangular lattice an itinerant hole actually helps to produce an \emph{antiferromagnetic} ground state.  They suggest that this phenomenon holds on all lattices with ``electronic frustration,'' defined as those for which the sign of the hopping amplitude around the lattice's smallest closed loop is negative. (Note that Nagaoka's theorem only applies to un-frustrated systems.) 

\begin{figure}[h]
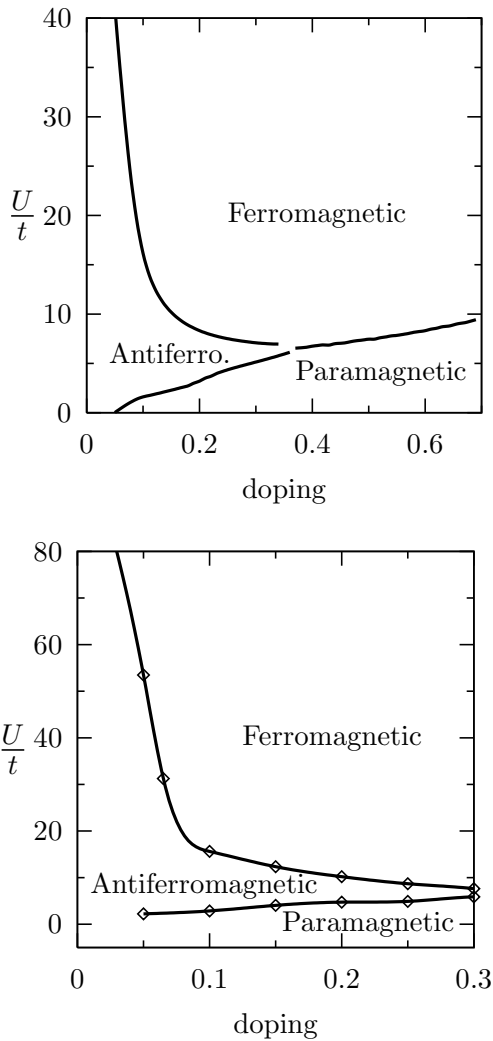

\begin{center}
\begin{tabular}{c}
\includegraphics[width=2.5in]{figs/sq10PhaseDiag.ps} \\
\\
\includegraphics[width=2.6in]{figs/sc8PhaseDiag3.ps}
\end{tabular}
\caption{Zero temperature mean-field theory phase diagram of the Hubbard model on a $10 \times 10$ square lattice (top) and $8 \times 8 \times 8$ (512 sites) simple cubic lattice (bottom).  Doping (horizontal axis) is defined as the number of extra electrons (above half-filling) per site.
\label{figMFTdiagram}}
\end{center}
\end{figure}

\subsection{Elusive Ferromagnetism}
A qualitative picture of the Hubbard model's magnetic behavior at zero temperature can be obtained by a mean-field analysis on square and simple cubic lattices, which results in the phase diagrams shown in Fig.~\ref{figMFTdiagram}.  The phase transitions were found by comparing the ground state energy and spin-spin correlations of self-consistent mean-field calculations that were initialized in paramagnetic, AF, FM, and random configurations.  
  Our analysis does not include the possibility of phase separation, \emph{e.g.}~the existence of polarons corresponding to ``carrier-rich" ferromagnetic and ``carrier-poor" antiferromagnetic regions.  If it occurs, phase separation could substantially alter\cite{EisenbergHuseAltshuler_2002} the simple phase diagrams given here.  Barbieri and Young construct phase diagrams for the large-$U$ Hubbard model in 2 and 3 dimensions using a variational Gutzwiller technique,\cite{BarbieriYoung_1991} and find phase separation occurs in both cases.  Dagotto et~al.,\cite{DagottoPhaseSep_1992} however, argue based on their results on 10- and 16-site square lattices that phase separation is generally absent in the Hubbard model, at least at short length scales.  Figure \ref{figMFTdiagram} also agrees with the extensive work by Hirsch\cite{HirschMFT_1985} in two dimensions.  

We focus on the region of low-doping and large $U/t$ (the top left of Fig.~\ref{figMFTdiagram}), where there is a FM-AF transition.  As expected, at zero doping (half-filling), the system is an antiferromagnet for all values of $U/t$ due to the effective exchange interaction and an absence of mobile carriers.  As the doping is increased from zero, it is clear from the mean field perspective that for large enough $U/t$ we expect, on some mesoscopic or macroscopic length scale, a transition to a FM ground state (even though its precise location in phase space depends on dimension as well as lattice structure, and requires more careful work).

Though the stability of the Nagaoka state has been studied extensively and is seen to exist in the Hubbard model, such ferromagnetism has not been observed experimentally.  In many Mott-insulator oxides and chacogenide systems this may be explained by an insufficient $U/t$ to allow for ferromagnetism (and finding a naturally occurring material with large enough $U/t$ seems unlikely).  However, in doped semiconductors at low dopant densities, $U/t$ is tunable over several orders of magnitude due to the exponential dependence of the hopping $t$ on the dopant spacing [\emph{e.g.}~$t(r)\sim\exp(-r/a_\mathrm{B})$ in the tight binding model].  This versatility makes doped semiconductors a promising candidate for Nagaoka ferromagnetism, as it allows $U/t$ to become large ($\sim 100-1000$), achieving for all practical purposes the limit $U/t \rightarrow \infty$ required by Nagaoka's theorem.  Despite this, the absence of ferromagnetism in experiments on a variety of doped semiconductors, both uncompensated\cite{AndresBhatt_1981,SasakiKinoshita_1968,Sasaki_1976,Ue_1971,Quirt_1973} and compensated,\cite{Hirsch_NoExpFerro_1992} is quite clear.  In these experiments, the nearest neighbor coupling, though distributed broadly, had a median value of 1-10K, and FM behavior was searched for down to much lower (mK) temperatures to probe the $\mathrm{T}=0$ behavior.  Even with the additional hope of alternative theories that predicted ferromagnetism of Anderson localized electrons,\cite{Kamimura_1978,Kamimura_1985} both uncompensated and compensated systems exhibited a significantly lower (by factors of 10-50) magnetic susceptibility compared to the high temperature paramagnetic Curie result, indicating that the systems were predominantly characterized by AF correlations, both at and slightly below half-filling.

To understand how this can be, let us return to the requirements of Nagaoka's theorem.  Placing dopants on a superlattice has become possible only very recently\cite{Schofield_2003} -- in naturally formed doped semiconductors, including those of all relevant experiments, the dopants are distributed randomly and an important hypothesis of Nagaoka's theorem is not met.  Adding such positional disorder to the Hubbard-like description of the system turns out to be an important ingredient.  (It is incorporated into the Hubbard model by setting $t \rightarrow t_{ij}$, which then depends on the separation $r_{ij}$, see section \ref{secHubbardForHydrogenic}, specifically Eq.~\eqref{eqnHubHamDisordered} below).  

After introducing positional disorder into the Hubbard model, is not clear whether any of the aforementioned theorems and arguments for uniform systems are still valid (or even relevant).  First, we expect a locally fluctuating carrier density, which may wash out any distinction between phase separation at macroscopic and mesoscopic length scales for such systems. Second, since the itinerancy of carriers depends on the local lattice geometry, when the geometry becomes spatially inhomogeneous the itinerancy might be suppressed, and at the very least the magnetic structure will show similar spatial inhomogeneity.


More precisely, it was found that the lack of low-temperature ferromagnetism in semiconductors can be explained by disorder localizing otherwise mobile carriers, thereby reducing the kinetic energy gain (previously $\sim t\delta$) and destroying ferromagnetism.  Bhatt and Lee\cite{BhattLee_JAP_1981,BhattLee_1982} gave insight into the true nature of the half-filled (uncompensated semiconductor) case using a perturbative renormalization group method tailored for the large amount of disorder present in the actual system.  They found that the randomness of the dopants results in what has been dubbed a valence-bond glass,\cite{Bhatt_1990,Bhatt_1988} random singlet,\cite{Fisher_1994} or Bhatt-Lee phase.\cite{Holcomb_SUSSP_1986,Paalanen_1988}  In such a state, spins pair up to form (spin zero) singlets (see also Ref.~\onlinecite{Bhatt_1986}) in a hierarchical fashion, and the resulting structure and behavior is qualitatively different from the antiferromagnet state predicted on a bipartite lattice.  There is no long-range AF order, and the magnetic susceptibility is strongly temperature dependent, even down to tens of millikelvin.  That compensated semiconductors show no evidence of ferromagnetism\cite{Hirsch_NoExpFerro_1992} can be attributed to the localization of holes on one (or a few) valence bonds, and their consequent inability to move long enough distances to disrupt the local magnetic arrangements.  As a result, holes are unable to gain the kinetic energy which favors a spin-polarized background.  Thus, even though doped semiconductors give one the ability to tune $U/t$ over several orders of magnitude, Nagaoka ferromagnetism remains elusive. 

\section{Hubbard model for hydrogenic systems \label{secHubbardForHydrogenic}}

\subsection{Overview and formulation}
An important question that can still be asked of a system with positional disorder is whether or not the ground state is spin polarized (resulting in macroscopic spin degeneracy). In the remainder of this paper, we attempt to answer an even more basic question -- does there exist, even on the nanoscale, large spin degeneracy in systems of hydrogenic centers, using an appropriate Hubbard-like description?  The paramount conclusion is that there \emph{does} exist a regime in doped semiconductors which is more amenable to Nagaoka ferromagnetism.  Interestingly, this regime is attainable in nanoscale quantum dots and heterostructures, but not accessible to bulk systems.  There we find Nagaoka-like ferromagnetism in the presence of disorder, at least at the nanoscale, and that this regime also possesses a higher likelihood of emerging on mesoscopic or macroscopic scales (\emph{e.g.}~in modulation doped systems).  In this section we introduce and motivate the generalized Hubbard model used to characterize the doped semiconductor problem.  




\subsection{Random Hubbard Model: positional disorder\label{subsecModel}}
As a first approximation, a system of $\Nsites$ randomly positioned donors can be modeled with the Hubbard Hamiltonian obtained by adding site-dependence to the hopping amplitude in Eq.~\eqref{eqnHubHamOriginal}.  Specifically, we make $t_{ij}$ a function of the site separation: $t_{ij}=t\left(|r_i-r_j|\right)$, resulting in the Hamiltonian:
\begin{equation}
\mathcal{H}_{rdm} = - \sum_{i,j,\sigma} \left( t_{ij}c^\dag_{i\sigma} c_{j\sigma} + \mbox{h.c.} \right) + U\sum_i n_{i\uparrow}n_{i\downarrow} \label{eqnHubHamDisordered}
\end{equation}
where $i,j=1\ldots \Nsites$.

This takes into proper account the random positioning of the donors, and, as we discuss below, should be a good model for both uncompensated and compensated \emph{bulk} semiconductors with $\le 1$ electron per donor site (in the latter case, a more rigorous treatment would additionally include random on-site energies reflecting the random fields generated by the (positively charged) acceptor sites). 

\subsection{Hubbard model generalization: occupation-dependent hopping\label{subsecModel2}}
 A shortcoming of $\mathcal{H}_{rdm}$ (Eq.~\eqref{eqnHubHamDisordered}), both for the lattice and random case, is that it does not account for a fundamental property of hydrogen: the two-electron wavefunction of the $H^-$ ion has much greater extent than the one-electron wavefunction of the $H$ atom.  This is reflected in the binding energy (the energy required to remove an electron) of $H^-$ being only 0.0555 \Ry, whereas $1\,\Ry$ is necessary to remove the electron of $H$.\cite{MottBook,BS_QMbook_1977}  Indeed, using that an effective Bohr radius $a^*$ scales as $1/\sqrt{E_{\mathrm{\scriptsize binding}}}$, we find that the ratio of Bohr radii for $H^-$ and $H$, $a^*_{H^-}/a^*_H = \sqrt{1.0}/\sqrt{0.0555} \approx 4$, showing that the wavefunction of $H^-$ is several times larger than that of $H$.  Variational treatments of the $H^-$ ion,\cite{BS_QMbook_1977} as well as an effective pseudopotential calculation,\cite{NielsenBhattTransport} determine the ratio to be in the range $2-4$.  This affects the Hubbard-description of the system because it is much easier for an electron on a doubly-occupied hydrogenic center to hop away than it is for the electron on a singly-occupied site to make a similar hop.  This implies that the hopping amplitude seen by an itinerant electron, hopping around in a background of singly-occupied sites, is larger than that seen by a hole in a similar background. The fact that the ratio of the two radii is substantial ($2-4$), and the hopping amplitude is exponentially dependent on the radius (in the low density regime), suggest that a doped semiconductor above half-filling is in a quite different regime of parameters than the conventional compensated semiconductor (a system below half-filling). Such a regime, while not obtainable in bulk doped semiconductors, should be realizable in semiconductor heterostructures, as well as quantum dots.  In Hubbard model parlance, near half-filling the hopping amplitude for an electron is much larger than for a hole.  At the very least, the different radii of the doubly- vs.~singly occupied sites suggest that we modify lattice Hubbard Hamiltonian (\ref{eqnHubHamOriginal}) to become:
\begin{equation}
\mathcal{H}^* = - \sum_{\langle i,j \rangle \sigma} \left( t(n_i,n_j)c^\dag_{i\sigma} c_{j\sigma} + \mbox{h.c.} \right) + U\sum_i n_{i\uparrow}n_{i\downarrow} \label{eqnHubHamOccDep}
\end{equation}
where $n_i$ is the total occupation of site $i$, and the hopping now has occupation dependence given by the piecewise function (the hopping corresponding to the different amplitudes $\tInner$ and $\tOuter$ is shown pictorially on the right):
\begin{displaymath}
t(n_i,n_j) = \hspace{2.5in}
\end{displaymath}
\begin{equation}
\left\{ \begin{array}{ccc}
\tOuter & \hspace{0.3cm} n_j=1, n_i=2 & \hspace{0.3cm} 
\parbox{1.4in}{\includegraphics[width=1.5in]{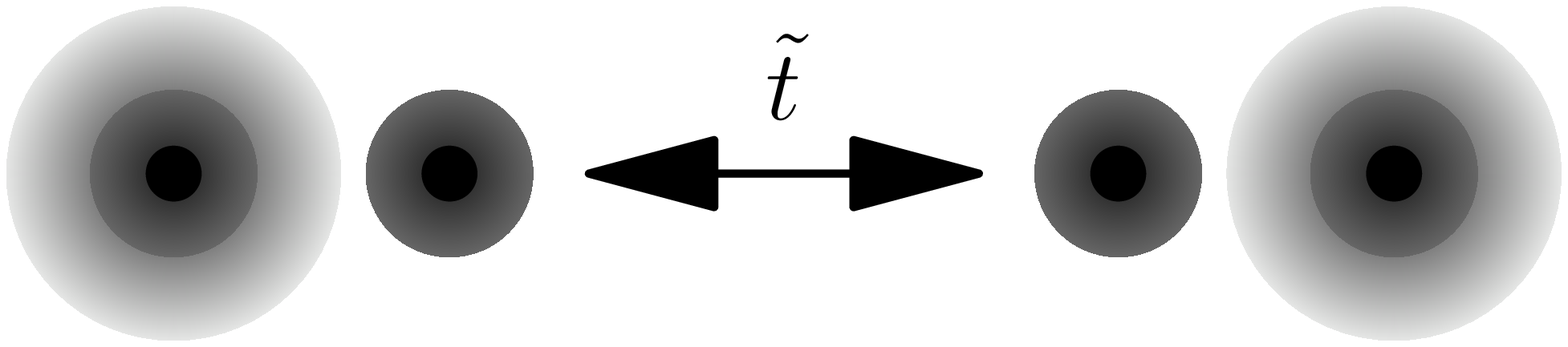}}  \\
\vspace{0.2cm} & & \\
\tInner & \hspace{0.3cm}\mbox{otherwise} & \hspace{0.3cm} 
\parbox{1.4in}{\includegraphics[width=1.5in]{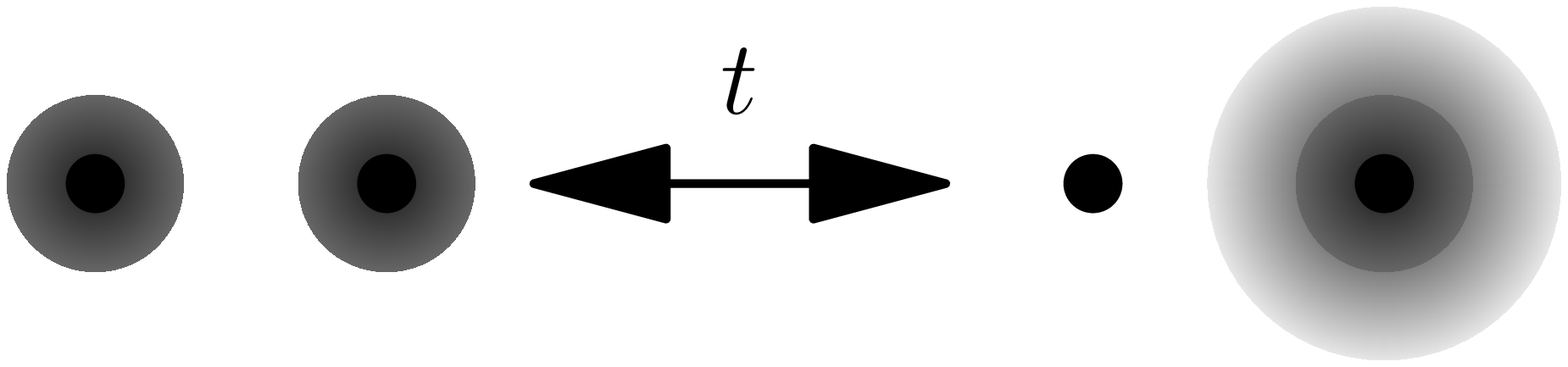} \\
 \includegraphics[width=1.5in]{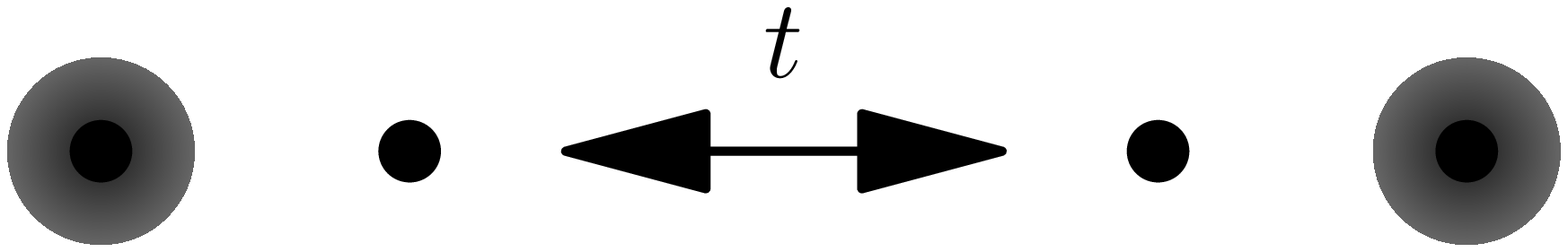}}
\end{array} \right. \label{eqnPiecewiseT}
\end{equation}
where $\tOuter$ is larger (and as we will see, can be much larger) than $\tInner$.\cite{ErikNanoscaleFM_2007}  This model enhances the hopping from a doubly-occupied site to an already singly occupied site (which will become doubly occupied after the hop).  One may question why the hopping from a doubly-occupied site to an empty site (the middle picture of Eq.~\eqref{eqnPiecewiseT}) is not also enhanced.  The primary reason is that the present formulation is the only way, within the single-band Hubbard model, to preserve the asymptotic spatial dependence of the effective exchange interaction: $J(r) \sim e^{-2r/\aB}$ (recall $J \sim t^2/U$ and $t \sim e^{-r/\aB}$).  This is of essential importance, since this relation for $J$ has been shown to be asymptotically exact.\cite{HerringFlicker_1964}  


Note that Eq.~\eqref{eqnHubHamOccDep} is in general \emph{not} electron-hole symmetric.  Only when $\tOuter = \tInner$ and the system is on a bipartite lattice is electron-hole symmetry preserved.\cite{FradkinBook_1991}  The general lack thereof is readily seen, since an itinerant hole hops with amplitude $\tInner$ whereas an itinerant electron hops with $\tOuter$.  Indeed, the effective low energy theory of the three-parameter Hubbard Hamiltonian when there is less than one electron per site (below half-filling), in the limit $U\gg \tInner$, is independent of $\tOuter$ and given by the familiar $t-J$ Hamiltonian:

\parbox{2in}{\begin{eqnarray} 
\mathcal{H}_{tJ} &=& - t \sum_{<ij>\sigma} \left( (1-n_{i\bar{\sigma}})c^\dag_{i\sigma}c_{j\sigma}(1-n_{j\bar{\sigma}}) + \mbox{h.c.}\right) \nonumber \\
& &  +\, J \sum_{<ij>} \left(\vec{S}_i\cdot \vec{S}_j - \frac{1}{4}n_i n_j \right) \nonumber
\end{eqnarray}} 
\hfill \parbox{.1cm}{\begin{eqnarray}  \label{tJEffModel} \end{eqnarray}}

\noindent where the AF exchange $J=4\tInner^2/U$, $c_{i\sigma}^\dag$ ($c_{i\sigma}$) is the electron creation (annihilation) operator, and the spin operator $\vec{S}_i$ is as previously defined.  When there is greater than one electron per site, however, the low energy spectrum (in the large $U/t$ limit) is given by a $\tOuter-J$ model, where $\tInner$ is replaced by $\tOuter$ in Eq.~\eqref{tJEffModel}, $(1-n_{i\sigma}$ is replaced by $n_{i\sigma}$, and where $J$ remains determined by the Hubbard $\tInner$ parameter, as one might expect.\cite{ChernyshevEffTheories_2004}  The Hilbert space restriction then excludes doubly-\emph{vacant} sites.  It is worth noting that in the usual $t-J$ model on a non-bipartite graph (defined as a set of sites and hopping links), excluding doubly-vacant sites is \emph{not} equivalent to excluding doubly-occupied sites.  This directly corresponds to the lack of electron-hole symmetry in the corresponding Hubbard problem.

It is important to remember that the electron creation and annihilation operators in these models act on a system with a fixed number and arrangement of sites.  In a semiconductor, each site corresponds to a dopant atom, and when we speak of adding electrons or holes to the system we mean addition or subtraction of carriers while \emph{leaving the underlying dopant configuration fixed}.  Thus, the electron-hole asymmetry here is \emph{not} an asymmetry between n-type and p-type semiconductors, but an asymmetry between a doped semiconductor which has more electrons than dopant atoms and one which has less electrons dopant atoms.

Hirsch has investigated a similar Hubbard model with occupation-dependent hopping, but in a different regime with its focus on superconducting pairs.\cite{HirschOccDepHopping_1995}  We proceed with semiconductors in mind, and to allow for the random placement of sites, we add positional dependence to the hopping amplitude in Eq.~\eqref{eqnHubHamOccDep}, similar to the modification yielding Eq.~\eqref{eqnHubHamDisordered} earlier, to arrive at: 
\begin{equation}
\mathcal{H}_{rdm}^* = - \sum_{i,j,\sigma} \left( t_{ij}(n_i,n_j)c^\dag_{i\sigma} c_{j\sigma} + \mbox{h.c.} \right) + U\sum_i n_{i\uparrow}n_{i\downarrow} \label{eqnHubHamDisorderedOccDep}
\end{equation}
where $n_i$ is the total occupation of site $i$, and $t_{ij}$ now has an occupation dependence given by:
\begin{equation}
t_{ij}(n_i,n_j) = \left\{ \begin{array}{cc}
\tOuter_{ij} & n_j=1 \,\, \mbox{and}\,\, n_i=2 \\
\tInner_{ij} & \mbox{otherwise} 
\end{array} \right. \label{eqnPiecewiseTDisordered}
\end{equation}

One way to view the manifest electron-hole asymmetry of models (\ref{eqnHubHamOccDep}), (\ref{tJEffModel}), and (\ref{eqnHubHamDisorderedOccDep}) is that systems above half-filling are effectively \emph{less random}, and hold greater hope for the Nagaoka phenomenon to take place.  This reasoning follows from electrons having more extended wavefunctions than holes, and the concomitant existence of two distinct length scales.  Because the electron wavefunctions average over much more of the disorder, systems with a small percentage of extra electrons experience a greatly reduced effect of the positional disorder when compared with corresponding hole-doped (\emph{i.e.}~compensated) systems, and so behave more like the uniform lattice.    

Hope for Nagaoka ferromagnetism in electron-doped semiconductors is also found by considering the relation of conventional doped semiconductors to diluted magnetic semiconductors (DMS), for which ferromagnetism does co-exist with disorder.  In one type of DMS (III-V), a transition metal atom acts as both a dopant and a local moment (coming from the unfilled d-shell of the atom).  For instance, in Ga$_{1-x}$Mn$_x$As,\cite{Ohno_1998,ChibaOhno_2003} the Mn atom acts as an acceptor (p-type) and local moment.  These systems also have substantial disorder (due to dopant positions and  anti-site defects in the semiconductor itself, \emph{e.g.}~As on Ga sites), but possess macroscopic ferromagnetism for temperatures up to 100K!\cite{Ohno_1998}  Thus, disorder by itself does not always destroy ferromagnetism; in fact, in some cases it may even enhance the ferromagnetic transition temperature.\cite{Berciu_DMS_2001,Kennett_2002}

One important difference between conventional ``non-magnetic'' doped semiconductors and DMS is that there exists in the latter two different length scales -- the Bohr radius of the Mn hole wavefunction ($\sim 10$ \AA) and the extent of the localized spin on the Mn ($\sim 1-2$ \AA).  Thus, each hole's wavefunction extends over several Mn spins, a phenomenon which is only accentuated as holes delocalize further at higher Mn density.  This allows the carrier-magnetic moment interaction to dominate, resulting in a FM ground state\cite{Berciu_DMS_2001,Berciu_DMS_2004}.  In the electron-doped semiconductor, the Bohr radius of the electrons that singly-occupy sites (which give rise to the effective AF exchange interaction $J\sim\tInner^2/U$), is much smaller than the radius of the electrons which doubly-occupy a site.  This dichotomy of length scales could similarly conspire to result in carrier hopping being dominant and ultimately a ferromagnetic (Nagaoka) ground state. (The other difference, of course, is the existence of multiple bands in DMS, which facilitates FM.) 

\subsection{Parameter ranges and calculation details\label{secCalcDetails}}
The first step in our analysis of the Hamiltonians (\ref{eqnHubHamOccDep}) and (\ref{eqnHubHamDisorderedOccDep}) is to find values (or ranges of values) appropriate for their parameters.  The models are described by the dimensionless ratios $\tOuter/\tInner$ and $U/\tInner$ (which depend on a pair of site indices in the case of Eq.~\eqref{eqnHubHamDisorderedOccDep}).  To find values of $U/t$ and $\tOuter/\tInner$ appropriate for doped semiconductors, we performed a calculation of the single particle states of donors placed on a simple cubic lattice.  Note that although much of our work deals with 2D systems, atomic hydrogen is intrinsically a 3D problem, and thus the calculation of realistic parameters for a system of many hydrogenic centers should likewise be in three dimensions.  We choose the simplest such 3D arrangement of centers, the simple cubic lattice.

As already stated, a hydrogen ($H$) atom binds its electron with a strength of 1 $\Ry$ and will bind a second electron with $0.0555\,\Ry$ to form a $H^-$ ion.  If all of the dopants are positioned on a superlattice, then these two levels broaden in the usual manner into two impurity bands.  The exact details of the particle bands depend on the spin configuration in the ground state.  Due to the $H^-$ ion's wavefunction being more spatially extended than that of the $H$ atom, the width of the upper impurity band is significantly greater than that of the lower band. 

\begin{figure}[h]
\begin{center}
\includegraphics[width=3in]{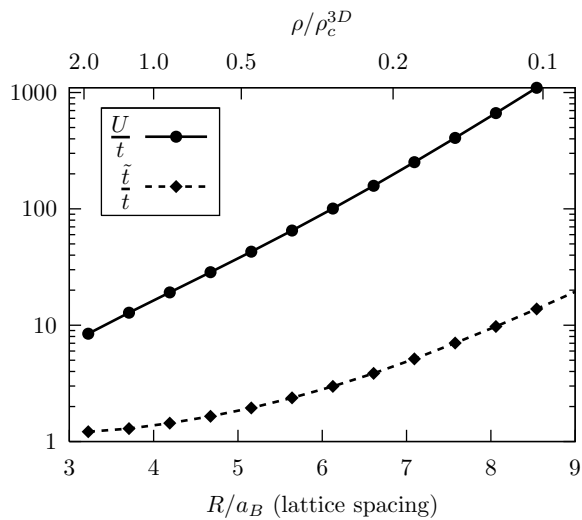}
\caption{Variation of ratios $U/t$ and $\tOuter/\tInner$ with the dopant spacing (related to the dopant density $\rho$ by $\rho = \frac{1}{R^3}$, so the metal-insulator occurs at $R_c/a_{\mathrm{B}} = 4$). \label{figParamRatiosVsLatSpacing}}
\end{center}
\end{figure}

We have calculated these bands for a ferromagnetic configuration of spins in the ground state of a filled lower band (\emph{i.e.}~the uncompensated case).  We follow Bhatt and Rice,\cite{BhattRice_1981} and use pseudopotentials and a sphericalized Wigner-Seitz (WS) method on a cubic superlattice.  Details of the band calculation can be found elsewhere.\cite{NielsenBhattTransport}  We then extract the dependence of $\tInner$ and $\tOuter$ on the impurity density (or equivalently, on the lattice constant) by fitting the calculated bandwidths to a tight binding model.  Using the well-known tight binding relationship between hopping parameter and bandwidth on un-frustrated lattices yields (where $z$ is the lattice coordination number): 
\begin{eqnarray}
 2z\tInner &=& \mbox{width of lower band} \nonumber \\
 2z\tOuter &=& \mbox{width of upper band} \label{eqnTightBindingFit} \\
 U &=& \mbox{band gap at zero density} \,. \nonumber
\end{eqnarray}
We find $U \approx 1 \Ry$ and, by matching the bandwidths for the 3D case, we obtain the tight binding parameters $\tInner(b)$, $\tOuter(b)$.  Figure \ref{figParamRatiosVsLatSpacing} shows the dependence of the dimensionless Hubbard parameter ratios on the superlattice spacing (lower axis) and impurity density (upper axis).  It shows clearly that the range of $U/t$ and $\tOuter/\tInner$ can be varied substantially in the doped semiconductors.  The large span of $U/t$ originates in the exponential dependence of the hopping parameter on the atomic spacing, and the variation of $\tOuter/\tInner$ from the relatively large size of the two-electron wavefunction appearing as a factor in this exponential. 

In the results that follow, we either use the exact parameter ratios found here or consider the effect of varying the parameter ratios within the ranges $U/t=[5,100]$ and $\tOuter/\tInner=[1,10]$, which are conservative when compared to the physically attainable ranges. 
After determining the parameter ranges of interest, we solve both Hubbard and $t-J$ models on finite systems.  We numerically find the ground state, and determine how its spin depends on $\tOuter/\tInner$, $U/\tInner$, system size, and system geometry.  Hamiltonians (\ref{eqnHubHamOccDep}) and (\ref{eqnHubHamDisorderedOccDep})  were solved using exact diagonalization, ultimately using a generalization of the Lanczos method.\cite{Calvetti_1994}  With four states allowed on a site, the Hilbert space grows exponentially in the number of sites, restricting the size of tractable systems significantly.  Several optimizations have been exploited to push back this computational barrier.  First, since both Hubbard and $t-J$ Hamiltonians commute with the $z$-component of total spin, it follows from the properties of the SU(2) group, that we can restrict the the Hilbert space to the minimal $S_z$ sector without reducing the support of the spectrum.  Second, all spatial symmetries are utilized via group theoretic techniques to divide the Hilbert space into sectors for which the Hamiltonian matrix is block diagonal.  Third, we factorize the action of the Hubbard Hamiltonian into ``up spin'' and ``down spin'' parts, allowing more efficient computation of the matrix elements.  In the $t-J$ model, this can be done only for the kinetic term.  


\section{Results for ground state spin in finite clusters\label{secHubbardResults}}
Here we present the results of solving our generalized Hubbard model on finite systems.  The results and discussion are divided into units based on the amount of structure present in the system, and what type of boundary conditions were used.  Section \ref{secFiniteLattices} considers systems with finite lattice structure and periodic boundary conditions.  Note that only nearest neighbor links are kept in the model (see Eq.~\eqref{eqnHubHamOccDep}), so that there is a single pair ($\tInner,\tOuter$) of kinetic parameters.  We refer to a lattice as being bipartite or non-bipartite if the corresponding Hubbard model with only nearest neighbor hopping is respectively bipartite or not.  Section \ref{secSelectedClusters} presents results from clusters with open boundary conditions and selected structures for which all nearest neighbors are equidistant (so there is again a single pair of kinetic parameters).  We use the term \emph{cluster} in this section to refer to a finite system possessing less symmetry than a finite lattice.  In section \ref{secGeomDistorted}, clusters constructed to have only two or three pairs of kinetic parameters are considered with open boundary conditions.  There we also describe a method of adding random perturbations to clusters, and present results for several cases. Finally, in section \ref{secFixedDensityClusters} we analyze ensembles of random clusters.  We generate these ensembles with a fixed density, and exact diagonalization results of the individual clusters are averaged to produce our final results. 

\begin{figure*} 
\begin{center}
\begin{tabular}{|c|c|} \hline
Square & \rule[-0.9in]{0in}{1.9in}\parbox{4.7in}{
\begin{tabular}{c}
\includegraphics[height=1.4in]{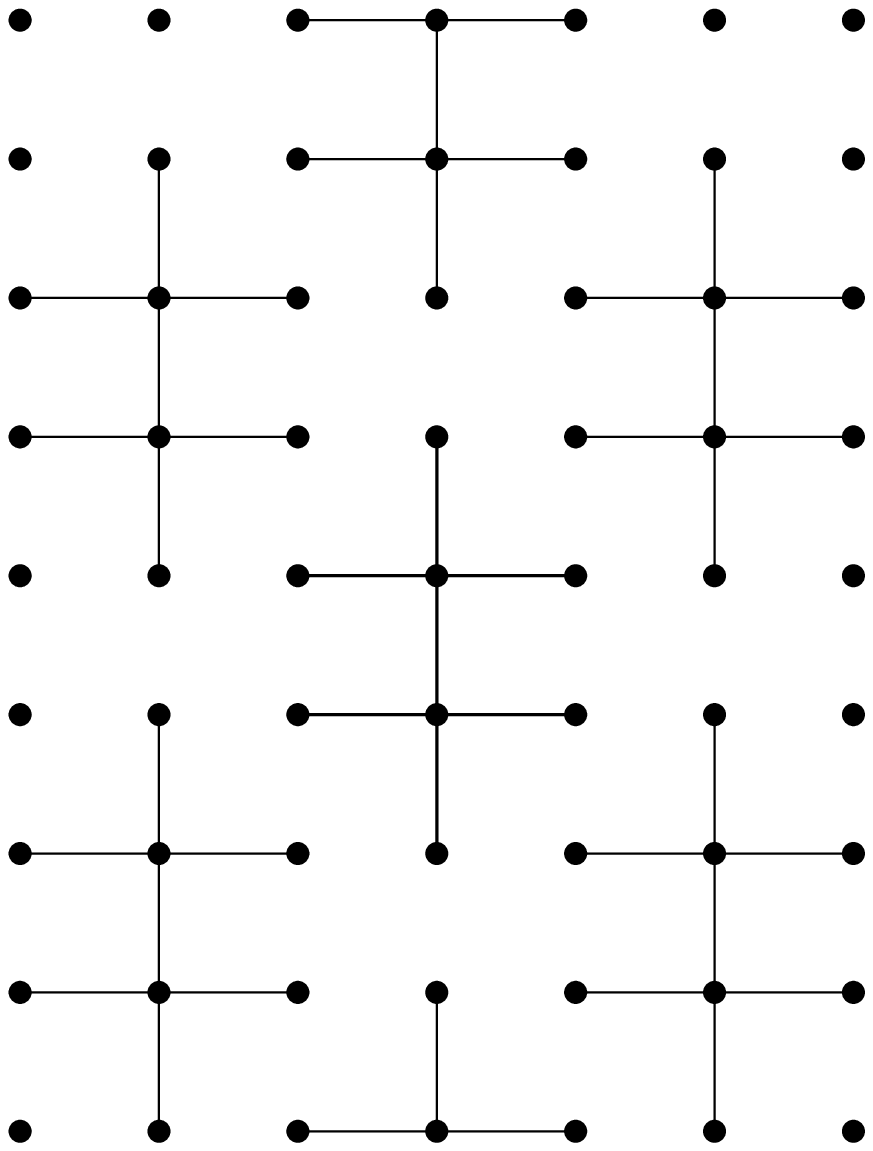}\\ 8 sites \end{tabular}
\hspace{0.5cm} 
\begin{tabular}{c}
\includegraphics[height=1.2in]{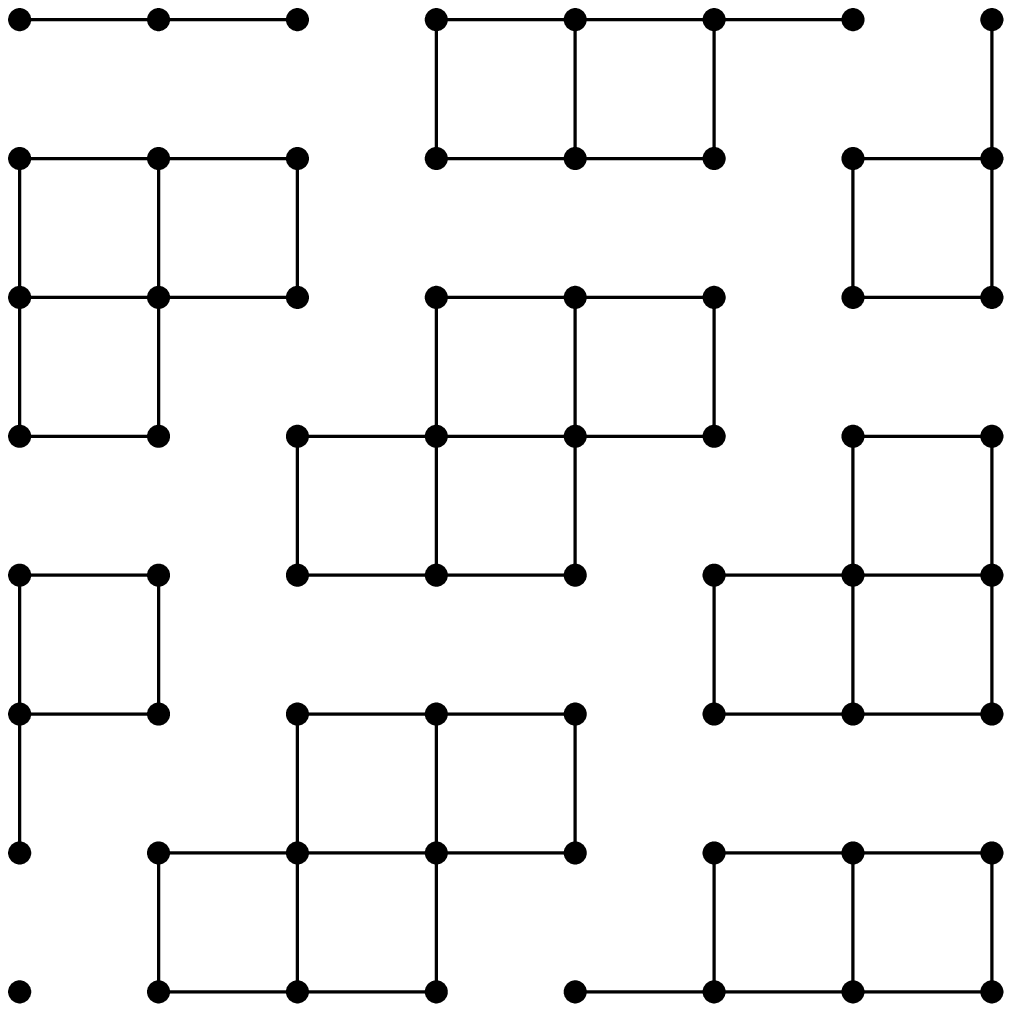}\\10 sites \end{tabular}
\hspace{0.5cm} 
\begin{tabular}{c}
\includegraphics[height=1.4in]{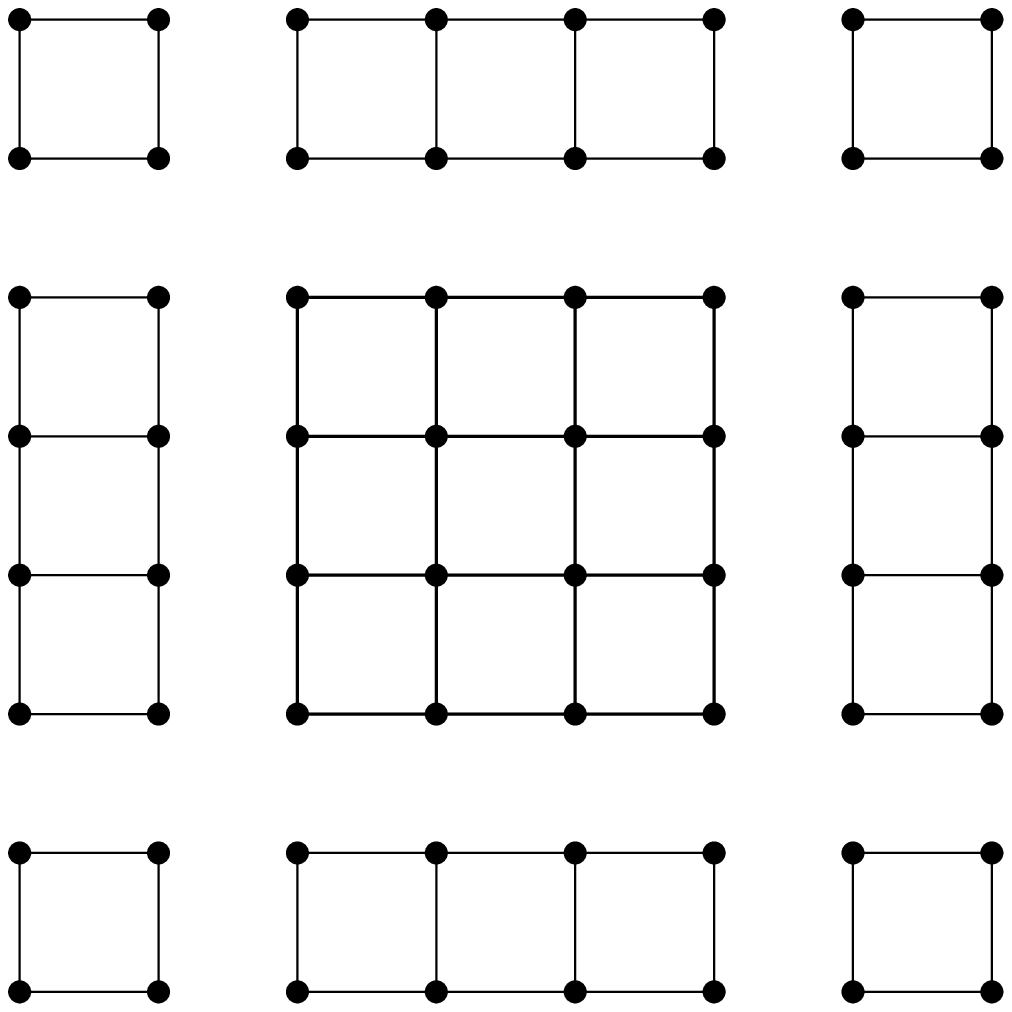}\\16 sites \end{tabular}}
 \\ \hline
Honeycomb & \rule[-0.8in]{0in}{1.7in}\parbox{4.5in}{
\begin{tabular}{c}
\includegraphics[height=1.2in]{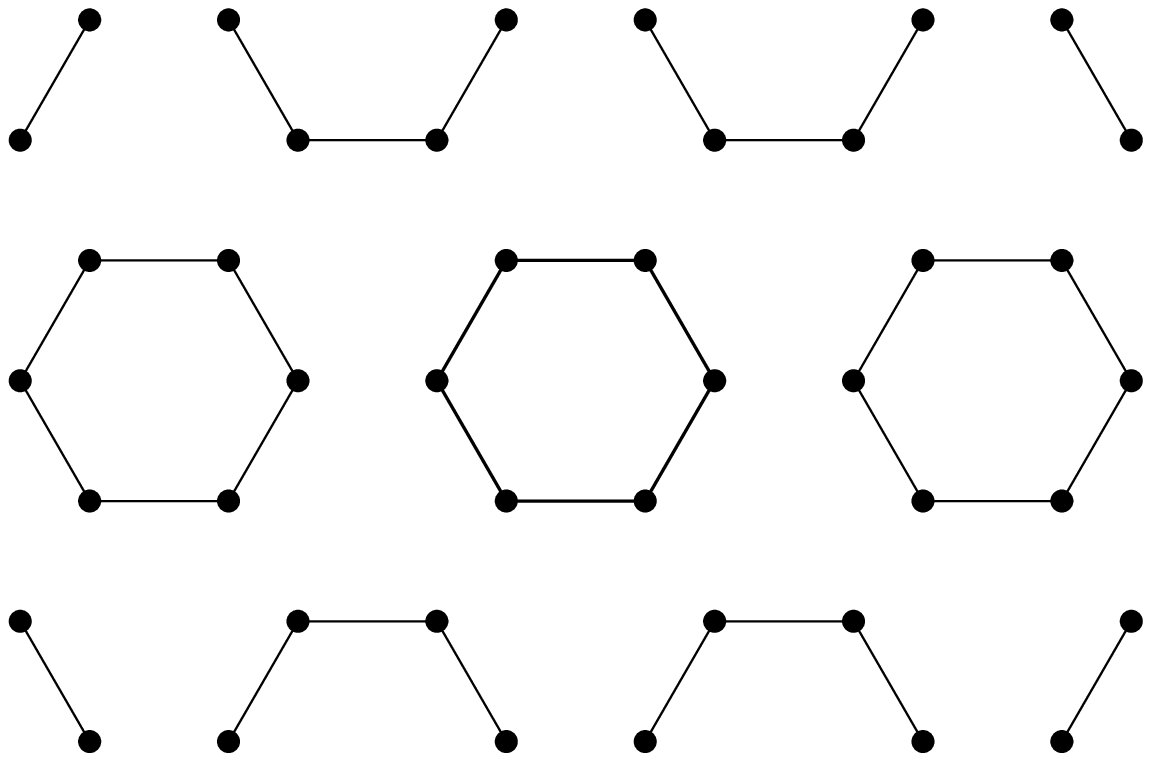}\\6 sites \end{tabular}
\hspace{1cm} 
\begin{tabular}{c}
\includegraphics[height=1.2in]{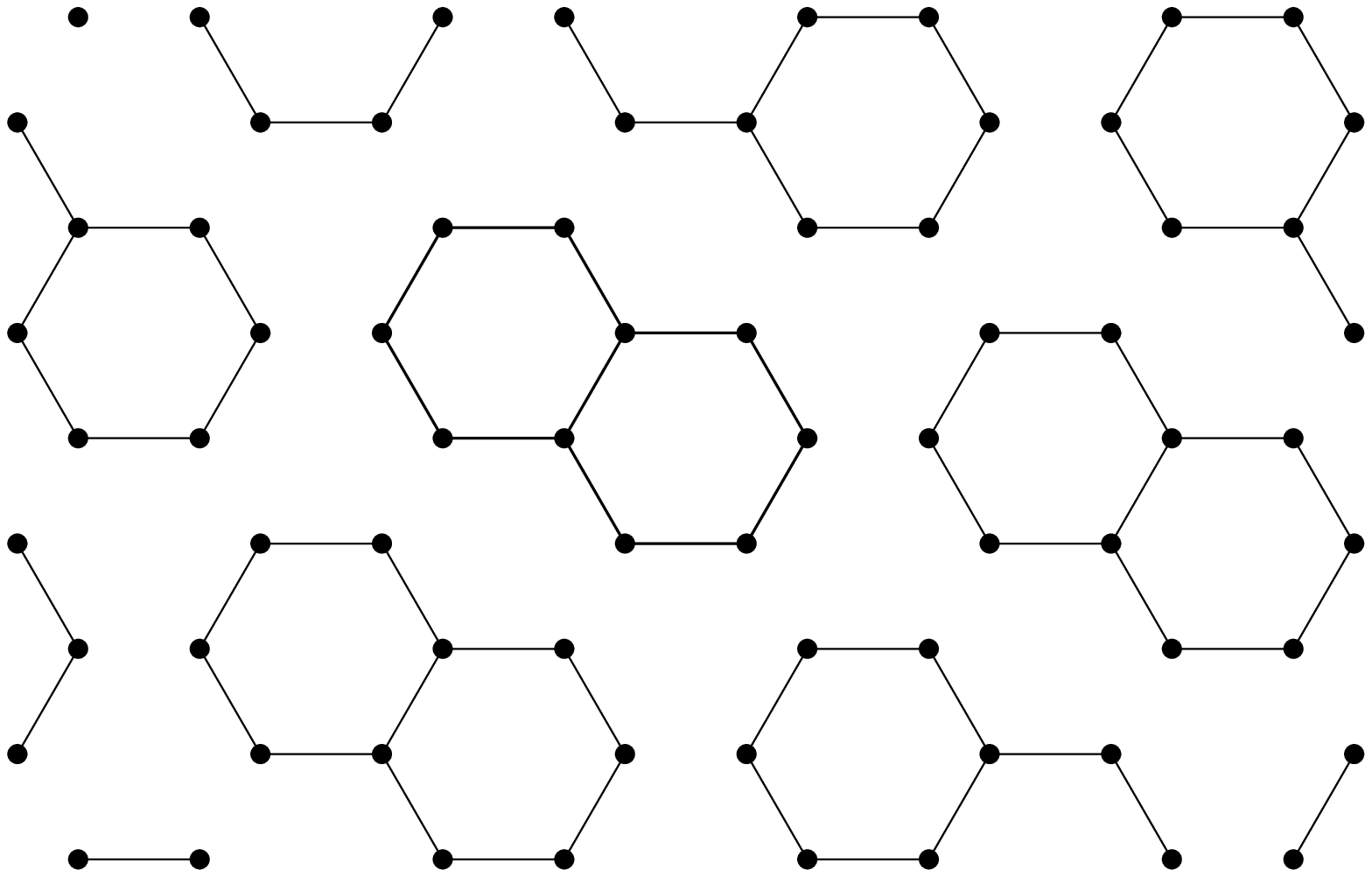}\\10 sites \end{tabular}}
 \\ \hline
Triangular & \rule[-0.8in]{0in}{1.7in}\parbox{4.5in}{
\begin{tabular}{c}
\includegraphics[height=1.2in]{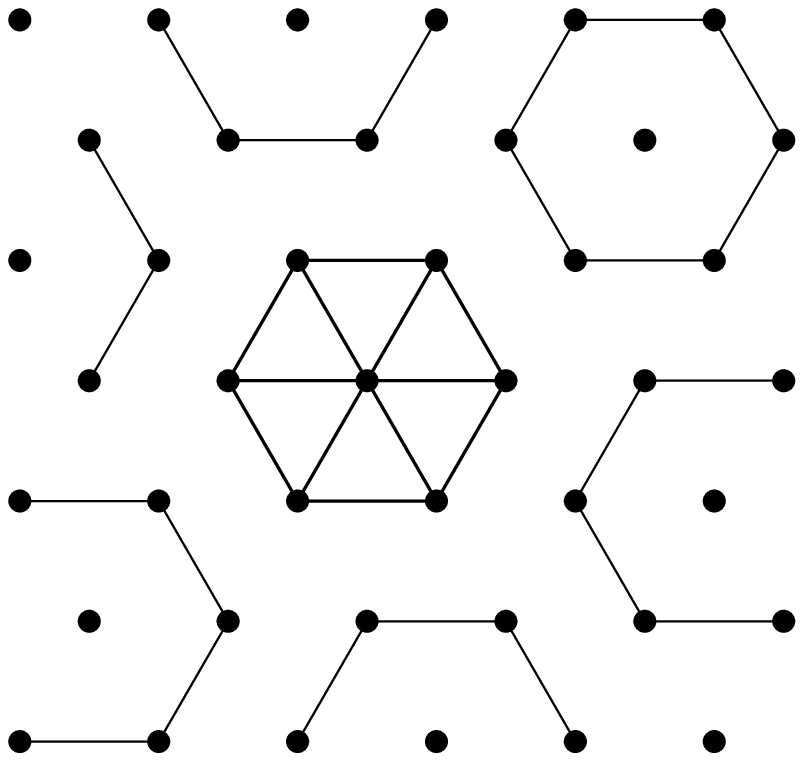}\\7 sites \end{tabular}
\hspace{1cm} 
\begin{tabular}{c}
\includegraphics[height=1.2in]{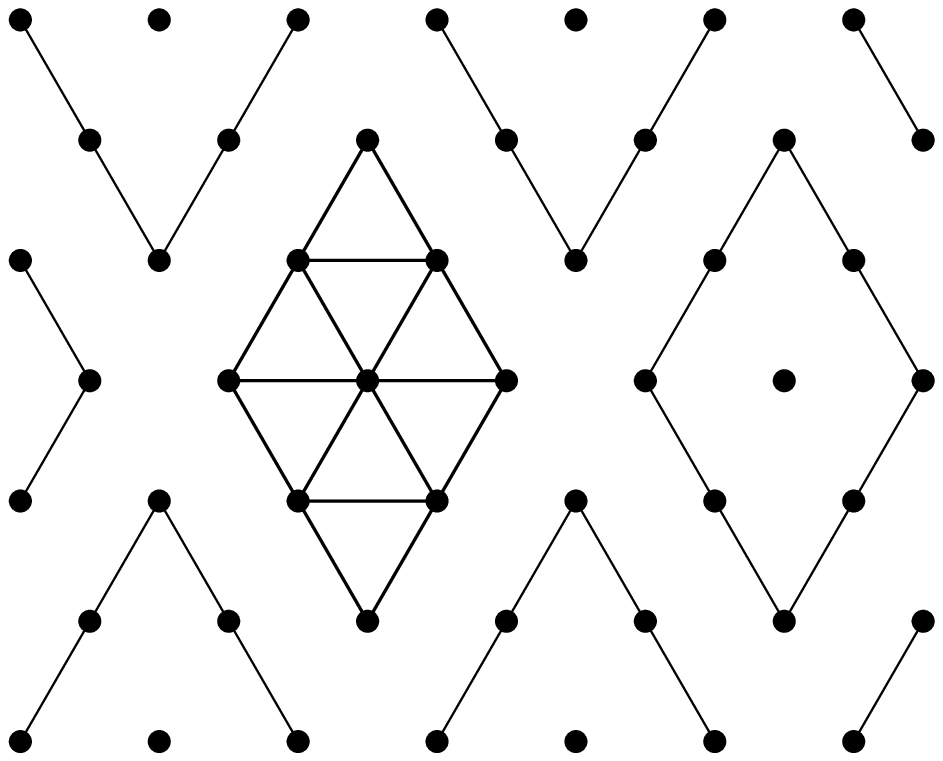}\\9 sites \end{tabular}}
 \\ \hline
\end{tabular}
\caption{Lattice geometries for the square, honeycomb and triangular lattices used in this section.  The lines connect sites of the finite lattice, which is repeated to show how periodic boundary conditions are implemented. \label{figLatticeGeometries}}
\end{center}
\end{figure*}

\subsection{Finite Lattices \label{secFiniteLattices}}

We have solved the nearest-neighbor Hubbard and corresponding $\tOuter-J$ models on finite square (8, 10, and 16 sites), honeycomb (6 and 10 sites), and triangular (7 and 9 sites) lattices.  These are shown in Fig.~\ref{figLatticeGeometries} with the sites of a single unit cell connected, so that the method of applying periodic boundary conditions in each case is clear.  Note the choice of unit cell for all of the bipartite lattices (square and honeycomb) allows a classical Neel state spin assignment, where all of a site's nearest neighbors have spin opposite to it.  This requirement is important since a finite bipartite lattice that is magnetically frustrated due to boundary conditions may have an exaggerated preference for FM.

  Each finite lattice, with periodic boundary conditions, was doped with up to two electrons or holes away from half-filling.  Denoting the number of electrons $\Nelec$, this means that $\Nsites - 2 \le \Nelec \le \Nsites + 2$.  The Hubbard model depends on the two dimensionless ratios $U/\tInner$ and $\tOuter/\tInner$, whereas the $\tOuter-J$ model depends only on $\tOuter/J = \frac{1}{4}(\tOuter/\tInner)(U/\tInner)$.  Thus, the value of $\tOuter/J$ marking the onset of the Nagaoka state defines a straight line in $\log U/\tInner$ vs.~$\log \tOuter/\tInner$ space with slope $-1$.  We consider each lattice in turn below.

\vspace{0.5in}

\begin{figure} 
\begin{center}
\includegraphics[width=3in]{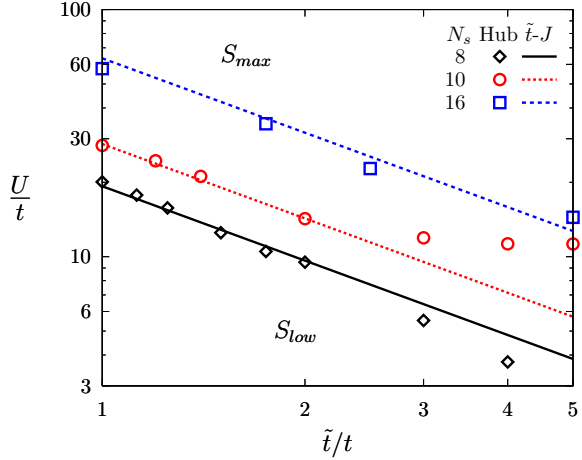}
\renewcommand{\baselinestretch}{1}\normalsize
\caption{(Color online) Ground state spin diagram resulting from the exact diagonalization of Eq.~\eqref{eqnHubHamOccDep} on 8-,10-, and 16-site square lattices (periodic b.c.) with 9, 11, and 17 electrons respectively.  Hubbard model results are displayed as open symbols.  Lines show the result of the corresponding $\tOuter-J$ model as described in the text. $S_{max}$ denotes the region of largest allowed spin (actual value depends on the lattice size), and $S_{low}$ marks the region of unsaturated (usually minimal) ground state spin. \label{figSqLatticeResults}}
\end{center}
\end{figure}

\subsubsection{Square Lattice}
\begin{figure}
\begin{center}
\includegraphics[width=2.7in]{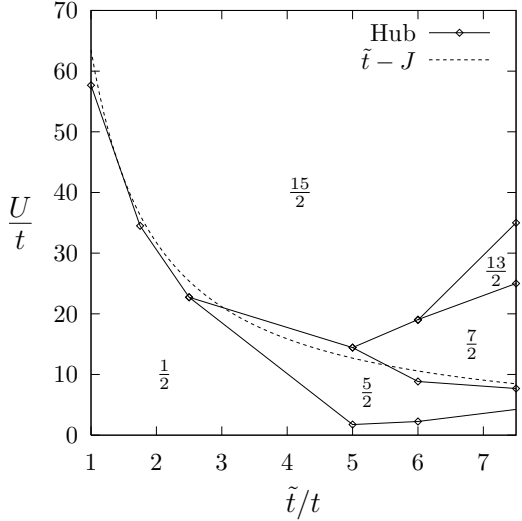} 
\renewcommand{\baselinestretch}{1}\normalsize
\caption{Detailed ground state spin diagram of the 16-site square lattice with 17 electrons.  Labels indicate the ground state's total spin.  As $\tOuter/\tInner$ increases beyond 3, the transition to a maximally polarized state is less abrupt and regions of partial spin polarization exist.  We find that the $\tOuter-J$ model gives a direct transition from $S=\frac{1}{2}$ to $S=\frac{15}{2}$, which is shown as the dashed line.\label{fig16siteIntermediateSpins}}
\end{center}
\end{figure}

The square lattice, the stereotypical 2D lattice, is bipartite and is itself a Bravais lattice.  Figure \ref{figSqLatticeResults} shows the ground state spin phase diagram for the 8-, 10-, and 16-site square lattices doped with one electron, up to $\tOuter/\tInner=5$.  One sees that an increase in $\tOuter/\tInner$ causes the region where the ground state attains its maximum spin to increase.  This confirms our intuition about the model, that a FM ground state is more likely when the carriers (an extra electron in this case) have greater hopping amplitude.  (Recall that a greater hopping amplitude increases the kinetic energy gain of a delocalized electron in a background of \emph{aligned} spins relative to the case when the background spins are in an AF or random arrangement.)  Up to $\tOuter/\tInner=5$, the minimal $U/t$ needed for a fully polarized ground state falls roughly as a power law with $\tOuter/\tInner$.  The $\tOuter-J$ model gives a fairly accurate fit to the Hubbard data (predicting a power law with exponent -1, shown by the lines in Fig.~\ref{figSqLatticeResults}).  The fit is especially good at low $\tOuter/\tInner$, which coincides with larger $U/t$ values and thus is where we expect the $\tOuter-J$ model to be most accurate.  Beyond $\tOuter/\tInner=5$, the same general trend is observed, but the phase diagram becomes more complicated as regions of intermediate polarization arise, making the transition from low spin to maximal spin less abrupt (and closer to a second order transition).  This behavior is shown in Fig.~\ref{fig16siteIntermediateSpins} for the 16-site square lattice with $\Nelec=17$.  The $\tOuter-J$ line in this case runs through the regions of intermediate spin, though the model itself gives a direct transition from minimal to fully saturated ground state spin.

\begin{figure}
\begin{center}
\includegraphics[width=3in]{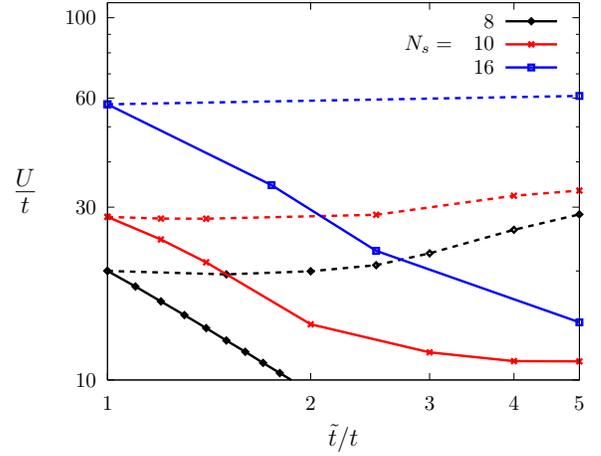}
\renewcommand{\baselinestretch}{1}\normalsize
\caption{(Color online) Ground state spin diagram for the 8-, 10-, and 16-site square lattices showing the asymmetry between doping with a single hole (dashed line) and a single electron (solid line).
\label{figSingleHoleVsElectron}}
\end{center}
\end{figure}

A comparison of these electron-doped systems with corresponding hole-doped systems reveals a pronounced electron-hole asymmetry.  This is expected from the model, since for $\tOuter \ne \tInner$ the Hamiltonian is not electron-hole symmetric: electrons hop with $\tOuter$ whereas holes hop with amplitude $\tInner$.  Figure \ref{figSingleHoleVsElectron} compares the Hubbard model with $\Nelec = \Nsites \pm 1$ (one extra electron or one hole) on finite square lattices.  In the larger 10- and 16-site lattices with one hole we see very little dependence of the ground state spin on $\tOuter/\tInner$, as would be naively expected.  [In the 8-site square lattice an increase in $\tOuter/\tInner$ actually hinders ferromagnetism, seen by an increase in the $U/t$ necessary to reach the totally spin-polarized state. This is most likely a finite size effect, but may have interesting ramifications in the context of finite clusters (see section \ref{secGeomDistorted} below)].  It is clear that the asymmetry between the electron- and hole-doped results originates from the electronic states having greater radius than the hole states, since for equal radii ($\tOuter = \tInner$) the square lattice is bipartite and the problem is electron-hole symmetric.  Figure \ref{figSingleHoleVsElectron} is the first of many that illustrate a central result of this thesis:  high-spin ground states are attained at \emph{much} lower $U/t$ in the electron-doped case than in the hole-doped case.

\begin{figure}[H]
\begin{center}
\includegraphics[width=3in]{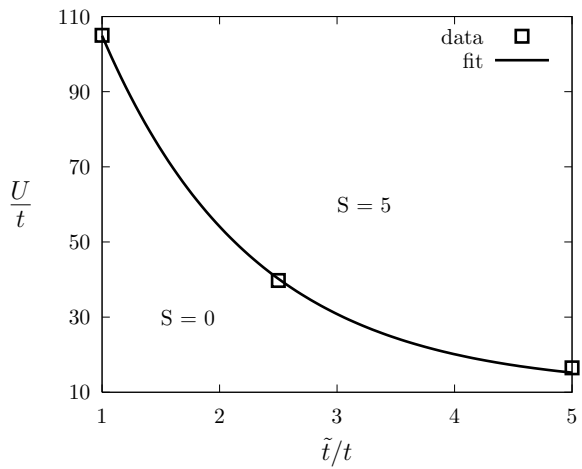}
\renewcommand{\baselinestretch}{1}\normalsize
\caption{Ground state phase diagram of the 16-site square lattice with 4 electrons above half-filling (20 electrons total).  The line is a spline fit, and is provided as a guide for the eye.
\label{fig20e}}
\end{center}
\end{figure}

The ground state spin of the Hubbard model on finite 2D square lattices with periodic boundary conditions is known\cite{RieraYoung_HubSq16_1989} to behave somewhat erratically as a function of the number of electrons ($\Nelec$), and techniques involving an average over varied boundary conditions have had some success as smoothing out, as well as explaining this behavior.\cite{Gros_1996}  We do not address these issues here; instead we focus on the square lattice at two dopings that are known to give high-spin ground states when used with periodic boundary conditions. In addition to the single electron or hole configurations already described, the 16-site square lattice with 4 electrons ($\Nelec=20$) is known to have a ground state spin of maximal value ($S=5$). Figure \ref{fig20e} shows the effect of varying $\tOuter/\tInner$ in this case, and we see, similarly to the case of a single carrier, that increasing $\tOuter/\tInner$ decreases the value of $U/t$ needed to attain the fully saturated ground state.  

\subsubsection{Honeycomb Lattice}
\begin{figure}
\begin{center}
\includegraphics[width=3in]{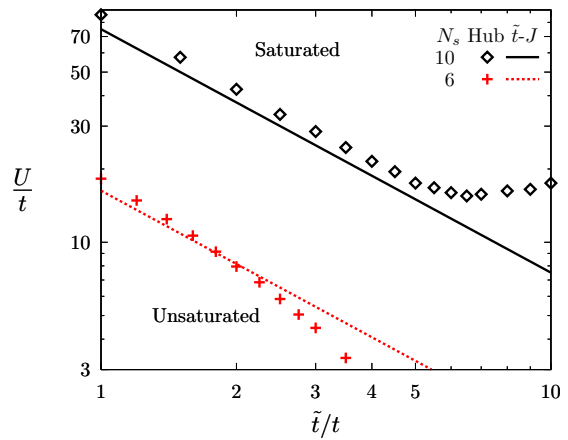} 
\caption{(Color online) Ground state spin diagram from the exact diagonalization of 6- and 10-site honeycomb lattices doped with a single electron (\emph{i.e.}~with 7 and 11 electrons respectively) showing the boundary of the region where there is a complete spin polarization.  In the 6-site lattice the transition is from S=5/2 to S=3/2, whereas in the 10-site lattice the transition is more abrupt, changing from S=9/2 to S=1/2 within the resolution used.
\label{figHoney1ExtraElec}}
\end{center}
\end{figure}

\begin{figure}
\begin{center}
\includegraphics[width=3in]{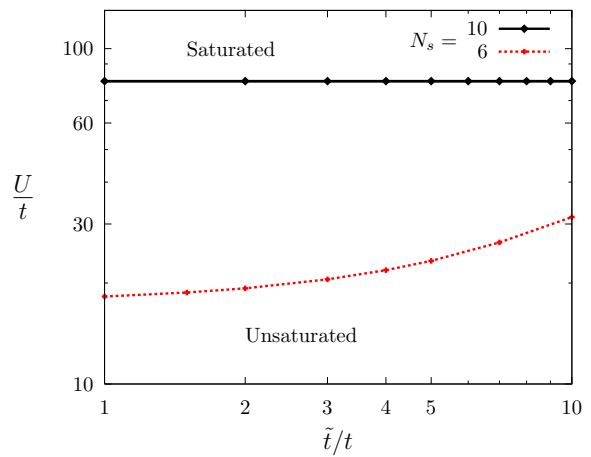}
\caption{(Color online) Exact diagonalization results showing the boundary of the fully spin polarized region on the 6- and 10-site honeycomb lattices doped with a single hole (\emph{i.e.}~with 5 and 9 electrons respectively).  In the 10-site case, the spin on the unsaturated side of the transition is $S=\frac{1}{2}$ except for a region of $S=\frac{5}{2}$ found at intermediate $U/\tInner$ for $\tOuter/\tInner > 10$; on the 6-site lattice the unsaturated state has uniform spin $\frac{3}{2}$.  Note that there is much less variation with respect to $\tOuter/\tInner$ when compared with Fig.~\ref{figHoney1ExtraElec}.
\label{figHoney1ExtraHole}}
\end{center}
\end{figure}

There has been a revived interest in the honeycomb lattice since the recent surge in graphene-related research.  Though it is not itself a Bravais lattice (it is a triangular lattice with a two-point basis), the honeycomb lattice is bipartite and thus the Hubbard model is electron-hole symmetric on it for $\tOuter=\tInner$.  The mean-field ground state phase diagram of the $\tOuter=\tInner$ Hubbard model for hole-doped systems shows the existence and stability of the Nagaoka phase at large $U/\tInner$ near half-filling.\cite{Peres_2004}  The magnetic ground state diagrams for Hamiltonian (\ref{eqnHubHamOccDep}) on 6- and 10-site honeycomb lattices with one electron or hole away from half-filling ($\Nelec = \Nsites \pm 1$) are shown in Figs.~\ref{figHoney1ExtraElec} and \ref{figHoney1ExtraHole} respectively.

We find similar qualitative behavior to that of the square lattices: for systems with $\Nelec = \Nsites + 1$, increasing $\tOuter/\tInner$ expands the region of phase space for which the spin is maximal.  Again, the $\tOuter-J$ model result agrees well with the Hubbard results for low $\tOuter/\tInner$.  In the case of single hole-doping ($\Nelec = \Nsites - 1$), there is little dependence on $\tOuter/\tInner$ in the 10-site lattice whereas there is the opposite $\tOuter/\tInner$ dependence in the smaller 6-site lattice, similar to the case of the 8-site square lattice.

\subsubsection{Triangular Lattice}
The triangular lattice is a Bravais lattice of particular interest, since it magnetically frustrated (not bipartite).  A recent study of the triangular lattice\cite{GhoshSingh_2008} using a many-body expansion technique finds that, at large $U/t$, a $120\,^{\circ}$-ordered AF phase is stable at and below half-filling, and becomes unstable above half-filling.  In past studies of finite clusters, it was likewise found that at half-filling antiferromagnetic states are optimal in \emph{non-}bipartite systems (due to the quantum fluctuations arising from what would be frustrated bonds in a static picture).\cite{PastorHirschMuhlschlegel_1994}

\begin{figure}
\begin{center}
\includegraphics[width=3in]{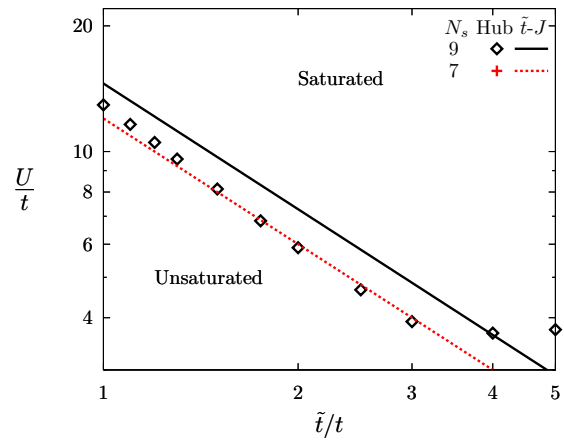} 
\caption {(Color online) Ground state spin diagram from the exact diagonalization of 7- and 9-site triangular lattices when doped with a single electron, showing the region of saturated spin.  On the 9-site lattice, the unsaturated region is predominantly $S=0$ except for a sliver of $S=2$ close to the transition. There is no transition on the Hubbard 7-site lattice, which has a maximally polarized ground state ($S=3$) for the entire plotted area.  In the corresponding $t-J$ model, however, the 7-site lattice has a transition from $S=3$ to $S=2$ near $\tOuter / J \approx 3.0$ (shown by the dotted line). 
\label{figTri1ExtraElec}}
\end{center}
\end{figure}

With a single extra electron ($\Nelec = \Nsites+1$), the Hubbard model on 7- and 9-site lattices displays saturated ferromagnetism very strongly (on the 9-site lattice with $\tOuter=\tInner$, $U/t\approx 15$ results in a spin polarized ground state).  Figure \ref{figTri1ExtraElec} shows our results for the Hubbard model on finite triangular lattices with one extra electron.  Classically, the observed dominance of ferromagnetism could be linked to a suppression of competing AF configurations (frustrated on the triangular lattice).  One must be careful, however, when applying this reasoning to quantum models, as studies have shown that antiferromagnetism is \emph{enhanced} on the triangular lattice with a single hole\cite{HaerterShastryAFTriangle_2005} due to the subtle interplay of quantum phases.  The regnancy of ferromagnetism may also be due to the large number of tight loops in the lattice. Pastor \emph{et al.}\cite{PastorHirschMuhlschlegel_1996}~have remarked that the presence of triangular or square loops coincides with ferromagnetism in finite clusters, and we reach similar findings in our study of clusters below (see sections \ref{secSelectedClusters} and \ref{secGeomDistorted}). The strong FM we see here suggests that this connection extends to lattices as well.

 The $\tOuter-J$ data for the triangular lattice fits the Hubbard data less well than in the previous bipartite lattices.  For the 9-site triangular lattice the $\tOuter-J$ result underestimates the region of saturated spin, and in the case of the 7-site triangular lattice, the Hubbard model does not even transition to the unsaturated state predicted by the $\tOuter-J$ model. The discrepancy is not an immediate cause for concern, and might even be expected, given the low $U/t$ values at which the the transitions occur.

\begin{figure}
\begin{center}
\includegraphics[width=3in]{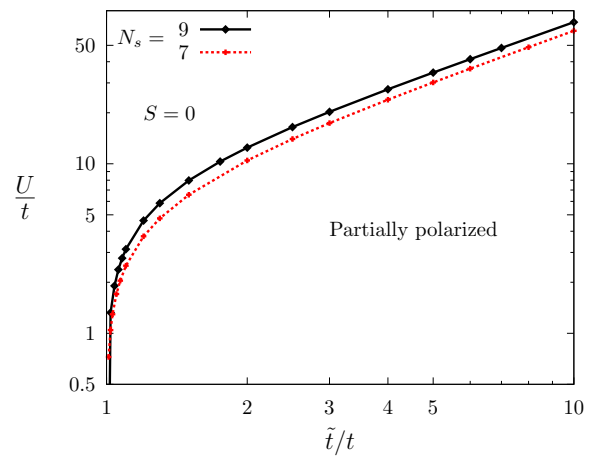} 
\caption {(Color online) Ground state spin diagram for the 7- and 9-site triangular lattices doped with a single hole.  Nowhere is the ground state spin saturated.  Instead, there is a region of minimal spin ($S=0$) at large $U/t$ which is encroached upon by a region of partial spin polarization ($S=2$ and $S=3$ for 7- and 9-sites respectively) as $\tOuter/\tInner$ increases.
\label{figTri1Hole}}
\end{center}
\end{figure}

Since the triangular lattice problem is not bipartite, there can be (and is) electron-hole asymmetry even when $\tOuter = \tInner$.  Figure \ref{figTri1Hole} shows the ground state phase diagram for single hole-doped 7- and 9-site triangular lattices ($\Nelec = \Nsites - 1$).  These plots are qualitatively different from those of the the hole-doped square and honeycomb lattices: the high-spin region is unsaturated and lies at \emph{lower} $U/\tInner$ than a minimal-spin region which dominates at large $U/\tInner$.  As $\tOuter/\tInner$ is increased, the partially polarized region expands up to larger $U/\tInner$ values.  The mechanism for this may be related to the ``kinetic antiferromagnetism'' studied by Haerter and Shastry,\cite{HaerterShastryAFTriangle_2005} which explains how the phase dependence of a single hole's motion enhances antiferromagnetism.

\subsection{Selected Symmetric Clusters \label{secSelectedClusters}}

\begin{figure}
\begin{center}
\includegraphics[width=3.3in]{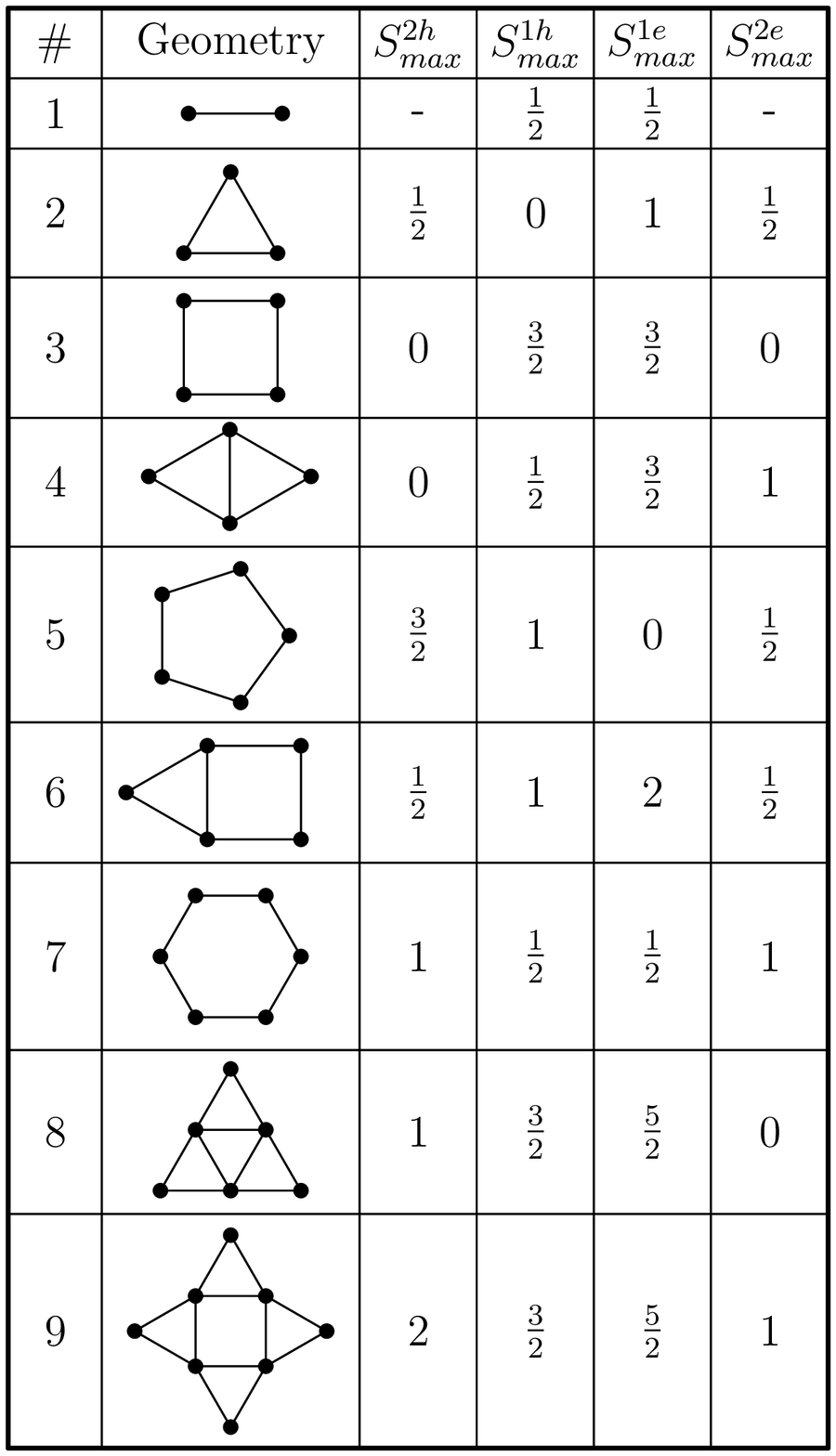}
\renewcommand{\baselinestretch}{1}\normalsize
\caption{ Summary of clusters that have a single pair of hopping parameters.  $S_{max}^x$ is the maximum spin obtained in the window $\tOuter/\tInner \in [1,10]$, $U/t \in [5,100]$ when the system has 1 or 2 holes or electrons away from half-filling ($x = 1h,2h,1e,2e$ respectively).  Note the correspondence of high-spin states with larger numbers of tight loops.
\label{figSingleHopSummary}}
\end{center}
\end{figure}

Next we consider a select group of two-dimensional Hubbard clusters that, like the finite lattices, have only a single pair of hopping amplitudes, $\tInner$ and $\tOuter$.  Unlike the lattices, these clusters are given \emph{open} boundary conditions.  This corresponds to the physical situation in which a small number of sites (dopants or quantum dots) are positioned in a plane such that every pair of nearest neighbors is equidistant.  Pastor \emph{et al.}\cite{PastorHirschMuhlschlegel_1994,PastorHirschMuhlschlegel_1996}~have studied the ordinary Hubbard model (Eq.~\eqref{eqnHubHamOriginal}) on all possible geometrically realizable clusters in two and three dimensions.  Our analysis of cluster structure here is not as exhaustive, but we calculate the phase diagram along the $\tOuter/\tInner$ axis.  Clusters are chosen to lie in the plane such as to retain some spatial symmetries, and their ground state spin is calculated for $1 \le \tOuter/\tInner \le 10$ and $5 < U/t < 100$ when doped with 1 or 2 electrons away from half-filling (in either direction).  Figure \ref{figSingleHopSummary} summarizes the results, giving each cluster's geometric structure and its maximal spin as a function of doping.  We see that in most cases, the highest spin is attained when doped with a single electron, following our expectation that a low density of extra electrons will favor spin polarization.  Indeed, clusters 1-4, 6, and 7, attain their \emph{maximal} ground state spin when doped with one electron.  In contrast, clusters 5 and 9 have greater spin polarization below half-filling, and that their polarization is maximal when doped with two holes.

\begin{figure*} 
\begin{center}
\begin{tabular}{|c|cc|} \hline
Geometry & \multicolumn{2}{c|}{Ground state phase diagrams}  \\ \hline

\includegraphics[width=0.5in]{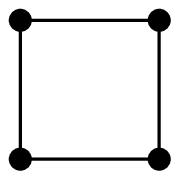} & 
\parbox{2in}{\vspace{.1cm}\includegraphics[width=2in]{figs/sq4_3_alt.ps}\vspace{.1cm}} &
\parbox{2in}{\vspace{.1cm}\includegraphics[width=2in]{figs/sq4_5_alt.ps}\vspace{.1cm}} \\ \hline

\includegraphics[width=0.65in]{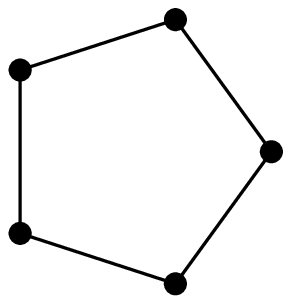} & 
\parbox{2in}{\vspace{.1cm}\includegraphics[width=2in]{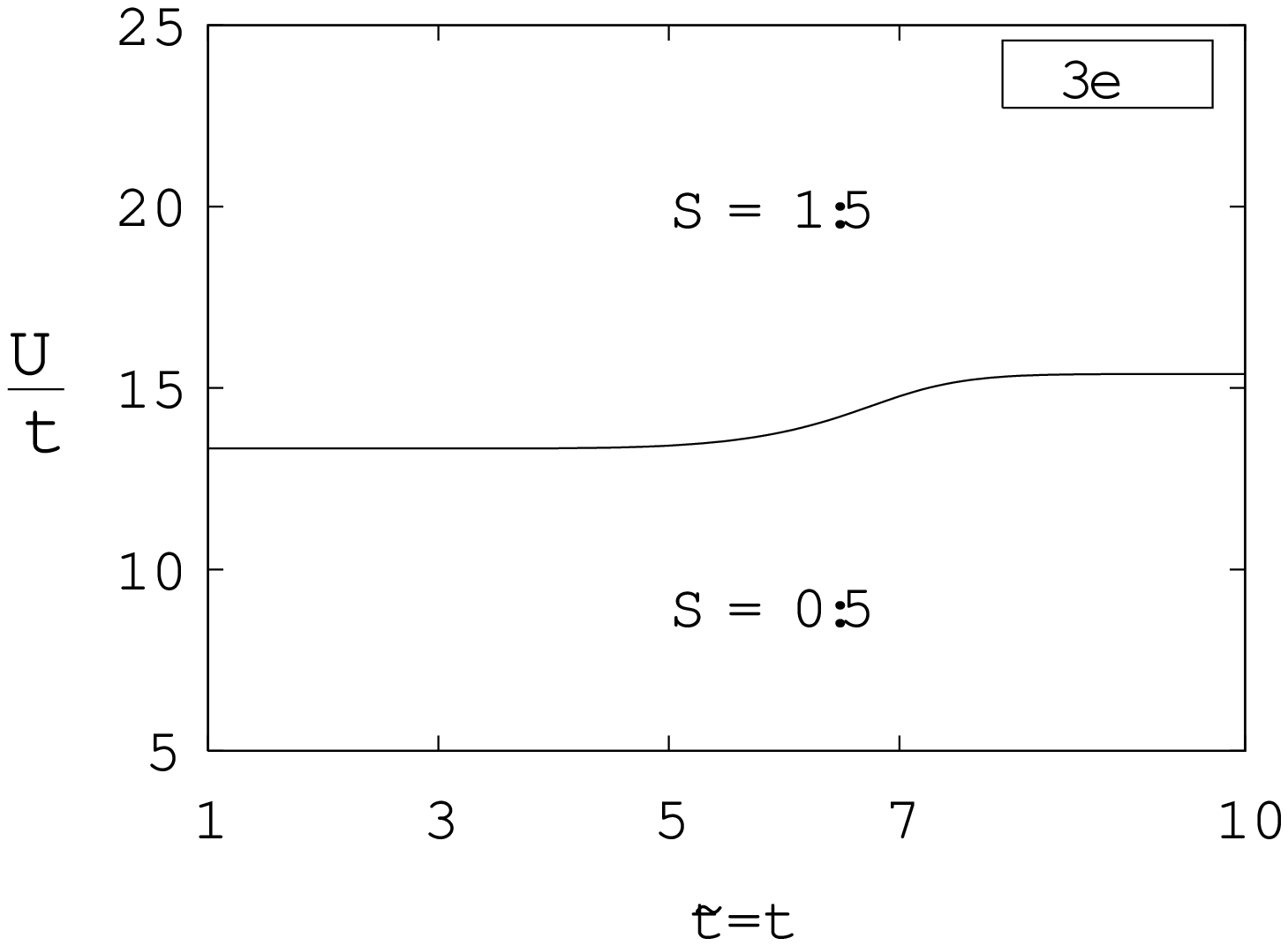}\vspace{.1cm}} & 
\parbox{2in}{\vspace{.1cm}\includegraphics[width=2in]{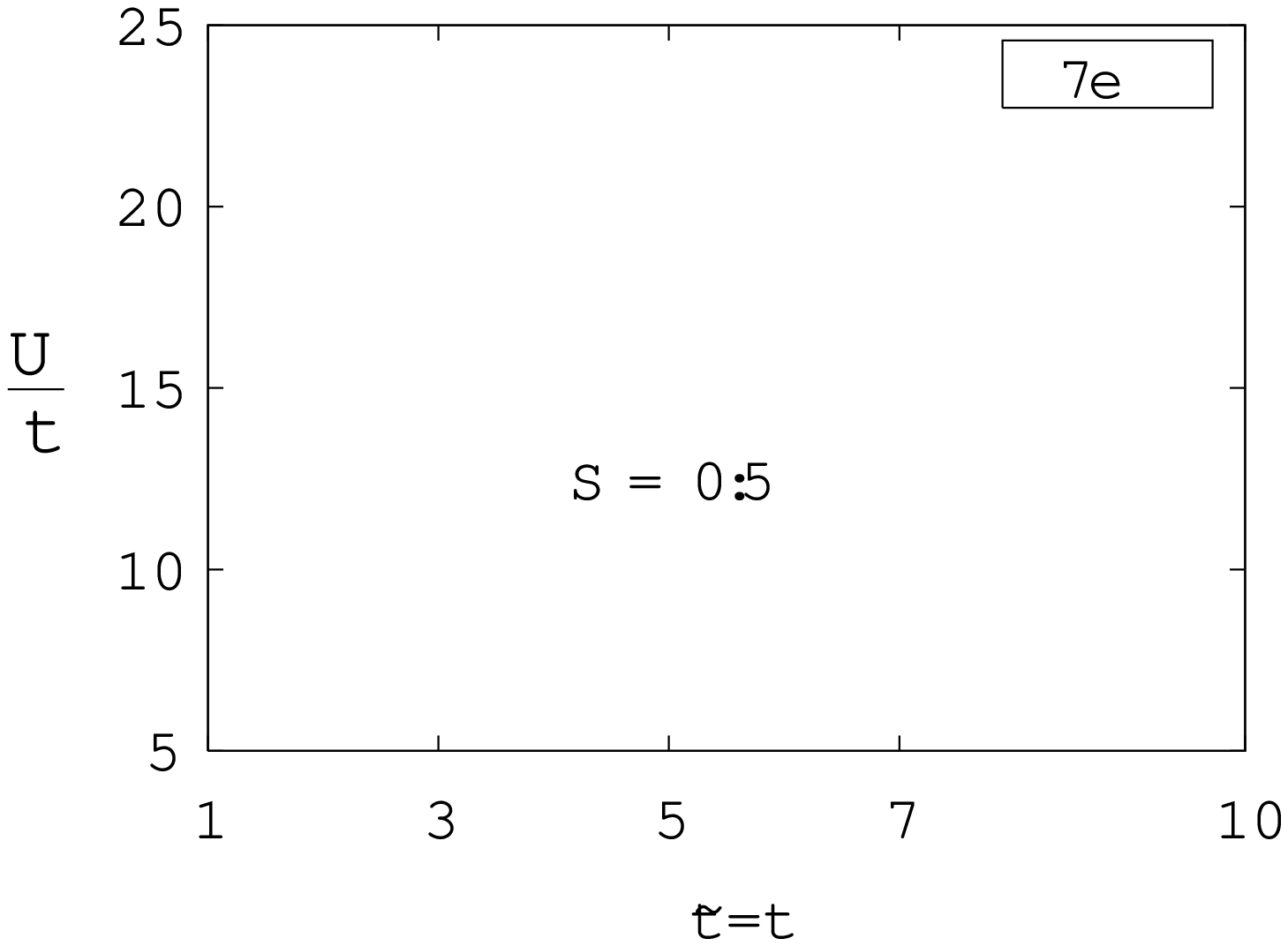}\vspace{.1cm}} \\ \hline

\includegraphics[width=0.65in]{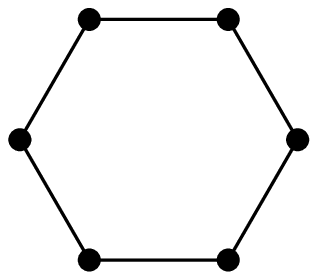} & 
\parbox{2in}{\vspace{.1cm}\includegraphics[width=2in]{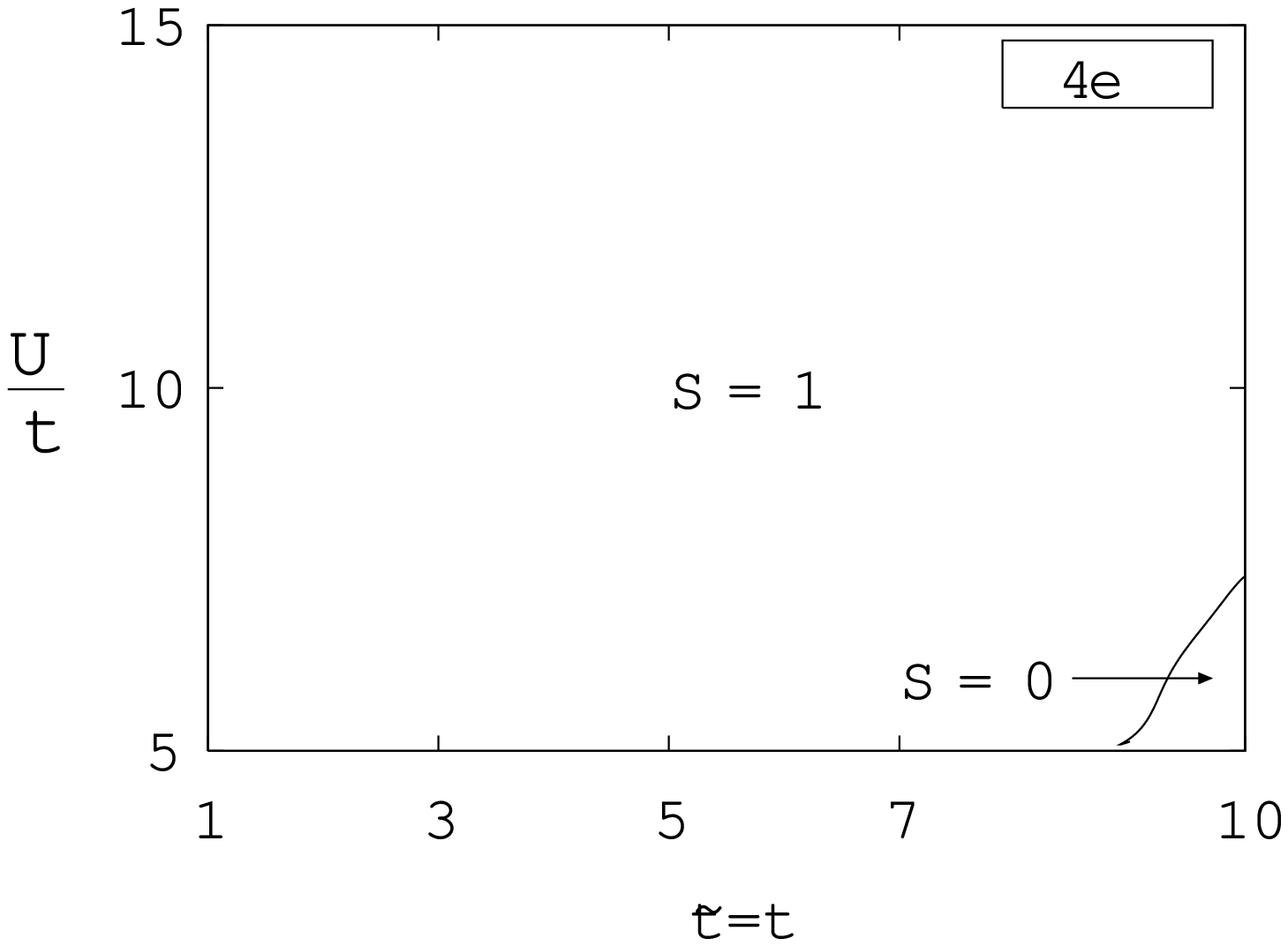}\vspace{.1cm}} & 
\parbox{2in}{\vspace{.1cm}\includegraphics[width=2in]{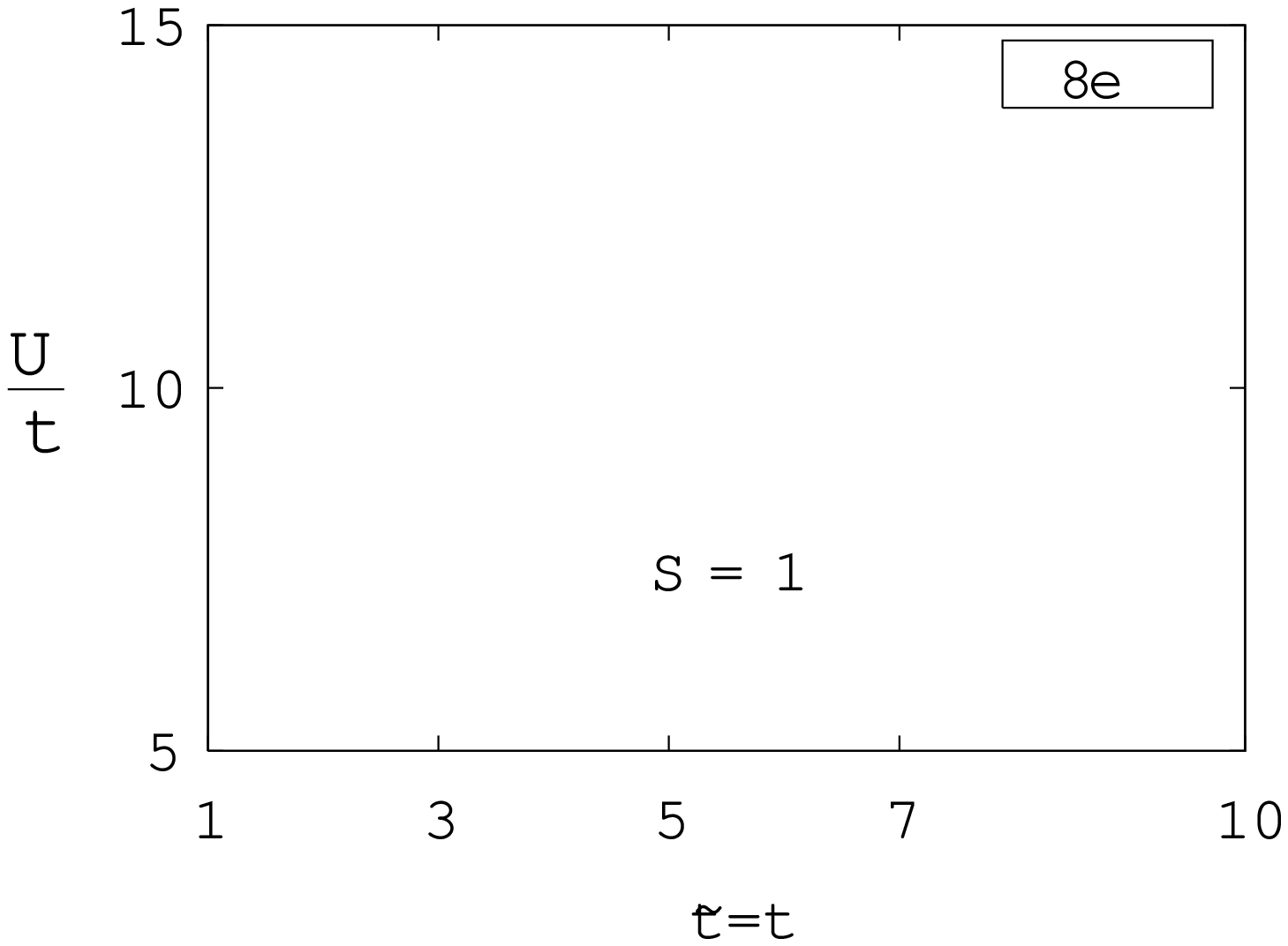}\vspace{.1cm}} \\ \hline
\end{tabular}
\caption{ Ground state spin (T = 0) phase diagrams in the $U/t - \tOuter/\tInner$ plane for clusters 3, 5, and 7 from Fig.~\ref{figSingleHopSummary}.  These 2D clusters are ``ring-like'' in the sense that they are equivalent to 1D chains with periodic boundary conditions.  The fixed electron number is given in the upper-right corner of each plot, and only selected non-trivial diagrams are shown. \label{figSingleHopRingDiagrams}}
\end{center}
\end{figure*}

Clusters 1-3, 5, and 7 we call ``ring-like'', since each is equivalent to a 1-dimensional chain of sites with periodic boundary conditions.  In the pair and triangle (clusters 1 and 2), the spin listed in Fig.~\ref{figSingleHopSummary} is the only spin found in the considered parameter range.  Figure \ref{figSingleHopRingDiagrams} compares the ground state phase diagrams of the remaining clusters with $|\Nelec^*-\Nsites|$ electrons above and below half-filling, for all values of $\Nelec^*$ such that the resulting phase diagrams are non-trivial (have at least two spin regions).  We see from Figs.~\ref{figSingleHopSummary} and \ref{figSingleHopRingDiagrams} that the triangle and square show the greatest percentage spin polarization above half-filling (both have maximally polarized ground states; for the square at large $\tOuter/\tInner$).  Also note the $\tOuter/\tInner$ dependence of the square with one hole vs.~with one electron, where we see behavior similar to that of the square lattices.  The pentagon is unusual in that it has higher ground state spin when hole-doped.  A fully-polarized ground state occurs at large $U/t$ when the system is doped with two holes.  Lastly, the hexagon shows very little $\tOuter/\tInner$ dependence, though with two holes ($4e^-$) larger $\tOuter/\tInner$ creates an interval in $U/t$ with low spin ($S=0$).  This behavior was also seen in the hole-doped bipartite lattices of section \ref{secFiniteLattices}.

The remaining (non-ring-like) clusters, 4, 6, 8, and 9 of Fig.~\ref{figSingleHopSummary}, are created by adjoining triangles and squares.  This was done with the hope of engineering clusters with a high-spin ground states, given the individual properties of the triangle and square.  Detailed ground state phase diagrams for these clusters are presented in Appendix \ref{appSingleHopDiagrams}.  We see in general that increasing $\tOuter/\tInner$ enlarges the high-spin region of the phase diagram for electron-doped clusters and, in this sense, indicates that the high-spin state has become more robust. In hole-doped systems we see a much weaker dependence on $\tOuter/\tInner$,  and in the clusters 8 and 9 we see the opposite behavior: as $U/t$ increases there is a transition to lower ground state spin.  Upon electron-doping, we find a correlation between structures that have a large number of triangular or square loops and those with high spin ground states.  This relationship has also been seen in previous work.\cite{PastorHirschMuhlschlegel_1996}  Though a precise reason for this correspondence has not been found, we believe it is due to such systems being electronically unfrustrated, allowing an electron to easily hop among all the sites and to be very effective at increasing the kinetic energy of the FM state.  Whatever the mechanism, a heuristic rule for constructing clusters with high spin ground states is that a large cluster with many tight loops (triangular or square) is likely to be strongly magnetic.  This has recently become relevant to experiment through the work of Schofield \emph{et al.},\cite{Schofield_2003} who are able to position phosphorous dopants within bulk silicon to nanometer accuracy using a scanning tunneling microscopy (STM) tip.  Such capability allows for the construction of cluster geometries made ``to order'', and opens an entirely new area of application for our work.  In particular, the ability to test for FM behavior (\emph{i.e.}~high spin ground states) in finite lattices of dopants would be very valuable.


\subsection{Distorted clusters \label{secGeomDistorted}}
\begin{figure*}
\begin{center}
\includegraphics[width=3in]{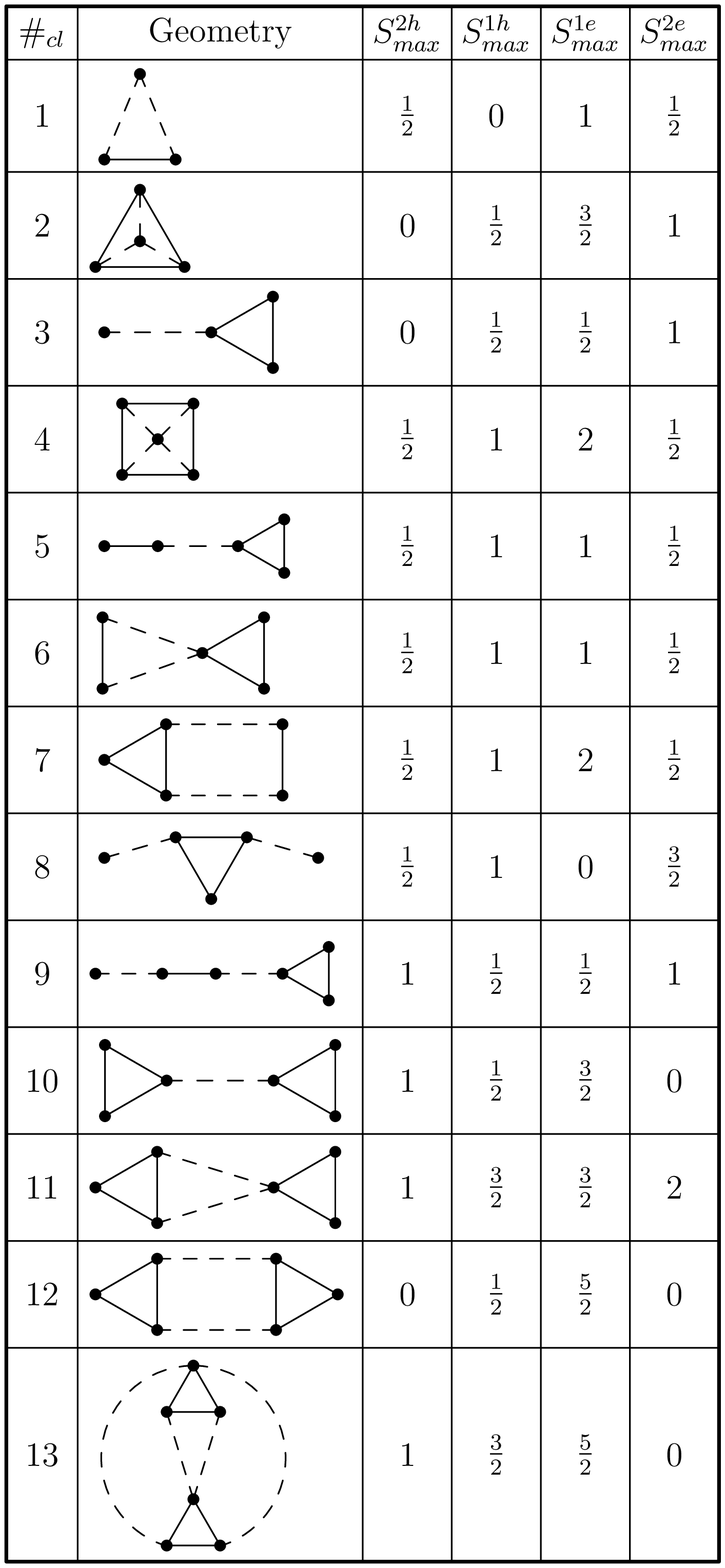}
\includegraphics[width=3in]{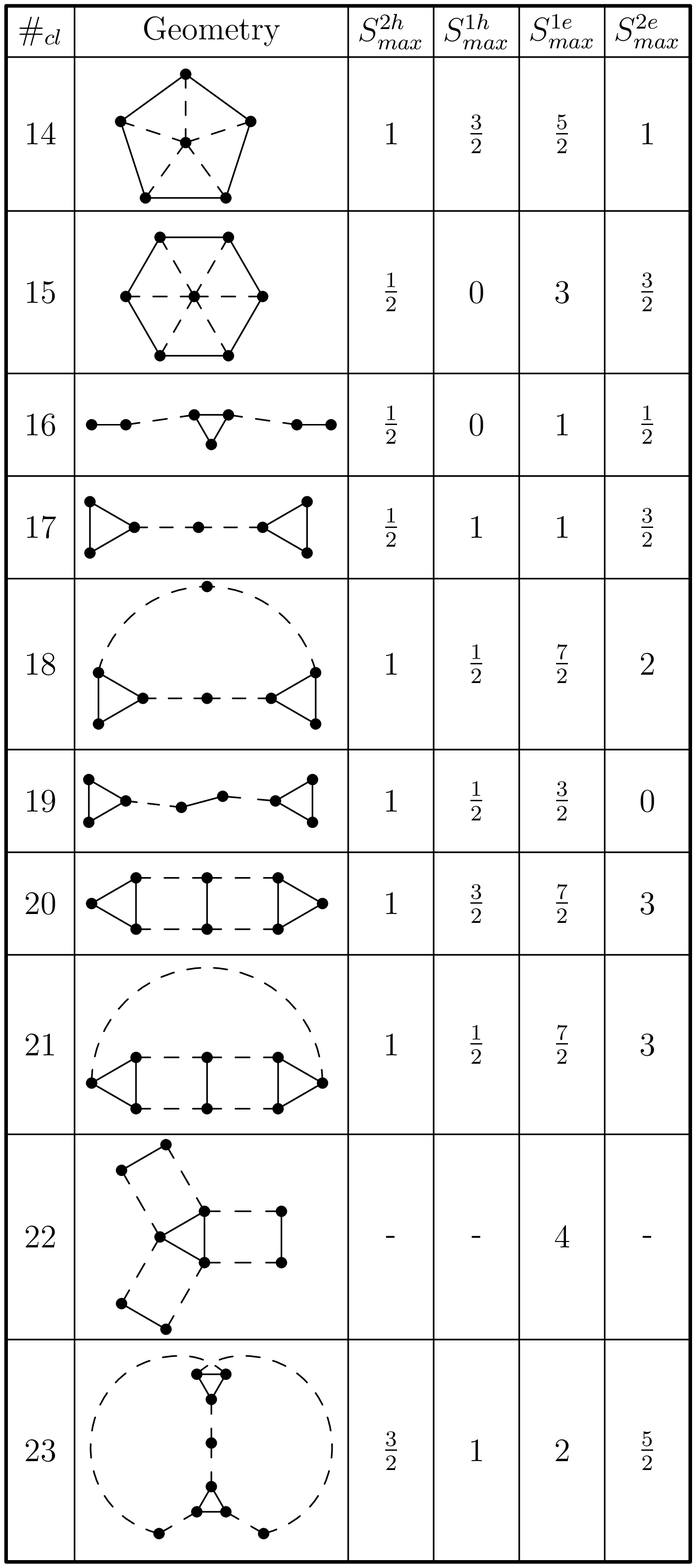}
\renewcommand{\baselinestretch}{1}\normalsize
\caption{Summary of maximum ground state spins for clusters that have two pairs of kinetic parameters (two distinct nearest neighbor distances).  Solid lines represent hopping amplitude $t_1$, and dashed lines $t_2$.  Cluster geometries are listed by size, and maximal spin is given for dopings of -2,-1,1, and 2 electrons away from half-filling.  Each cluster is identified by a number, \#$_{cl}$, and the maximum is taken over the region $t_2/t_1 \in [1,10]$, $t_1/U \in [0.01,0.5]$ for $\tOuter_i/\tInner_i$ uniformly set $= 1$, $5$, and $10$.\label{figDoubleHopSummary}}
\end{center}
\end{figure*}

More complex 2D clusters are obtained by allowing more than one pair of hopping parameters (\emph{i.e.}~hopping is allowed between sites of different separation distances).  In this section we consider clusters with two and three pairs of distinct hopping parameters $\left\{ (\tInner_i,\tOuter_i) \,:\, i \in (1,2,3) \right\}$.  Some of these can be viewed as geometric perturbations of clusters in the last section, while many are new geometries not possible under the restriction of equidistant nearest neighbors.  For a select group of clusters with two pairs of hopping parameters, we consider the ground state spin as a function of $\tInner_2/\tInner_1$ and $U/\tInner_1$ at a uniform fixed $\tOuter_i / \tInner_i$, $i=1,2$.  Our analysis is done over the substantial region of phase space: $t_2/t_1 \in [1,10]$, $t_1/U \in [0.01,0.5]$.  (Note that this extends to $U/t < 10$, outside the physical range found earlier, but in the direction that favors non-ferromagnetic behavior.)  The results are summarized in Fig.~\ref{figDoubleHopSummary}, which show for each geometry the maximal spin achieved with a doping of up to two electrons or holes (the maximum is taken over the region of phase space stated above).  Again we find that most clusters attain their highest spin when doped with $1e^-$ (clusters 1, 2, 4, 7, 10, 12, 14, 15, 18, 20, and 22).  Some of the larger clusters also have high spins when doped with two electrons (clusters 11, 18, 20, and 23), since their density is still low enough to favor FM.  Although in most cases the maximal spin is greater for electron-doping than hole-doping, there are some which attain high spins even when hole-doped (\emph{e.g.}~clusters 8, 9, 11, and 15).

\begin{figure*}
\begin{tabular}{|c|c|c|c|c|} \hline
 \#$_{cl}$ & Geometry & Q & $\tOuter/\tInner=1$ & $\tOuter/\tInner=5$ \\
\hline
\raisebox{0cm}{11}
& \raisebox{-3.3cm}[0cm][0cm]{\includegraphics[height=1.2in]{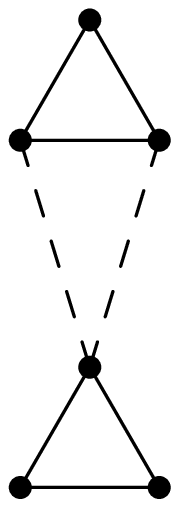}}
& +1
& \parbox{2.3in}{\vspace{0.1cm}\includegraphics[width=2.2in]{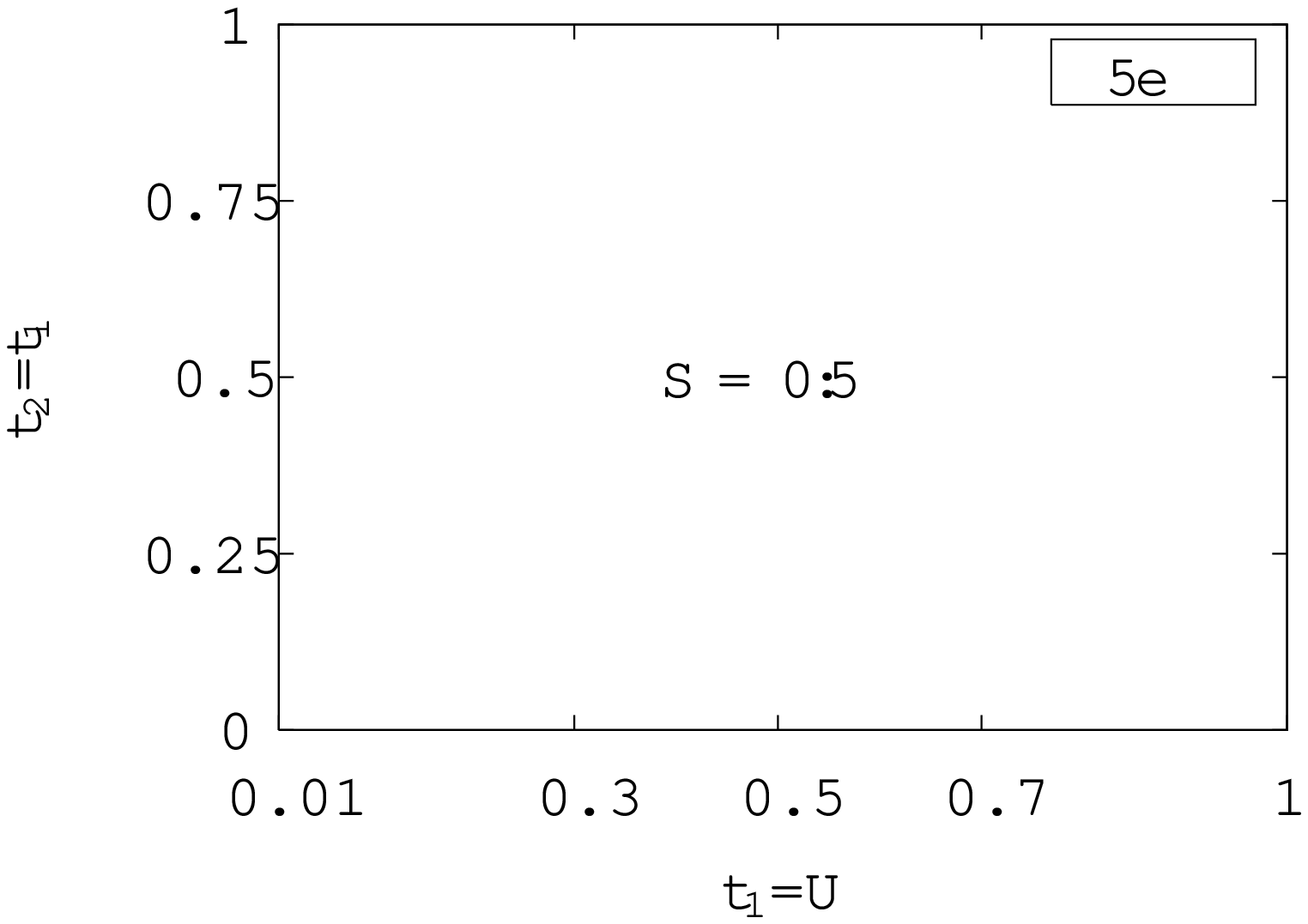}\vspace{0.1cm}}
& \parbox{2.3in}{\vspace{0.1cm}\includegraphics[width=2.2in]{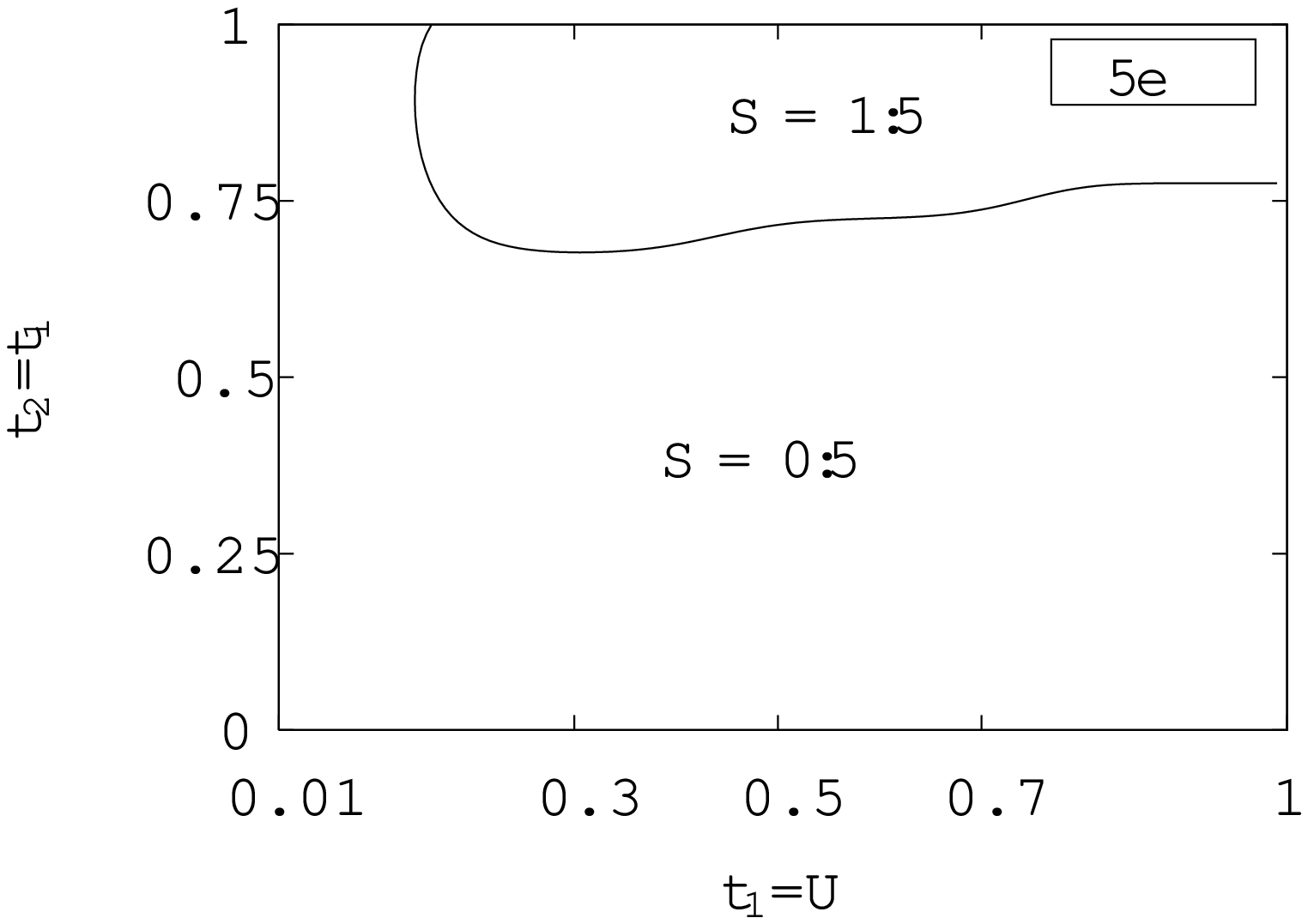}\vspace{0.1cm}} \\ \cline{3-5}
& 
& -1
& \parbox{2.3in}{\vspace{0.1cm}\includegraphics[width=2.2in]{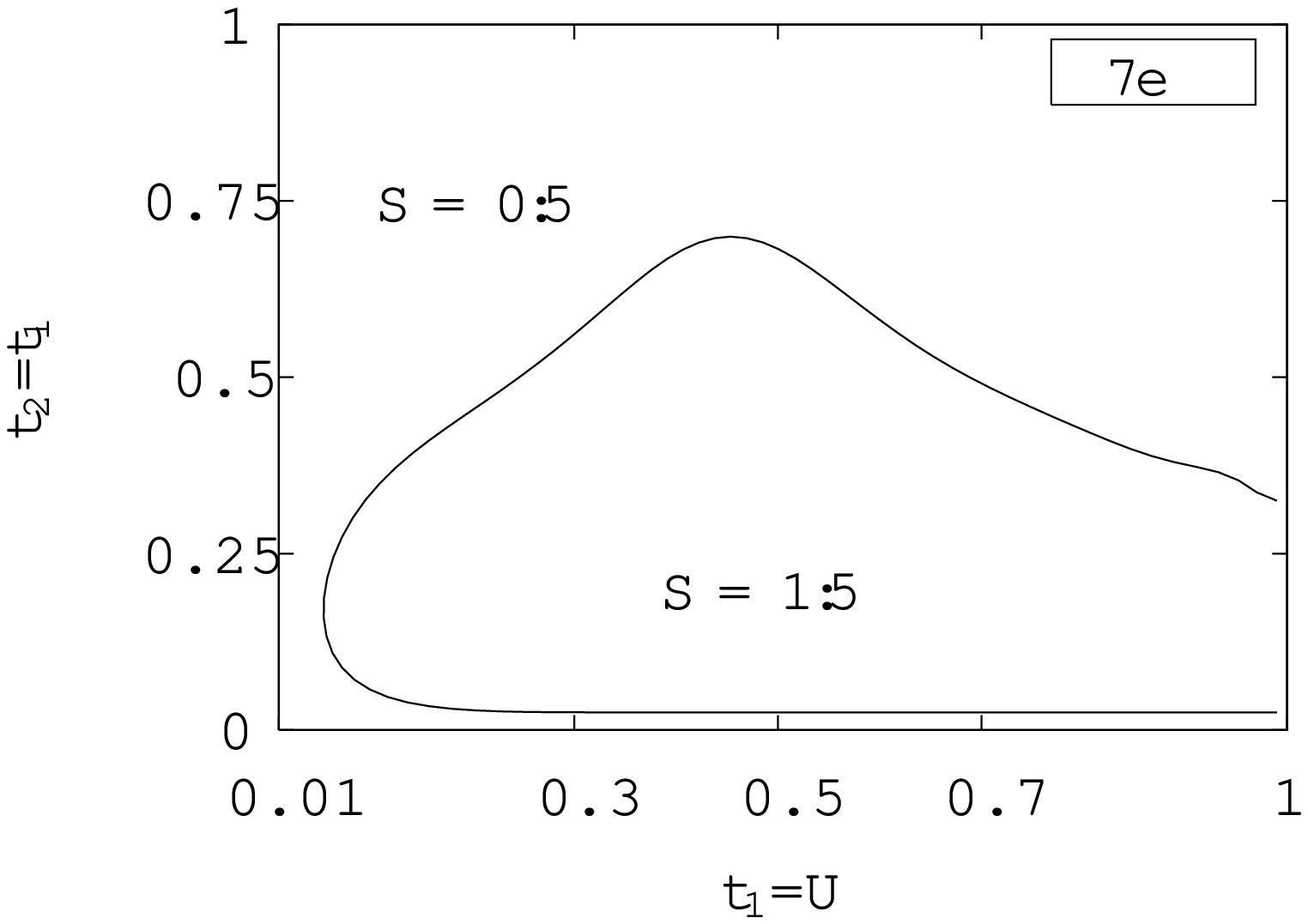}\vspace{0.1cm}}
& \parbox{2.3in}{\vspace{0.1cm}\includegraphics[width=2.2in]{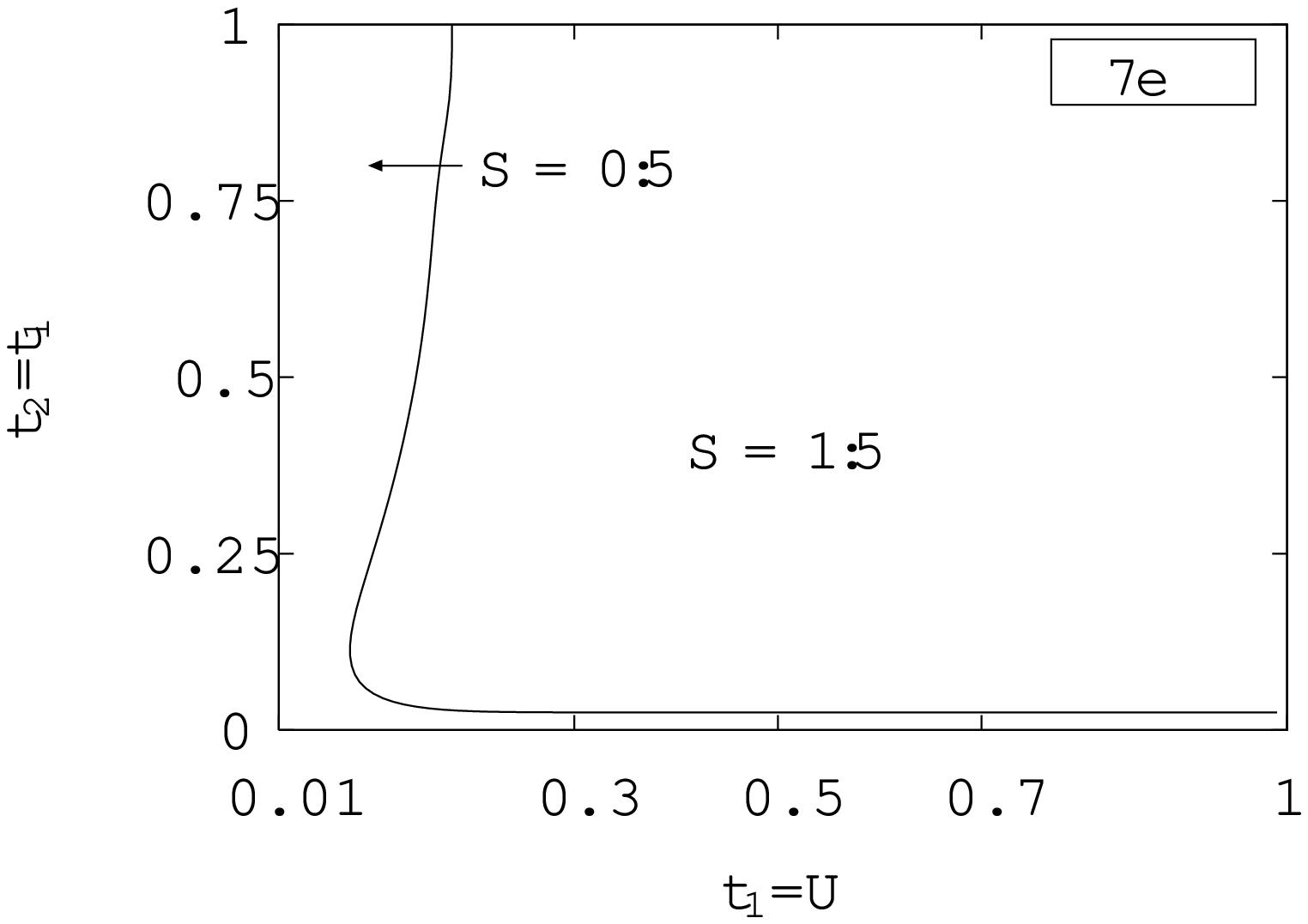}\vspace{0.1cm}} \\ \hline
\raisebox{0cm}{12}
& \raisebox{-1.3cm}{\includegraphics[height=1.2in]{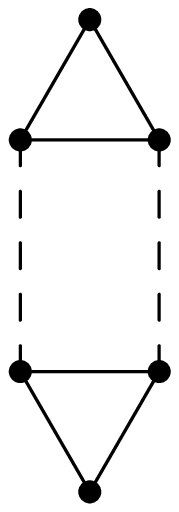}}
& -1
& \parbox{2.3in}{\vspace{0.1cm}\includegraphics[width=2.2in]{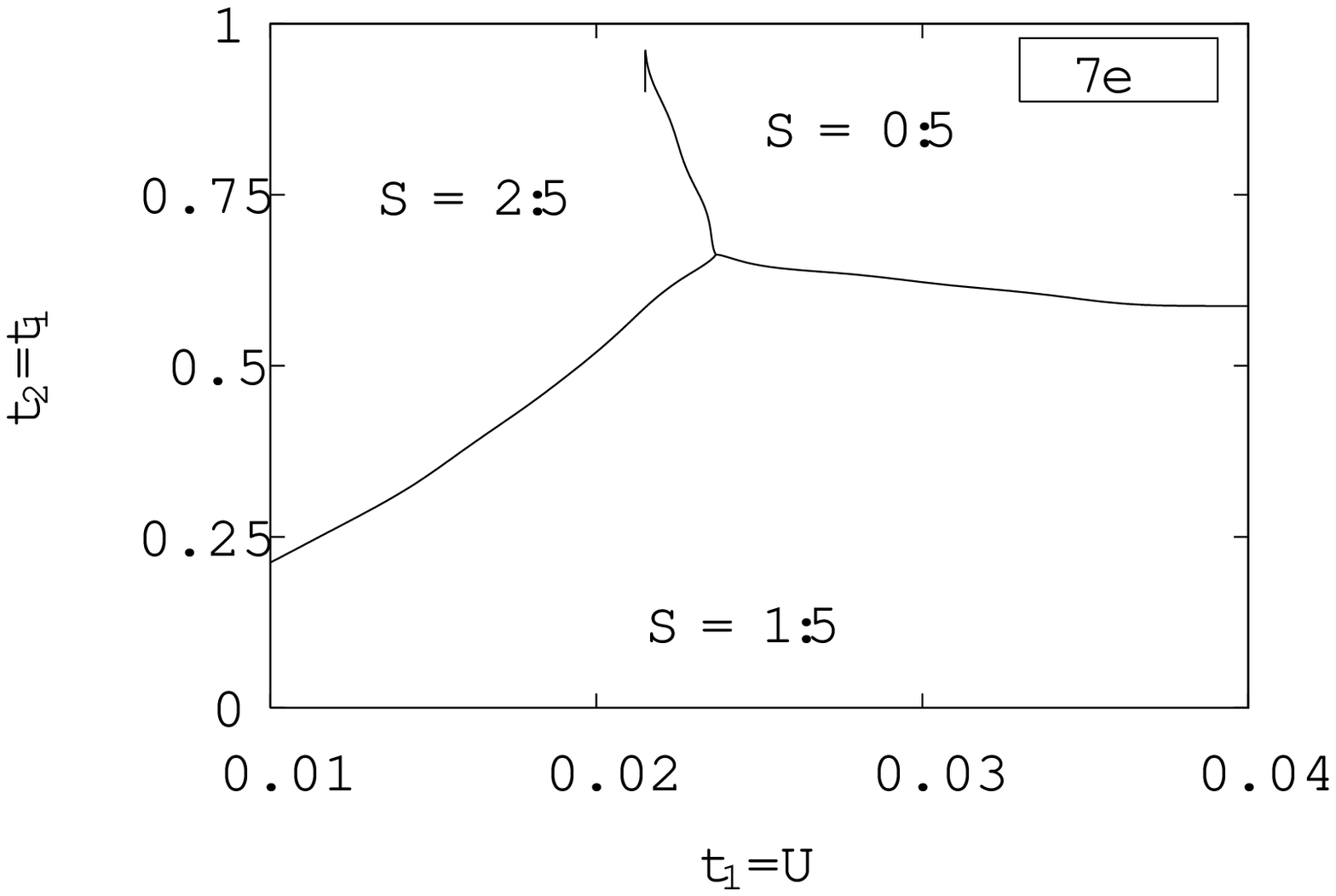}\vspace{0.1cm}}
& \parbox{2.3in}{\vspace{0.1cm}\includegraphics[width=2.2in]{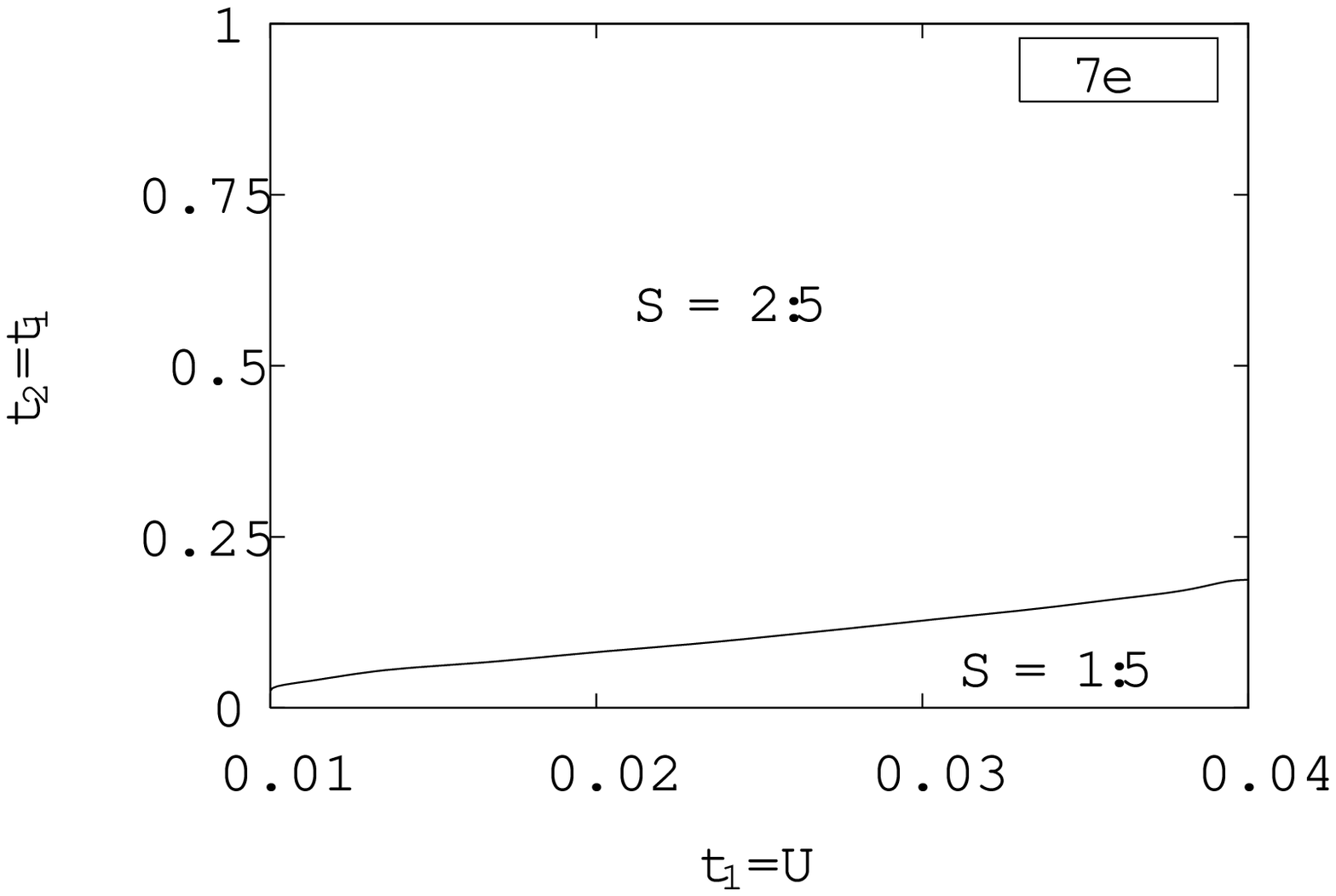}\vspace{0.1cm}} \\ \hline
\raisebox{0cm}{20}
& \raisebox{-3.3cm}[0cm][0cm]{\includegraphics[height=1.2in]{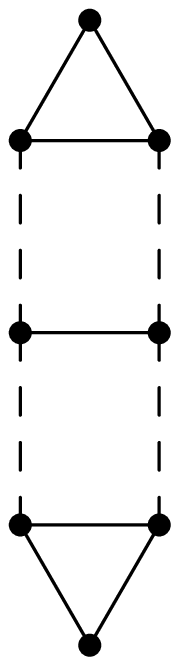}} 
& +1
& \parbox{2.3in}{\vspace{0.1cm}\includegraphics[width=2.2in]{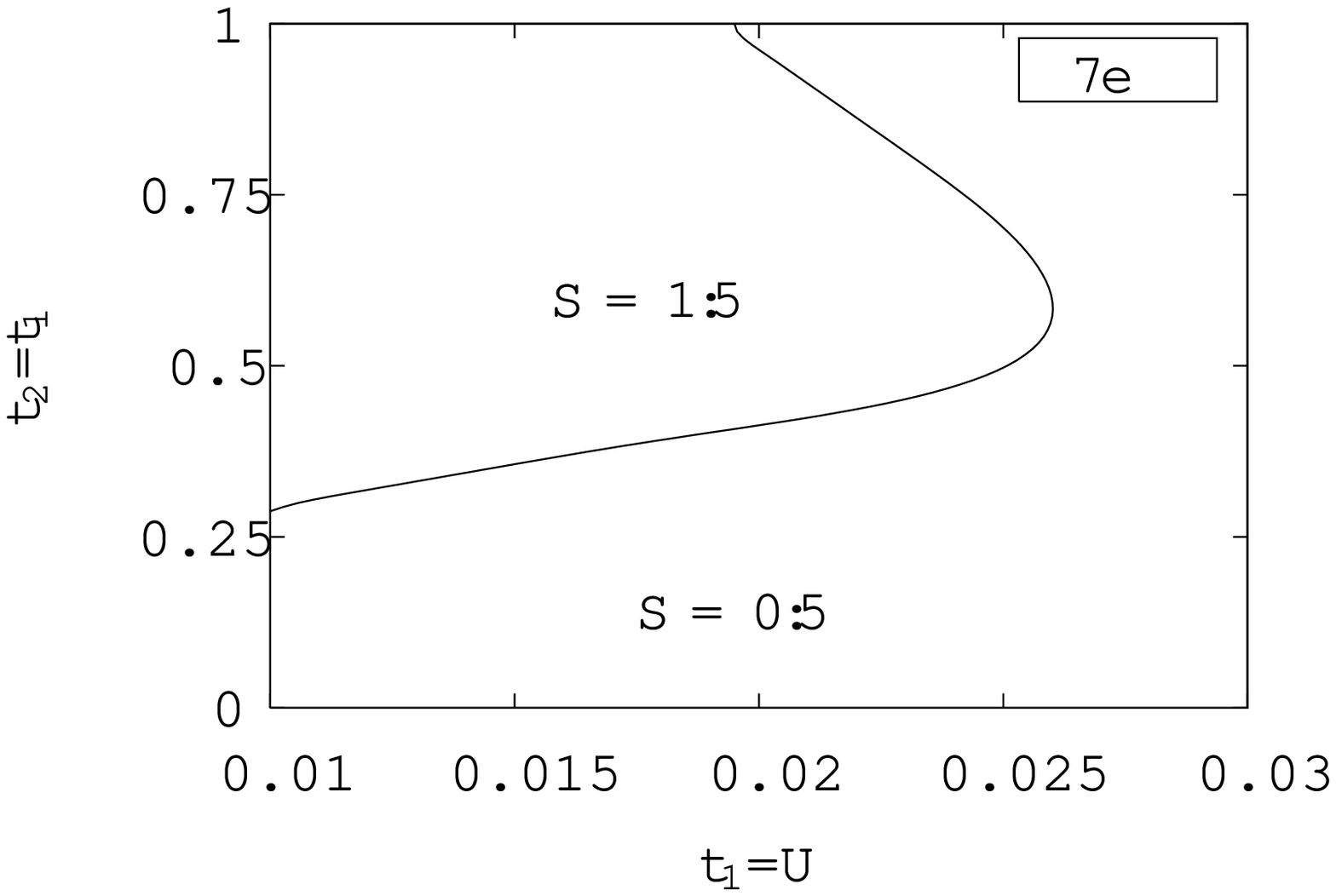}\vspace{0.1cm}}
& \parbox{2.3in}{\vspace{0.1cm}\includegraphics[width=2.2in]{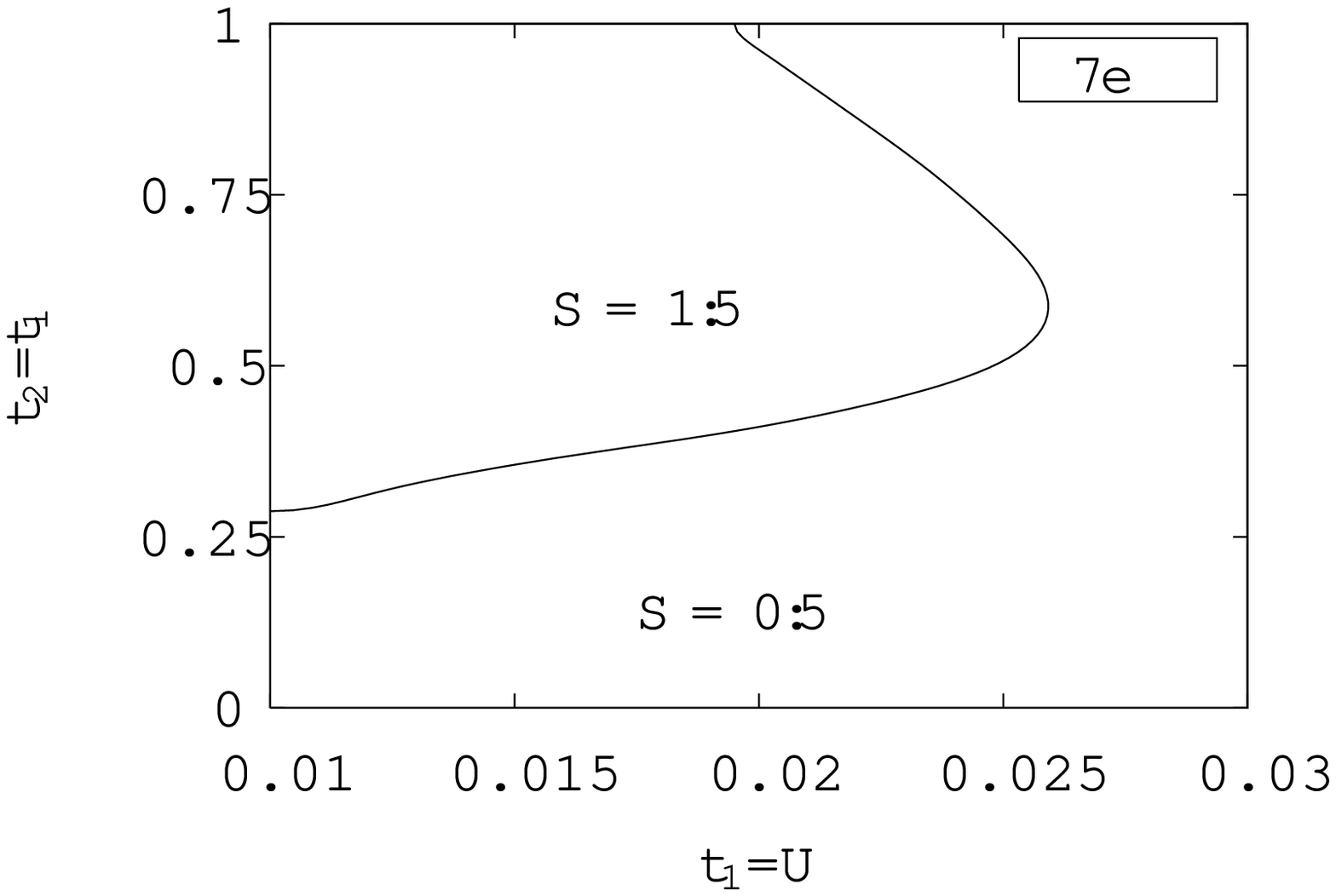}\vspace{0.1cm}} \\ \cline{3-5}

& 
& -1
& \parbox{2.3in}{\vspace{0.1cm}\includegraphics[width=2.2in]{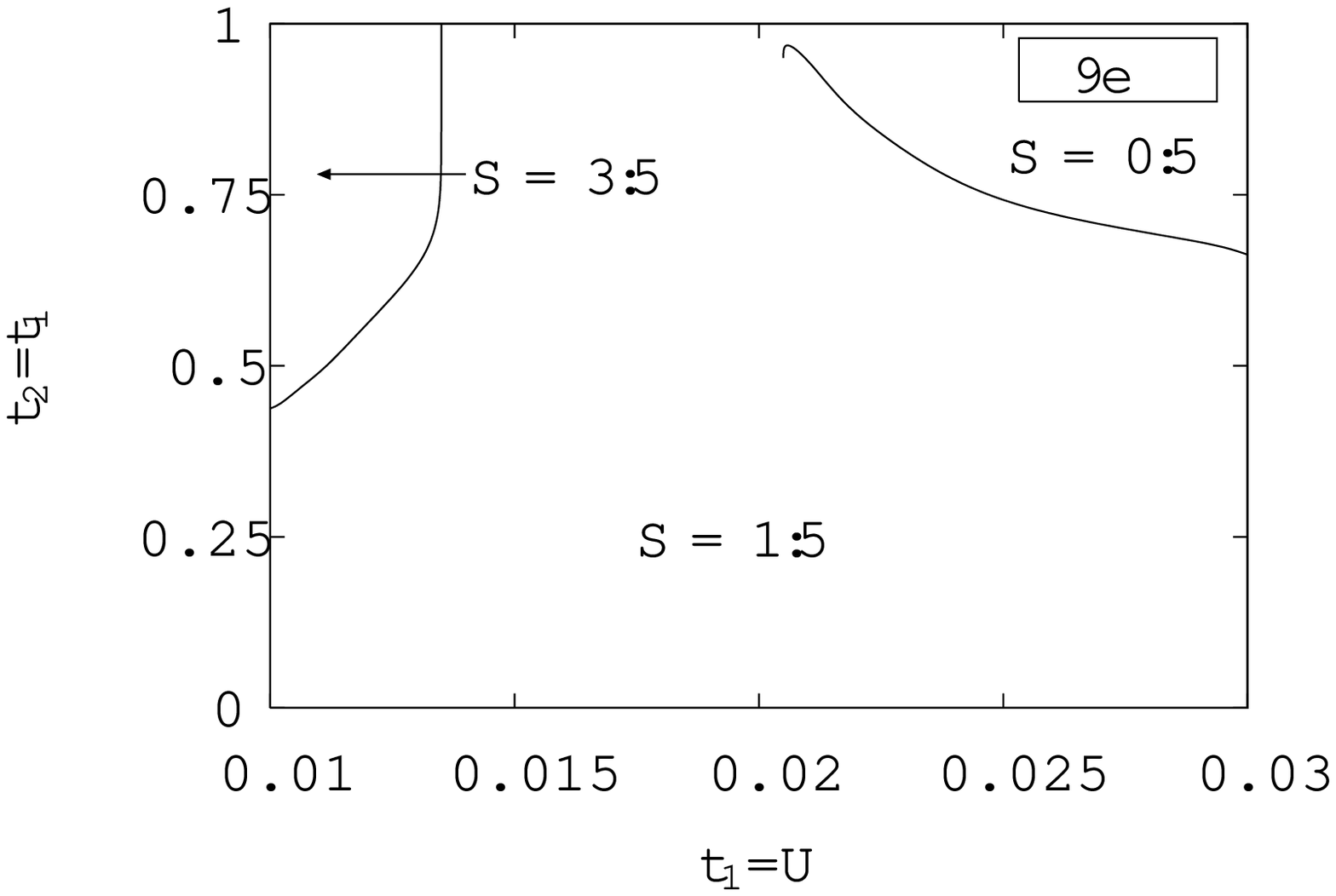}\vspace{0.1cm}}
& \parbox{2.3in}{\vspace{0.1cm}\includegraphics[width=2.2in]{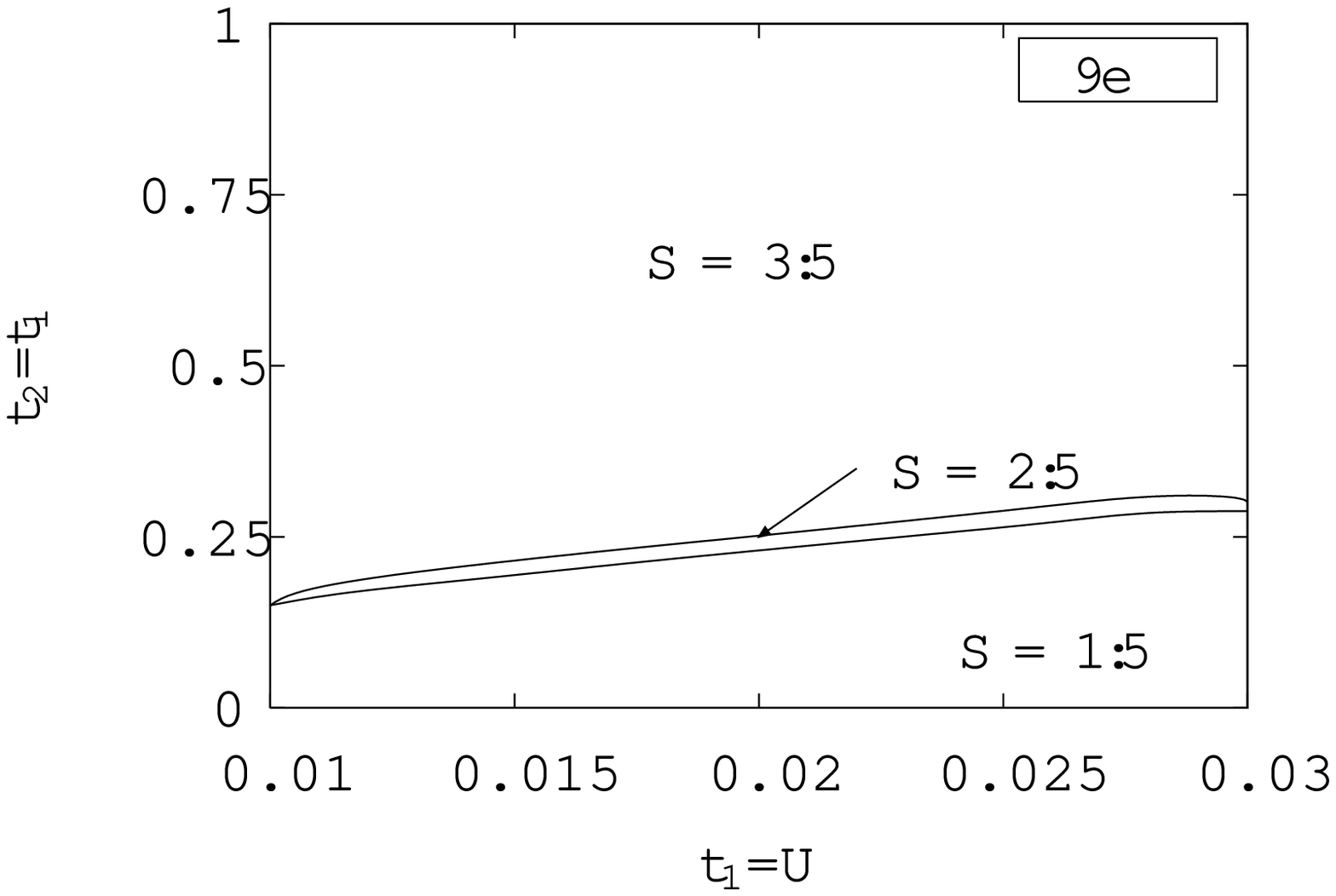}\vspace{0.1cm}} \\ \hline
\end{tabular}
\caption{Ground state spin diagrams for selected clusters from Fig.~\ref{figDoubleHopSummary}. \label{figDoubleHopDiagrams}}
\end{figure*}

We focus on the ground state spin behavior of three clusters from Figs.~\ref{figDoubleHopSummary}: 11, 12, and 20.  Ground state phase diagrams showing the spin for these clusters are in the Fig.~\ref{figDoubleHopDiagrams}.  Each row of the table shows the geometry and two ground state phase diagrams of a cluster with a fixed number of sites $\Nsites$ and electrons $\Nelec$.  The two diagrams correspond to $\tOuter/\tInner = 1$ and $5$, as indicated by the column headings.  The charge of the cluster $Q = \Nsites - \Nelec$ (the negative of its doping relative to half filling) is given in the third column.  For each selected cluster, phase diagrams are only shown for $Q=\pm 1$.  The transition lines in these plots are found by finding the ground state spin on a grid in parameter space, then fitting the transitions between grid points with smooth curves.  Detailed phase diagrams of \emph{all} non-trivial cases are given Appendix \ref{appDoubleHopClusters}.


Clusters 11, 12, and 20 have fully spin-polarized ground states when doped with one electron, and for this reason will be used as starting points in later perturbation schemes.  The movement of ground state spin boundaries as $\tOuter/\tInner$ is increased in steps ($\tOuter/\tInner = 1,5,10$) is seen in each row of the table.  In cluster 11 the region of $\tInner_2/\tInner_1$ vs.~$U/\tInner_1$ space with maximal spin expands for both electron- and hole-doped cases as $\tOuter/\tInner$ increases, which is interesting since the effect of a larger $\tOuter/\tInner$ on a hole-doped system is expected to be relatively minor.  In cluster 12, a similar increase in polarization with larger $\tOuter/\tInner$ is only seen in the single electron-doped case (the $Q = -1$ case is all that is shown, since all other dopings have minimal spin throughout the plotted region; see Fig.~\ref{figDoubleHopSummary}).  Cluster 20 behaves very much like we naively expect:  in the hole-doped case ($Q = +1$) the diagram is almost insensitive to changing $\tOuter/\tInner$, while for $Q=-1$, the region of maximal polarization clearly expands at the expense of other lower-spin regions.  We note that in all cases high-spin ground states occur when $\tInner_2/\tInner_1$ is close to 1, that is, when the dotted hopping links in the tables are nearly as strong as the solid links and the triangles and pairs that make up cluster are more strongly coupled.

Several major conclusions may be drawn from this data.  First, there are many instances of high-spin ground states among these clusters, many of which can be thought  of as a weak coupling ($t_2$) between triangles and pairs with a stronger internal coupling ($t_1$).  In a real system, where the broad distribution of inter-site distances due to positional randomness creates exponentially strong and weak bonds, these results give some hope that the spin-polarization seen in the isolated triangle, for example, will survive in the presence of perturbation due to other sites, and that this interaction may even lead to spin polarization on longer length scales.  Second, it is found almost universally that increasing $\tOuter/\tInner$ leads to greater spin polarization in \emph{electron-doped} clusters, just as in the finite lattices (section \ref{secFiniteLattices}) and single-hopping parameter clusters (section \ref{secSelectedClusters}).  In electron-doped clusters, we continue to see a correlation between the number of triangular loops in a cluster and that cluster's maximal spin.  For instance, compare clusters 5 and 7 with clusters 14 and 15 of Fig.~\ref{figDoubleHopSummary} (the latter are much more magnetic).  
 In hole-doped systems we generally find lower spin values, and often there is a high-spin region at \emph{low} $U/t_1$.  This inverted relationship in clusters below half-filling was also found in section \ref{secSelectedClusters} and on the 8-site square lattice.  Lastly, we note that although there is potential for high-spin states, there are many clusters that have large regions of minimal ground state spin.  We find overall that the Nagaoka-like ferromagnetic effect we observe is very sensitive to geometry, though the sensitivity decreases at large $\tOuter/\tInner$.

\begin{figure} 
\begin{center}
\includegraphics[width=3in]{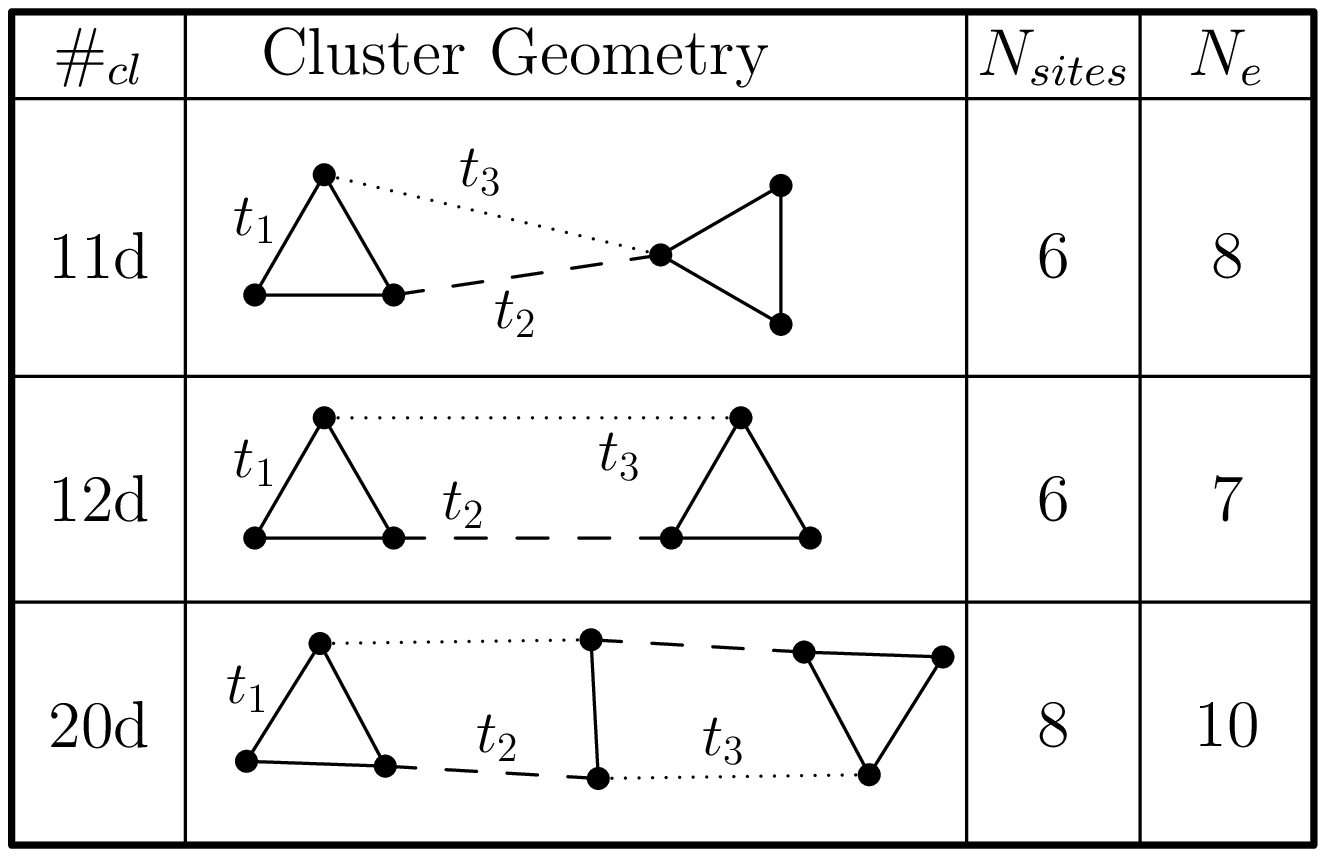}
\caption{Geometries of clusters obtained by geometric distortion of clusters 11, 12, and 20 of Fig.~\ref{figDoubleHopSummary}, with three pairs of kinetic parameters $\tInner_1 > \tInner_2 > \tInner_3$. \label{figDistortedGeometries}}
\end{center}
\end{figure}

Next, we test the stability of a select few of the high-spin ground states found above.  For clusters 11, 12, and 20 of Fig.~\ref{figDoubleHopSummary}, we further reduce the spatial symmetry by additional geometric distortion, as shown in Fig.~\ref{figDistortedGeometries}.  The distortion introduces a third pair of hopping amplitudes  ($\tInner_3,\tOuter_3$), and the ratio $\tInner_3/\tInner_2$ measures the amount of distortion.  

\begin{figure}[H]
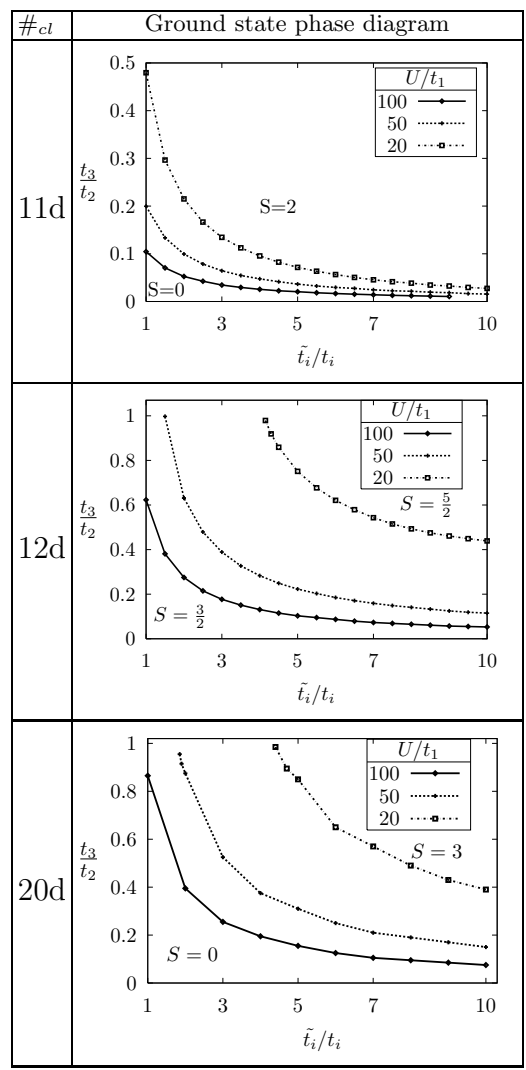

\begin{center}
\begin{tabular}{|l|c|}
\hline
\#$_{cl}$ & Ground state phase diagram \\ \hline
\large{11d} & \parbox{2.3in}{\vspace{0.2cm}\includegraphics[width=2.2in]{figs/triangle2d.ps}\vspace{0.2cm}}\\ \hline
\large{12d} & \parbox{2.3in}{\vspace{0.2cm}\includegraphics[width=2.2in]{figs/triangle2e.ps}\vspace{0.2cm}}\\ \hline
\large{20d} & \parbox{2.3in}{\vspace{0.2cm}\includegraphics[width=2.2in]{figs/pairLinkedTriangles2_d2.ps}\vspace{0.2cm}}\\ \hline
\end{tabular}
\caption{Ground state spin diagram for distorted clusters.\label{figDistortedPhaseDiagrams}}
\end{center}
\end{figure}

We fix $\tInner_2/\tInner_1$ at a value for which the undistorted ($\tInner_3 = \tInner_2$) cluster has a high-spin ground state, and determine the amount of distortion that can be applied (\emph{i.e.}~the lowest value $\tInner_3/\tInner_2$ can attain) before the cluster loses its high spin state.  The value of $\tOuter_i/\tInner_i$ is fixed (\emph{i.e.}~in each run, all of the links forming the cluster have the same $\tOuter/\tInner$ ratio), and the resulting ground state phase diagrams as a function of $\tInner_3/\tInner_2$ and $U/\tInner_1$  are shown in Fig.~\ref{figDistortedPhaseDiagrams}.  There are two key points resulting from this data.  First, as $\tOuter/\tInner$ becomes larger, the high-spin ground states become more robust to the geometric fluctuation considered here: regions with high-spin ground states persist to lower values of $\tInner_3/\tInner_2$ as $\tOuter_i/\tInner_i$ is raised.  (Recall that lower $\tInner_3/\tInner_2$ corresponds to larger geometric distortion.)  Second, the high-spin ground states are more robust at larger $U/\tInner_1$, since the curves for fixed $U/\tInner_1$ move to lower values of $\tInner_3/\tInner_1$ as $U$ increases (\emph{e.g.}~$U=100$ curve lies below the $U=50$ and $U=20$ curves).

\begin{widetext}

\begin{figure*}[h]
\begin{tabular}{c|ccc}
  & Cluster 11 & Cluster 12 & Cluster 20 \\ \hline
$\tOuter/\tInner=1$ & \parbox{2.3in}{\includegraphics[width=1.8in,angle=270]{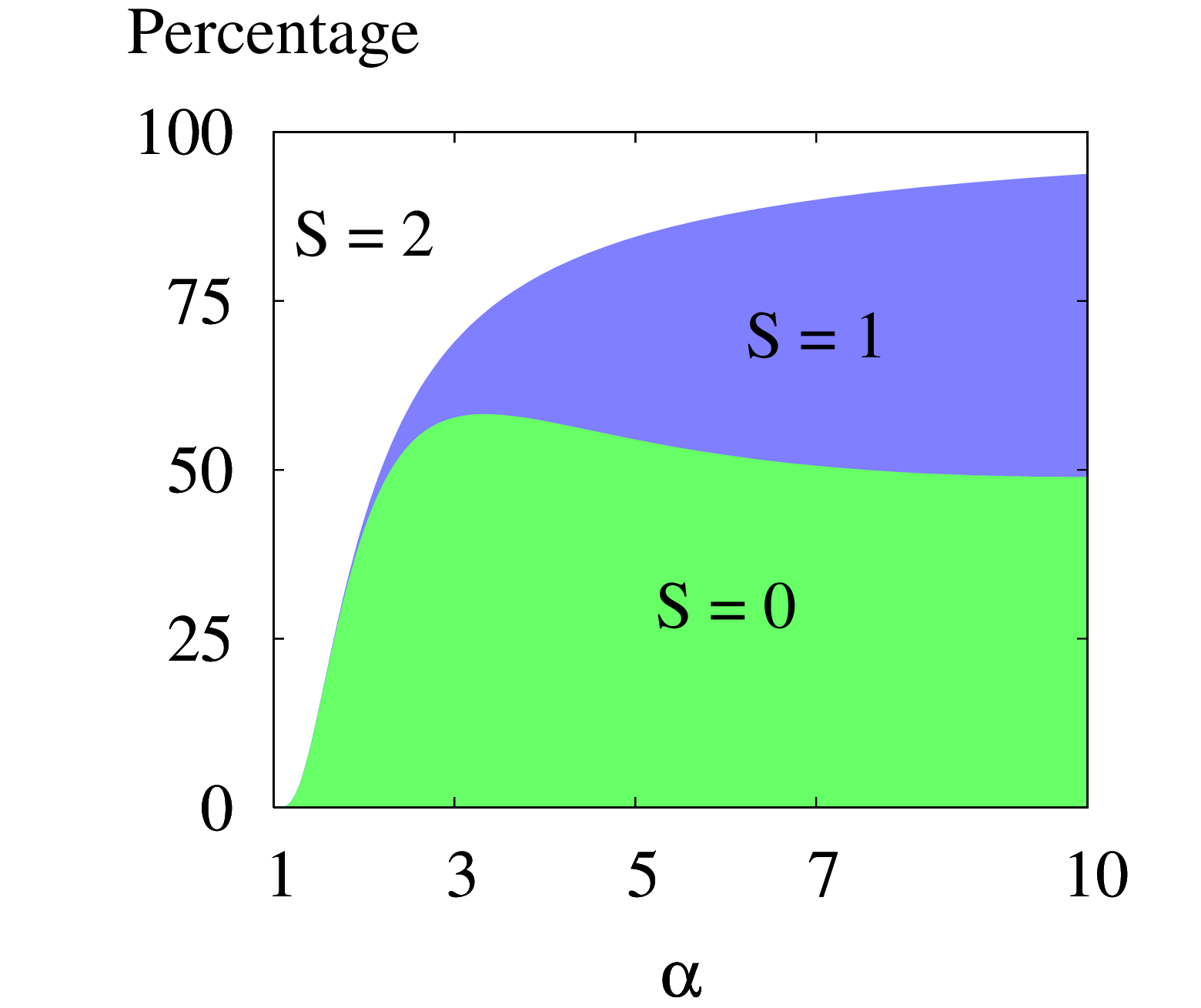}} &
\hspace{-1cm}\parbox{2.3in}{\includegraphics[width=1.8in,angle=270]{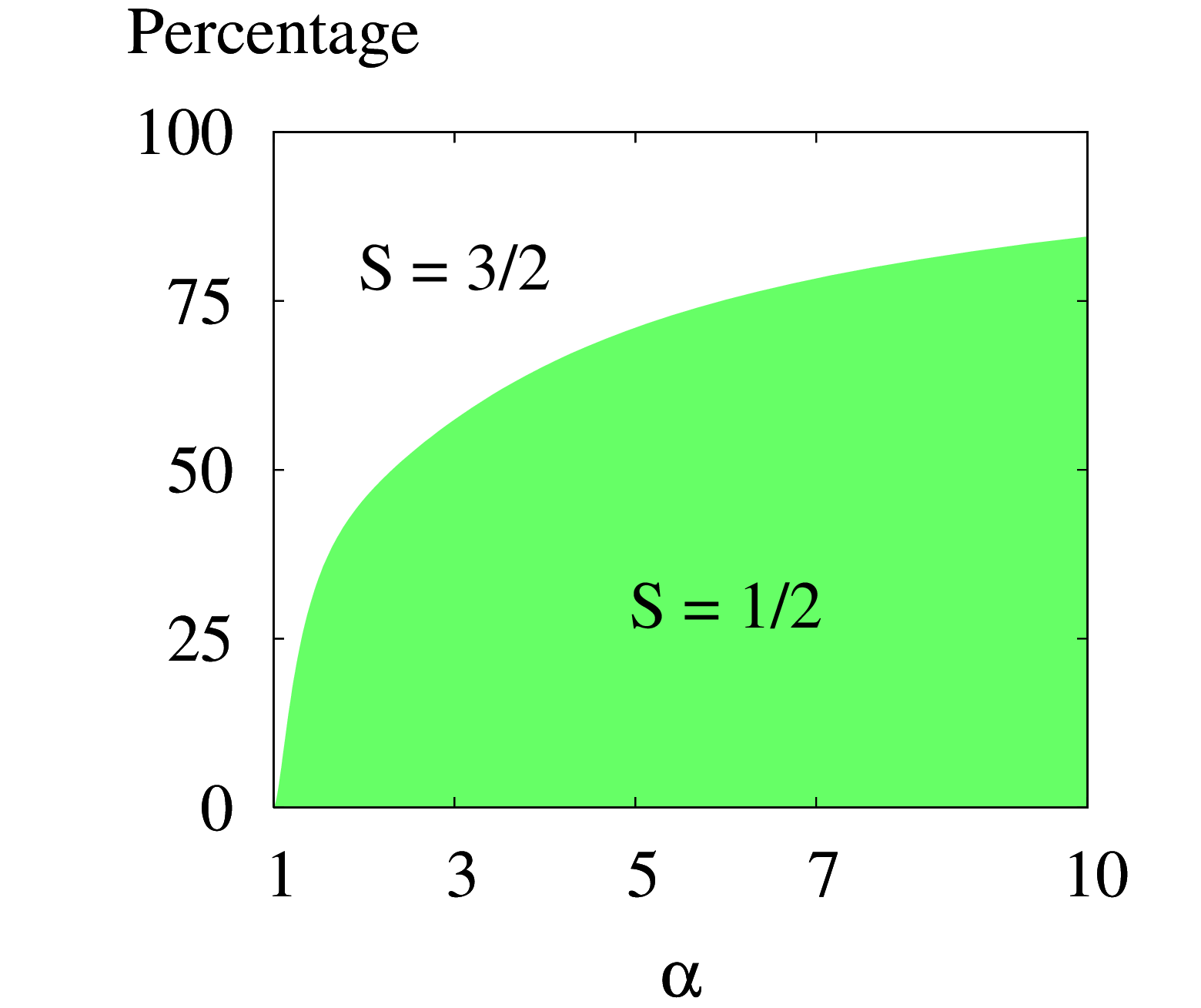}} &
\hspace{-1cm}\parbox{2.3in}{\includegraphics[width=1.8in,angle=270]{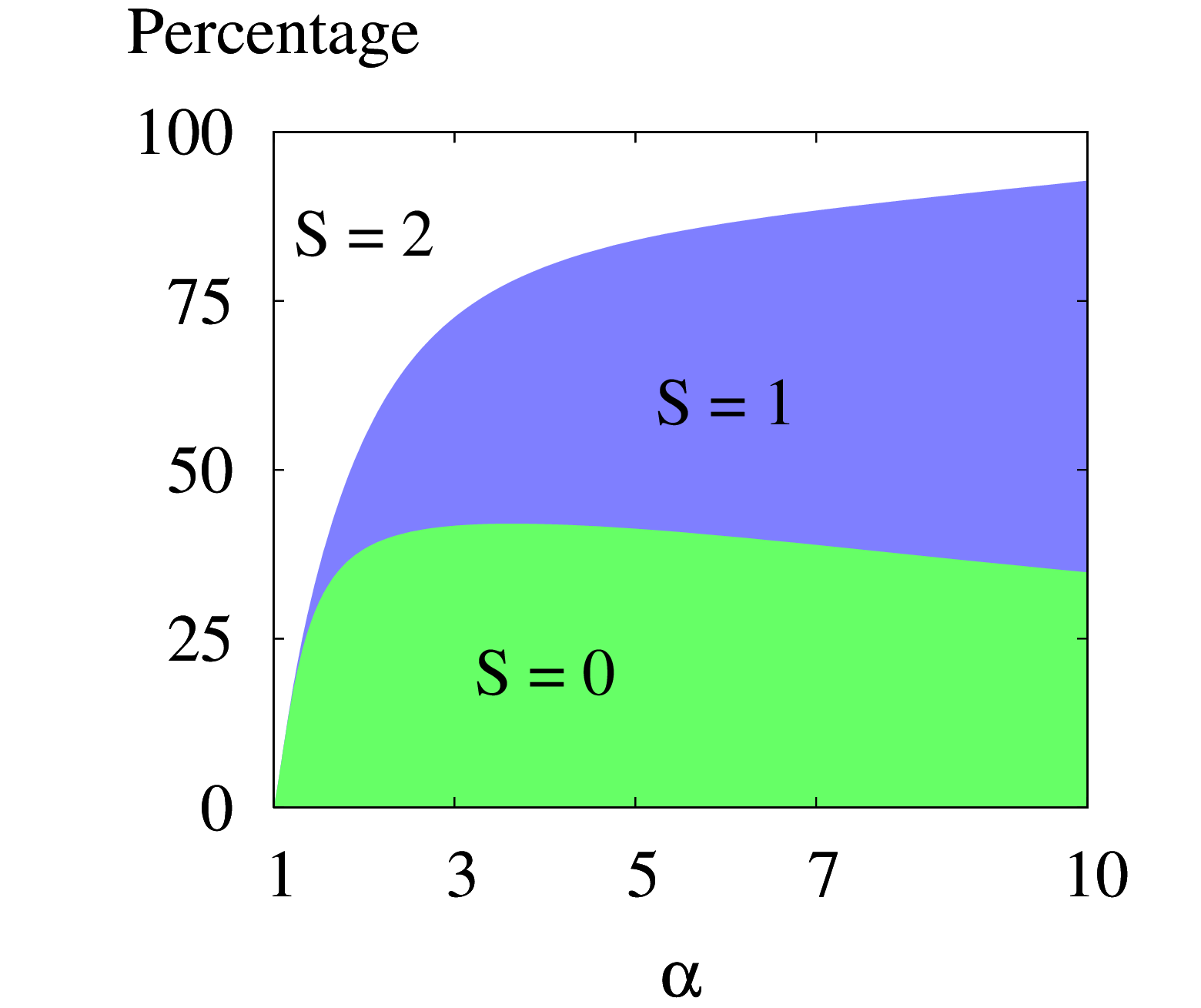}} \\ 
$\tOuter/\tInner=2.5$ & \parbox{2.3in}{\includegraphics[width=1.8in,angle=270]{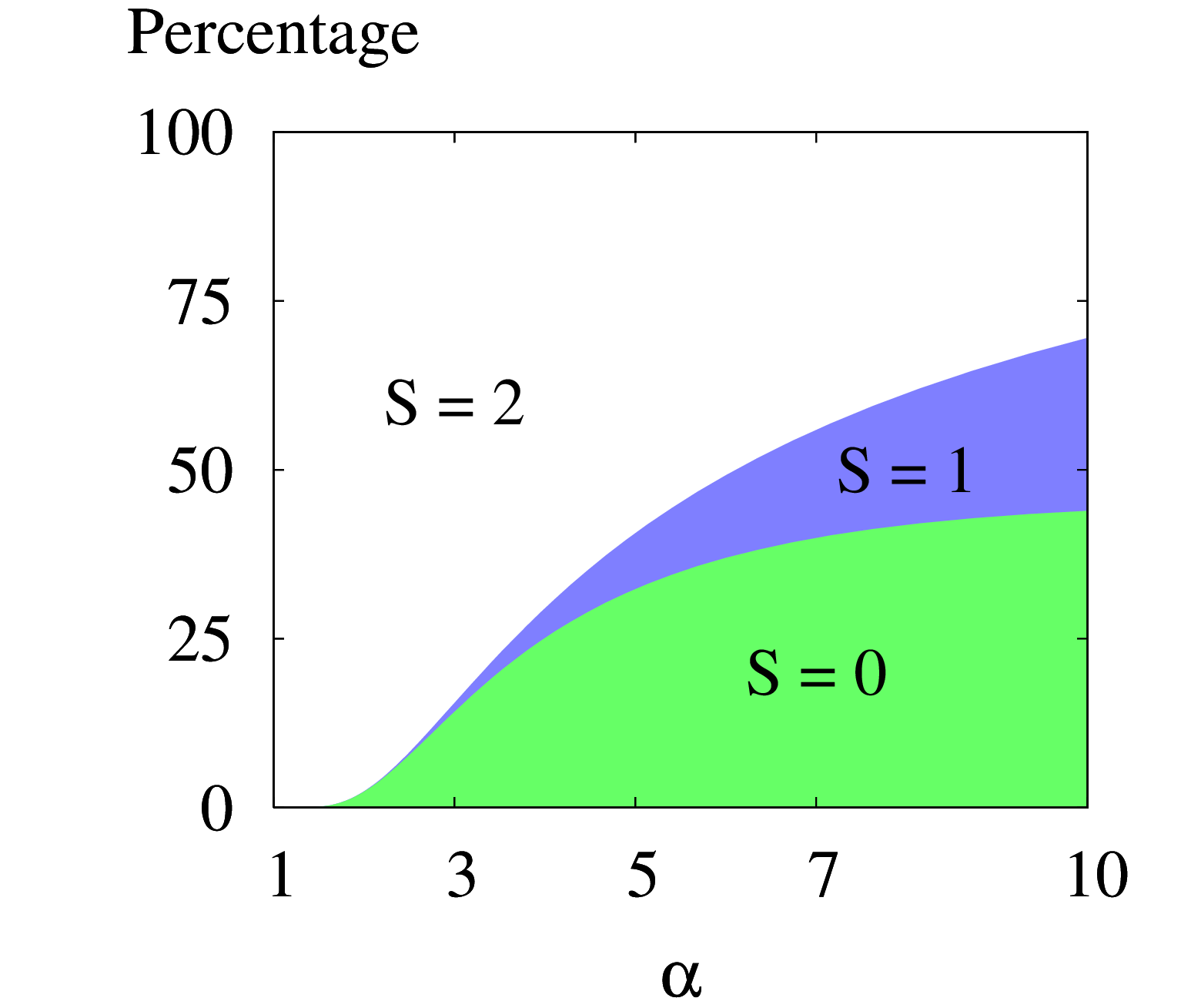}} &
\hspace{-1cm}\parbox{2.3in}{\includegraphics[width=1.8in,angle=270]{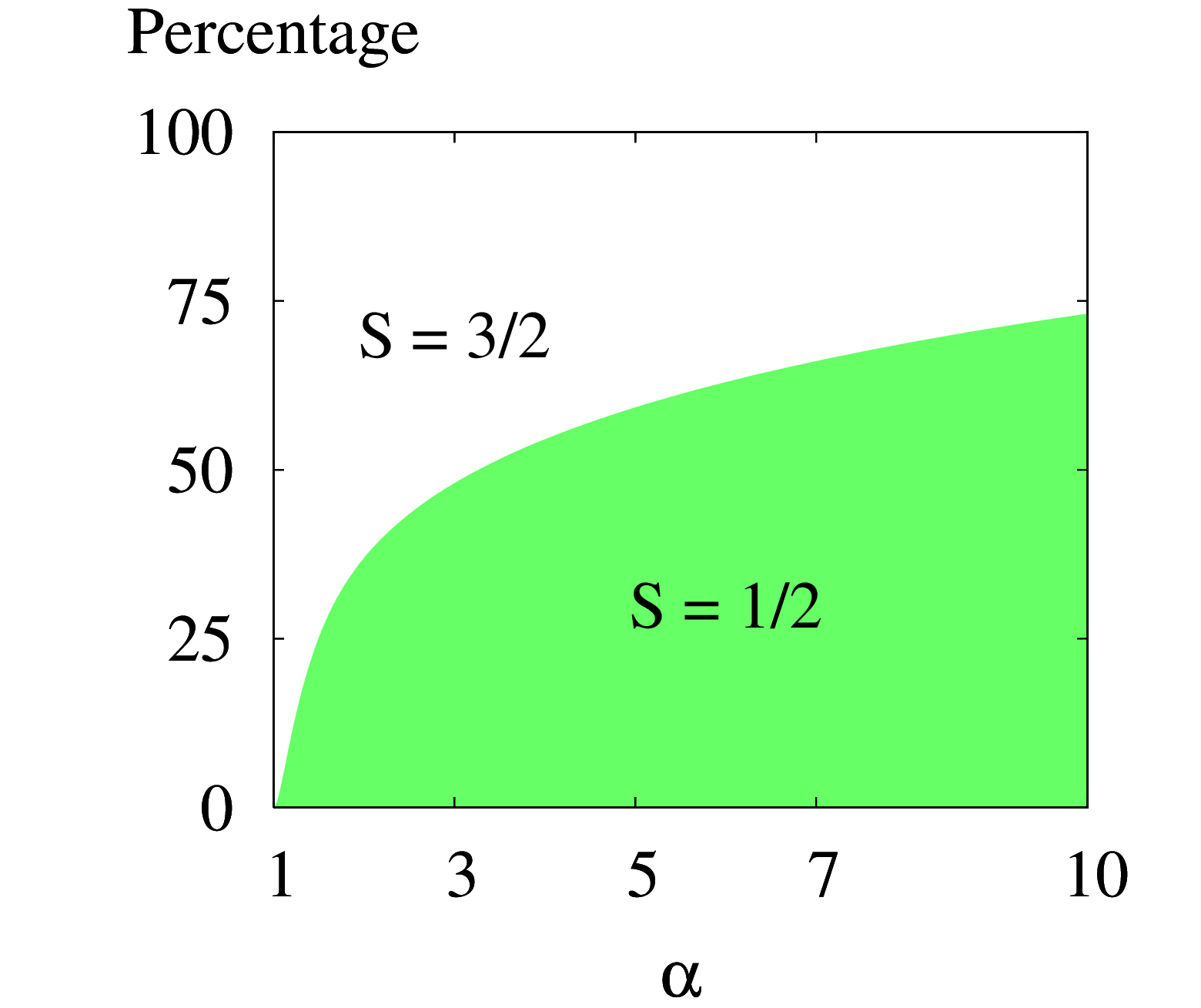}} &
\hspace{-1cm}\parbox{2.3in}{\includegraphics[width=1.8in,angle=270]{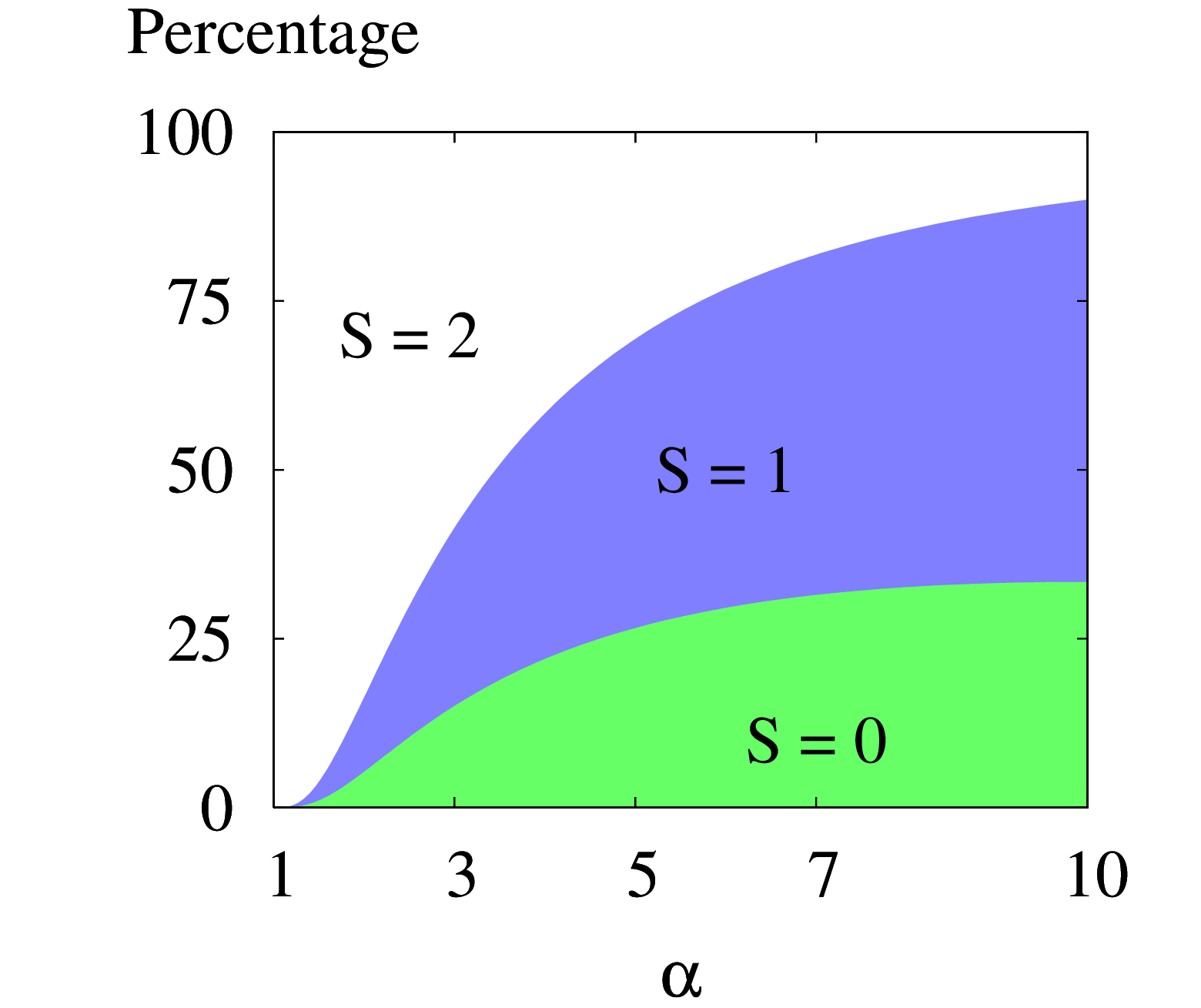}} \\ 
$\tOuter/\tInner=5$ & \parbox{2.3in}{\includegraphics[width=1.8in,angle=270]{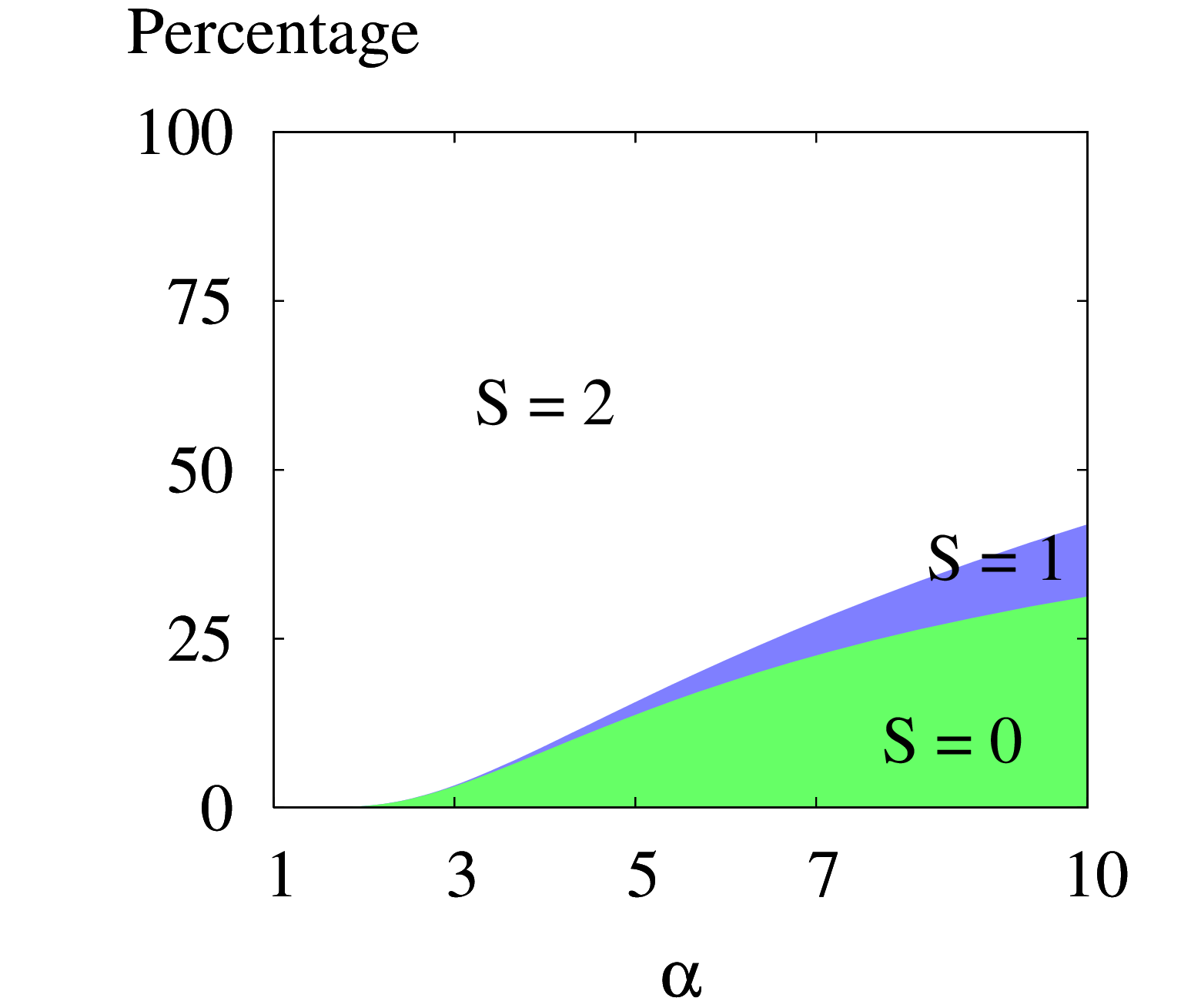}} &
\hspace{-1cm}\parbox{2.3in}{\includegraphics[width=1.8in,angle=270]{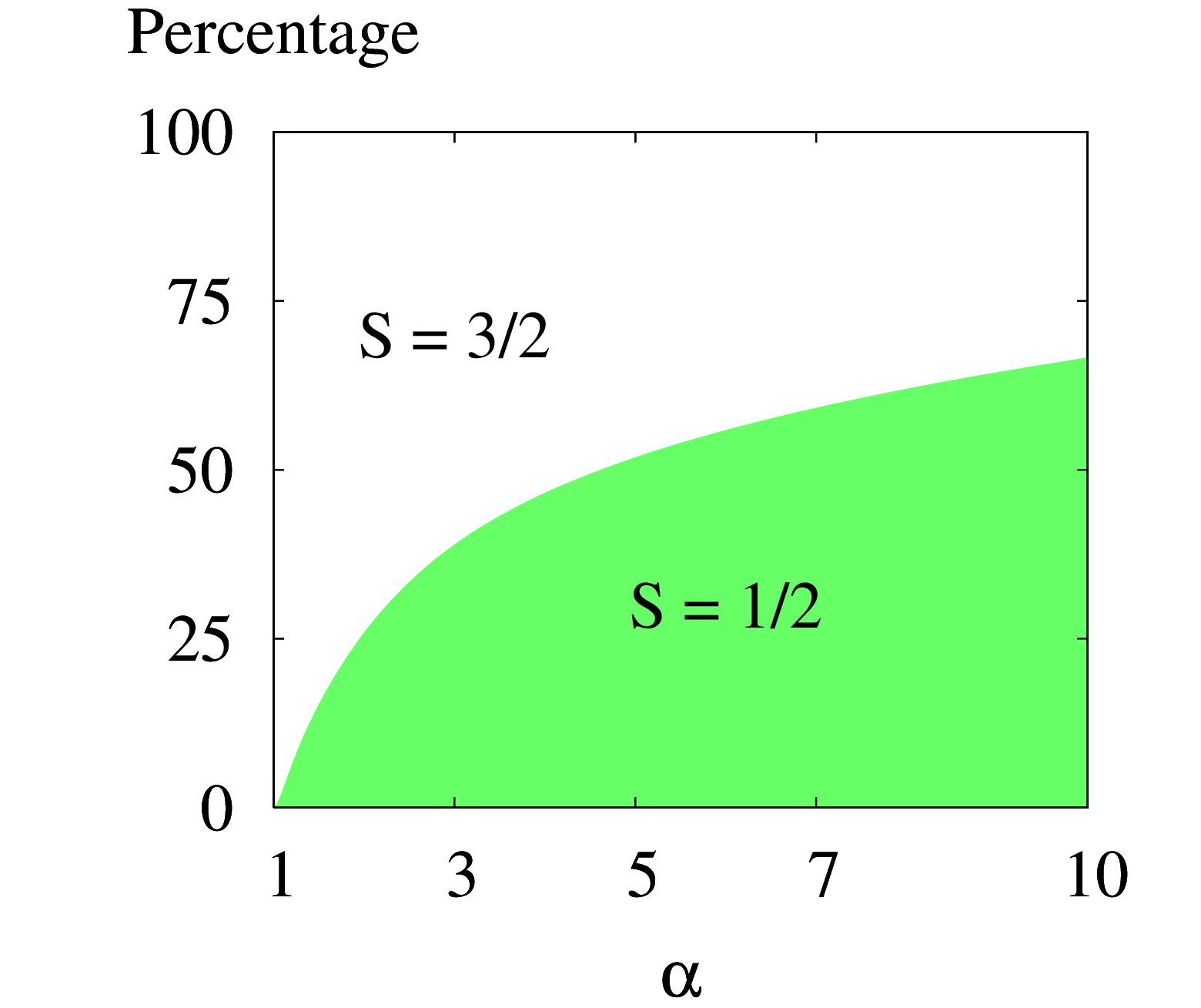}} &
\hspace{-1cm}\parbox{2.3in}{\includegraphics[width=1.8in,angle=270]{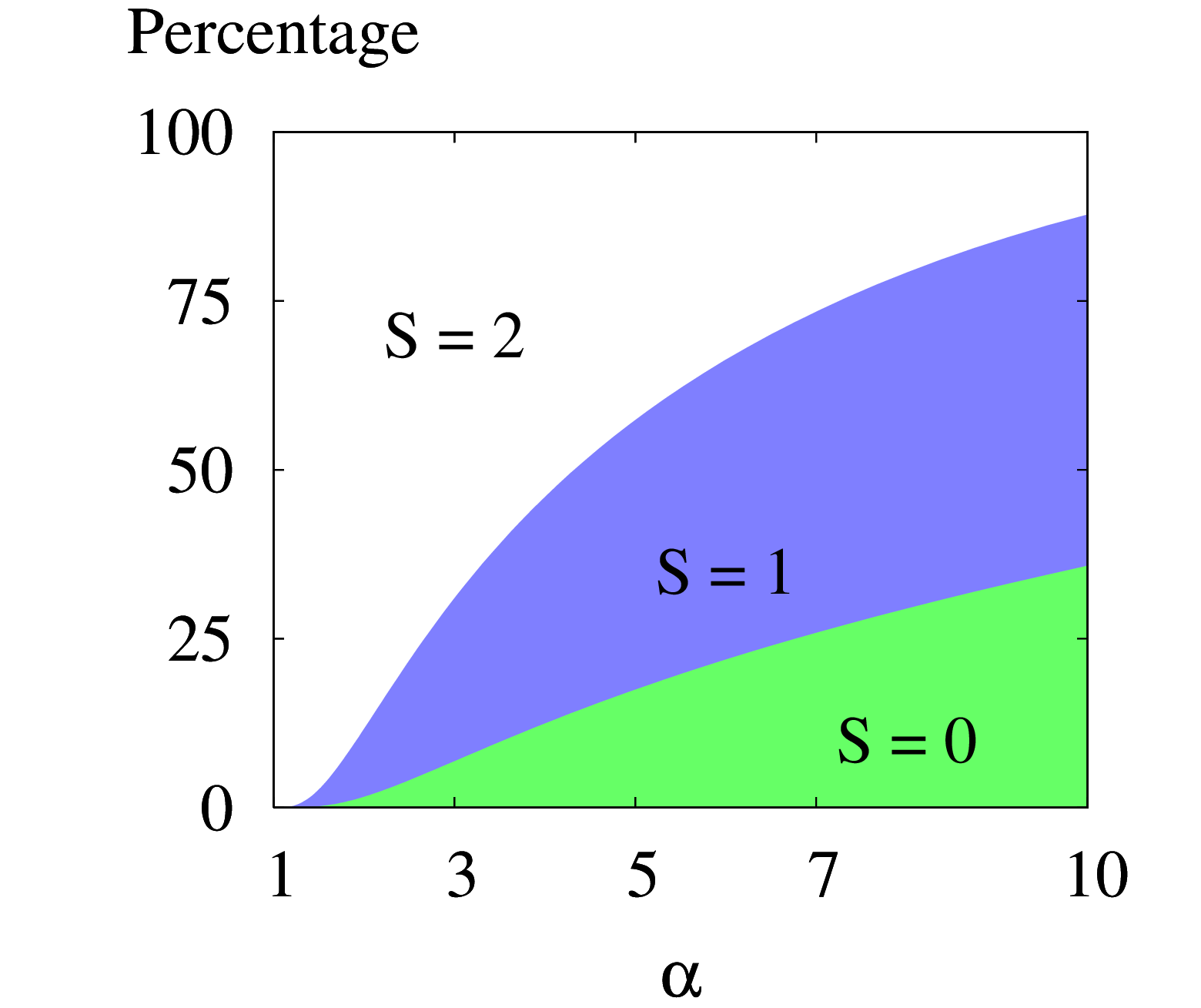}}  
\end{tabular}
\caption{(Color online) Result of randomizing clusters 11, 12, and 20 of Fig.~\ref{figDoubleHopSummary}.  For clusters 11 and 12, $U/t_1=20$ and $t_2/t_1 = 0.1$; for cluster 20, $U/t_1=100$ and $t_2/t_1 = 0.3$.  We set $\tOuter_i/\tInner_i$ to 1, 2.5, and 5 as indicated by the row headers.  
\label{figRandomized}}
\end{figure*}

\end{widetext}

To further probe the robustness of a given cluster's high spin ground state, we consider multiplying each $t_i$ of the cluster by a random factor $\lambda$ whose \emph{logarithm} is chosen from the box distribution $P(\log\lambda)=1/(2\log\alpha)$, $\log\lambda \in [-\log\alpha,+\log\alpha)$ where $\alpha \ge 1$.  Thus, when $\alpha=1$ the system is unperturbed, and for $\alpha \ge 1$ each hopping amplitude is independently multiplied by a different random number between $1/\alpha$ and $\alpha$.  Compared to the specific geometrical distortions analyzed in the preceding paragraph, this method of introducing randomness more accurately characterizes the fluctuations we expect in a real system, since the hopping is \emph{exponentially} dependent on the inter-site distance and we do not expect the fluctuations to preserve any symmetry present in the cluster.  We take as our starting point a cluster known to have a high spin ground state and average over 1-5 thousand of the just described random perturbations.  Then, we tabulate the percentage of the randomly perturbed clusters possessing each possible value of the ground state spin.  The shaded regions in plots of Fig.~\ref{figRandomized} show how these percentages vary as a function of $\alpha$, with the different figures corresponding to initial clusters 11, 12, and 20 of Fig.~\ref{figDoubleHopSummary}.  The boundaries of the regions are spline fits to the data.  We set $U/t=20$, a relatively low value for doped semiconductors, to more clearly see the effect of $\alpha$ (at larger $U/t$ the high-spin ground state becomes increasingly robust).  We see in all cases the general movement, in a probabilistic sense, of the clusters from high to low spin as $\alpha$ is increased, but that this effect is significantly mitigated by raising $\tOuter/\tInner$.  As $\tOuter/\tInner$ becomes larger, the percentage of the clusters that retain the high-spin ground state of the original ($\alpha=1$) cluster grows substantially.  Thus, we again find that increasing $\tOuter/\tInner$ makes high spin ground states significantly more robust to random geometric fluctuations, this time to fluctuations similar to those we expect in an actual doped semiconductor.  This result gives additional hope for the viability of constructing magnetic clusters using an STM tip (described above), where there would inevitably be slight errors in the dopant positions.

\subsection{Randomly distributed finite clusters of fixed density\label{secFixedDensityClusters}}
In sections \ref{secSelectedClusters} and \ref{secGeomDistorted}, we solved generalized Hubbard and $\tOuter-J$ models on a variety of clusters that were constructed to have some spatial symmetries and at most a few pairs of hopping parameters ($\tInner_i$,$\tOuter_i$).  This section and the next give an extensive analysis of clusters with completely random structures and several types of boundary conditions.  Also, instead of considering a range of $\tOuter/\tInner$ values, we use only the parameters given by our realistic band calculation described in section \ref{secCalcDetails}.  In $d$-dimensions, clusters with $\Nsites$ sites and fixed density $\rho$ are generated by randomly placing $\Nsites$ sites within a $d$-dimensional hypercube of side length $L$ such that $\rho (\aB)^{-d}=\Nsites/L^d$.  We fix $U=1\,\Ry$ and determine the hopping parameters $\tInner_{ij}$ by setting $\tInner_{ij}=t(|\vecr_i-\vecr_j|)$, where $t(r)$ is given by the lattice calculation described earlier (see Fig.~\ref{figParamRatiosVsLatSpacing}).  We consider three different models, each corresponding to a different method of setting $\tOuter_{ij}$:
\begin{enumerate}
\item $\tOuter_{ij} = \tInner_{ij}$.
\item Analogous to $\tInner_{ij}$, using $\tOuter(r)$: $\tOuter_{ij}=\tOuter(|\vecr_i-\vecr_j|)$, where $\tOuter(r)$ is obtained from the broadening of the upper impurity band, referred to as the $D^-$ band in semiconductor literature.
\item Set $\tOuter_{ij} \equiv C$, where $C$ is a constant. The value of $C$ is chosen to be $U/2$.
\end{enumerate}

The first case is the regular Hubbard model for randomly distributed sites, and does not take into account the special property of hydrogenic centers.  Model 2 takes into account the larger extent of the $D^-$ state.  Model 3 is to simulate a situation when the radius of the $D^-$ state becomes very large, to see how big an effect that would have on the possibility of Nagaoka ferromagnetism.  We choose $C=U/2$ since this is close $\tOuter(r)$ when $r=\aB$, the smallest separation for which the tight binding model could apply.  Since $\tOuter(r)$ increases with decreasing $r$, $\tOuter(\aB) \approx U/2$ is of order the maximal $\tOuter$ found in the entire system. 

Given a fixed cluster size and density, we exactly solve many (between $10^4$ and $10^6$) clusters and construct a histogram of ground state spin values.  Results obtained using each of the three models are compared to assess the effect of the nature of the doubly-occupied state.  We have calculated the spin histograms for clusters in two and three dimensions with sizes from $\Nsites=4-7$ and for densities $\rho=\frac{1}{1600}$, $\frac{1}{160}$, and $\frac{3}{160}$ in 2D, (corresponding to $\approx 0.005$, $0.05$, and $0.15$ times the Mott metal-insulator transition density) and $\rho=\frac{1}{6400}$, $\frac{1}{640}$, and $\frac{3}{640}$ in 3D (corresponding to 0.01, 0.1 and 0.3 times the Mott density). Further, we have considered open as well as periodic boundary conditions.  In an actual macroscopic sample, clusters will be connected to other clusters of different local densities.  Thus, the physical situation will be intermediate between the cases of open (where each cluster is surrounded by no others) and periodic (where each cluster is effectively surrounded by others of the same density) boundary conditions.  The latter is closer to the actual case at high densities, the former at low density.  

\subsubsection{Hopping set by band calculation: $\tOuter_{ij} = \tOuter(r_{ij})$}
Here we present results for two- and three-dimensional random clusters.  Clusters have all inter-site links present (\emph{i.e.}~hopping is not restricted to be between nearest neighbor sites only).  We find the distribution of ground state spin values for ensembles of clusters with fixed size $\Nsites$, density $\rho$, doping (either one extra electron or one hole), and model for determining $\tOuter_{ij}$.

Raw spin distribution data, shown by tables containing the percentage of clusters with each possible spin, are given for two-dimensional clusters, with open and periodic boundary conditions, in Appendix \ref{appRandomClusterData}.  Corresponding results for three-dimensional clusters could not be included in this paper due to length considerations, and can be found in Ref.~\onlinecite{NielsenThesis}.

Here, we summarize the data by plotting the average spin and the percentage of magnetic clusters (those with greater than minimal spin) as a function of doping (zero doping = half-filled).  We show only the results for 2D clusters with open boundary conditions; similar plots for periodic boundary conditions can be found in Appendix \ref{appRandomClusterData}.  Figure \ref{figAvgSpin2D} shows the average spin of such clusters.  There is some variation in the average spin due to even-odd asymmetry: clusters with an odd number of electrons have minimum spin $S_{min} = \frac{1}{2}$, while those with an even number have $S_{min}=0$.  To remove this effect, Fig.~\ref{figAvgSpin2Dr} shows the average spin \emph{relative to $S_{min}$} (\emph{i.e.}~0.5 is subtracted from cases of odd electron number).  A second measure of a systems magnetic behavior is the percentage of clusters with above minimal spin.  We define any cluster with greater than minimal ground state spin (equivalently, spin $\ge 1$ since the minimal spin is either 0 or 1/2) as a \emph{magnetic cluster}, and Fig.~\ref{figPcMag2D} shows this quantity as a function of doping for the different cluster sizes (2D clusters with open b.c.).  Although both the average spin and percentage of magnetic clusters provide less detailed information than the spin distribution data (Appendix \ref{appRandomClusterData}), they also suffer less from finite size effects and give a more concise picture of the results. 

\begin{figure*}
\begin{center}
\begin{tabular}{|c|c|} \hline
$\rho$ & \textbf{2D \ : \ Average Spin \ : \ open b.c.}\\ \hline
$\frac{1}{1600}$ & \parbox{4in}{
\includegraphics[width=2in, angle=270]{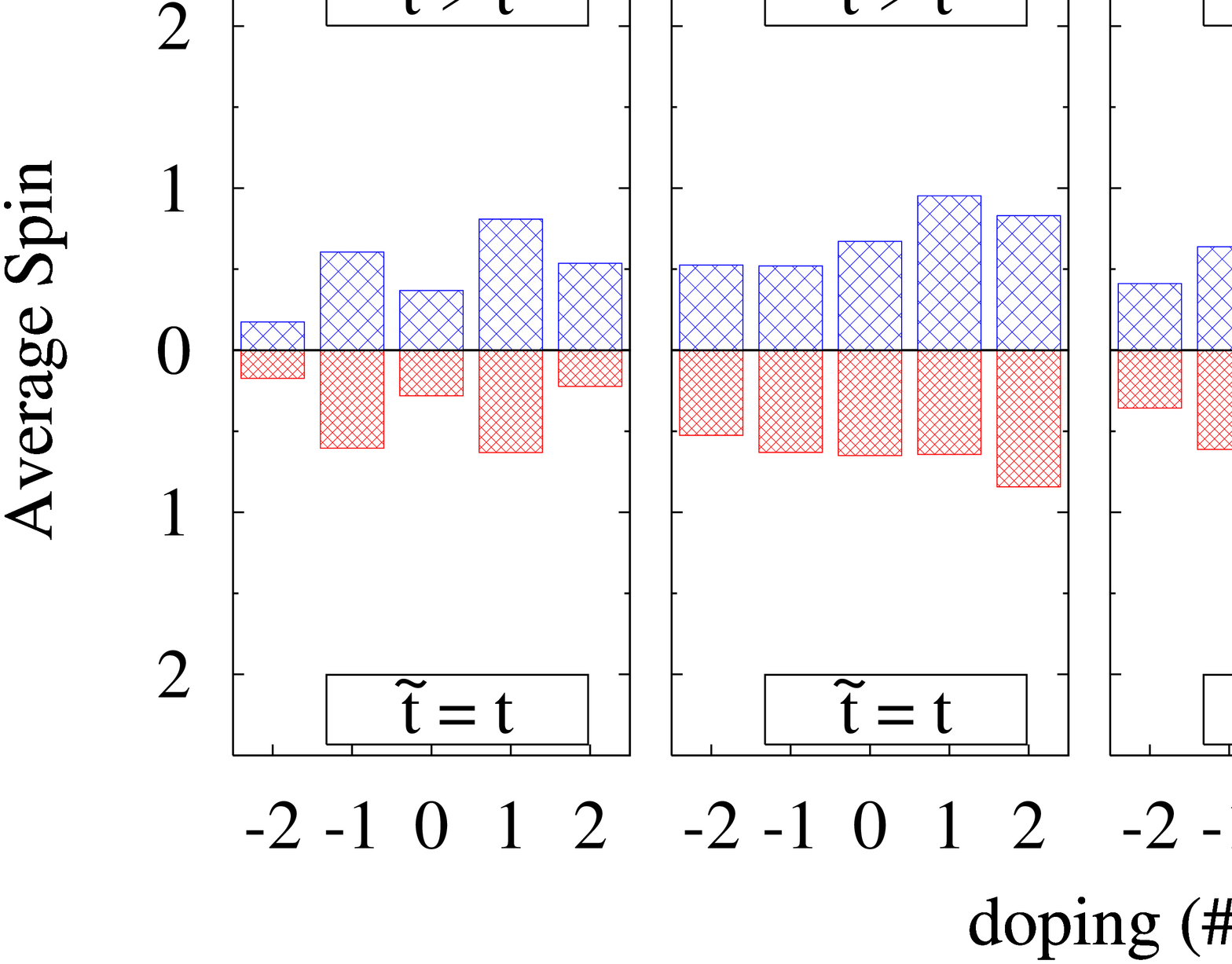}} \\ \hline
$\frac{1}{160}$ & \parbox{4in}{
\includegraphics[width=2in, angle=270]{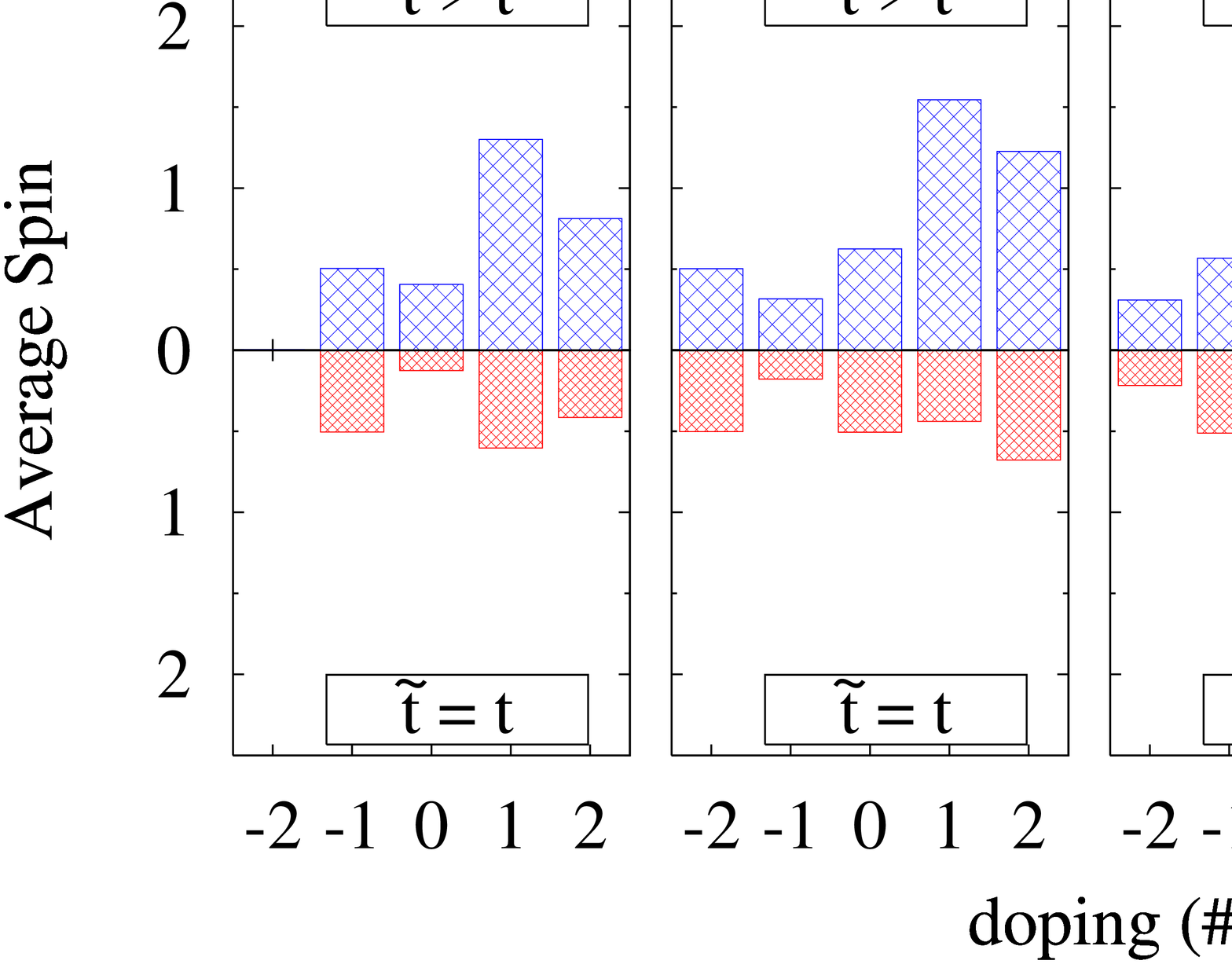}} \\ \hline
$\frac{3}{160}$ & \parbox{4in}{
\includegraphics[width=2in, angle=270]{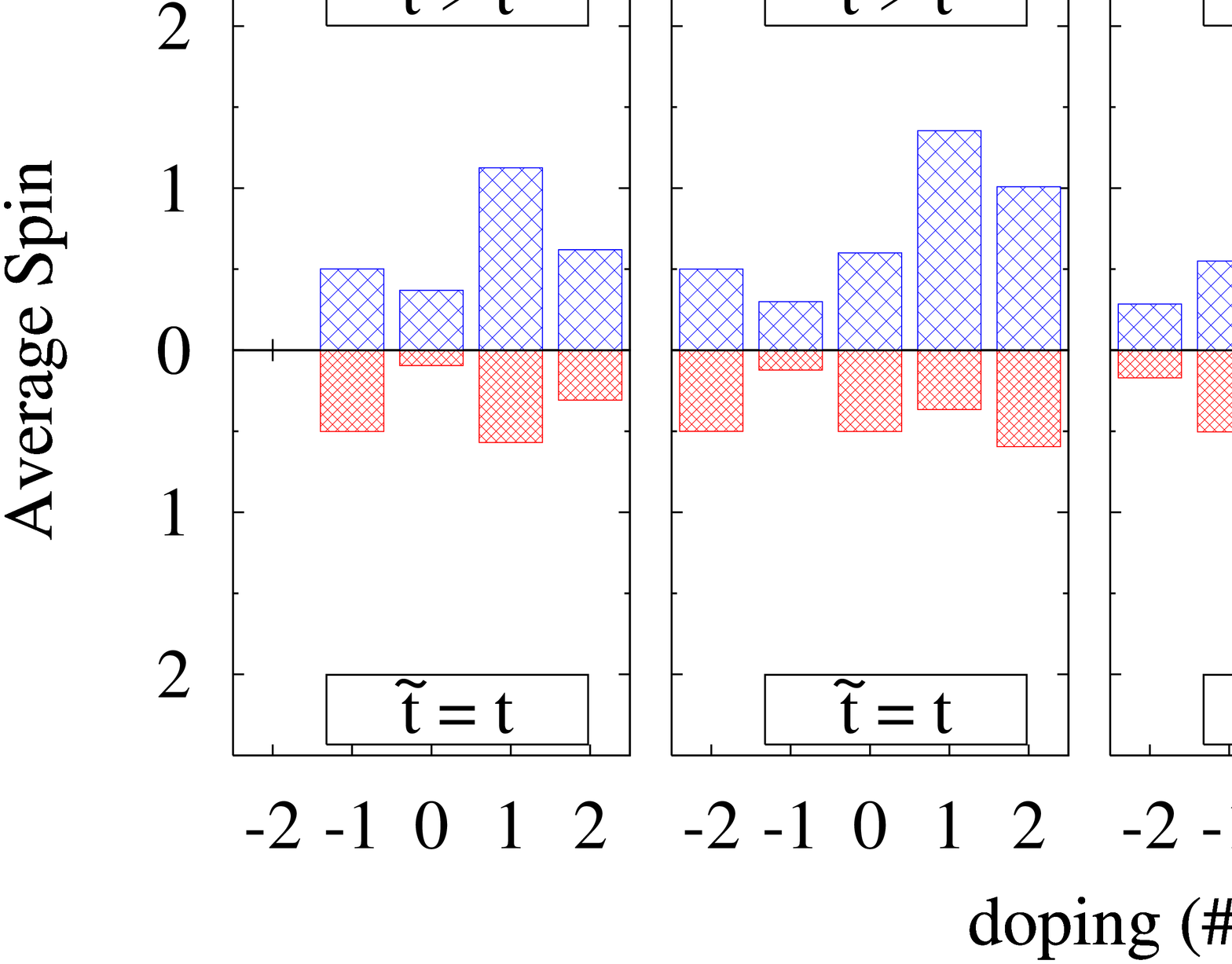}} \\ \hline
\end{tabular}
\caption{(Color online) Ground state average spin of 2D random clusters with fixed size and density, and \emph{open boundary conditions}, as a function of electron-doping (negative = hole-doping). The lower half of plots are the result of setting $\tOuter_{ij}=\tInner_{ij}$, determined by the bandwidth of the lower Hubbard band.  The upper half use $\tOuter_{ij}$ determined by the bandwidth of the upper Hubbard ($D^-$) band. \label{figAvgSpin2D}}
\end{center}
\end{figure*}

\begin{figure*}
\begin{center}
\begin{tabular}{|c|c|} \hline
$\rho$ & \textbf{2D \ : \ Average Spin - $\mathbf{S_{min}}$\ : \ open b.c.}\\ \hline
$\frac{1}{1600}$ & \parbox{4in}{
\includegraphics[width=2in, angle=270]{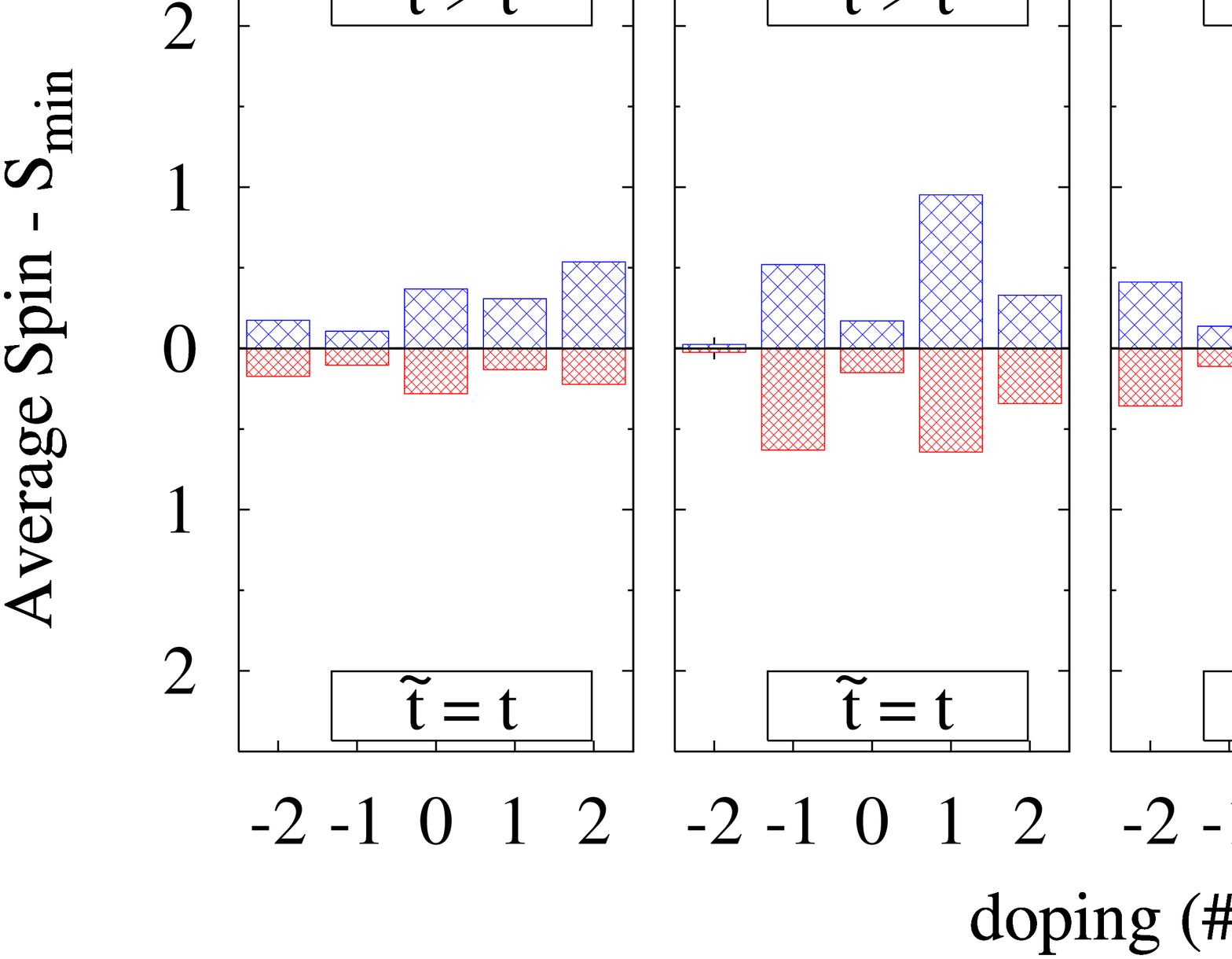}} \\ \hline
$\frac{1}{160}$ & \parbox{4in}{
\includegraphics[width=2in, angle=270]{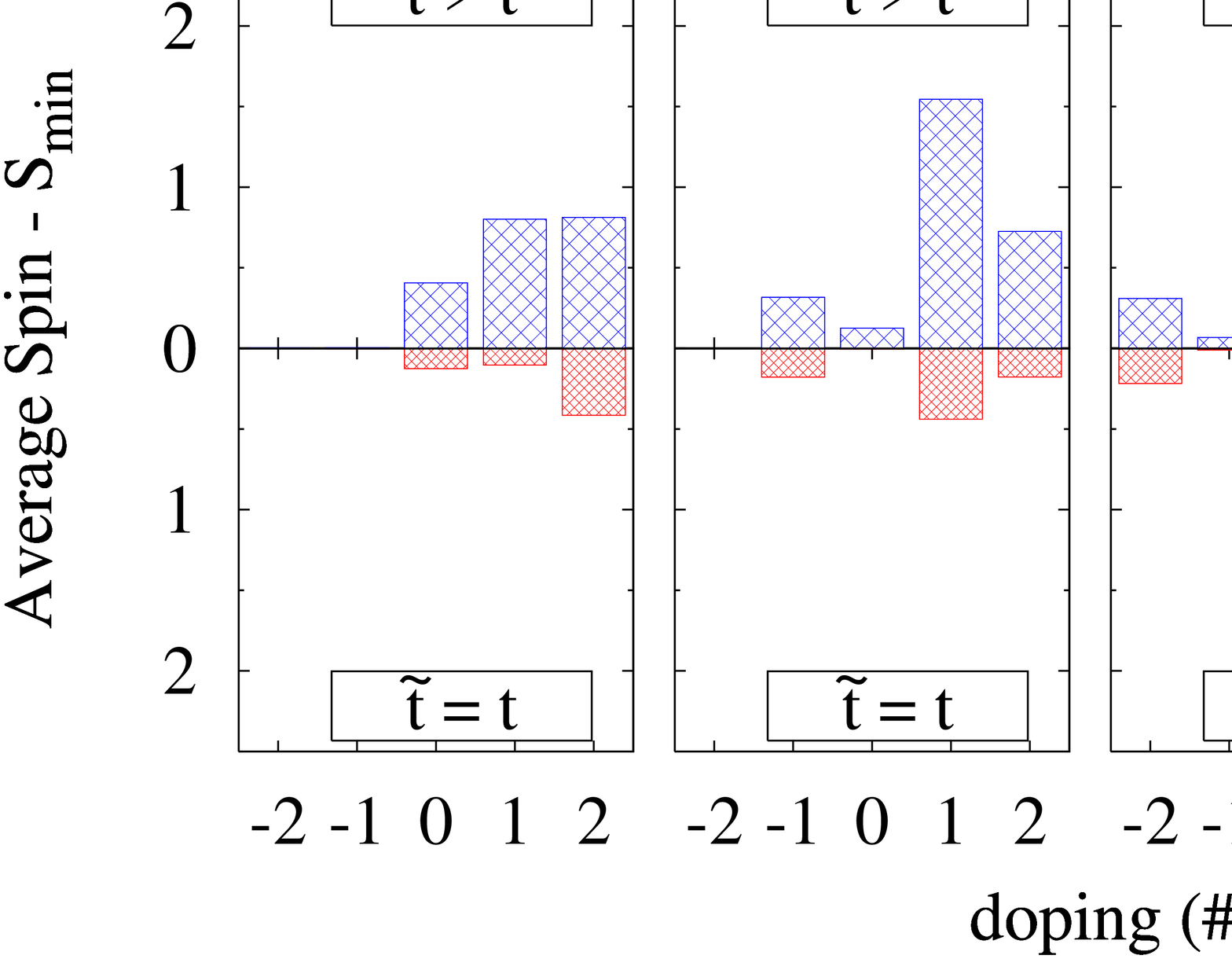}} \\ \hline
$\frac{3}{160}$ & \parbox{4in}{
\includegraphics[width=2in, angle=270]{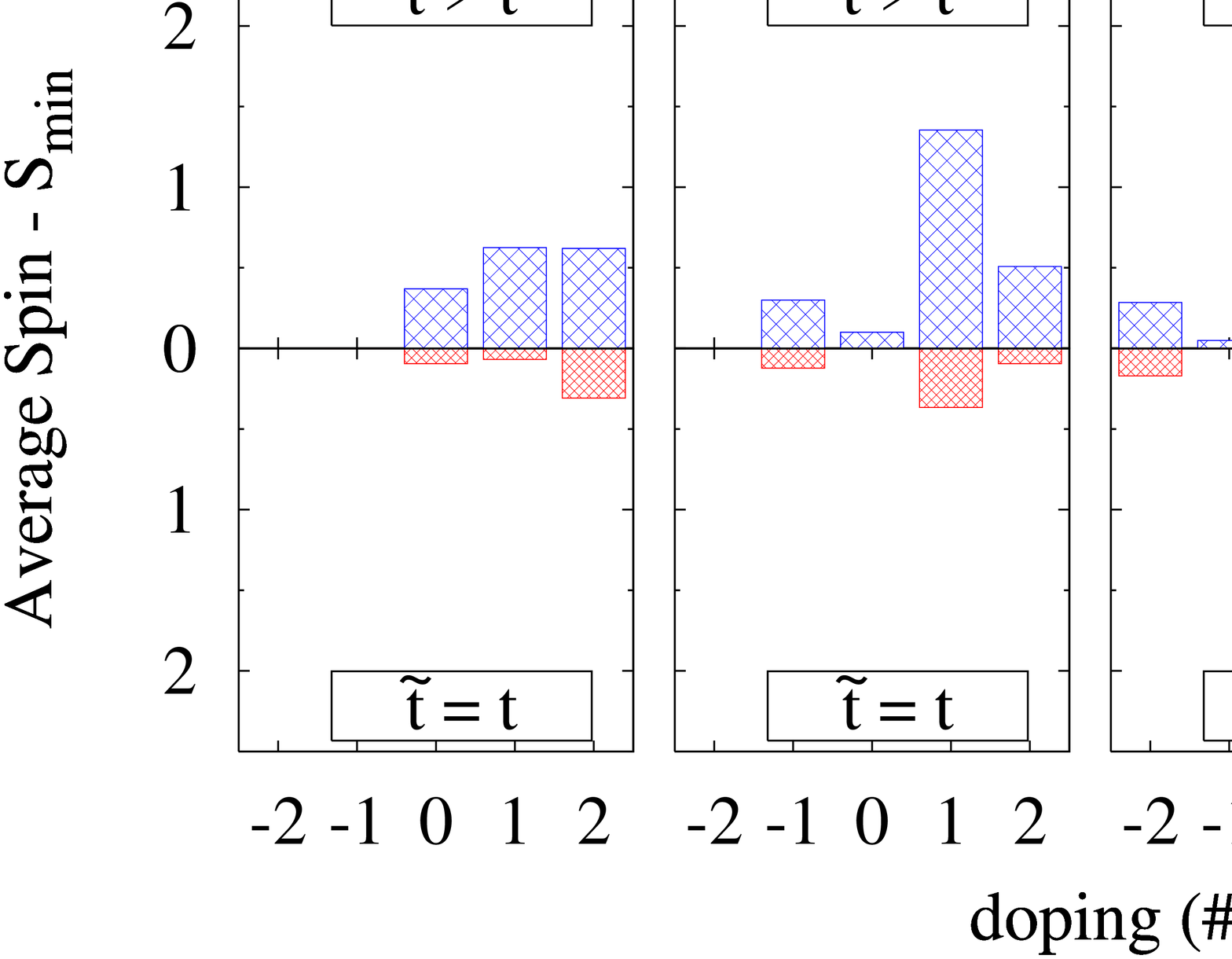}} \\ \hline
\end{tabular}
\caption{(Color online) Ground state average spin \emph{relative to minimum spin} of 2D random clusters with fixed size and density, and \emph{open boundary conditions}, as a function of electron-doping (negative = hole-doping). The lower half of plots are the result of setting $\tOuter_{ij}=\tInner_{ij}$, determined by the bandwidth of the lower Hubbard band.  The upper half use $\tOuter_{ij}$ determined by the bandwidth of the upper Hubbard ($D^-$) band. \label{figAvgSpin2Dr}}
\end{center}
\end{figure*}

\begin{figure*}
\begin{center}
\begin{tabular}{|c|c|} \hline
$\rho$ & \textbf{2D \ : \ \% magnetic clusters \ : \ open b.c.}\\ \hline
$\frac{1}{1600}$ & \parbox{4in}{
\includegraphics[width=2in, angle=270]{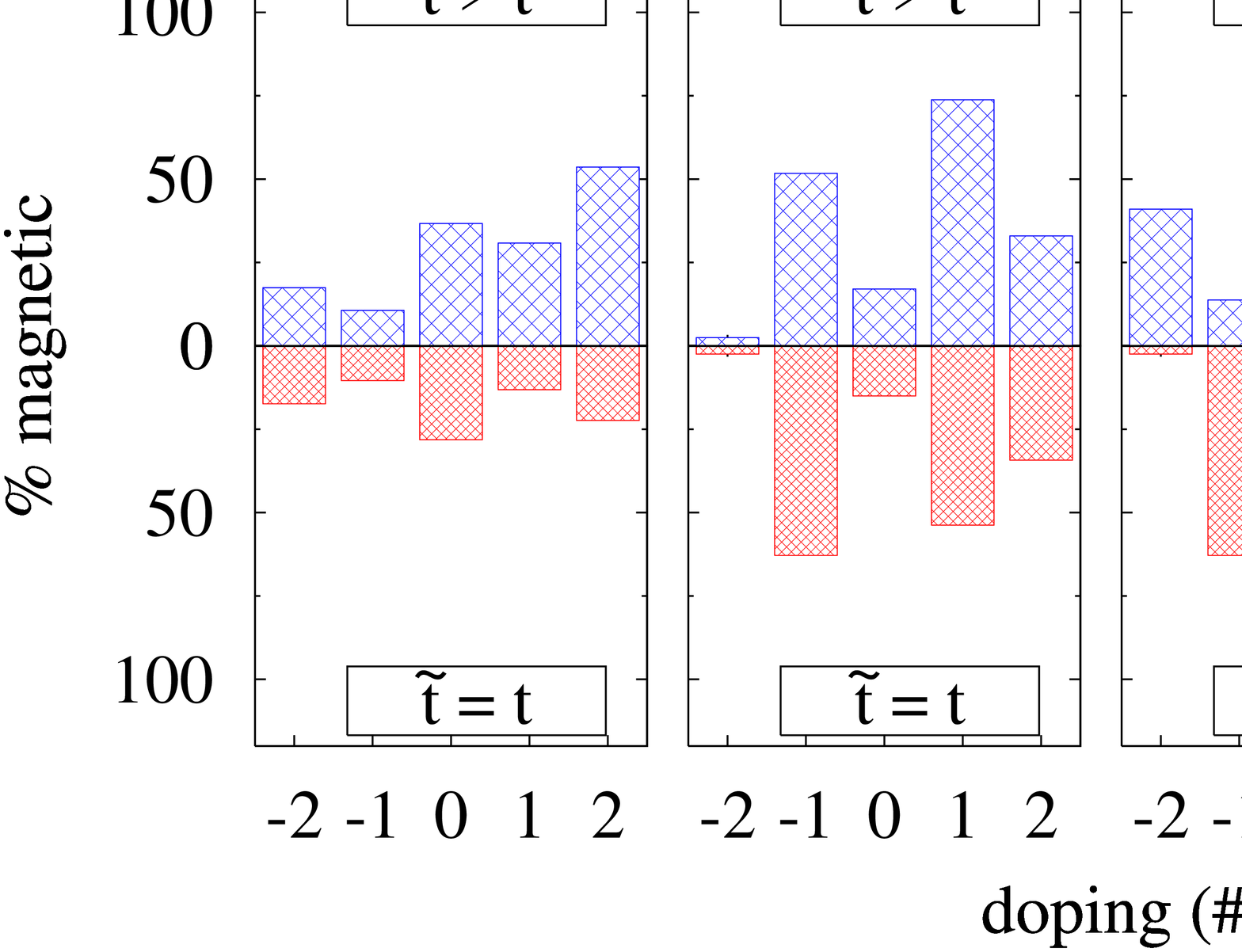}} \\ \hline
$\frac{1}{160}$ & \parbox{4in}{
\includegraphics[width=2in, angle=270]{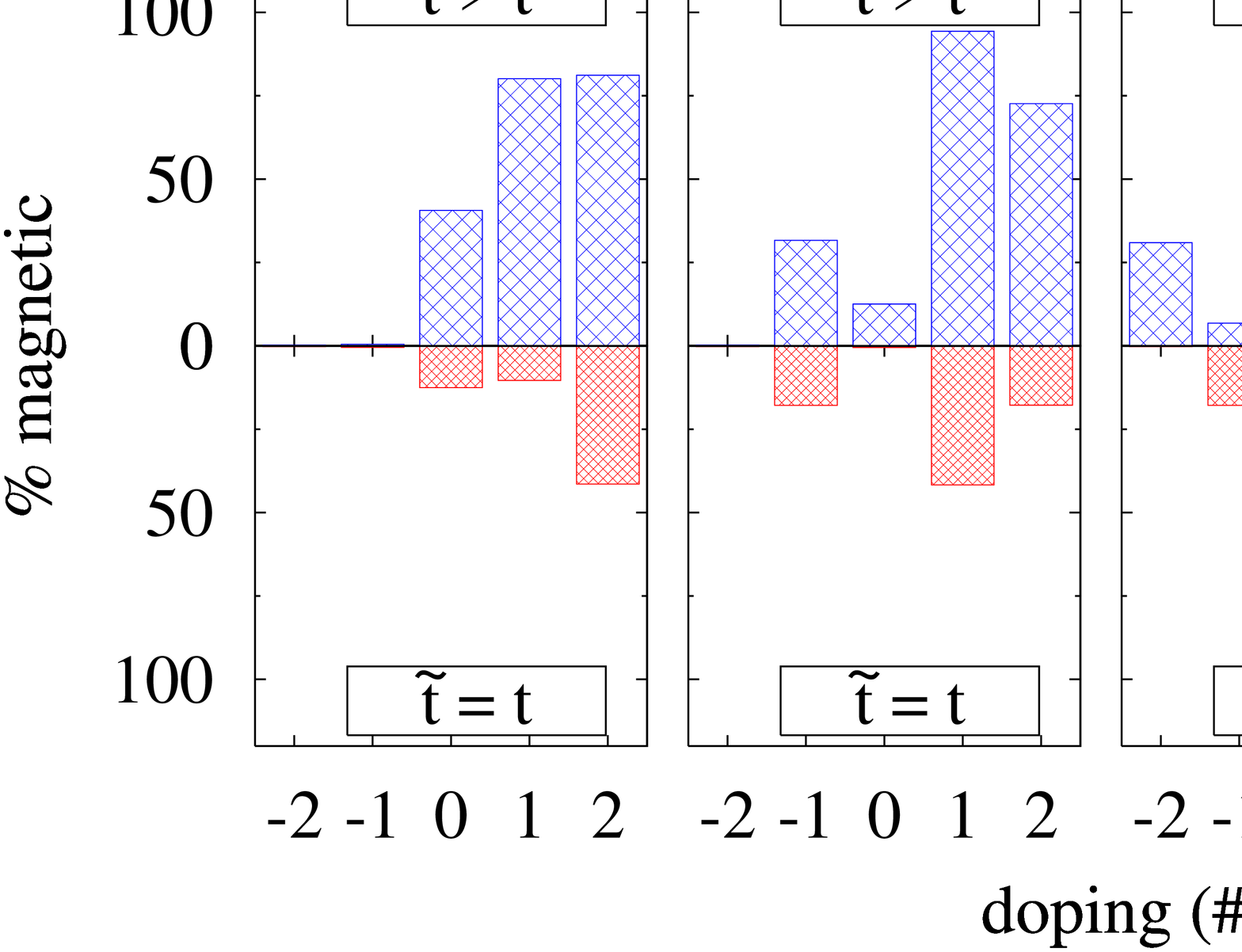}} \\ \hline
$\frac{3}{160}$ & \parbox{4in}{
\includegraphics[width=2in, angle=270]{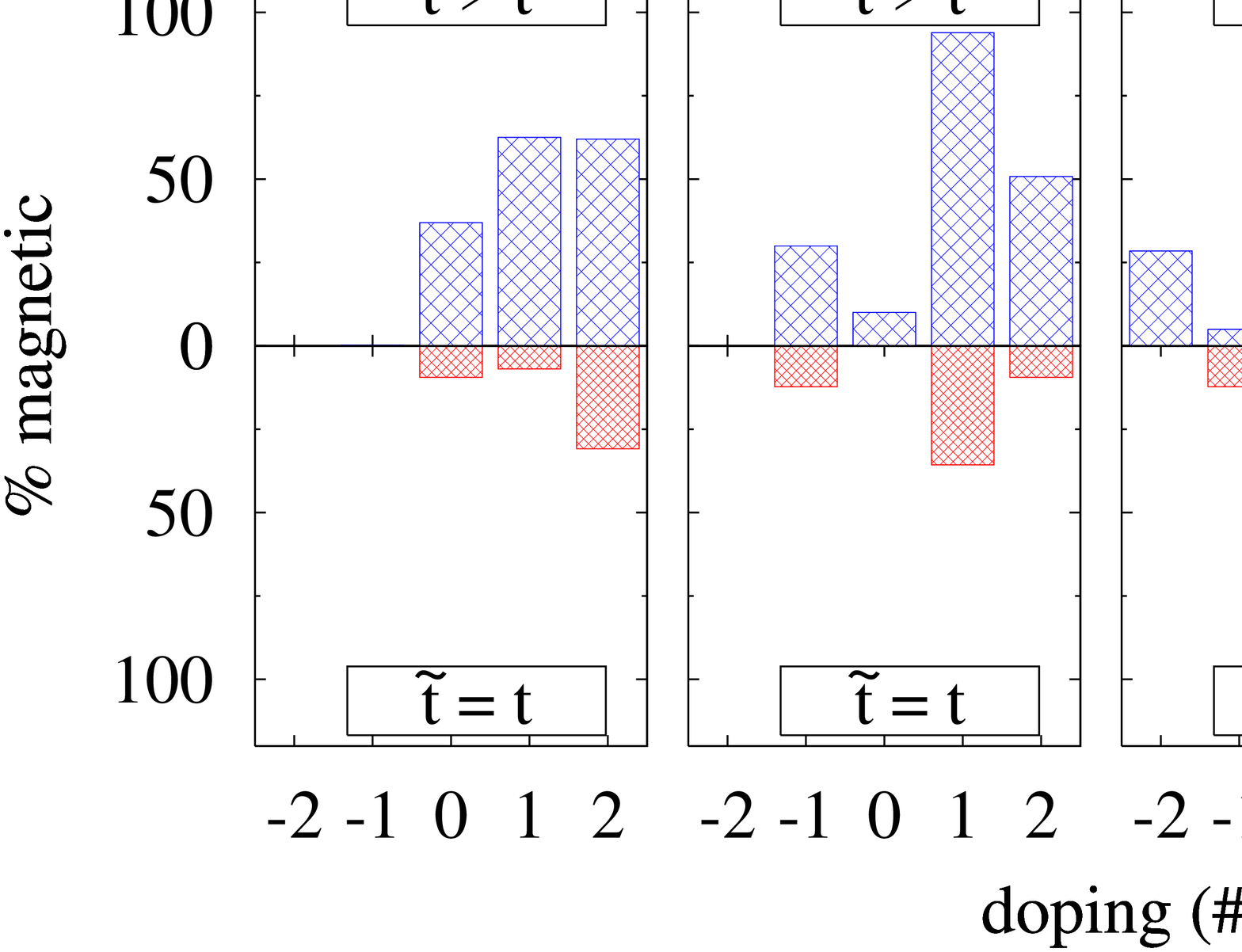}} \\ \hline
\end{tabular}
\caption{(Color online) Percentage of magnetic clusters (spin 1 or greater) in an ensemble of 2D random clusters with fixed size and density, and \emph{open boundary conditions}, as a function of electron-doping (negative = hole-doping). The lower half of plots are the result of setting $\tOuter_{ij}=\tInner_{ij}$, determined by the bandwidth of the lower Hubbard band.  The upper half use $\tOuter_{ij}$ determined by the bandwidth of the upper Hubbard ($D^-$) band.\label{figPcMag2D}}
\end{center}
\end{figure*}

The study of Figs.~\ref{figAvgSpin2D}-\ref{figPcMag2D}, and the more extensive data of Appendix \ref{appRandomClusterData} and Ref.~\onlinecite{NielsenThesis} reveals several trends.  First, clusters with periodic boundary conditions tend to have a larger total spin than those of the same size and density but with open boundary conditions (see Appendix \ref{appRandomClusterData}, Table \ref{tabRandClusters2D_diff}).  This is especially true for large clusters ($\Nsites = 6,7$) and at lower density.  This may be due to the increased connectedness of clusters with periodic boundary conditions compared to those with open boundary conditions.  In a system that is more connected (\emph{i.e.}~where there are more nonzero hopping amplitudes $t_{ij}$), electrons can more easily move among the sites and their kinetic energy (which favors FM) is a stronger contribution to the total energy.  Seen another way, the application of periodic boundary conditions to a cluster with open boundary conditions effectively raises the density of the cluster's environment, since the cluster then appears to be surrounded by other clusters of the same density.  

A comparison between odd-$\Nsites$ and even-$\Nsites$ clusters shows that clusters with an odd number of sites (which have integer spin for $\pm1e^-$ away from half-filling) generally have greater average spin relative to the minimum possible spin (zero for $\pm1e^-$).  This difference is not great, however, and their absolute average spin (\emph{e.g.}~in Fig.~\ref{figAvgSpin2D}) is comparable to that of the even-$\Nsites$ clusters, which have a minimum spin of 1/2 as opposed to 0.

Cluster size is a third point of comparison, where we find that larger clusters usually have ground states with higher spin, and higher average spin overall.  One should keep in mind, however, that larger clusters are able to have higher spin values just by virtue of having more sites (and \emph{total} electrons).  (Indeed, we find that smaller clusters have larger average spin \emph{relative to their maximal allowed spin}.)  The rise in average spin and the existence of higher spin ground states as cluster size increases is greater and more consistently true of electron-doped clusters.  In this case the dependence of average spin on cluster size is particularly significant:  we find a substantial percentage of maximally polarized clusters for all sizes (4-7) investigated, showing that the spin polarization induced by extra electrons persists to larger random systems yielding large spins (up to $S=3$).  We also see that the polarization of larger clusters (6-7 sites) remains (and sometimes increases) when there are two electrons above half-filling.  The average spin of hole-doped clusters shows a much weaker shift toward larger spin values with cluster size than the electron-doped case, which again highlights our central argument that electron-doping is very different from hole-doping.

Fourth, we see that with increasing density there are usually fewer high-spin clusters in all categories except for clusters with one extra electron that have $\tOuter_{ij}$ set by method 2 above ($\tOuter > \tInner$).  In this case the distribution with highest weight on large spins occurs at \emph{intermediate} density ($\rho = \frac{1}{160}$ in 2D, $\frac{1}{640}$ in 3D), a result also seen in the ensembles of section \ref{secVaryDensityClusters} below.  This suggests that there exists an optimal density for finding high-spin states in doped semiconductors above half-filling.  We generally expect low density to be most favorable for FM, since this corresponds to large $U/t$, and believe this is the reason why all but the aforementioned case show this behavior.  In the exceptional case, when there is one extra electron and $\tOuter_{ij}$ is set by model 2, the additional parameter $\tOuter/\tInner$, will play a significant role, and the dependence of the pair ($U/\tInner$,\,$\tOuter/\tInner$) on the density could result in an optimal density for FM that is greater than zero.

Lastly, the most striking trend we find is by comparing electron-doped and hole-doped clusters.  When $\tOuter = \tInner$ the clusters with one extra electron have a spin distribution shifted to substantially higher spin values than those with one less electron (\emph{i.e.}~one hole).  When $\tOuter_{ij}$ is determined by our band calculation (\emph{i.e.}~$ > \tInner_{ij}$), this effect increases dramatically (particularly at intermediate density, as mentioned earlier).  This effect is expected, since in our model an extra electron hops with amplitudes $\tOuter_{ij}$ while an extra hole hops with amplitudes $\tInner_{ij}$.  Recall that the motivation for the model comes from the special properties of the hydrogen atom which result in mobile electrons having spatially larger wavefunctions than mobile holes.  These cluster results show that even in strongly disordered systems a Nagaoka-like ferromagnetism can emerge at least on the nanoscale, and one of the ideal conditions for this FM is an electron-doped system.  Compared to those below half-filling, systems above half-filling also hold greater promise for spin polarization on longer length scales, since this would most likely arise from many aligned high-spin clusters.

\begin{table*}
\begin{center}
\begin{tabular}{|c|c|c||c|c|c|c||c|c|c|c||c|c|c|c||} \hline 
 & \multicolumn{2}{c||}{Dim \& b.c.} & \multicolumn{12}{c||}{2D, open b.c}\\ \cline{2-15}
$N_s$ & \multicolumn{2}{c||}{\rule[-3mm]{0mm}{8mm}$\rho$} & \multicolumn{4}{c||}{$\frac{1}{1600} \approx 0.005 \rho_c^{2D}$} & \multicolumn{4}{c||}{$\frac{1}{160} \approx 0.05 \rho_c^{2D}$} & \multicolumn{4}{c||}{$\frac{3}{160} \approx 0.15 \rho_c^{2D}$}\\ \cline{2-15}
 & \multicolumn{2}{c||}{spin} & \textbf{0.5} & \textbf{1.5} & \textbf{2.5} & \textbf{3.5} & \textbf{0.5} & \textbf{1.5} & \textbf{2.5} & \textbf{3.5} & \textbf{0.5} & \textbf{1.5} & \textbf{2.5} & \textbf{3.5}\\ \hline\hline 
\raisebox{-0.75cm}[0pt][0pt]{4} & \raisebox{-0.3cm}[0pt][0pt]{$ \tilde{t} > t$} & 1h & 89 & 11 & 0 & 0 & 100 & 0 & 0 & 0 & 100 & 0 & 0 & 0\\ \cline{3-15} 
  &  & 1e & 69 & 31 & 0 & 0 & 20 & 80 & 0 & 0 & 37 & 63 & 0 & 0\\ \cline{2-15} 
  & \raisebox{-0.3cm}[0pt][0pt]{$ \tilde{t} = U/2$} & 1h & 90 & 10 & 0 & 0 & 100 & 0 & 0 & 0 & 100 & 0 & 0 & 0\\ \cline{3-15} 
  &  & 1e & 0 & 100 & 0 & 0 & 3 & 97 & 0 & 0 & 10 & 90 & 0 & 0\\ \hline 
\raisebox{-0.75cm}[0pt][0pt]{6} & \raisebox{-0.3cm}[0pt][0pt]{$ \tilde{t} > t$} & 1h & 86 & 14 & 0 & 0 & 93 & 7 & 0 & 0 & 95 & 5 & 0 & 0\\ \cline{3-15} 
  &  & 1e & 60 & 32 & 8 & 0 & 16 & 43 & 41 & 0 & 17 & 59 & 24 & 0\\ \cline{2-15} 
  & \raisebox{-0.3cm}[0pt][0pt]{$ \tilde{t} = U/2$} & 1h & 0 & 100 & 0 & 0 & 0 & 100 & 0 & 0 & 1 & 99 & 0 & 0\\ \cline{3-15} 
  &  & 1e & 0 & 1 & 99 & 0 & 0 & 6 & 94 & 0 & 1 & 16 & 83 & 0\\ \hline 
 & \multicolumn{2}{c||}{spin} & \textbf{0} & \textbf{1} & \textbf{2} & \textbf{3} & \textbf{0} & \textbf{1} & \textbf{2} & \textbf{3} & \textbf{0} & \textbf{1} & \textbf{2} & \textbf{3}\\ \hline\hline 
\raisebox{-0.75cm}[0pt][0pt]{5} & \raisebox{-0.3cm}[0pt][0pt]{$ \tilde{t} > t$} & 1h & 48 & 51 & 0 & 0 & 68 & 32 & 0 & 0 & 70 & 30 & 0 & 0\\ \cline{3-15} 
  &  & 1e & 26 & 52 & 21 & 0 & 6 & 34 & 60 & 0 & 6 & 52 & 41 & 0\\ \cline{2-15} 
  & \raisebox{-0.3cm}[0pt][0pt]{$ \tilde{t} = U/2$} & 1h & 20 & 80 & 0 & 0 & 0 & 100 & 0 & 0 & 1 & 99 & 0 & 0\\ \cline{3-15} 
  &  & 1e & 0 & 1 & 99 & 0 & 0 & 5 & 95 & 0 & 0 & 13 & 86 & 0\\ \hline 
\raisebox{-0.75cm}[0pt][0pt]{7} & \raisebox{-0.3cm}[0pt][0pt]{$ \tilde{t} > t$} & 1h & 38 & 58 & 4 & 0 & 66 & 33 & 1 & 0 & 54 & 45 & 1 & 0\\ \cline{3-15} 
  &  & 1e & 29 & 54 & 15 & 2 & 7 & 26 & 41 & 26 & 6 & 30 & 51 & 13\\ \cline{2-15} 
  & \raisebox{-0.3cm}[0pt][0pt]{$ \tilde{t} = U/2$} & 1h & 0 & 0 & 100 & 0 & 0 & 0 & 100 & 0 & 0 & 2 & 98 & 0\\ \cline{3-15} 
  &  & 1e & 0 & 0 & 0 & 100 & 0 & 0 & 6 & 94 & 0 & 3 & 18 & 79\\ \hline 
\end{tabular}

\caption{Comparison of large $\tOuter=U/2$ and band calculation $\tOuter$ distributions of ground state spin values for 2D random clusters with \emph{open boundary conditions}.  Table entries give the percentage of clusters with the ground state spin specified in the column header.  Results are the ensemble average of many clusters with fixed size $\Nsites$, density $\rho$, and doping = one electron (1e) or hole (1h). Estimated error $\pm0.5\%$.\label{tabRandClusters2D_cband}}
\end{center}
\end{table*}

\subsubsection{Large $\tOuter$ case: $\tOuter = U/2$\label{secCBand}}
In model 3, the hopping $\tOuter_{ij}$ is set to a constant $C=U/2$, a value near the maximum of $\tOuter(r)$ (used in model 2).  This corresponds qualitatively to the case when the wavefunction on doubly-occupied sites is extended across the system (as if, for instance, the state had merged with conduction band states).  One may access this regime experimentally if the binding energy of the $D^-$ state can be tuned (\emph{e.g.}~in many-valley semiconductors or by an applied field).  Here we focus on the case of 2D clusters with open boundary conditions, for which the raw spin distribution data comparing models 2 and 3 is shown in Fig.~\ref{tabRandClusters2D_cband}.  Data  for 3D clusters can be found in Appendix \ref{appRandomClusterData}.  Two trends found in our discussion of models 1 and 2 above also appear in the $\tOuter = U/2$ results:  odd-$\Nsites$ clusters have greater spin polarization relative to their minimum spin, and spin polarization increases with cluster size.  Unlike the results of method 2 with one electron (1e), where the intermediate density was optimal for FM, the results of method 3 show spin polarization increasing with decreasing density (as in method 1).  This fits with our belief that the optimal density found when method 2 was used is due to the interplay of \emph{two} density-dependent Hubbard model parameters (in method 3 there is only one, $U/t$, as in method 1).  The electron-hole asymmetry found when $\tOuter = U/2$  is qualitatively similar to when $\tOuter_{ij} = \tOuter(r_{ij})$ (method 2), but with higher spin values (for both electron- and hole-doped systems).  This is expected in the single electron-doped case (1e), since the second electrons are even more weakly bound, causing correspondingly stronger spin polarization.  In the hole-doped case two aspects are particularly noteworthy.  First, we find that the large $\tOuter = U/2$ results in clusters with higher spin than those of method 2, opposite to the trend seen in hole-doped bipartite lattices (cf.~Figs.~\ref{figSingleHoleVsElectron} and \ref{figHoney1ExtraHole}).  Second, the largest spin in the distribution saturates at a value of one below the maximal allowed spin, denoted $S_{max}$ (for instance, in 2D clusters with $\Nsites=6$, the spin distribution in nearly 100\% $S=1.5$).  This behavior is somewhat similar to the hole-doped triangular lattice (Fig.~\ref{figTri1Hole}), which has a partially-polarized ground state (with spin = $S_{max}-1$) which covers larger intervals of $U/t$ as $\tOuter/\tInner$ is increased.  In summary, large $\tOuter$ results in almost 100\% of (single) electron-doped clusters having ground state spin $S_{max}$, and almost 100\% of (single) hole-doped clusters having ground state spin $S_{max}-1$.

\section{Cluster analysis of large samples\label{secVaryDensityClusters}}
We now study the viability of ferromagnetism in a macroscopic sample.  For this section, because of its relevance to hydrogenic n-doped semiconductors, we present results for $\tOuter_{ij}$ and $\tInner_{ij}$ from the band calculation only.  Our strategy will be to consider a large two- or three-dimensional system of random sites and divide it into clusters that can be approximately treated as independent as far as the Hubbard part of the Hamiltonian is concerned.  Choice of the number of carriers in each cluster involves long-range Coulomb forces and is treated in a classical approximation described later.  We solve the clusters individually and then analyze the resulting distribution of their ground state spins.  The analysis of section \ref{secFixedDensityClusters} characterized random clusters with a fixed density; here the average density of a large system is fixed while the local density of individual clusters is free to vary.  

\subsection{Decomposition into clusters}
\begin{figure} [h] 
\begin{center} 
\begin{tabular}{c|c|} \cline{2-2}
a) & \parbox{2.5in}{ \includegraphics[width=2.4in]{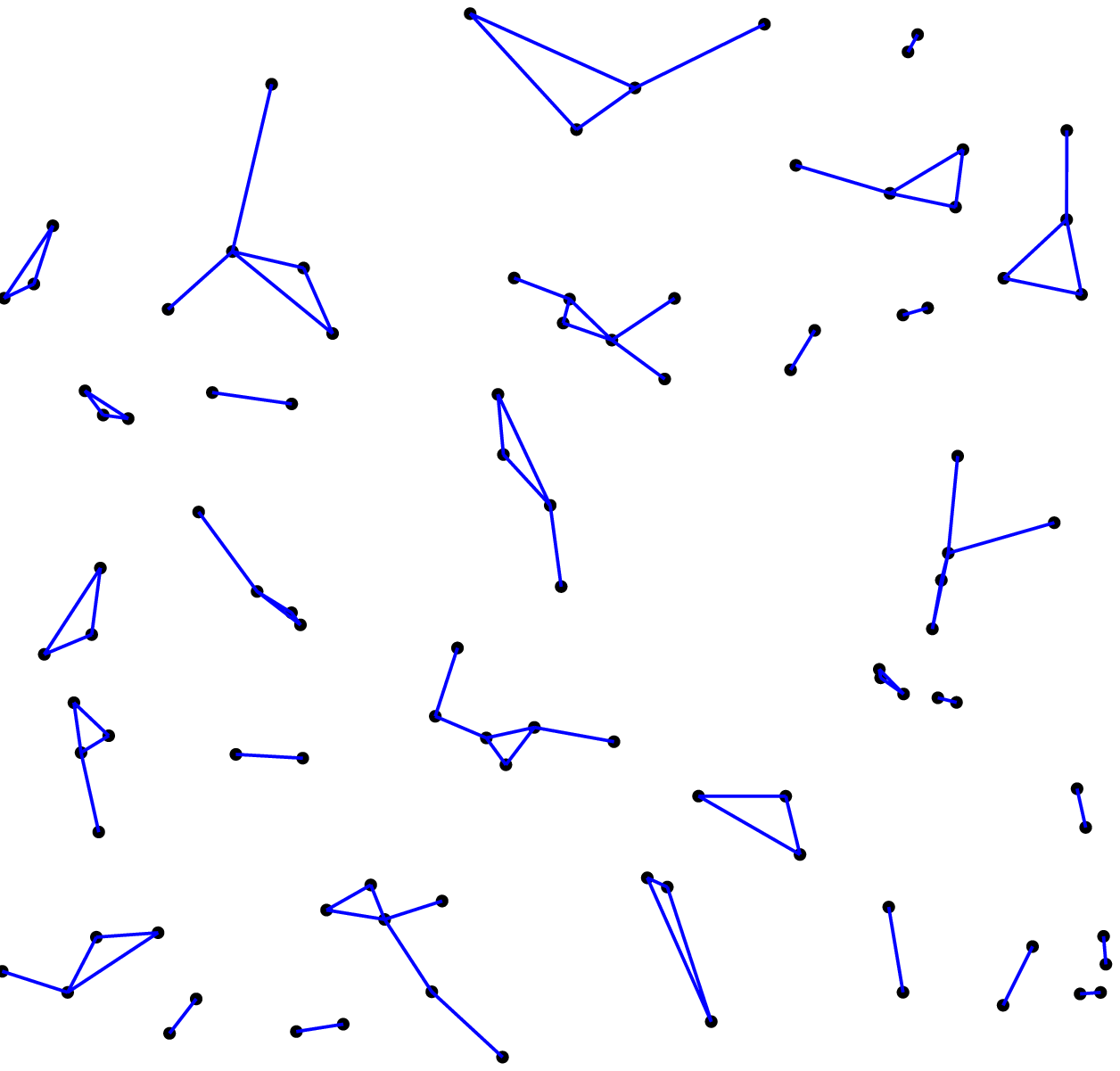} } \\ \cline{2-2}
\multicolumn{2}{c}{ } \\  \cline{2-2}
b) & \parbox{2.5in}{ \includegraphics[width=2.4in]{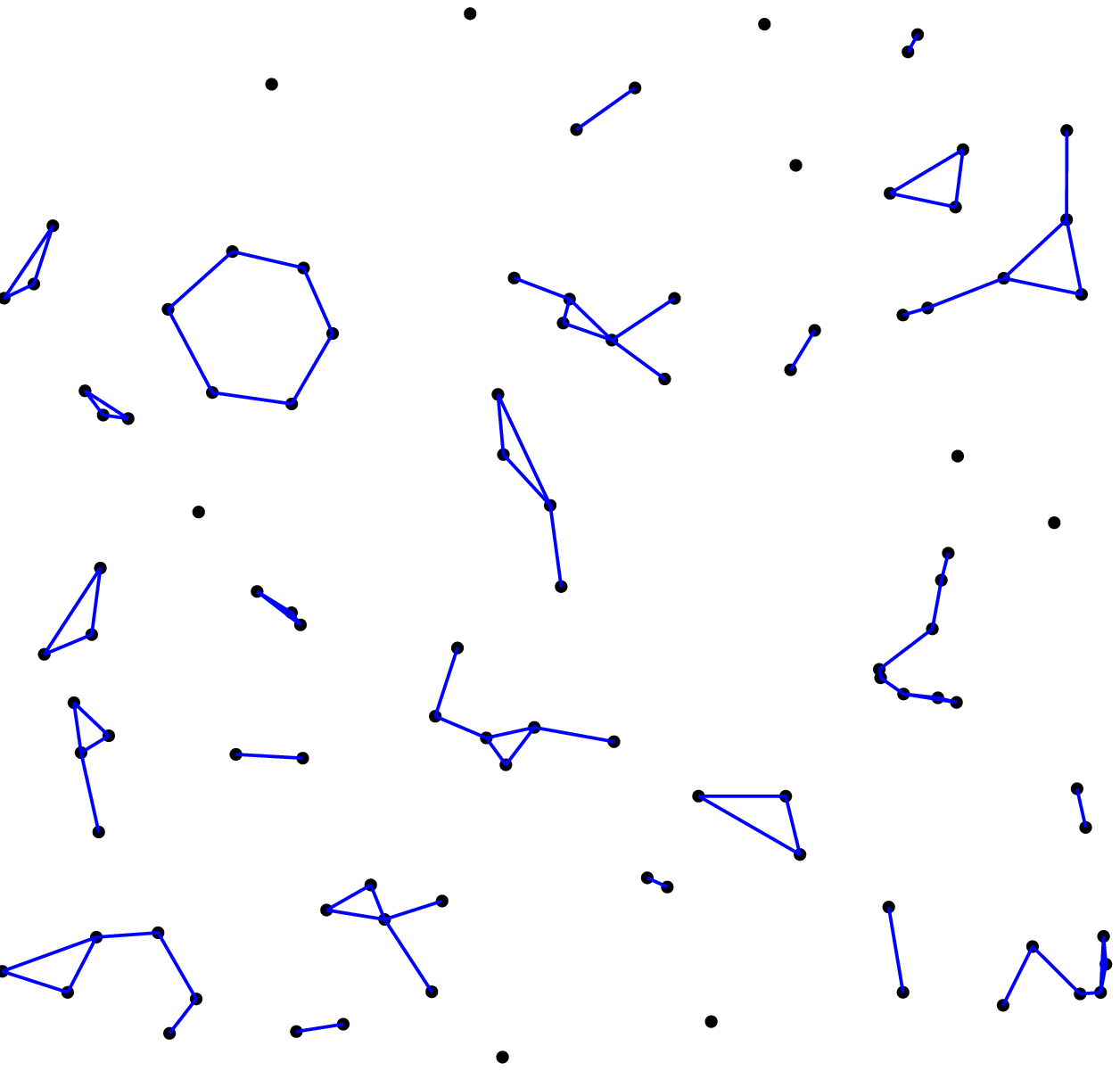} } \\ \cline{2-2}
\end{tabular}
\renewcommand{\baselinestretch}{1}\normalsize
\caption{(Color online) Example of decomposing a 100-site system into clusters. Part (a) uses the nearest-neighbor method, and part (b) the threshold method (both described in the text).  The blue lines link points in the same clusters (not all hopping links between the points are shown). \label{figClusteringMethods}}
\end{center}
\end{figure}

We begin with a set of $\Nsystem$ randomly positioned points with some average density $\bar{\rho}$ where $\Nsystem$ is typically 10,000 to 1,000,000.  We then divide the points into approximately isolated clusters, solve the cluster generalized Hubbard Hamiltonian exactly, and consider their ground state statistics. We choose to divide the large set of points into clusters using a simple algorithm that proceeds as follows:
\begin{enumerate}
\item Initially each point is a single cluster, and all points are ``unused''.
\item Choose any unused point $p$, and find its nearest neighbor $q$.
\item Merge the cluster containing $p$ with the cluster containing $q$.
\item Set point $p$ to ``used'' status.
\item Repeat at step 2 until no unused points remain.
\end{enumerate}
In this way we form the smallest clusters such that each point belongs to the same cluster as its nearest neighbor (\emph{i.e.}~the point most strongly coupled to it).  Note also that the minimum cluster size is 2.  The advantage of this ``nearest-neighbor'' method is that it always keeps nearest neighbor points in the same cluster, which is desirable from a perturbation theory standpoint.  It does not, however, guarantee that the clusters include all the hopping amplitudes of the original system above some threshold.  We show in Fig.~\ref{figClusteringMethods}(a) the decomposition of a 2D system into clusters using the algorithm.  A weakness of the nearest neighbor method is that it will form separate clusters of strongly-coupled pairs even when they are nearby other clusters, and it is clearly seen from Fig.~\ref{figClusteringMethods}(a) that some of the neglected bonds are stronger than other bonds that are kept.  On the same set of random sites, the result of an alternate algorithm that keeps all hopping amplitudes greater than a certain threshold (chosen so that the size of the clusters is not too large) is shown in Fig.~\ref{figClusteringMethods}(b).  This technique removes the problem of isolating strongly coupled pairs/triangles from other nearby sites, but it has the disadvantage of being very sensitive to the threshold, adding another degree of arbitrariness.  We find that both methods give reasonable decompositions into clusters, and the choice of algorithm not unique.  In this work, we use the nearest-neighbor method outlined above, and leave a more detailed assessment and comparison of clustering methods for later work.

We first determine, for fixed average densities $\bar{\rho}=\frac{1}{1600}$, $\frac{1}{160}$ and $\frac{3}{160}$, the distribution of cluster sizes which converges to the density-independent values shown in Table \ref{tabClusterSizeDist}.  By considering clusters with $< 8$ sites, which are within the reach of exact diagonalization techniques, we can account for over $97\%$ of the sites.  The remaining large clusters are converted into smaller clusters ($<8$ sites) by removing the smallest number of weakest links.

\begin{table}
\begin{center}
\begin{tabular}{|c|c|c|} \hline
   & \multicolumn{2}{c|}{Percentage of clusters} \\ \hline
  $\Nsites$ & \hspace{0.5cm}2D\hspace{0.5cm} & \hspace{0.5cm}3D\hspace{0.5cm} \\
  \hline
    2 & 22.9 & 20.9 \\
    3 & 28.2 & 25.0 \\
    4 & 22.0 & 20.6 \\
    5 & 13.7 & 14.7 \\
    6 &  7.2 &  8.6 \\
    7 &  3.5 &  4.8 \\
    8 &  1.5 &  2.7 \\
    9 &  0.6 &  1.3 \\
   10 &  0.3 &  0.6 \\
$>$10 &  0.1 &  0.8 \\
 \hline
\end{tabular}
\caption{Distribution of cluster sizes in a large 2D or 3D system of random sites with a fixed average density.  Clusters are formed from smallest sets of sites such that each site is in the same set as its nearest neighbor.  Since this criterion does not depend on the value of the average density, this table is valid for all fixed average densities. \label{tabClusterSizeDist}}
\end{center}
\end{table}

We can estimate the local density, $\rholoc$, of an $\Nsites$-site $d$-dimensional cluster with sites at positions $\vec{r}_i,\,i=1...N$ from the formula:
\begin{equation}
\rho_{\mbox{\scriptsize loc}} = \left\{
\begin{array}{ccc}
\frac{\Nsites}{\pi R_{cl}^2} & \quad & \mbox{in 3D} \\
\frac{\Nsites}{(4/3)\pi R_{cl}^3} & \quad & \mbox{in 2D}
\end{array} \right.
\end{equation}
where $R_{cl}$, the average radius of the cluster, is given by
\begin{eqnarray}
R_{cl} &=& \sqrt{\sum_{i=1}^{\Nsites} \left( \vec{r}_i-\vec{r}_0 \right)^2} \\
\vec{r}_0 &=&  \frac{1}{\Nsites}\sum_{i=1}^{\Nsites} \vec{r}_i 
\end{eqnarray}
\begin{figure}
\begin{center}
\includegraphics[width=2in,angle=270]{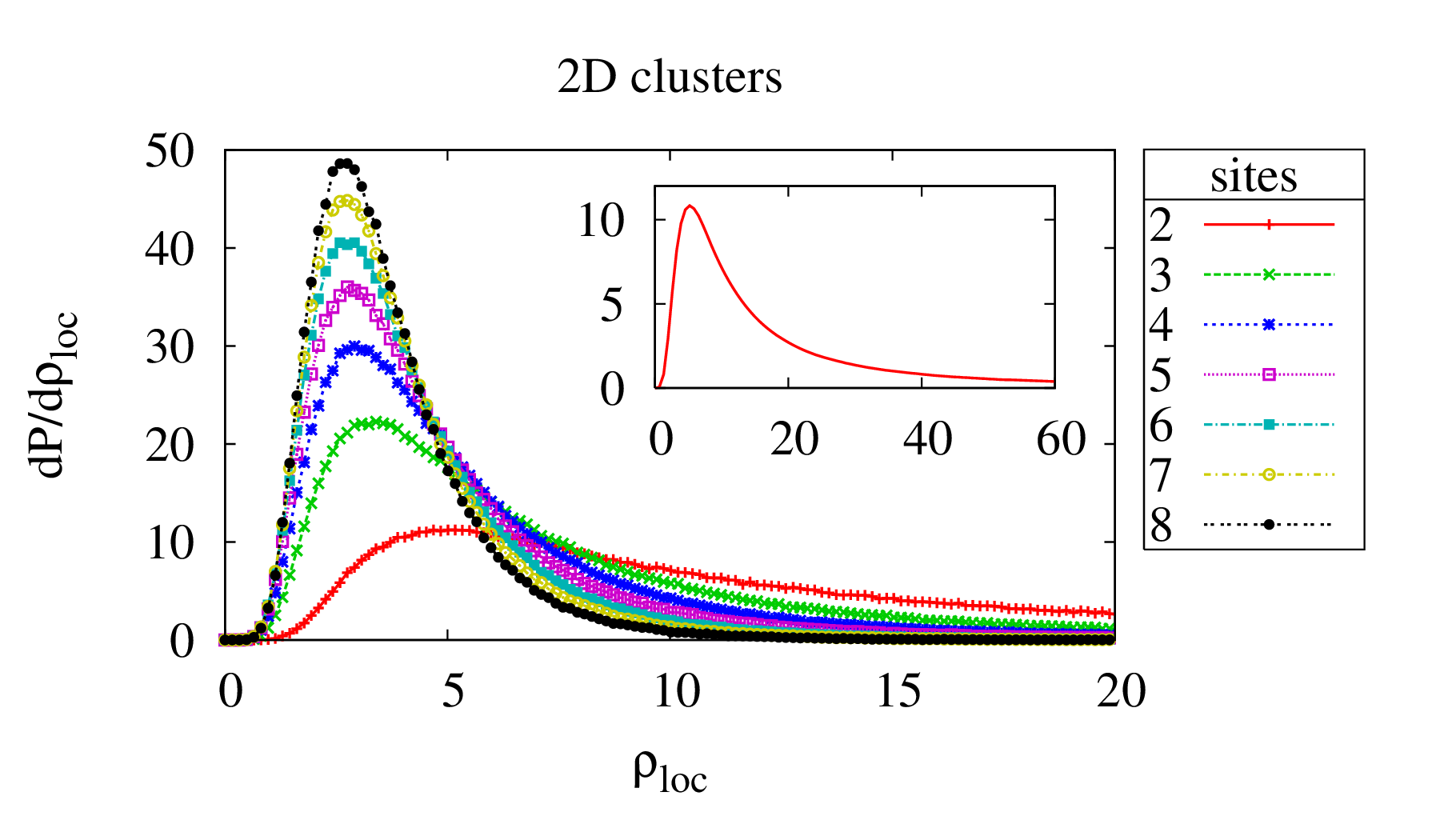}
\includegraphics[width=2in,angle=270]{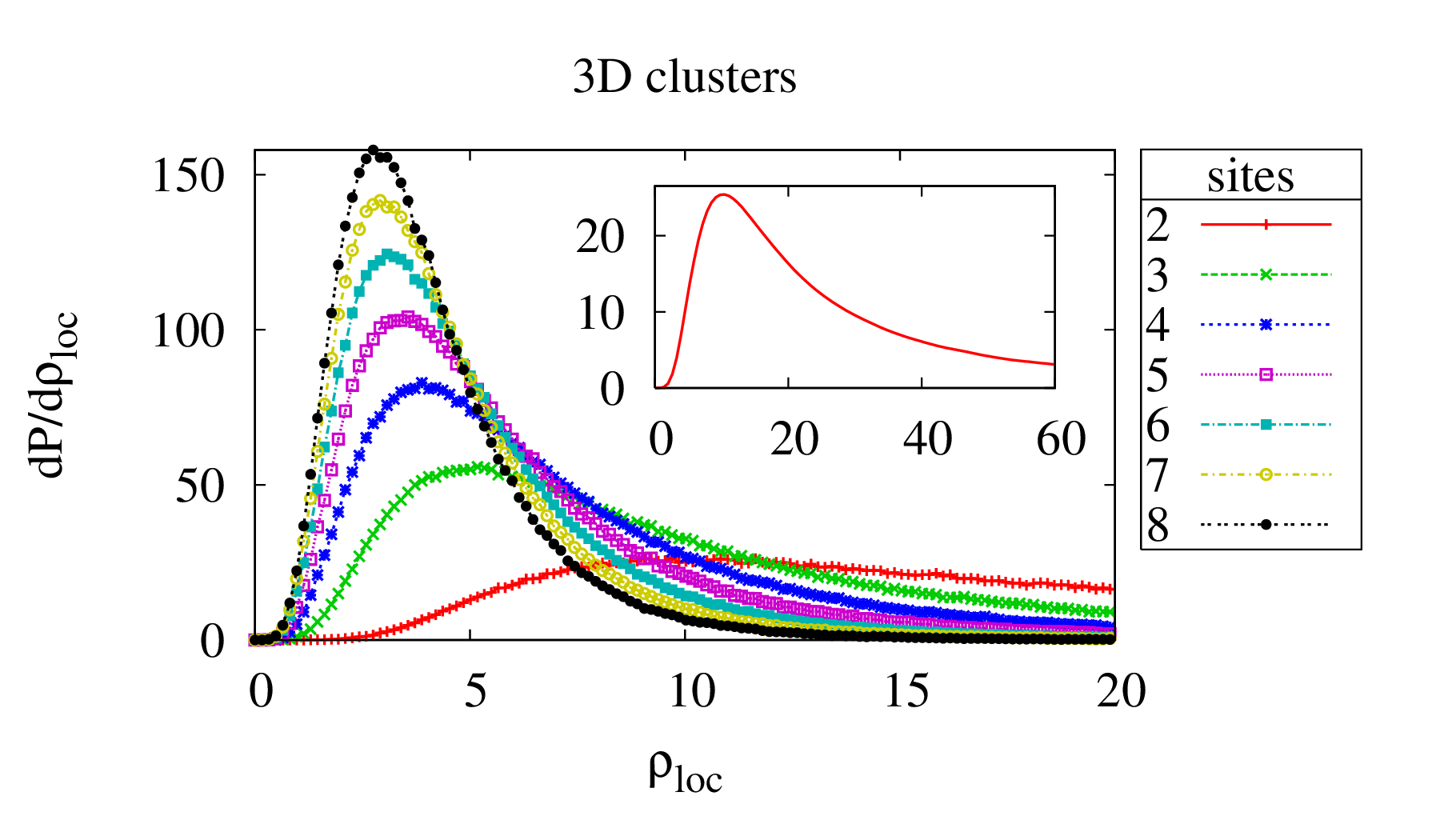}
\caption{(Color online) Individual density distributions for 2- to 8-site clusters in two and three dimensions when the average density $\bar{\rho}=1.0$.  Note that the majority of the weight falls above $\bar{\rho}$, indicating that the clusters chosen are significantly more dense than the average.  The inset shows the long tail of the 2-site cluster curve, indicating the existence of strong pairs.\label{figLocalDenDists}}
\end{center}
\end{figure}
For clusters of a given size $\Nsites$ and electron number $\Nelec$, the local density will also have some variation about its mean.  We plot the local density distribution of clusters with 2-7 sites for normalized global average density $\bar{\rho}=1.0$ in Fig.~\ref{figLocalDenDists}.  We find that clusters of larger size have a lower mean density, that is, at lower local densities the process of following nearest neighbors links has greater probability of connecting together a larger number of sites.  The reason for this trend is due to the suppressed probability of finding a group of mutual nearest-neighbors at low densities.  Let us consider the simplest case of two sites and compare the probability of finding a nearest neighbor at distance $r$ corresponding to local density $\rho_{\mbox{\scriptsize loc}} = r^{-d}$ with the probability of finding a nearest neighbor at this distance that also has the original site as its nearest neighbor.  We call such points ``mutual nearest neighbors,'' and the differential probability distribution is found by multiplying the probability of finding a nearest neighbor by the probability that the second site does not have a NN closer than the first site.  We thus define the differential probability of finding a mutual NN at distance $r$ by $p_{\mbox{\scriptsize mutualNN}}(r) = p_{nn}(r)*(1 - P_{nn}(r))$, where 
\begin{eqnarray}
P_{nn}(r) &=& 1 - \exp\left(-\frac{\pi^{d/2}nr^d}{\Gamma(\frac{d}{2}+1)}\right) \label{eq_prob_nn}\\ \nonumber\\
p_{nn}(r) &=& \left(\frac{2\pi^{d/2}}{\Gamma(\frac{d}{2})}nr^{d-1}\right) \exp\left(-\frac{\pi^{d/2}nr^d}{\Gamma(\frac{d}{2}+1)}\right) \,. \label{eq_diffProb_nn}\\ \nonumber
\end{eqnarray}
The function $p_{nn}(r)$ is the probability of finding a site's nearest neighbor between $r$ and $r+dr$, and $P_{nn}(r)=\int_0^r p_{nn}(r') dr'$ is the probability of finding a pair with length less than or equal to $r$.  As shown in Fig.~\ref{figMutualPnn} for 2D, at large $\rho_{\mbox{\scriptsize loc}}$ the distributions $p_{nn}(r)$ and $p_{\mbox{\scriptsize mutualNN}}$ approach one another, indicating that most nearest-neighbor links form mutual NN pairs.  However, at lower $\rho_{\mbox{\scriptsize loc}}$, due to the more rapid decrease of $p_{\mbox{\scriptsize mutualNN}}(r)$ at large $r$, the distributions separate and there is greater probability that a nearest neighbor will not be mutual, and thus lead to a larger (at least size 3) cluster.  The peak in $p_{\mbox{\scriptsize mutualNN}}$ near 3.0 coincides with the peak in the probability of 2D 2-site clusters in Fig.~\ref{figLocalDenDists} as one expects.  Clusters of greater than 2 sites will have greater probability density at lower $\rho_{\mbox{\scriptsize loc}}$, closer to the peak in $p_{nn}(r) - p_{\mbox{\scriptsize mutualNN}}$ near 1.75 (also shown in Fig.~\ref{figMutualPnn}). The distribution $p_{\mbox{\scriptsize mutualNN}}(r)$ decreases much more rapidly at large $r$ (small $\rho_{\mbox{\scriptsize loc}}$) than $p_{nn}(r)$ does, as shown in Fig.~\ref{figMutualPnn}. 

\begin{figure}
\begin{center}
\includegraphics[width=3in]{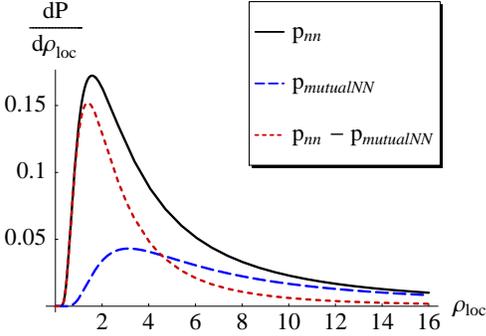}
\caption{(Color online) Probability density for finding a NN compared to that of finding a mutual NN (a NN that also has the initial site as its NN) vs. local spatial density $\rho_{\mbox{\scriptsize loc}}$.  The short-dashed line (the difference) shows the probability that a nearest neighbor is \emph{not} a mutual nearest neighbor, and thus will lead to a cluster of $>2$ sites.  The average density is set to unity.\label{figMutualPnn}}
\end{center}
\end{figure}

We diagonalize all the cluster Hamiltonians individually, and compile the resulting data to arrive at the distribution of spin values from the ensemble of clusters (obtained from many different large system realizations).  In this case, there is substantial fluctuation in the local density of clusters.  Even the mean local density for clusters of different size is different; only the average density of the \emph{entire} system is fixed.  The results show the same general trends as the clusters with fixed local density described earlier.  For comparison, the average spin and percentage of magnetic clusters in two and three dimensions are presented in Appendix \ref{appRandomClusterData} and Ref.~\onlinecite{NielsenThesis} respectively.   

\begin{figure} 
\begin{center} 
\hspace{-0.75in}\includegraphics[width=1.7in,angle=270]{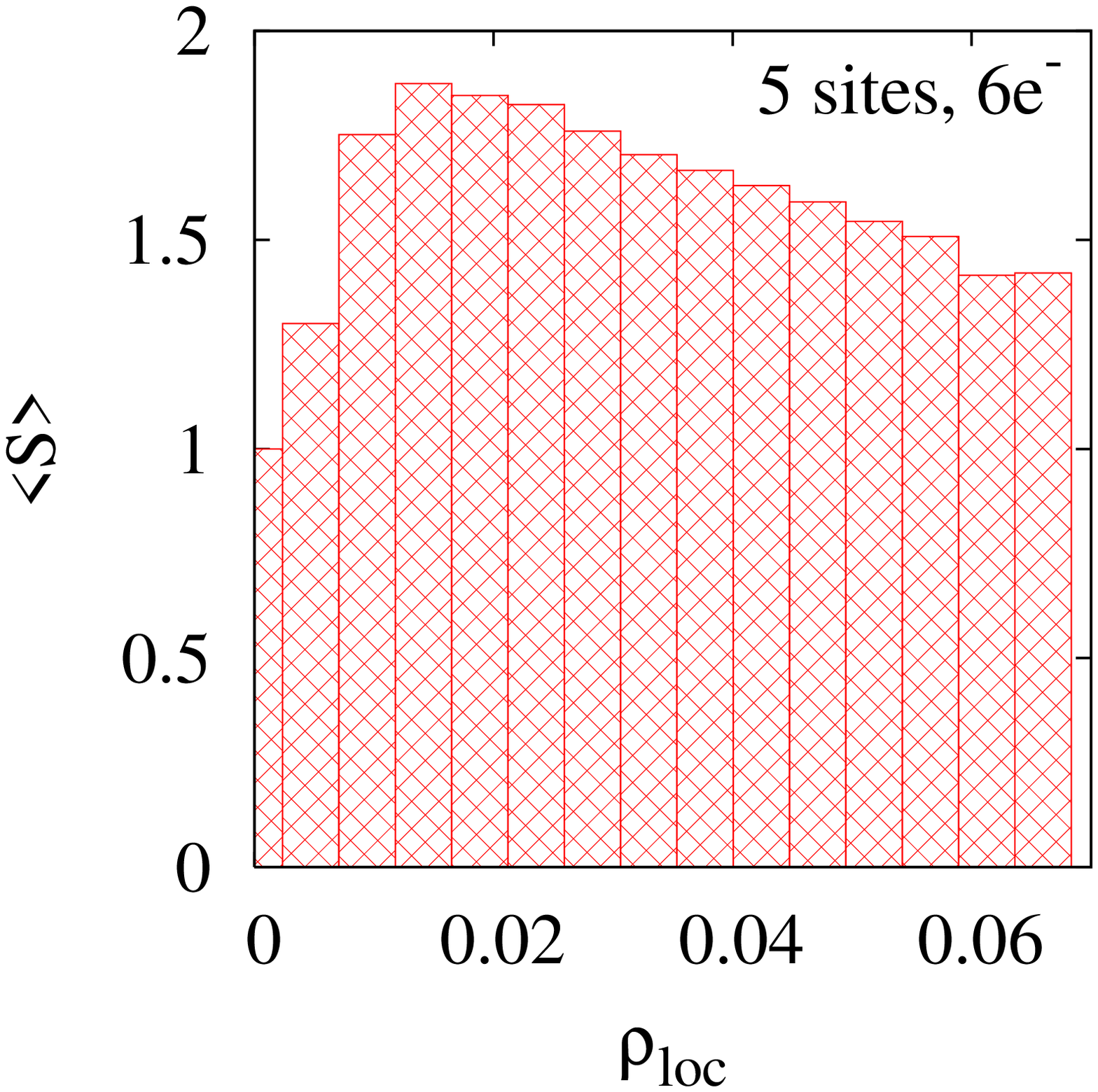} \hspace{-0.8in}
\includegraphics[width=1.7in,angle=270]{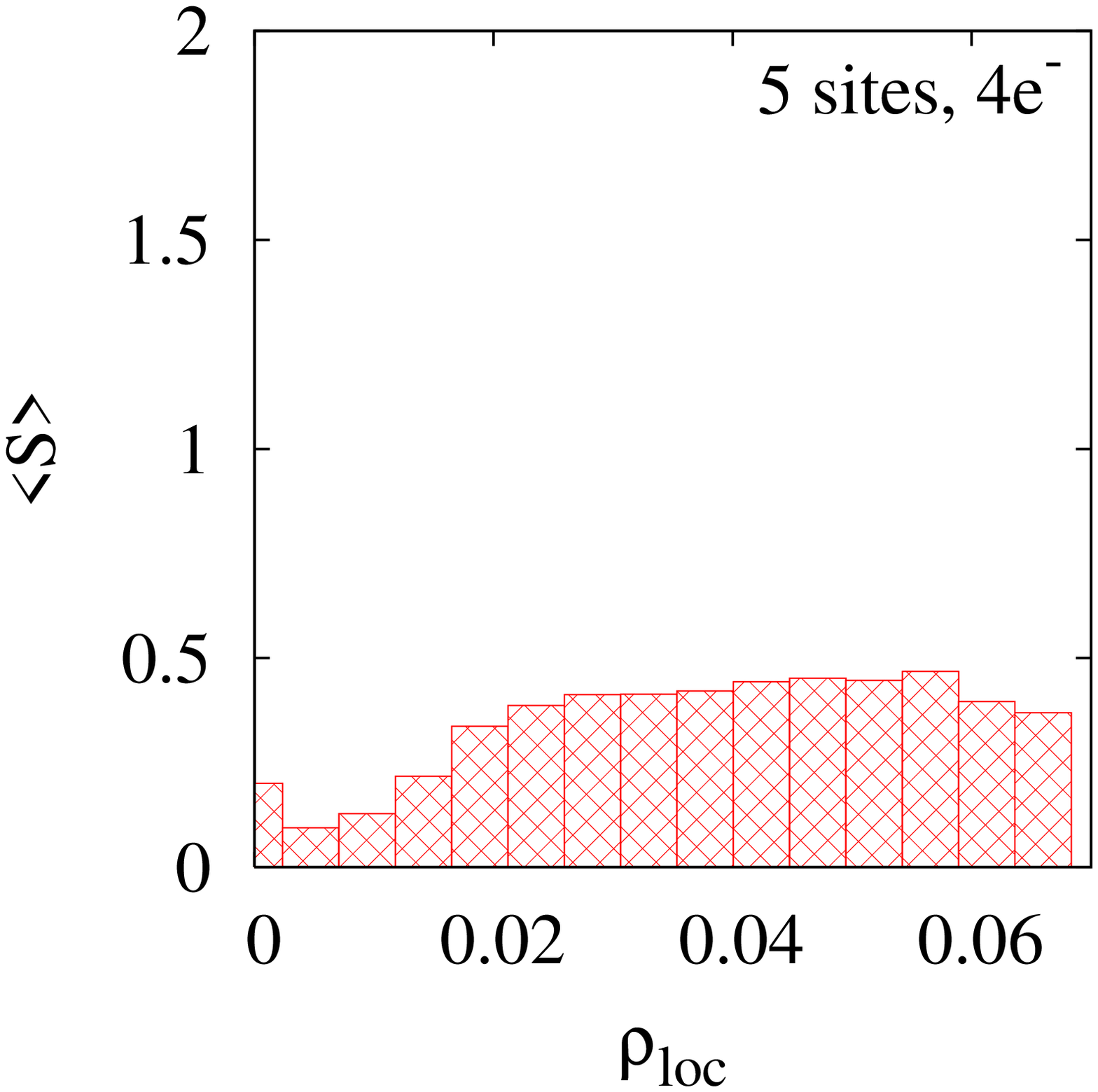} \hspace{-1.2in}
\caption{(Color online) Average ground state spin vs.~local density of 2D 5-site clusters.  The pertinent range of local densities is divided into bins, and bar heights indicate the average ground state spin of the 5-site clusters whose density falls within the corresponding density bin.  This data is from $\bar{\rho}=\frac{1}{160}$ clusters, but the behavior is typical (see text).\label{figLocalSpin}}
\end{center}
\end{figure}

We also find that there is a weak correlation between local density ($\rholoc$) and average ground state spin $\langle S \rangle$. We observe quite generally that 2D clusters with one extra electron ($\Nelec = \Nsites + 1$) have a peak in $\langle S \rangle$ near $\rholoc \approx 0.015$ while those with one hole ($\Nelec = \Nsites - 1$) have relatively smaller values of $\langle S \rangle$ that are less sensitive to changes in $\rholoc$.  Figure \ref{figLocalSpin} shows this typical behavior for 5-site clusters with $\bar{\rho} = \frac{1}{160}$ and $\Nelec = \Nsites \pm 1$.  Similar qualitative behavior is found for other clusters sizes $4 \le \Nsites \le 7$ and from systems with $\bar{\rho} = \frac{1}{1600},\frac{3}{160}$, though $\langle S \rangle$ tends to be higher for larger size clusters.  The location of the peak at $\rholoc \approx 0.015$ is important to our consideration of different large-system densities $\bar{\rho}$, since the density-independent histogram of local density given in Fig.~\ref{figLocalDenDists} shows that clusters with $\rholoc / \bar{\rho} \in [2,4]$ are most prevalent.  In the case $\bar{\rho} = \frac{1}{160} = 0.00625$, $\rholoc = 0.15$ corresponds to $\rholoc / \bar{\rho} = 2.4$, whereas for $\bar{\rho} = \frac{1}{1600} = 0.000625$ and $\bar{\rho} = \frac{3}{160} = 0.01875$ the corresponding values of $\rholoc / \bar{\rho}$ are $24$ and $0.8$ respectively.  This suggests that the $\bar{\rho} = \frac{1}{160}$ case will show the greatest overall magnetism, an inference that was seen in the fixed density clusters of section \ref{secFixedDensityClusters}, and is supported by the further investigation below (see section \ref{secElecDistNoCoulomb}).

So far we have focused on characterizing the ground state spin distribution for clusters with fixed size and electron number (but with varying densities).  We now turn to the spin distribution of the large systems from which we have taken the clusters.  Consider a large system with a fixed number of sites $\Nsystem$ and doping (fixed total electron number $\Ntelec$).  The system is partitioned into clusters of size $\Nsites=2-7$, which are approximated as being independent.  It only remains to determine how the electrons will be distributed among the clusters -- after the number of electrons on each cluster is known, the clusters can be independently solved and their ground-state spin tabulated.  We calculate the electron distribution using three different methods, two of which ignore Coulomb interactions and one which takes them into account using a classical approximation.  In the following sections we consider \emph{only} 2D systems, since our interest is primarily in 2D heterostructures and we can obtain better statistics in 2D than in 3D.

\subsection{Electron distribution without Coulomb interactions\label{secElecDistNoCoulomb}}

As a first attempt to find the distribution of electrons among the clusters, we ignore Coulomb interactions entirely and minimize the total energy, which is then just the sum of the cluster energies.  To accomplish this, we must compute the ionization energy and electron affinities of (half-filled) clusters and minimize the system's total energy to determine where the electrons will reside.  We pursue this goal in two ways. 

\paragraph{Average energy method\label{secAvgEnergyMethod}} 
The first, more approximate, calculation finds the (ensemble) \emph{average} energy required to remove or add an electron to a half-filled clusters of each considered size.  These values are shown in Table \ref{tabIonAffinity}.  The averages are over very broad distributions, however, with standard deviations comparable to the mean value shown in Table \ref{tabIonAffinity}.  This reveals a major shortcoming of this technique: it approximates broad energy distributions by their mean.  Its advantages lie in the simplicity and speed of its calculation, and that it applies to thermodynamically large systems.  We continue the analysis, knowing that results are to be treated only as a first approximation.

\begin{table}
\begin{center}
\begin{tabular}{|c||c|c||c|c||c|c|}
  \hline
  & \multicolumn{2}{|c|}{$\rho=\frac{1}{1600}$} & \multicolumn{2}{|c|}{$\rho=\frac{1}{160}$} &
   \multicolumn{2}{|c|}{$\rho=\frac{3}{160}$} \\ \hline
  \Nsites & $+1e^-$ & $-1e^-$ & $+1e^-$ & $-1e^-$ & $+1e^-$ & $-1e^-$ \\
  \hline
  1 & 1.000  &  0.000  & 1.000  &  0.000  & 1.000  &  0.000  \\
  2 & 0.994  & -0.002  & 0.971  & -0.011  & 0.971  & -0.009  \\
  3 & 0.990  & -0.004  & 0.951  & -0.035  & 0.933  & -0.090  \\
  4 & 0.988  & -0.005  & 0.940  & -0.049  & 0.914  & -0.133  \\
  5 & 0.986  & -0.006  & 0.931  & -0.060  & 0.900  & -0.167  \\
  6 & 0.985  & -0.007  & 0.924  & -0.069  & 0.887  & -0.193  \\
  7 & 0.984  & -0.007  & 0.917  & -0.082  & 0.875  & -0.217  \\
  \hline
\end{tabular}
\caption{Average energy (in units of $U \approx 0.945\Ry$) required to add (+1) or remove (-1) an electron from a half-filled cluster of size \emph{sites} in a large 2D system with total average density $\rho$.  We have used the $\tOuter(r)$ and $\tInner(r)$ (with $\tOuter > \tInner$) of our band calculation. \label{tabIonAffinity}} 
\end{center}
\end{table}

\begin{table}
\begin{center}
\begin{tabular}{|c|ccccccc|}
  \hline
  \Nsites & 1 & 2 & 3 & 4 & 5 & 6 & 7 \\
  \hline
  1 & 1.0 & .991 & .910 & .867 & .833 & .807 & .783 \\
  2 & .971 & .962 & .881 & .838 & .804 & .778 & .754 \\
  3 & .933 & .924 & .843 & .800 & .766 & .740 & .716 \\
  4 & .914 & .905 & .824 & .781 & .747 & .721 & .697 \\
  5 & .900 & .891 & .810 & .767 & .733 & .707 & .683 \\
  6 & .887 & .878 & .797 & .754 & .720 & .694 & .670 \\
  7 & .875 & .866 & .785 & .742 & .708 & .682 & .658 \\
  \hline
\end{tabular}
\caption{Net average energy (in units of $U$) required to transfer an electron between a half-filled cluster of the size specified by a column to a half-filled of the size specified by the row.  This data is for clusters in a large 2D system with total average density $\bar{\rho}=\frac{3}{160}$ and with $\tOuter(r)$ and $\tInner(r)$ set by our band calculation.\label{tabEnergyTransfer}}
\end{center}
\end{table}

\begin{figure}
\begin{center}
\includegraphics[width=3in]{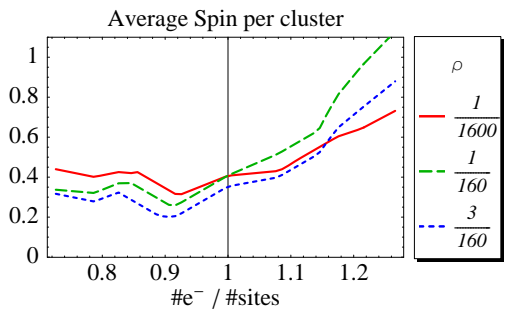}
\caption{(Color online) Average spin per cluster as a function of filling (number of electrons per site; half-filled corresponds to 1.0), where the energy optimizing electron distribution is used at each filling. \label{figOptimalAvgSvsDoping}}
\end{center}
\end{figure}
\begin{figure}
\begin{center}
\includegraphics[width=3in]{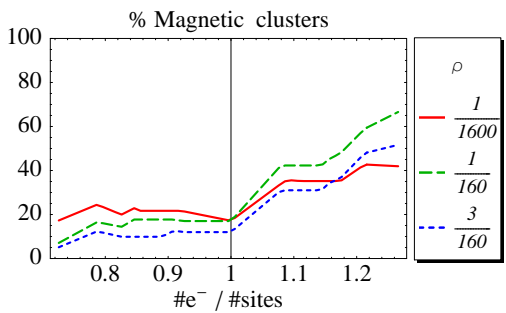}
\caption{(Color online) Percentage of magnetic clusters (those having above minimal ground state spin) as a function of filling (number of electrons per site; half-filled corresponds to 1.0), where the energy optimizing electron distribution is used at each filling. \label{figOptimalPcMagVsDoping}}
\end{center}
\end{figure}

By subtracting pairs of the values in Table \ref{tabIonAffinity}, we find the \emph{average} net energy gained (or lost) when transferring an electron from one cluster to another, shown in Table \ref{tabEnergyTransfer}.  The fact that all transfer energies are positive implies the stability of the electron configuration in which each cluster is exactly half-filled in this approximation.  Note, however, that Coulomb interactions (which lower the energy of two charged cluster system) may alter this picture substantially.  Using the average affinities and ionization energies, we determine the optimal distribution of electrons among the clusters.  Let $x_n^q$ be the fraction of the total clusters that have $n$ sites and charge $q$.  In our calculation, $n=2 \ldots 7$ and $q\in\{-1,0,+1\}$ (clusters are allowed at most one electron or hole on them), so there are 18 variables in all.  The optimal $x_n^q$ are found by minimizing the total energy, $E^{tot}(\{x_n^q\})$, subject to constraints.  The energy is written: 
\begin{equation}
E^{tot}(\{x_n^q\}) = \sum_{n,q=\pm 1} \alpha_{n,q} x_n^q
\end{equation}
where $\alpha_{n,-1}$ is the energy required to add an electron to a $n$-site cluster, $\alpha_{n,+1}$ is the energy required to remove an electron from a $n$-site cluster.  Constraints on the problem are:
\begin{itemize}
\item $x_n^q \ge 0$ for all $n,q$.
\item $\sum_{q=-1}^1 x_n^q = f_n$, where $f_n$ is the fraction of total clusters with size $n$ (found from Table \ref{tabClusterSizeDist}).
\item $N_e^{tot} = \sum_{n,q} (n+q) x_n^q$, where the sum ranges over all $n=2 \ldots 7$ and $q\in\{-1,0,+1\}$ (note that $n+q$ is the total number of electrons on $n$-site clusters with charge $q$).
\end{itemize}
Since the energy and all constraints are linear, this is a linear programming problem and can be solved with standard numerical routines.  The minimization is carried out at a fixed doping, and results in the optimal placement the electrons in a thermodynamically large system.  We determine the average spin per cluster of the entire system as a function of doping (summing the average spin of a cluster with $n$ sites and charge $q$ multiplied by $x_n^q$) in Fig.~\ref{figOptimalAvgSvsDoping}.  We also consider the percentage of clusters that have above-minimal spin (again using the average results for fixed-size clusters), classified above as magnetic clusters.  From the plot of the percentage of magnetic clusters vs.~doping in Fig.~\ref{figOptimalPcMagVsDoping} we see that when a system is doped 10-20\% above half-filling, nearly half of the clusters have greater than minimal ground state spin.  This suggests that at such doping some kind of percolation might be possible that would induce magnetic order on a mesoscopic or even macroscopic length scale.  


%

\begin{figure}
\begin{center}
\begin{tabular}{l}
\includegraphics[width=2in, angle=270]{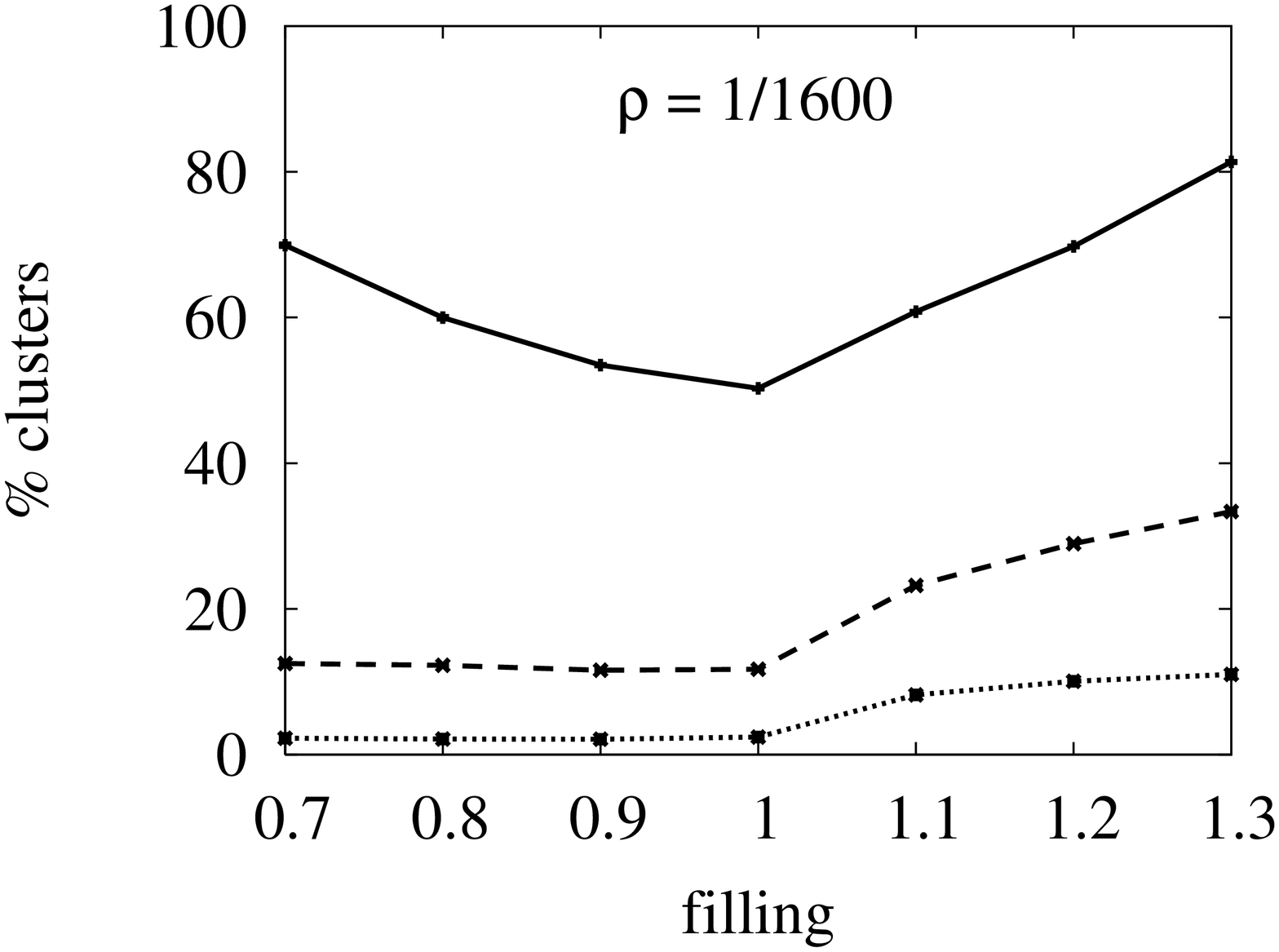} \\
\includegraphics[width=2in, angle=270]{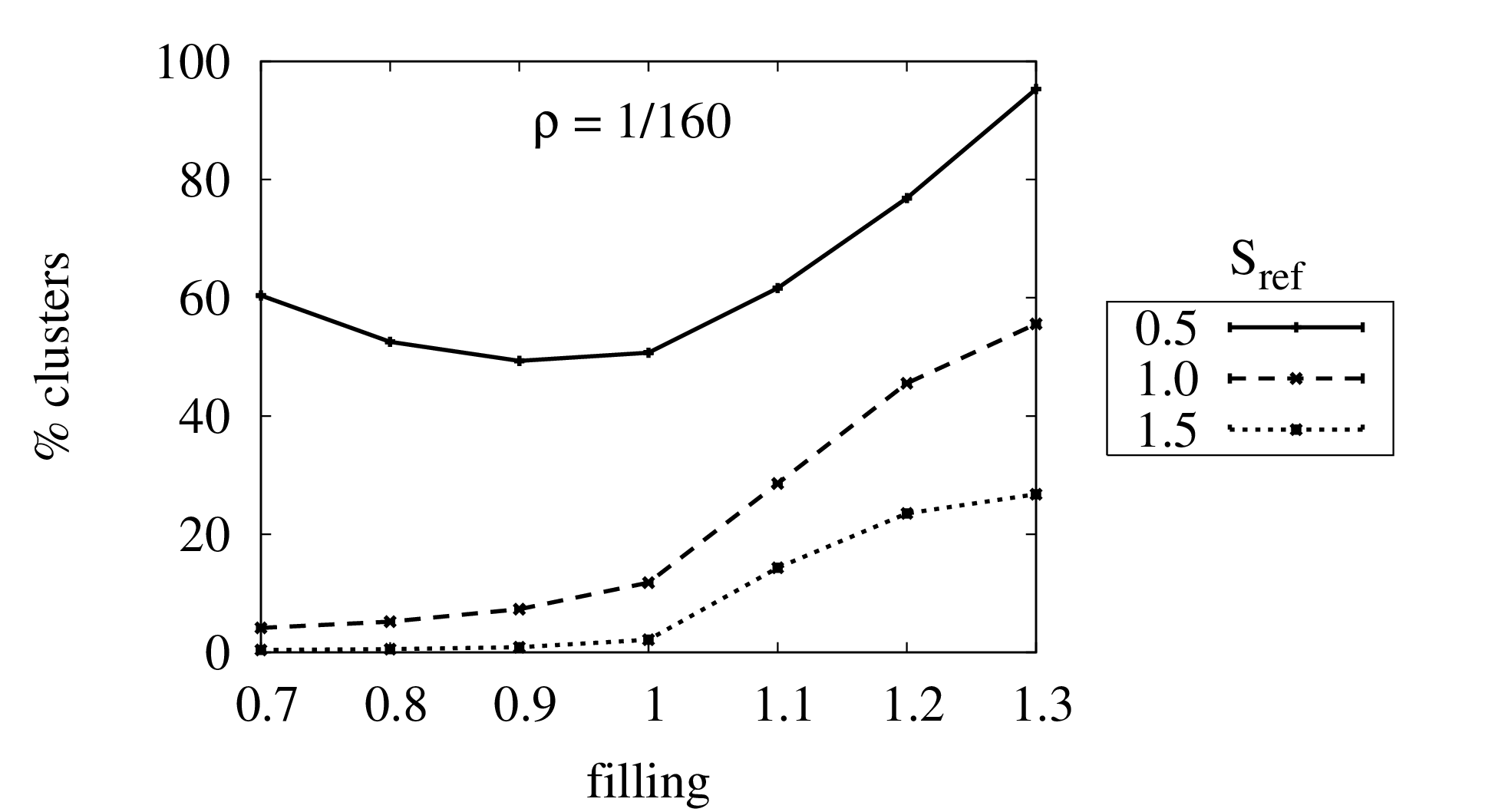} \\
\includegraphics[width=2in, angle=270]{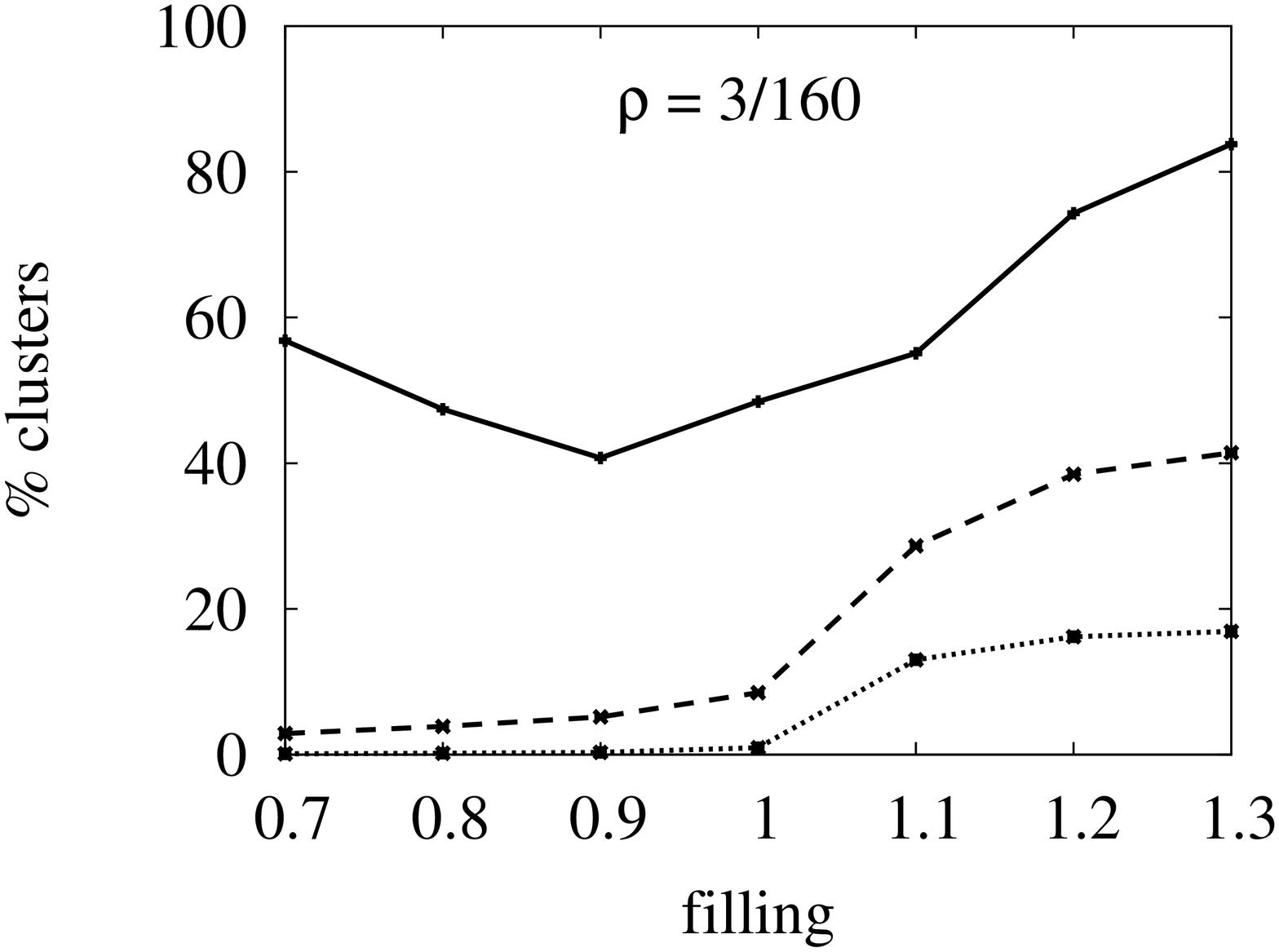} \\
\end{tabular}
\caption{Percentage of clusters with total spin greater than the reference value $S_{ref}$, specified in the key, as a function of filling (1.0 = half-filling).  Plots correspond to densities $\rho = \frac{1}{1600}$, $\frac{1}{160}$, and $\frac{3}{160}$ as indicated in their titles.\label{figElecGlassNoCoulomb}}
\end{center}
\end{figure}

\begin{figure}
\begin{center}
\includegraphics[width=3in]{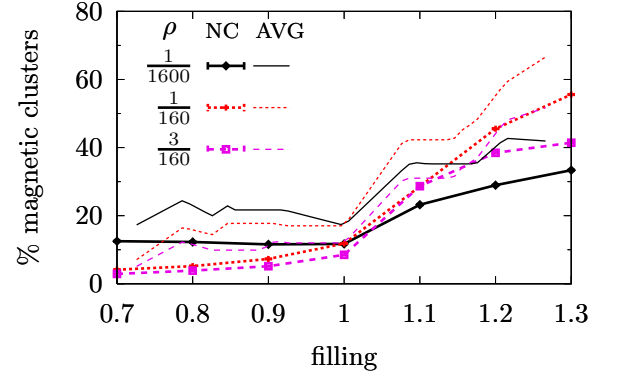}
\caption{(Color online) Comparison of the percentage of magnetic clusters (those with greater than minimal ground state spin) using the average energy method (AVG, thin lines) of section \ref{secAvgEnergyMethod} and the method using actual clusters without Coulomb interactions (NC, thick lines) of section \ref{secElecGlassNoCoulomb}. The style of the line (solid, dashed, or long-dashed) indicates the density. \label{figElecGlassCompareNoCoulomb}}
\end{center}
\end{figure}

\paragraph{Full cluster method\label{secElecGlassNoCoulomb}}
A more straightforward way of calculating the optimal electron distribution in the absence of Coulomb interactions is to consider an ensemble of large random systems. For each system, after separating the sites into clusters, we minimize the energy by repeatedly testing if the movement of an electron from one site to another lowers the total energy.  Specifically, the algorithm we use is as follows:
\begin{enumerate}
\item Initialize the system by placing electrons (if above half-filling) or holes (if below half-filling) on random clusters.
\item Randomly choose two clusters $i$ and $j$, and attempt to move an electron from $i$ to $j$.  If the resulting change in total system energy (just the sum of all cluster energies since there are no Coulomb interactions) is negative, accept the transfer.  If not, do not make the transfer.
\item Repeat the above step until the total energy converges.
\end{enumerate}
Knowing the electron distribution, the ground state spin of each cluster in the large system is then calculated.  Finally, we compute the distribution of cluster ground state spin values, and average them over an ensemble of large systems.  Figure \ref{figElecGlassNoCoulomb} shows how the percentage of clusters with spin greater than or equal to a reference spin $S_{ref}$.  For $S_{ref} = 1$, these percentages correspond to our earlier definition of ``magnetic clusters'', and we compare in Fig.~\ref{figElecGlassCompareNoCoulomb} the results of this section with those obtained earlier using the average energy method (section \ref{secAvgEnergyMethod}).  We see that the latter overestimates the number of high-spin clusters, particularly in the electron-doped case.  The above analysis neglected the effect of long-range Coulomb interactions, which we consider next.

\subsection{Electron distribution including Coulomb interactions\label{secElecGlassCoulomb}}
In low density insulating systems, Coulomb interactions between charged centers (or clusters) are not screened effectively and, because of their slow ($1/r$) fall-off, cannot be neglected.\cite{EfrosShklovskii_1975} 
 Therefore, a more accurate way to calculate the electron distribution is to include Coulomb interactions.  The approach described in this section accounts for the Coulomb interactions between charged clusters in an approximate way.  We begin in a fashion similar to the preceding analysis, solving each cluster for range of total electrons near half-filling.  We then determine the minimum energy electron distribution by solving a generalized electron glass problem\cite{EfrosShklovskii_1975,Efros_1976,DaviesLeeRice_1982,DaviesLeeRice_1984} which accounts for the differences in ground state energy of the clusters \emph{and} the Coulomb energy between charged clusters, as described below.

The generalized electron glass problem we solve consists of a set of two-dimensional clusters lying in a large two-dimensional space, indexed by $i=1\ldots N_{cl}$.  Each cluster is treated as an effective site, and is assigned a position $\vec{R}_i$ (given by the average positions of all of its points), and a dimensionless charge $q_i$.  The charge naturally corresponds to the occupation of the cluster (relative to half-filling), and is restricted in our analysis to be +1, 0, or -1.  
The problem is to minimize the classical Hamiltonian 
\begin{equation}
\mathcal{H}_{eg} = \sum_i \phi_i^{q_i} + \frac{e^2}{2\epsilon} \sum_{ij, i\ne j} \frac{q_i q_j}{r_{ij}}
\end{equation}
where $\epsilon$ is the dielectric constant, $r_{ij} = |\vec{R}_i - \vec{R}_j|$, and $\phi_i^{q_i}$ is the ground state energy of cluster $i$ when it has charge $q_i$.  The first term gives the on-site (or, more accurately, on-cluster) energy contribution, and the second term supplies the Coulomb interaction between clusters.  The minimization is performed with respect to the variables $q_i$ which must obey the constraint $\sum_i q_i = \Ntelec-\Nsystem$, where $\Ntelec$ is the total number of electrons in the $\Nsystem$-site system.  The details of the minimization are a generalization of the procedure outlined by Baranovskii \emph{et al.},\cite{Baranovskii_1979} divided into three steps:
\begin{enumerate}
\item Initialize the $\{q_i\}$ by starting them all equal to zero and randomly choosing clusters to add an electron to (if $\Ntelec > \Nsystem$) or remove an electron from (if $\Ntelec < \Nsystem$) until $\sum \limits_i q_i = \Ntelec-\Nsystem$.
\item Calculate all single-cluster energies
\begin{equation}
E_i^q = \phi_i^q + \frac{e^2}{\epsilon}\sum_{j \ne i} \frac{q_i q_j}{r_{ij}}
\end{equation}
and check that 
\begin{equation}
\Big(E_i^{q_i}-E_i^{q_i-1}\Big) < \Big(E_j^{q_j+1}-E_j^{q_j}\Big) \label{eqEGlassIneq1}
\end{equation}
 for all $i,j$ such that $q_i > -1$, $q_j < 1$, and $i \ne j$.  The left hand side of the inequality is the cost of having the last-placed electron on site $i$, which should be less than the cost of placing an electron on site $j$.  Otherwise, we can lower the system's energy (disregarding the Coulomb interaction for now) by moving an electron from $i$ to $j$.  In practice, we consider the pair $i,j$ that for Eq.~\eqref{eqEGlassIneq1} makes the left side maximal and right side minimal. If inequality (\ref{eqEGlassIneq1}) is not satisfied we move an electron from $i$ to $j$ and repeat the step from the beginning.  If the inequality is satisfied, we proceed to the next step.  This is analogous to the $\mu$-sub routine referred to by earlier work.\cite{Baranovskii_1979,DaviesLeeRice_1984}
\item Calculate the energies $E_i$, and iterate through all pairs $(i,j)$ such that $q_i > -1$, $q_j < 1$, and $i \ne j$ and check that each satisfies
\begin{equation} 
\Big(E_j^{q_j+1}-E_j^{q_j}\Big) - \Big(E_i^{q_i}-E_i^{q_i-1}\Big) - \frac{e^2}{\epsilon r_{ij}} > 0 \,.
\end{equation}
If a pair $(i,j)$ is found that does not satisfy the inequality, we move an electron from cluster $i$ to cluster $j$ and repeat the step (recalculate the $E_i$ and check again).
\end{enumerate}
This process results in a set $\{ q_i \}$ that is stable with respect to any single electron moving between clusters.  Further conditions (and steps) could be added that would make the final distribution of electrons stable against any two electrons simultaneously moving between sites, but previous work on the electron glass problem\cite{DaviesLeeRice_1982} has shown that these additional constraints do not significantly affect the result.  Therefore, we do not implement this additional step.

\begin{figure}
\begin{center}
\begin{tabular}{l}
\includegraphics[width=2in, angle=270]{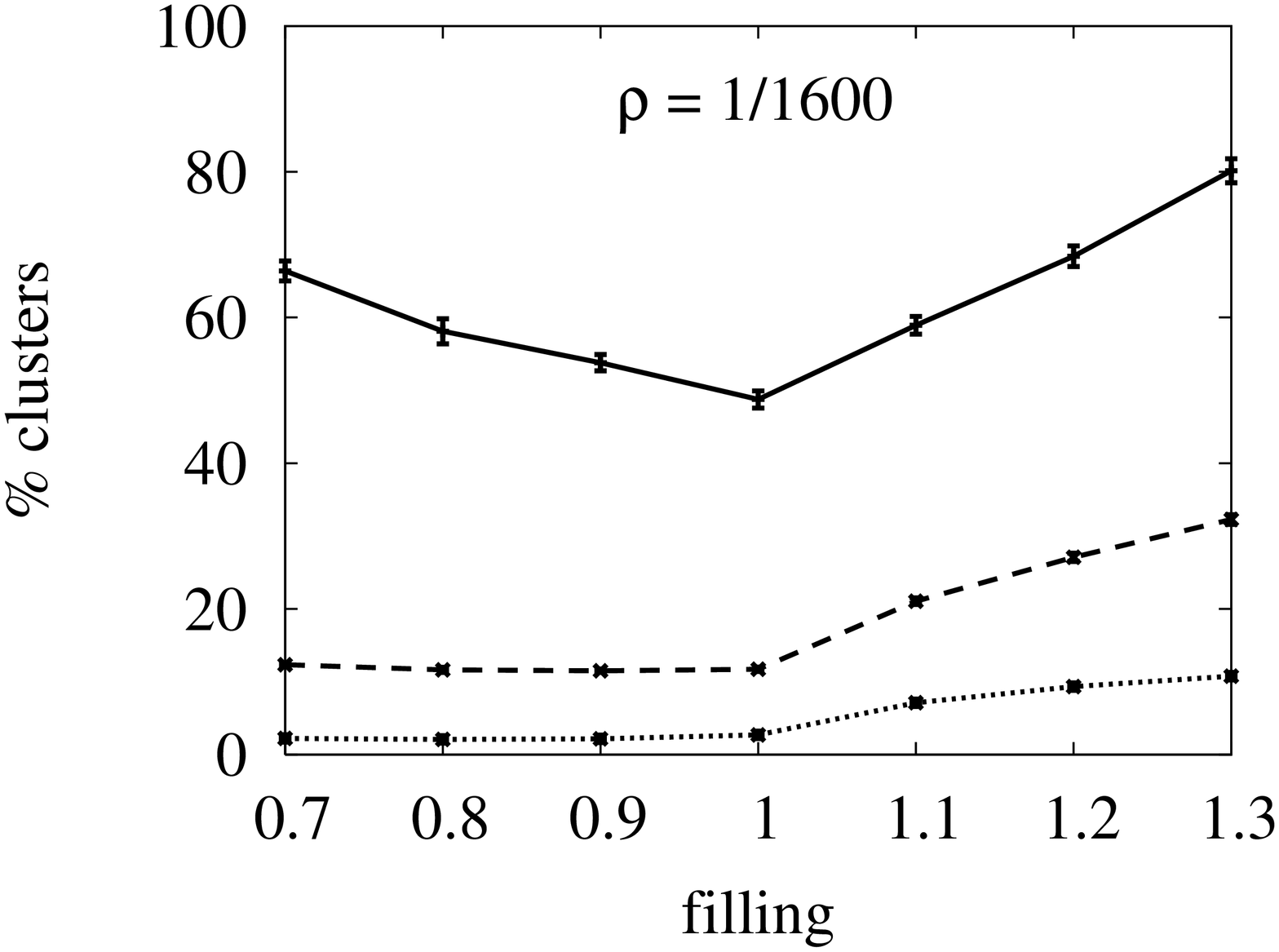} \\
\includegraphics[width=2in, angle=270]{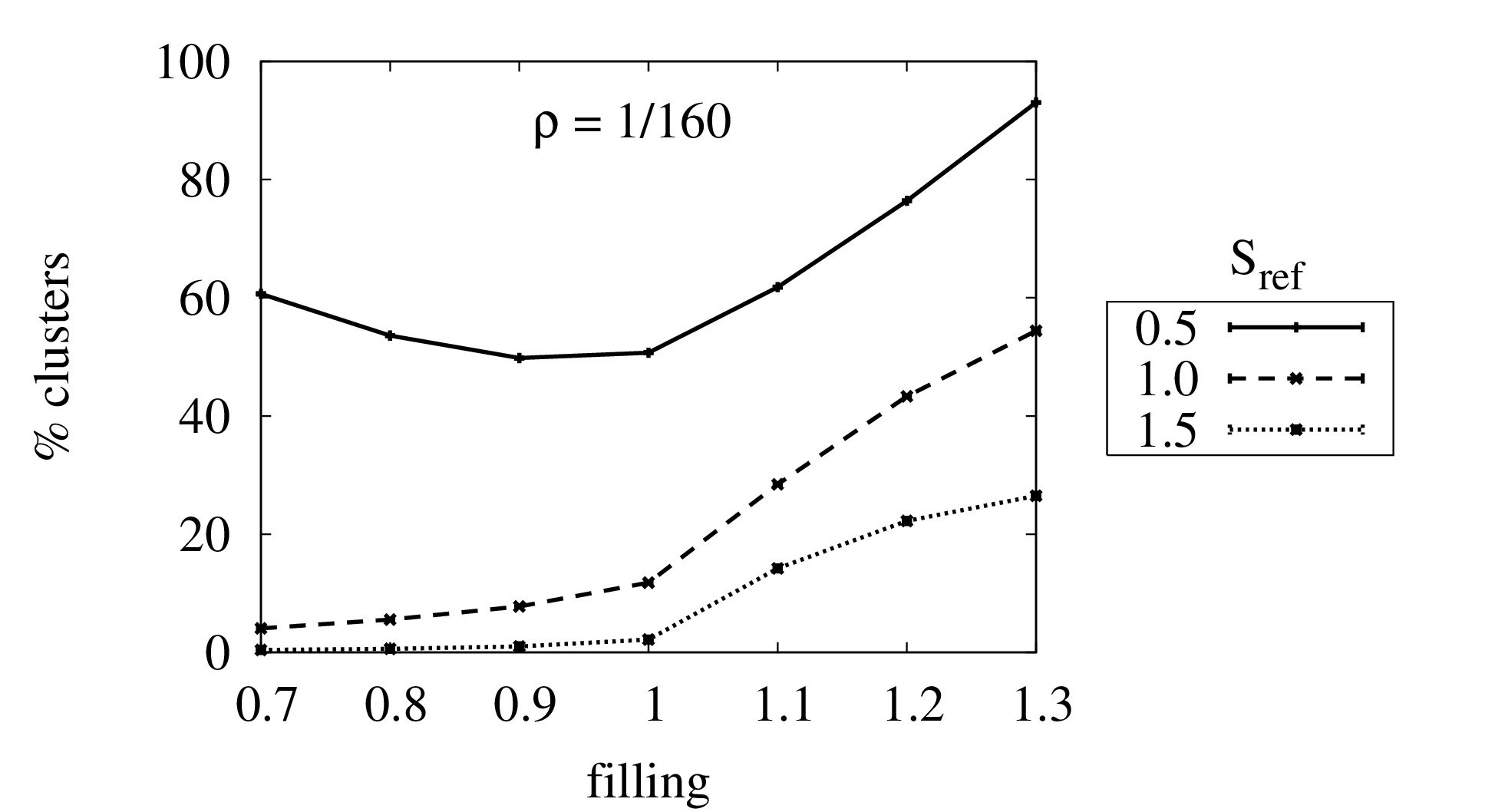} \\
\includegraphics[width=2in, angle=270]{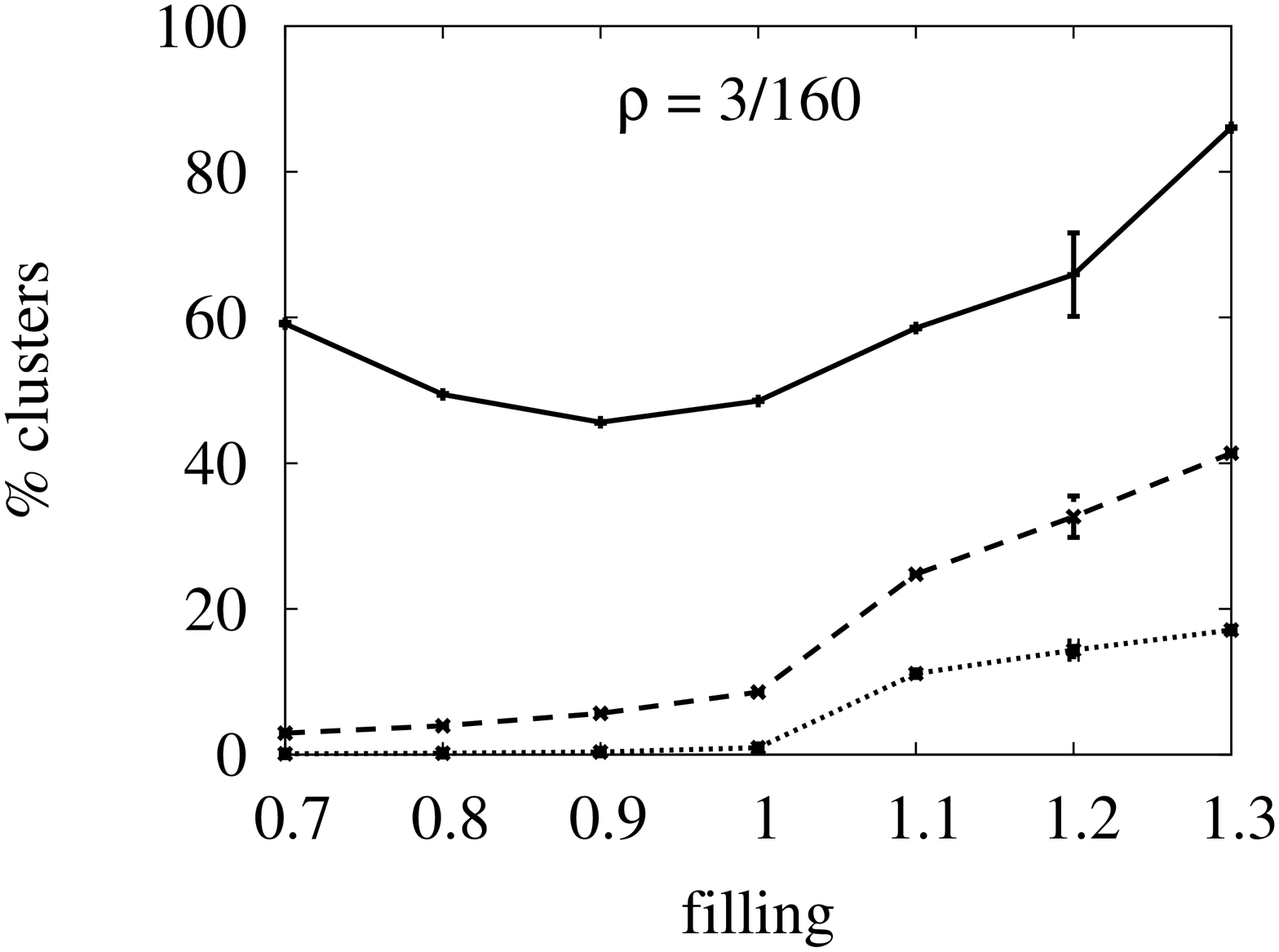} \\
\end{tabular}
\caption{Percentage of clusters with total spin greater than the reference value $S_{ref}$, specified in the key, as a function of filling (1.0 = half-filling).  Plots correspond to densities $\rho = \frac{1}{1600}$, $\frac{1}{160}$, and $\frac{3}{160}$ as indicated in their titles.\label{figElecGlassSeparateDensities}}
\end{center}
\end{figure}

\begin{figure}
\begin{center}
\includegraphics[width=3in]{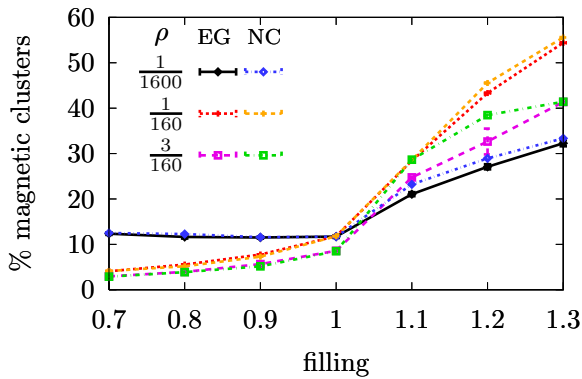}
\caption{(Color online) Comparison of the percentage of magnetic clusters (those with greater than minimal ground state spin) when Coulomb interactions are ignored (section \ref{secElecGlassNoCoulomb}) or accounted for (section \ref{secElecGlassCoulomb}, using a generalized Coulomb glass analysis.  The plot shows, for densities $\rho = \frac{1}{1600}$, $\frac{1}{160}$, and $\frac{3}{160}$, percentages of the no-Coulomb (NC) case and electron glass (EG).\label{figElecGlassCompareBoth}}
\end{center}
\end{figure}

Once we have determined the distribution of electrons among the many clusters of the large system, we compute the percentage of clusters with spin greater than a given reference spin $S_{ref}$.  This quantity is averaged over many random realizations of the large cluster system. 
The (ensemble-averaged) percentage of clusters with spin $\ge S_{ref}$ for $S_{ref} = \frac{1}{2}$, $1$, and $\frac{3}{2}$ is shown in Fig.~\ref{figElecGlassSeparateDensities} for our standard densities $\rho = \frac{1}{1600}$, $\frac{1}{160}$, and $\frac{3}{160}$.  For $S_{ref} = 1$, these percentages correspond to our earlier definition of ``magnetic clusters'', and Fig.~\ref{figElecGlassCompareBoth} compares the results with those of the previous section (\ref{secElecGlassNoCoulomb}) which neglects Coulomb interactions but is otherwise identical to the calculation performed here.  We see that Coulomb interactions slightly deplete the number of high-spin clusters, particularly in the electron-doped case.  This also shows that even in the presence of long-range Coulomb interactions there is a sizable percentage of magnetic clusters at modest electron-doping.  In order for the magnetic clusters to percolate in a strictly 2D system, they must account for 50\% of the system, which is only attained at large filling factors ($\approx 1.2$ in the best case of $\rho=\frac{1}{160}$).  In 3D, however, the percolation threshold is much lower, so a parallel calculation in a 3D or thick 2D system (which behaves as a 3D system on short length scales) may yield even more promising results. We also remark that as the impurity density is increased at fixed doping the average number of magnetic clusters has a non-monotonic behavior. There is an optimal impurity density (nearest to $\rho = \frac{1}{160}$ in our data) that results in an on-average maximal number of magnetic clusters.  Altogether, the presence of many high-spin clusters provides a necessary ingredient for ferromagnetism on macroscopic, or even mesoscopic, length scales.

\section{Conclusions\label{secConclusion}}

We have formulated a Hubbard model appropriate for doped semiconductors, which has an occupation-dependent hopping term and therefore intrinsic electron-hole asymmetry characteristic of the hydrogenic center.  This generalized disordered Hubbard model is numerically solved using exact diagonalization on 2D finite lattices, selected symmetric clusters, and completely random clusters in two and three dimensions.  We summarize the results of each in turn.

Our results on finite (periodic) lattices, as well as selected clusters and distorted/randomized versions of them, lead us to several important conclusions.  First, high-spin ground states generally occur at large $U/t$ (low impurity density).  On a bipartite lattice one carrier away from half-filling, Nagaoka's theorem guarantees a maximal spin state in the limit $U/t \rightarrow \infty$.  In the finite lattices that satisfy Nagaoka's theorem, we find maximal spin states at large but finite $U/t$. In clusters (with less symmetry than a lattice), high-spin ground states are found to be quite sensitive to the cluster geometry, though they all exist at large $U/t$.
 Second, the properties of the hydrogen atom give rise to a crucial difference between the electron-doping and hole-doping of hydrogenic systems.  In lattices as well as clusters we see a greatly enhanced occurrence of spin-polarization in electron-doped (above half-filling) systems.  In systems above half-filling we also find that increasing $\tOuter/\tInner$ can significantly increase the likelihood of this nanoscale ferromagnetism.  These results confirm our expectation that the greater the spatial extent of a doubly-occupied site's wavefunction (relative to the wavefunction of a singly-occupied site), the more favorable spin polarization becomes. 
 Lastly, we remark on the resilience of the high-spin ground states.  By perturbing a cluster geometry that has a high-spin ground state, we find that larger values of $U/t$ and $\tOuter/\tInner$ make the state more stable to geometric and random fluctuations in the hopping amplitudes.  An assessment of high-spin state robustness is relevant to situations in which sites are individually positioned within some tolerance. Overall, we have identified a regime where nanoscale FM exists (with some robustness) in finite lattices and artificially-made clusters of hydrogenic centers.

The analysis of ground state spin behavior in completely random clusters reveals several of the same conclusions we found for the selected symmetric clusters.  Namely, we find that electron-doping and a larger $\tOuter/\tInner$ favor spin polarization in random clusters as well.  The electron-hole asymmetry found in all of the random ensembles implies that in real semiconductor systems there is a significant difference between doping above and below half-filling.  Spin-polarization is much more prevalent in systems above half-filling, an effect which we again emphasize as arising from the physical properties of the dopant atom.   Unlike in the case of selected clusters, where ferromagnetism is generally more prevalent at larger $U/t$,  we find that within the low-density range considered (well below the metal-insulator transition), there is an optimal density for finding high-spin (random) clusters.  This interesting observation is likely due to clusters breaking up into separate, effectively disconnected, pieces at very low densities, which hinders carrier movement and thereby the alignment of spins in the ground state.

We also study the problem of distributing electrons onto the cluster components of a large system. Of particular interest is the relatively small effect of Coulomb interactions (between charged clusters) on the electron distribution.  Coulomb interactions slightly \emph{decrease} the number of clusters with above-minimal spin.  This effect was unexpected since Coulomb interactions reduce the energy cost of charged clusters, which generally have higher ground state spin than un-charged clusters.  A detailed look at differences in the electron distributions with and without Coulomb interactions may help to explain the reason for the small observed effect, and is left for future work. 

Taking into account all our data on finite systems,  we expect high spin clusters to be observable in systems with a low density (large Hubbard $U/t$) of centers and a small \emph{excess} of electrons.  The latter requirement is difficult to realize in 3D bulk systems, but could be met in doped quantum dots and 2D heterostructures.  For example, doped quantum dots with dopant number $N_d = 6-15$ and a small excess of electrons $N_e-N_d=1-2$ would be ideal systems for finding high-spin ground states.  Also, in modulated structures with dopants in both quantum wells and barrier regions, regions of excess electrons can be achieved, unlike in true bulk doped semiconductors.  We also note that the artificial cluster geometries studied in section \ref{secSelectedClusters} have real world applications through recently developed technology which allows precise placement of phosphorous donors in silicon.\cite{Schofield_2003} 

The same regime (low density, electron-doping) is also the most likely region for the possible appearance of true macroscopic ferromagnetism, as our calculations on the cluster constituents of large systems (in section \ref{secVaryDensityClusters}) reveal.  Obtaining a conclusive answer to this question numerically, however, requires going beyond the small sizes possible with exact diagonalization methods.  A possible route is to use more approximate methods such as density matrix or perturbative renormalization group methods in combination with other numerical techniques.  Even if true ferromagnetism on the macroscopic scale is absent, we have shown that there should be a significant asymmetry between the magnetic response of systems with excess electrons above the half-filled (uncompensated) case, and those with a deficit of electrons from the half-filled case, ({\it i.e.}~traditional compensated): the former should have a larger susceptibility in the paramagnetic phase at low temperatures.  This prediction can be experimentally tested by an experiment that uses gates to tune the electron density in a 2D layer.  Also, if ferromagnetism is attained on large enough length scales, it may show up as hysteresis in transport measurements due to magnetic domains. Clearly the temperature scales at which these ferromagnetic tendencies will manifest temselves will be much below the scales for diluted magnetic semiconductors like Galium Manganese Arsenide. This is due to several factors - (i) the energy scale for shallow impurities is low; (ii) the ferromagnetism
occurs only for large $U/\tInner$, \emph{i.e.}~low dopant and carrier densities where $J \sim \tInner^2/U$ is very small; and (iii) Nagaoka ferromagnetism in a Hubbard band is a much subtler effect involving two competing terms, compared with ferromagnetism arising out of a double exchange
mechanism. Nevertheless, the demonstration of high spin states and possible ferromagnetism in semiconductors doped with so-called "non-magnetic" shallow impurities would suggest that magnetism in semiconductors is not limited to semiconductors with transition metal elements, but is possible in a wider range of semiconductor based materials.

\section{AKNOWLEDGEMENTS}
This research was supported by NSF-MRSEC, Grant DMR-0213706 and DMR-0819860.

\appendix

\section{Clusters built from squares and triangles (one hopping parameter): detailed phase diagrams\label{appSingleHopDiagrams}}

The following table shows the ground state (T=0) phase diagrams of ``non ring-like'' clusters 4, 6, 8, and 9 of Fig.~\ref{figSingleHopSummary}.  The fixed electron number is given in the upper-right corner of each plot, and only non-trivial diagrams and their electron/hole pair are shown (\emph{i.e.} if the 1-electron diagram is non-trivial, the 1-hole diagram is shown for comparison).


\begin{widetext}
\begin{center}
\begin{tabular}{|c|cc|} \hline
Geometry & \multicolumn{2}{c|}{Non-trivial phase diagrams}  \\ \hline

\includegraphics[width=0.25in,angle=90]{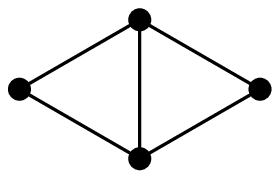} &
\parbox{2in}{\vspace{.1cm}\includegraphics[width=2in]{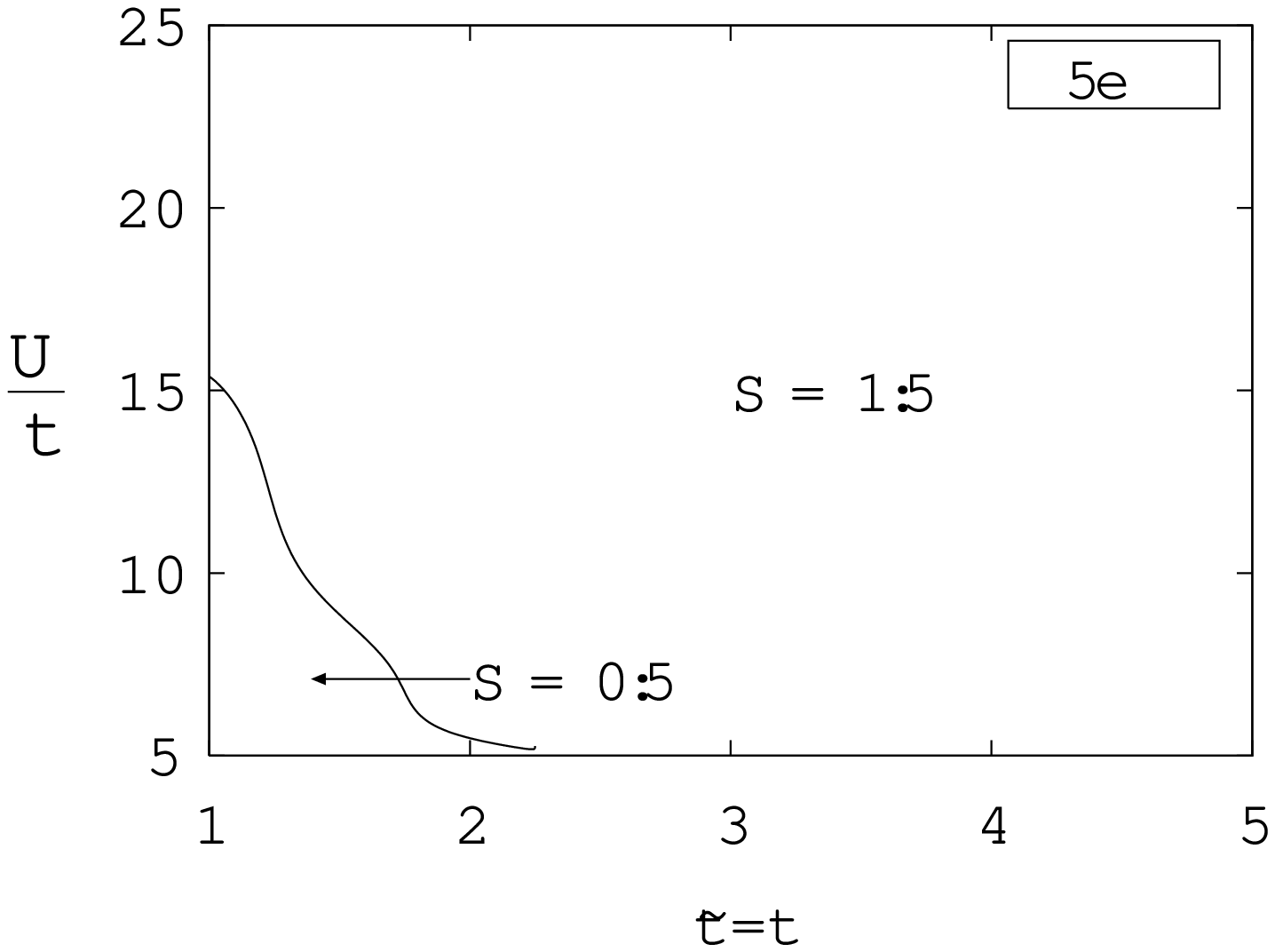}\vspace{.1cm}} &
\parbox{2in}{\vspace{.1cm}\includegraphics[width=2in]{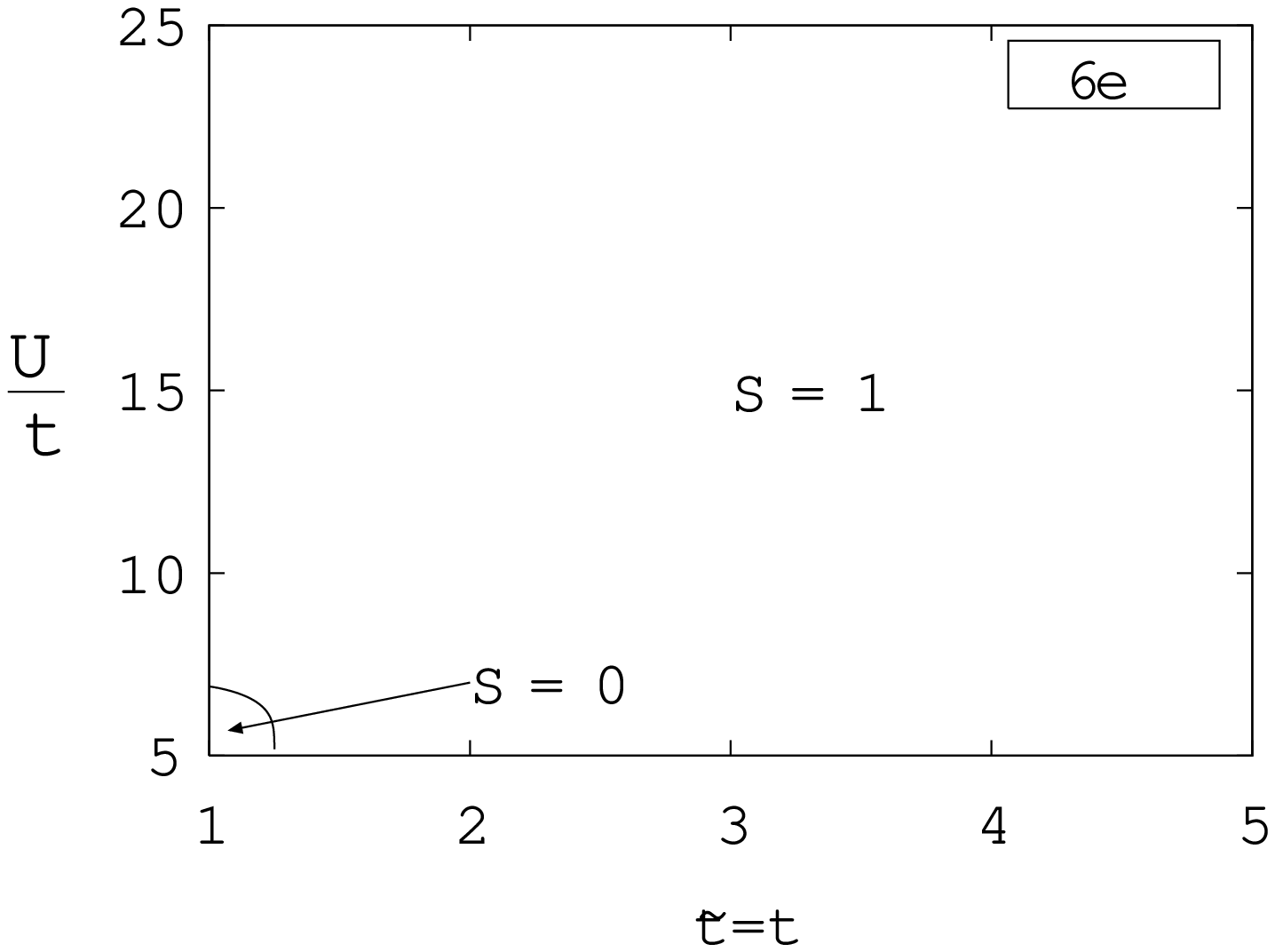}\vspace{.1cm}} \\ \hline

\raisebox{-1cm}[0cm][0cm]{\includegraphics[width=0.25in]{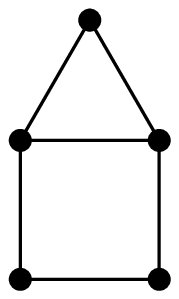}} &
\parbox{2in}{\vspace{.1cm}\includegraphics[width=2in]{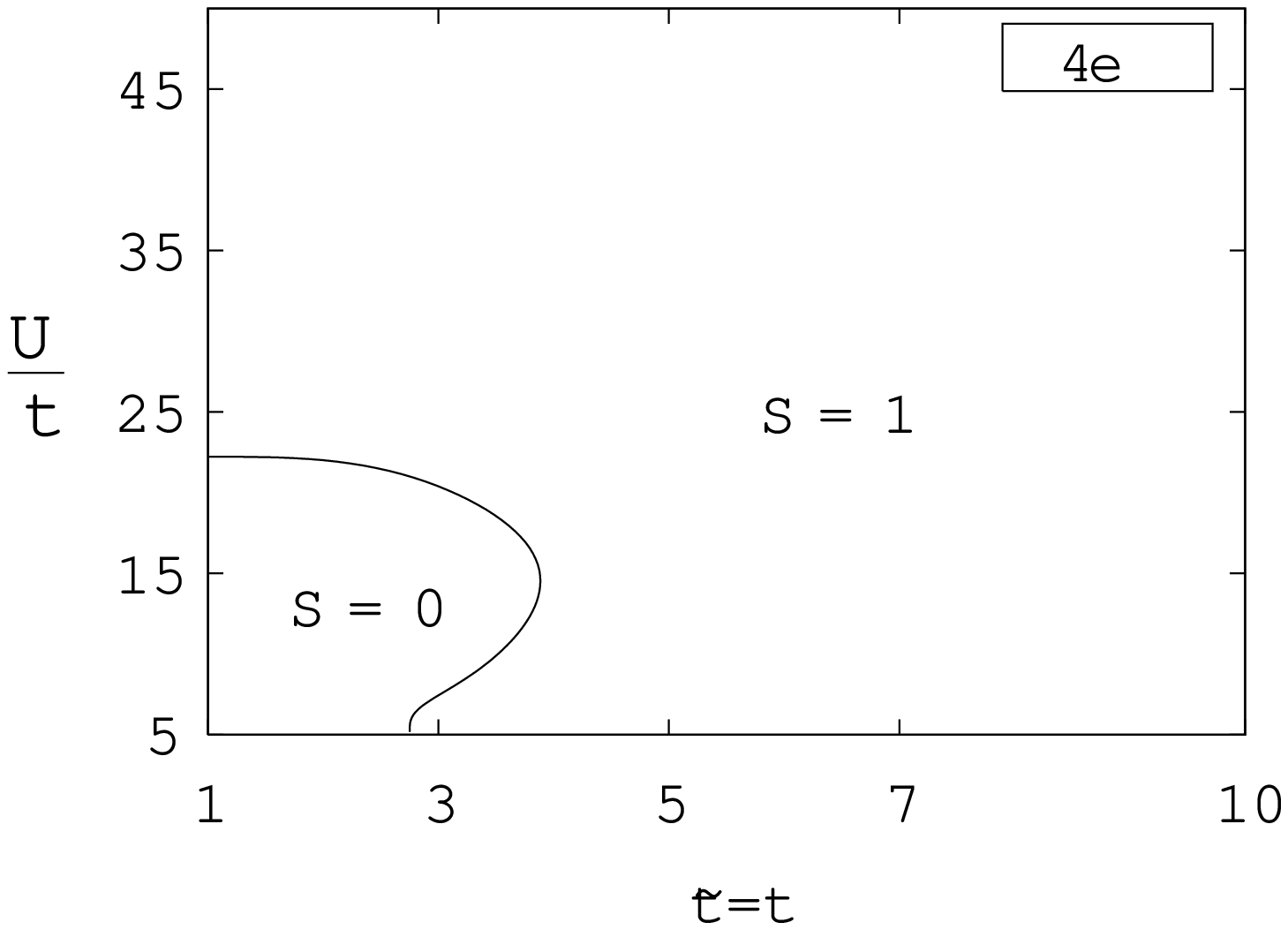}\vspace{.1cm}} &
\parbox{2in}{\vspace{.1cm}\includegraphics[width=2in]{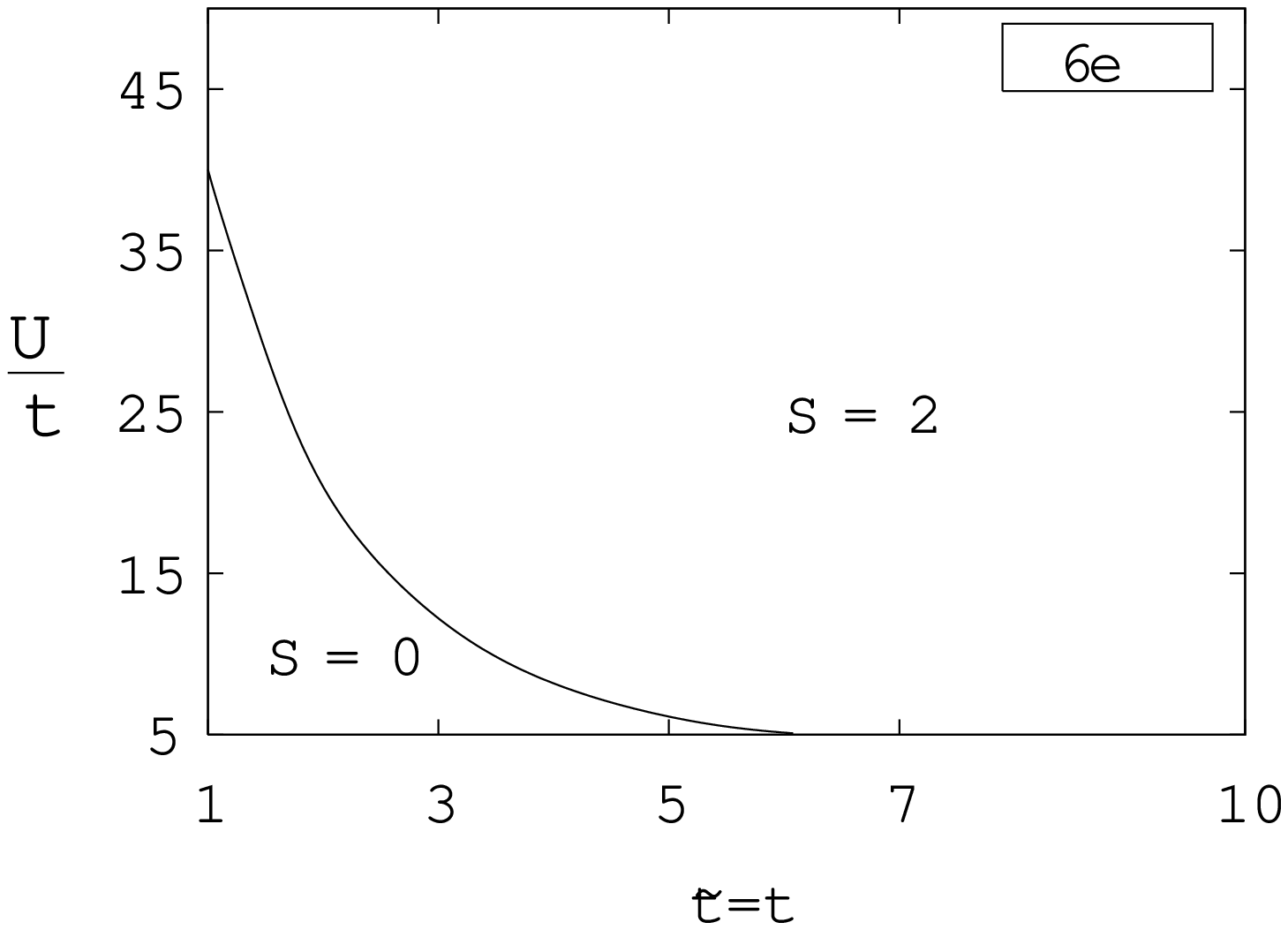}\vspace{.1cm}} \\ \hline

\raisebox{-1cm}{\includegraphics[width=0.5in]{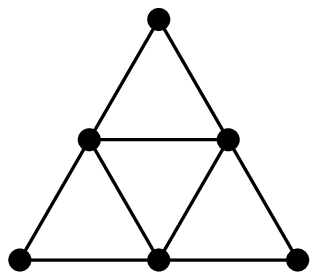}} &
\parbox{2in}{\vspace{.1cm}\includegraphics[width=2in]{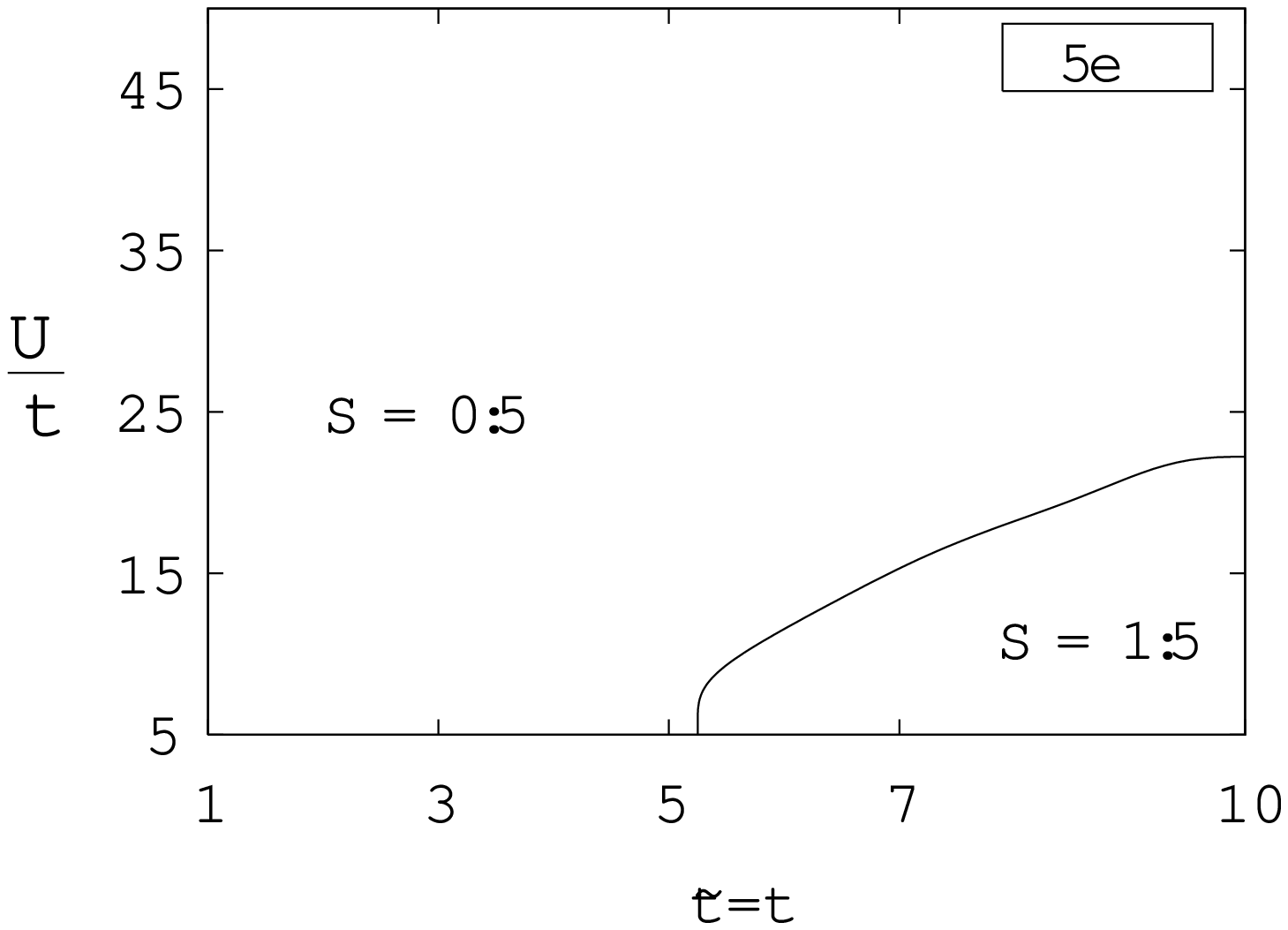}\vspace{.1cm}} &
\parbox{2in}{\vspace{.1cm}\includegraphics[width=2in]{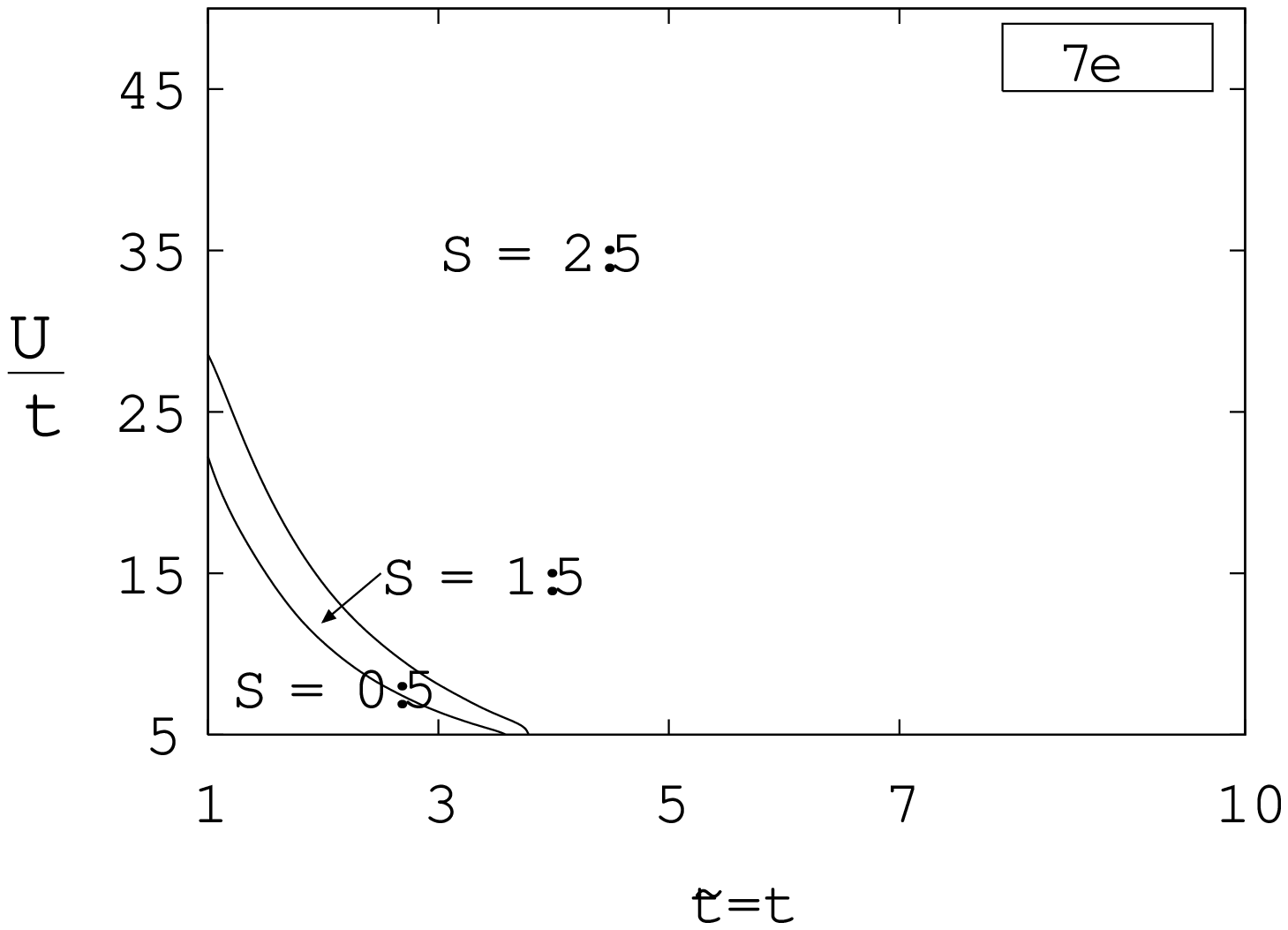}\vspace{.1cm}} \\ \hline

%
\end{tabular}

\begin{tabular}{|c|cc|} \hline
Geometry & \multicolumn{2}{c|}{Non-trivial phase diagrams}  \\ \hline
\raisebox{-3cm}[0cm][0cm]{\includegraphics[width=0.6in]{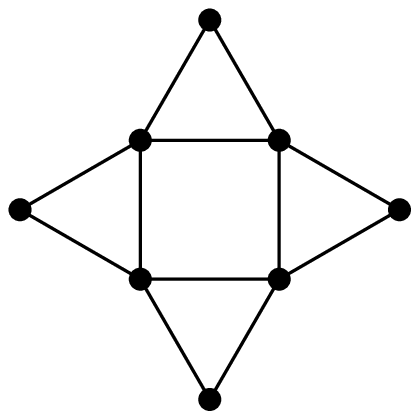}} &
\parbox{2in}{\vspace{.1cm}\includegraphics[width=2in]{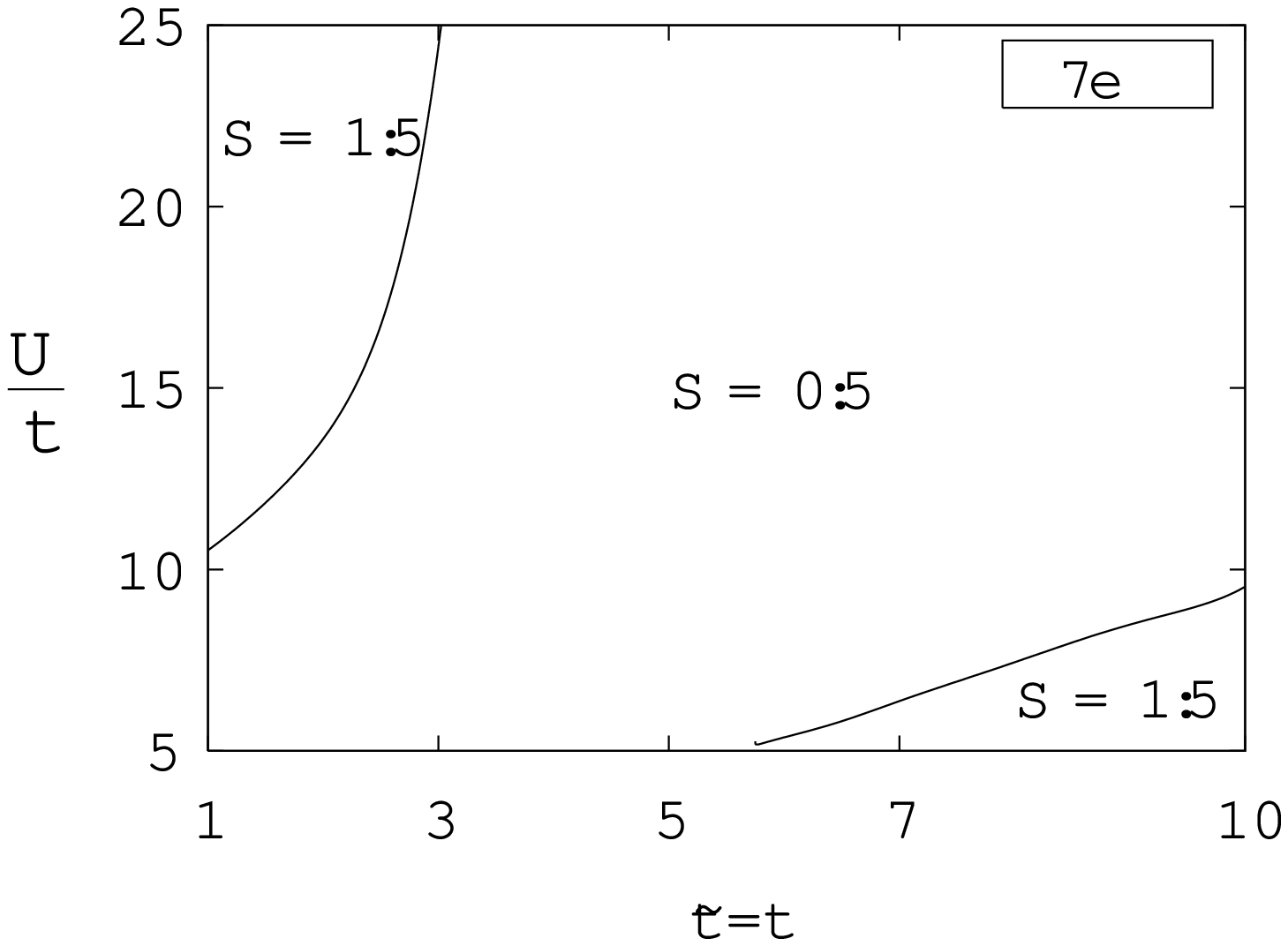}} &
\parbox{2in}{\vspace{.1cm}\includegraphics[width=2in]{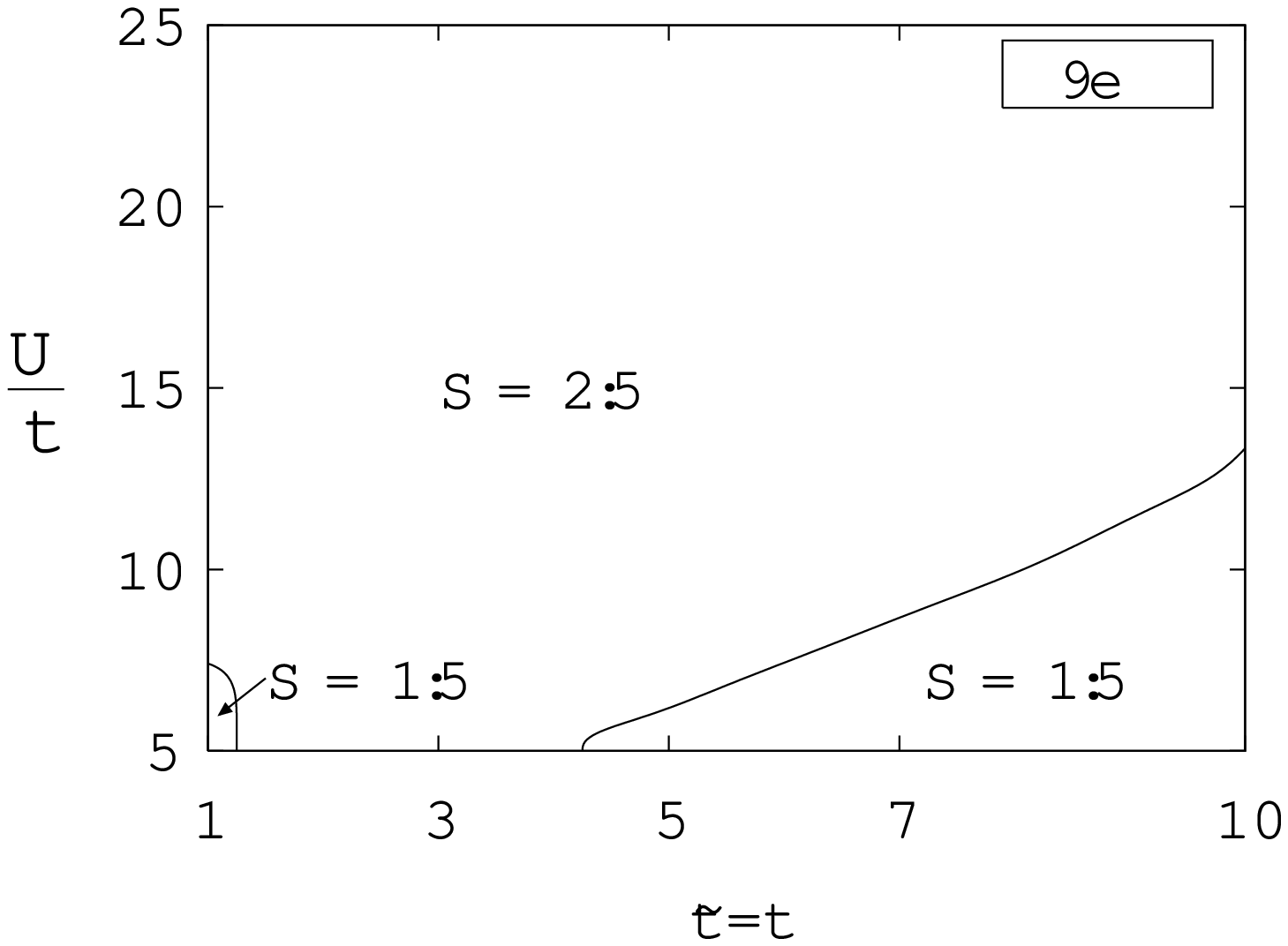}} \\
 
& \parbox{2in}{\includegraphics[width=2in]{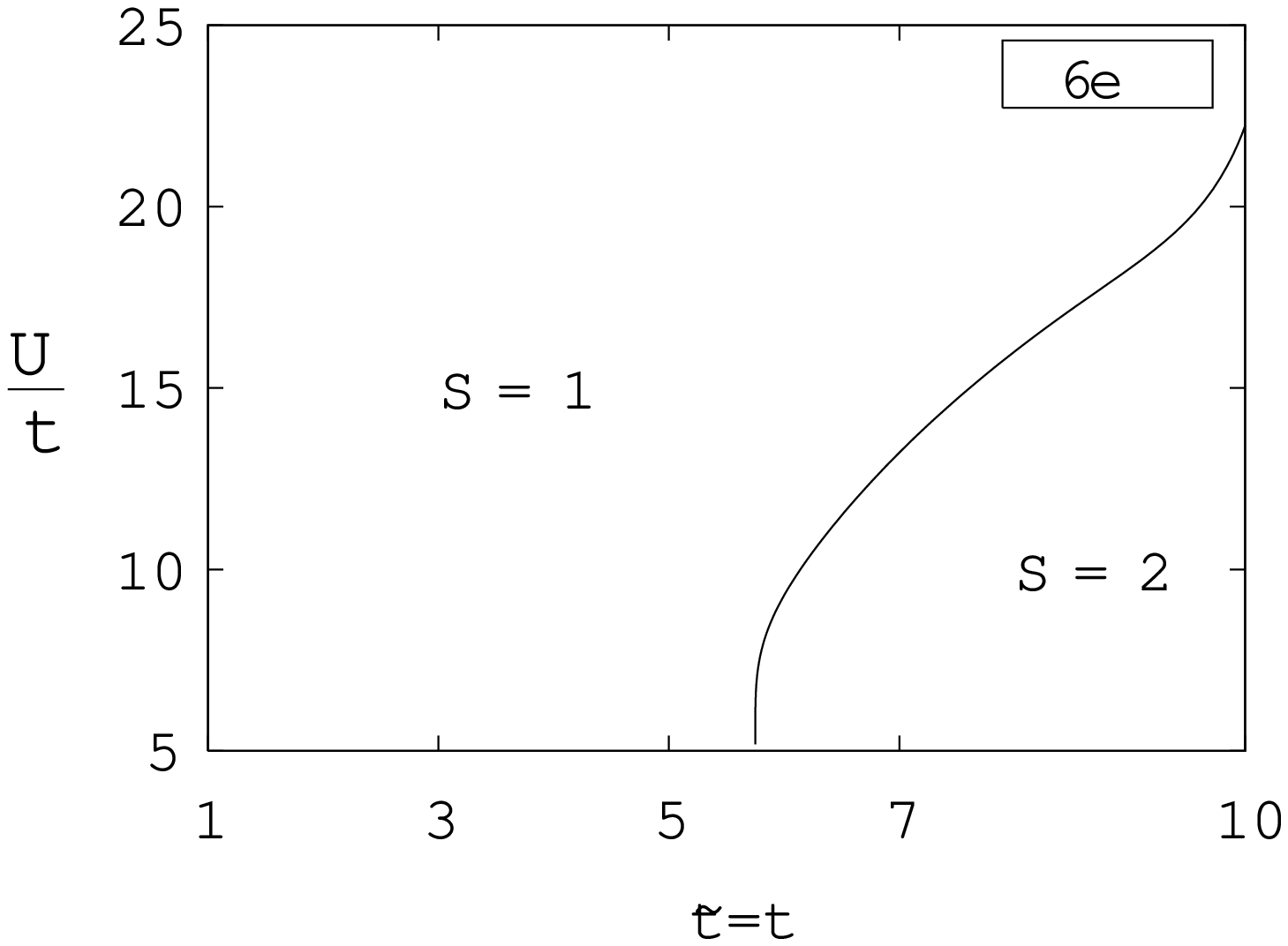}\vspace{.1cm}} &
  \parbox{2in}{\includegraphics[width=2in]{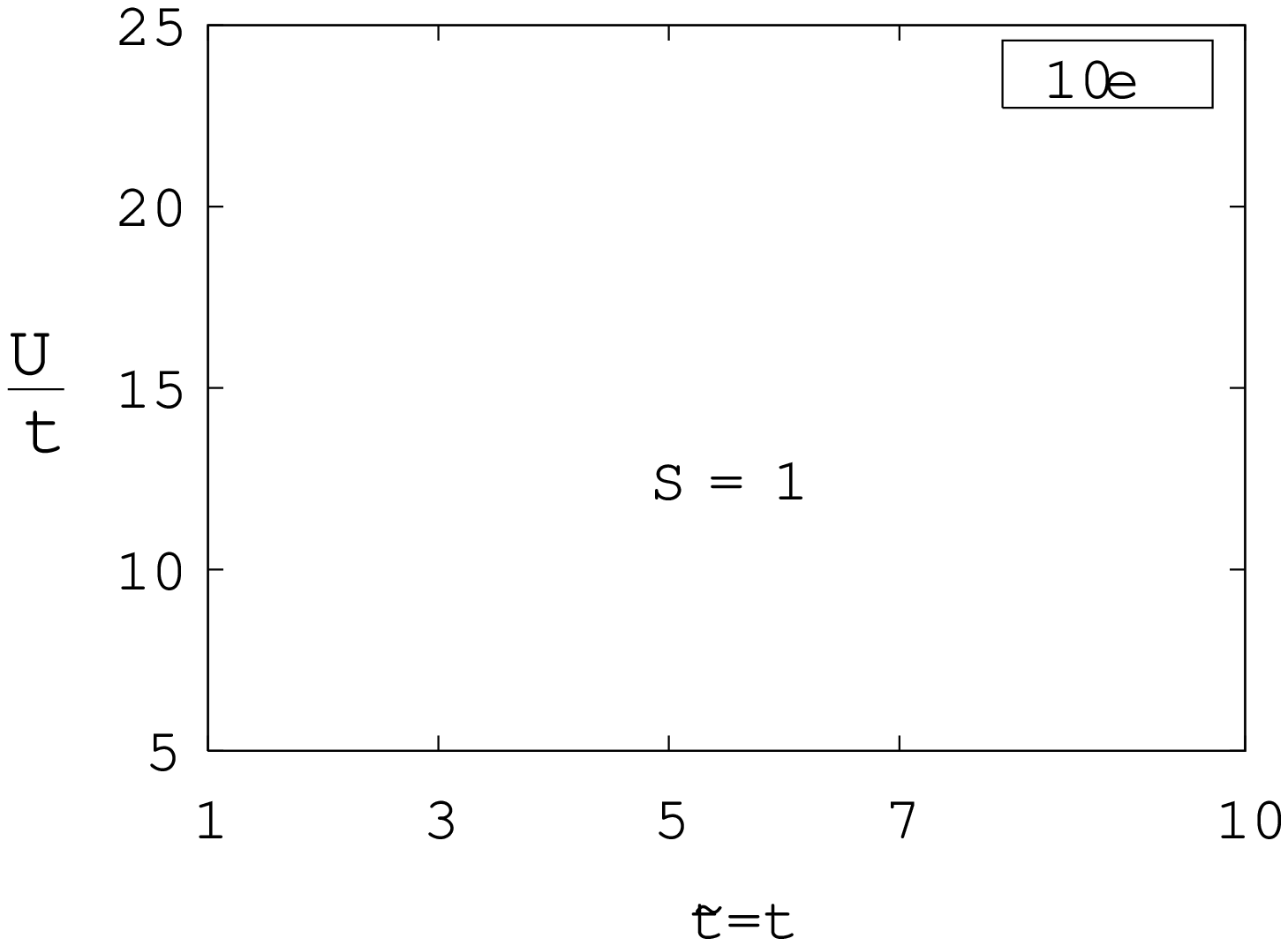}\vspace{.1cm}} \\ \hline
\end{tabular}

\end{center} 

\end{widetext}

\section{Clusters with two hopping parameters: detailed phase diagrams\label{appDoubleHopClusters}}

Here we present details of the ground state phase diagrams for the clusters listed in Fig.~\ref{figDoubleHopSummary}.  For each cluster with 0-2 electrons away from half filling (in either direction), we have computed phase diagrams for the parameter range $t_2/t_1 \in [1,10]$, $t_1/U \in [0.01,0.5]$, and  $\tOuter/\tInner = 1$, $5$, and $10$. Due to the large number of diagrams, we have divided them into three roughly defined categories:  trivial, simple, and complex.  \emph{Trivial} diagrams have no phase transitions (changes in ground state spin) within the considered parameter range.  \emph{Simple} diagrams have an abrupt transition between a single pair of ground state spin values, with very little parameter space where there are intermediate spin values (effectively, these diagrams contain a single phase boundary).  \emph{Complex} diagrams have substantial parameter space where the ground state spin ranges three or more values, so there is a sizeable region of itermediate spin.  

Diagrams categorized as either simple or complex are listed in table \ref{tblDiagramDescriptions} by cluster number (see Fig.~\ref{figDoubleHopSummary}) and charge $Q = \Nsites - \Nelec$ ($-Q$ is the doping relative to half filling).  The absence of data (for any cluster number 1-23 and $Q=-2,-1,0,+1,+2$) implies that the phase diagram is trivial, and the constant ground state spin can be found from Fig.~\ref{figDoubleHopSummary}.  For each simple diagram, table \ref{tblDiagramDescriptions} gives its two predominant spin values (min and max) and the equation of a line with approximates the phase boundary.  Full diagrams for these cases are given in Ref.~\onlinecite{NielsenThesis}.  Diagrams categorized as complex are indicated by a ``C'' in table \ref{tblDiagramDescriptions}, and are shown in their entirety in Figs.~\ref{figComplexDiagrams1}-\ref{figComplexDiagrams3}.


\begin{widetext}

\begin{table}
\hspace{-11cm}\begin{minipage}[b]{0.4\linewidth}
\begin{tabular}{|c|c|c|c|c|c|} \hline
\#$_cl$ & Q & $\tOuter/\tInner$ & $S_{min}$ & $S_{max}$ & Region where $S=S_{max}$ \\
\hline
1 & -1 & 1  & 0 & 1 & $t_2/t_1 > 0.0005$ \\ \cline{3-6}
  &    & 5  & 0 & 1 & $t_2/t_1 > 0.0002$ \\ \cline{3-6}
  &    & 10 & 0 & 1 & $t_2/t_1 > 0.0002$ \\ \cline{3-6}
\hline
2 & -1 & 1  & 0.5 & 1.5 & $t_2/t_1 > 0.0005$ \\ \cline{3-6}
  &    & 5  & 0.5 & 1.5 & $t_2/t_1 > 0.00025$ \\ \cline{3-6}
  &    & 10 & 0.5 & 1.5 & $t_2/t_1 > 0.0004$ \\ \cline{3-6}
\hline
4 & +1 & 1  & 0 & 1 & $t_2/t_1 < -4.6(t_1/U)+1.2$ \\
  &    &    &   &   & and $t_2/t_1 > 3.6(t_1/U)-0.2$ \\ \cline{2-6}
  & -1 & 1  & 1 & 2 & $t_1/U < 0.7$ \\ \cline{3-6}
  &    & 10 & 0 & 2 & $t_2/t_1 > 5(t_1/U) - 0.55$ \\
\hline
5 & +1 & 5  & 0 & 1 & $t_1/U > 0.8$ \\ \cline{3-6}
  &    & 10 & 0 & 1 & $t_1/U > 0.05$ \\
\hline
6 & +1 & 5  & 0 & 1 & $t_1/U > 0.15$ \\ \cline{3-6}
  &    & 10 & 0 & 1 & $t_1/U > 0.07$ \\ \cline{2-6}
  & -1 & 1  & 0 & 1 & $t_2/t_1 < 0.5$ \\ \cline{3-6}
  &    & 5  & 0 & 1 & $t_2/t_1 < 0.55$ \\ \cline{3-6}
  &    & 10 & 0 & 1 & $t_2/t_1 < 0.4$ \\
\hline
7 & +1 & 1  & 0 & 1 & $t_2/t_1 > 17(t_1/U) + 0.28$ \\ \cline{3-6}
  &    & 5  & 0 & 1 & $t_2/t_1 > 15(t_1/U) + 0.30$ \\ \cline{3-6}
  &    & 10 & 0 & 1 & $t_2/t_1 > 12(t_1/U) + 0.33$ \\ \cline{2-6}
  & -1 & 1  & 0 & 2 & C \\ \cline{3-6}
  &    & 5  & 1 & 2 & $t_2/t_1 > 0.3$ \\ \cline{3-6}
  &    & 10 & 1 & 2 & $t_2/t_1 > 0.2$ \\
\hline
8 & +1 & 5  & 0 & 1 & $t_1/U > 0.1$ \\ \cline{3-6}
  &    & 10 & 0 & 1 & $t_1/U > 0.04$ \\ \cline{2-6}
  & -2 & 1  & 0.5 & 1.5 & $t_1/U < 0.03$ \\
\hline
9 & +2 & 5  & 0 & 1 & $t_1/U > 0.18$ \\ \cline{3-6}
  &    & 10 & 0 & 1 & $t_1/U > 0.07$ \\
\hline
10 & +2 & 10 & 0 & 1 & $t_1/U > 0.07$ \\
\hline
11 & +2 & 5  & 0 & 1 & $t_1/U > 0.4$ \\ \cline{3-6}
   &    & 10 & 0 & 1 & $t_1/U > 0.05$ \\ \cline{2-6}
   & +1 & 5  & 0.5 & 1.5 & M \\ \cline{3-6}
   &    & 10 & 0.5 & 1.5 & $t_1/U > 0.03$ \\ \cline{2-6}
   & -1 & 1  & 0.5 & 1.5 & M \\ \cline{3-6}
   &    & 5  & 0.5 & 1.5 & M \\ \cline{3-6}
   &    & 10 & 0.5 & 1.5 & $t_1/U > 0.1$ \\ \cline{2-6}
   & -2 & 1  & 0 & 2 & C \\
\hline
12 & -1 & 1  & 0.5 & 2.5 & M \\ \cline{3-6}
   &    & 5  & 1.5 & 2.5 & M \\ \cline{3-6}
   &    & 10 & 1.5 & 2.5 & $t_2/t_1 > 0.12$ \\
\hline
13 & +2 & 10 & 0 & 1 & $t_1/U > 0.06$ \\ \cline{2-6}
   & +1 & 10 & 0.5 & 1.5 & $t_1/U > 0.05$ \\ \cline{2-6}
   & -1 & 1  & 0.5 & 2.5 & C \\ \cline{3-6}
   &    & 5  & 1.5 & 2.5 & $t_2/t_1 > 3.5(t_1/U) - 0.03$ \\ \cline{3-6}
   &    & 10 & 1.5 & 2.5 & $t_2/t_1 > 3(t_1/U) - 0.03$ \\ \cline{3-6}
\hline
14 & +1 & 5  & 0.5 & 1.5 & $t_2/t_1 < 0.5$ and $t_1/U > 0.06$ \\ \cline{3-6}
   &    & 10 & 0.5 & 1.5 & $t_1/U > 0.03$ \\ \cline{2-6}  
   & -1 & 1  & 0.5 & 2.5 & $t_2/t_1 > 14(t_1/U) + 0.16$ \\ \cline{3-6}
   &    & 5  & 0.5 & 2.5 & $t_2/t_1 > 3.5(t_1/U) + 0.1$ \\ \cline{3-6}
   &    & 10 & 0.5 & 2.5 & $t_2/t_1 > 2(t_1/U) + 0.08$ \\
\hline
15 & +2 & 5  & 0.5 & 1.5 & $t_2/t_1 < 0.25$ and $0.06 < t_1/U < 0.14$ \\ \cline{3-6}
   &    & 10 & 0.5 & 1.5 & $t_2/t_1 > 0.5$ and $t_1/U > 0.07$ \\ \cline{2-6} 
   & +1 & 5  & 0 & 1 & $t_2/t_1 < 0.25$ and $0.05 < t_1/U < 0.12$ \\ \cline{3-6}
   &    & 10 & 0 & 2 & C \\ 
\hline
\end{tabular}
\end{minipage}

\vspace{-20.55cm}

\hspace{8cm}\begin{minipage}[b]{0.4\linewidth}
\begin{tabular}{|c|c|c|c|c|c|} \hline
\#$_cl$ & Q & $\tOuter/\tInner$ & $S_{min}$ & $S_{max}$ & Region where $S=S_{max}$ \\
\hline
  
   & -1 & 1  & 0 & 3 & C \\ \cline{3-6}
   &    & 5  & 1 & 3 & $t_2/t_1 > 5.5(t_1/U) + 0.07$ \\ \cline{3-6}
   &    & 10 & 1 & 3 & $t_2/t_1 > 6.6(t_1/U) + 0.004$ \\
\hline
16 & -1 & 10 & 0 & 1 & $t_1/U > 0.13$ \\
\hline
17 & +1 & 5  & 0 & 1 & $t_1/U > 0.11$ \\ \cline{3-6}
   &    & 10 & 0 & 1 & $t_1/U > 0.05$ \\ \cline{2-6}
   & -1 & 5  & 0 & 1 & $t_2/t_1 < 0.1$ and $t_1/U > 0.1$ \\ \cline{3-6}
   &    & 10 & 0 & 1 & $t_2/t_1 < 0.2$ and $t_1/U > 0.05$ \\
\hline
18 & +2 & 5  & 0 & 1 & $t_1/U > 0.14$ \\ \cline{3-6}
   &    & 10 & 0 & 1 & $t_1/U > 0.05$ \\ \cline{2-6}
   & -1 & 5  & 0.5 & 3.5 & C \\ \cline{3-6}
   &    & 10 & 0.5 & 3.6 & C \\ \cline{2-6}
   & -2 & 5  & 0 & 2 & C \\ \cline{3-6}
   &    & 10 & 0 & 2 & C \\
\hline
19 & +2 & 5  & 0 & 1 & $t_1/U > 0.2$ \\ \cline{3-6}
   &    & 10 & 0 & 1 & $t_1/U > 0.1$ \\ \cline{2-6}
   & -1 & 1  & 0.5 & 1.5 & $t_2/t_1 < -1.7(t_1/U) + 1.1$ \\ 
\hline
20 & +2 & 1  & 0 & 1 & $t_2/t_1 > 15(t_1/U) + 0.15$ \\ \cline{3-6}
   &    & 5  & 0 & 1 & $t_2/t_1 > 15(t_1/U) + 0.15$ \\ \cline{3-6}
   &    & 10 & 0 & 1 & $t_2/t_1 > 10(t_1/U) + 0.2$ \\ \cline{2-6}
   & +1 & 1  & 0.5 & 1.5 & M \\ \cline{3-6}
   &    & 5  & 0.5 & 1.5 & M \\ \cline{3-6}
   &    & 10 & 0.5 & 1.5 & $t_2/t_1 > 0.4$ and $t_1/U < 0.025$ \\ \cline{2-6}
   & -1 & 1  & 0.5 & 3.5 & M \\ \cline{3-6}
   &    & 5  & 0.5 & 3.5 & M \\ \cline{3-6}
   &    & 10 & 0.5 & 3.5 & $t_2/t_1 > 0.2$ \\ \cline{2-6}
   & -2 & 1  & 0 & 3 & C \\ \cline{3-6}
   &    & 5  & 0 & 3 & C \\ \cline{3-6}
   &    & 10 & 0 & 3 & $t_2/t_1 > 0.1$ \\
\hline
21 & +2 & 1  & 0 & 1 & $t_2/t_1 < 23(t_1/U) + 0.13$ \\ \cline{3-6}
   &    & 5  & 0 & 1 & $t_2/t_1 < 23(t_1/U) + 0.13$ \\ \cline{3-6}
   &    & 10 & 0 & 1 & $t_2/t_1 < 18(t_1/U) + 0.17$ \\ \cline{2-6}
   & -1 & 1  & 0.5 & 3.5 & C \\ \cline{3-6}
   &    & 5  & 1.5 & 3.5 & $t_2/t_1 > 0.2$ \\ \cline{3-6}
   &    & 10 & 1.5 & 3.5 & $t_2/t_1 > 0.15$ \\ \cline{2-6}
   & -2 & 1  & 0 & 1 & $t_2/t_1 > 15(t_1/U) + 0.2$ \\ \cline{3-6}
   &    & 5  & 0 & 3 & C \\ \cline{3-6}
   &    & 10 & 0 & 3 & C \\
\hline
22 & -1 & 1  & 0 & 4 & C \\ \cline{3-6}
   &    & 5  & 1 & 4 & C \\ \cline{3-6}
   &    & 10 & 0 & 4 & C \\
\hline
23 & +2 & 1  & 0.5 & 1.5 & $0.3 < t_2/t_1 < 0.6$ and $t_1/U < 0.17$ \\ \cline{3-6}
   &    & 5  & 0.5 & 1.5 & $t_2/t_1 < -6(t_1/U) + 1.15$ \\
   &    &    &     &     & and $t_1/U < 1.3$ \\ \cline{3-6}
   &    & 10 & 0.5 & 1.5 & $t_1/U < 0.05$ \\ \cline{2-6}
   & +1 & 1  & 0 & 1 & $t_2/t_1  < 0.3$ and $t_1/U < 0.04$ \\ \cline{3-6}
   &    & 5  & 0 & 1 & $t_1/U < 0.05$ and $t_1/U > 0.14$ \\ \cline{3-6}
   &    & 10 & 0 & 1 & $t_2/t_1 > 0.17$ \\ \cline{2-6}
   & -1 & 1  & 0 & 2 & C \\ \cline{3-6}
   &    & 5  & 0 & 3 & C \\ \cline{3-6}
   &    & 10  & 0 & 3 & C \\ \cline{2-6}
   & -2 & 1  & 0.5 & 2.5 & C \\ \cline{3-6}
   &    & 5  & 0.5 & 3.5 & C \\ \cline{3-6}
   &    & 10  & 0.5 & 3.5 & C \\
\hline

\end{tabular}
\end{minipage}
\caption{Description of all non-trivial phase diagrams for clusters in Fig.~\ref{figDoubleHopSummary}. Cluster geometry is referenced by \#$_{cl}$ (see Fig.~\ref{figDoubleHopSummary}), and $Q$ is the clusters charge.  $\tOuter/\tInner$ is the ratio of both pairs of kinetic parameters, \emph{i.e.}~$\tOuter/\tInner = \tOuter_1/\tInner_1 = \tOuter_2/\tInner_2$.  In each case, the minimal and maximal spin, $S_{min}$ and $S_{max}$ are given, as well as a \emph{rough} approximation of the region of maximal spin.  If there is a substantial region of the phase space under consideration (see text) where the spin lies between $S_{min}$ and $S_{max}$, then a ``C'' (for ``complex'' classification) is placed in the final column, and the diagram is shown in Figs.~\ref{figComplexDiagrams1}-\ref{figComplexDiagrams3}.  If the phase diagram is given in the main text, regardless of its complexity, an ``M'' is placed in the final column.\label{tblDiagramDescriptions}}
\end{table}
 
\begin{figure}[b]
\begin{tabular}{cc}

\parbox[t]{0.5\linewidth}{
\begin{tabular}{|c|c|c|c|} \hline
 \#$_{cl}$ & Q & $\tOuter/\tInner$ & Phase Diagram \\
\hline
7 & -1 & 1 & \parbox{2.3in}{\vspace{0.1cm}\includegraphics[width=2.2in]{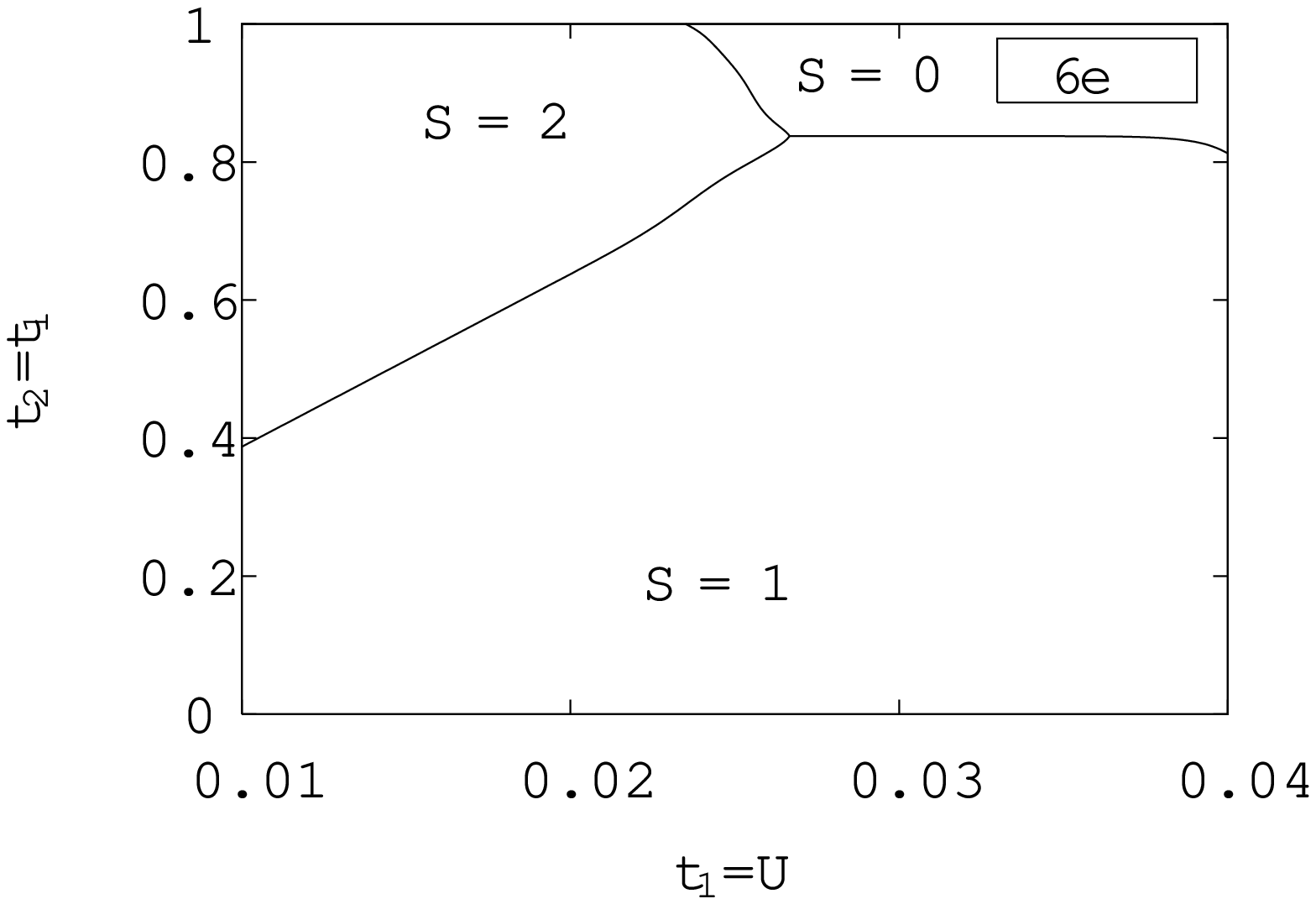}\vspace{0.1cm}} \\ \hline
11 & -2 & 1 & \parbox{2.3in}{\vspace{0.1cm}\includegraphics[width=2.2in]{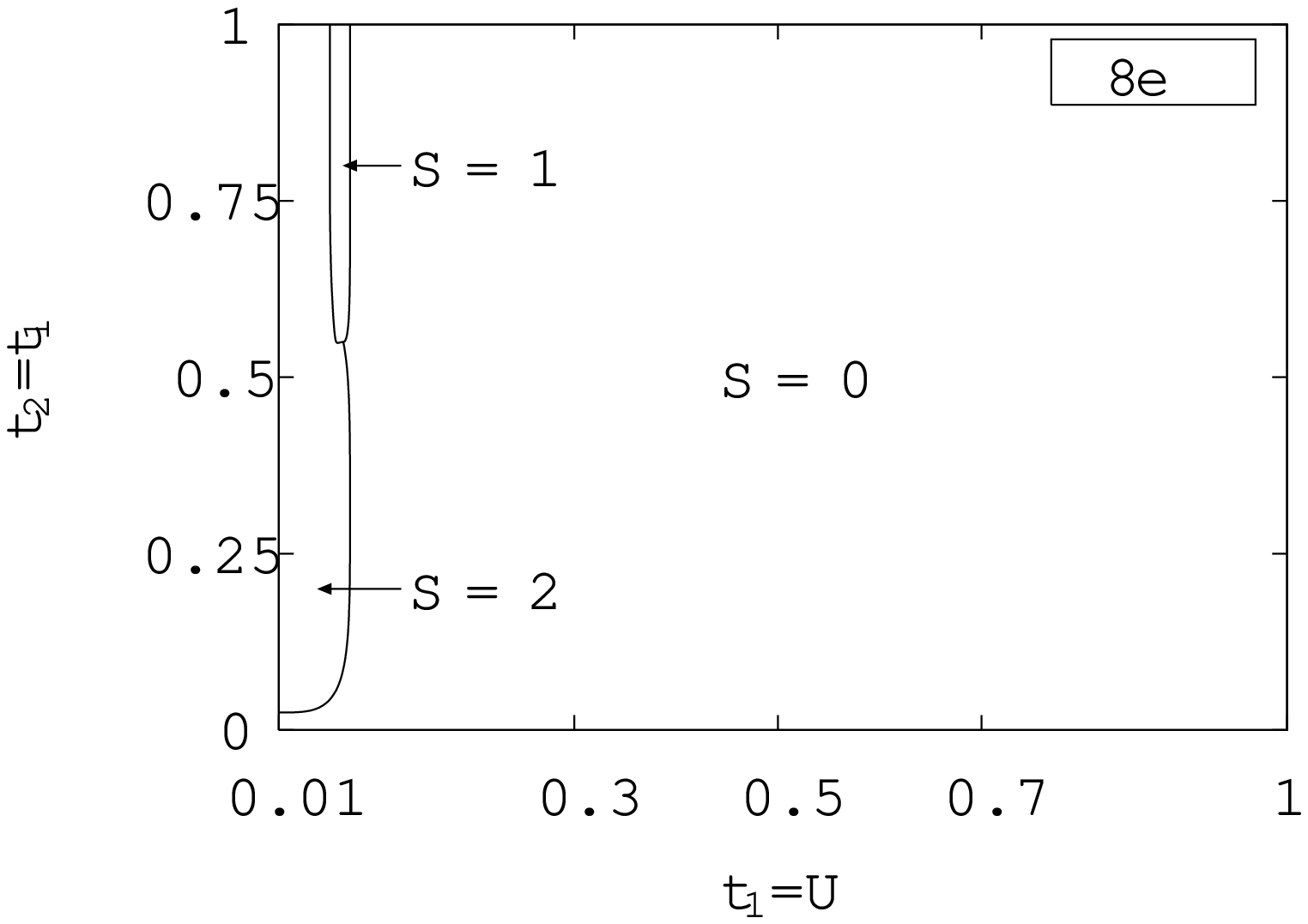}\vspace{0.1cm}} \\ \hline
13 & -1 & 1 &  \parbox{2.3in}{\vspace{0.1cm}\includegraphics[width=2.2in]{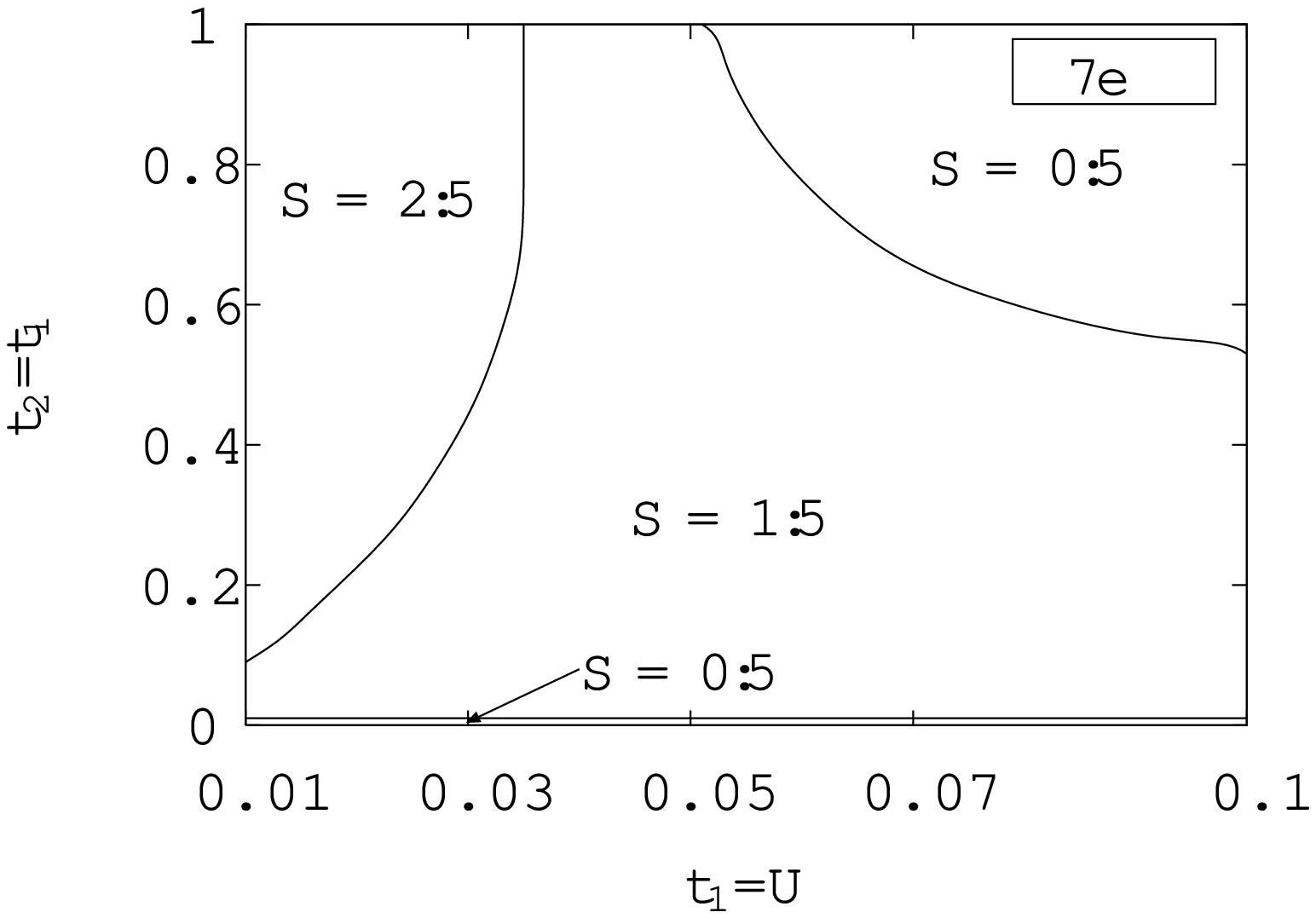}\vspace{0.1cm}} \\ \hline
15 & +1 & 10 & \parbox{2.3in}{\vspace{0.1cm}\includegraphics[width=2.2in]{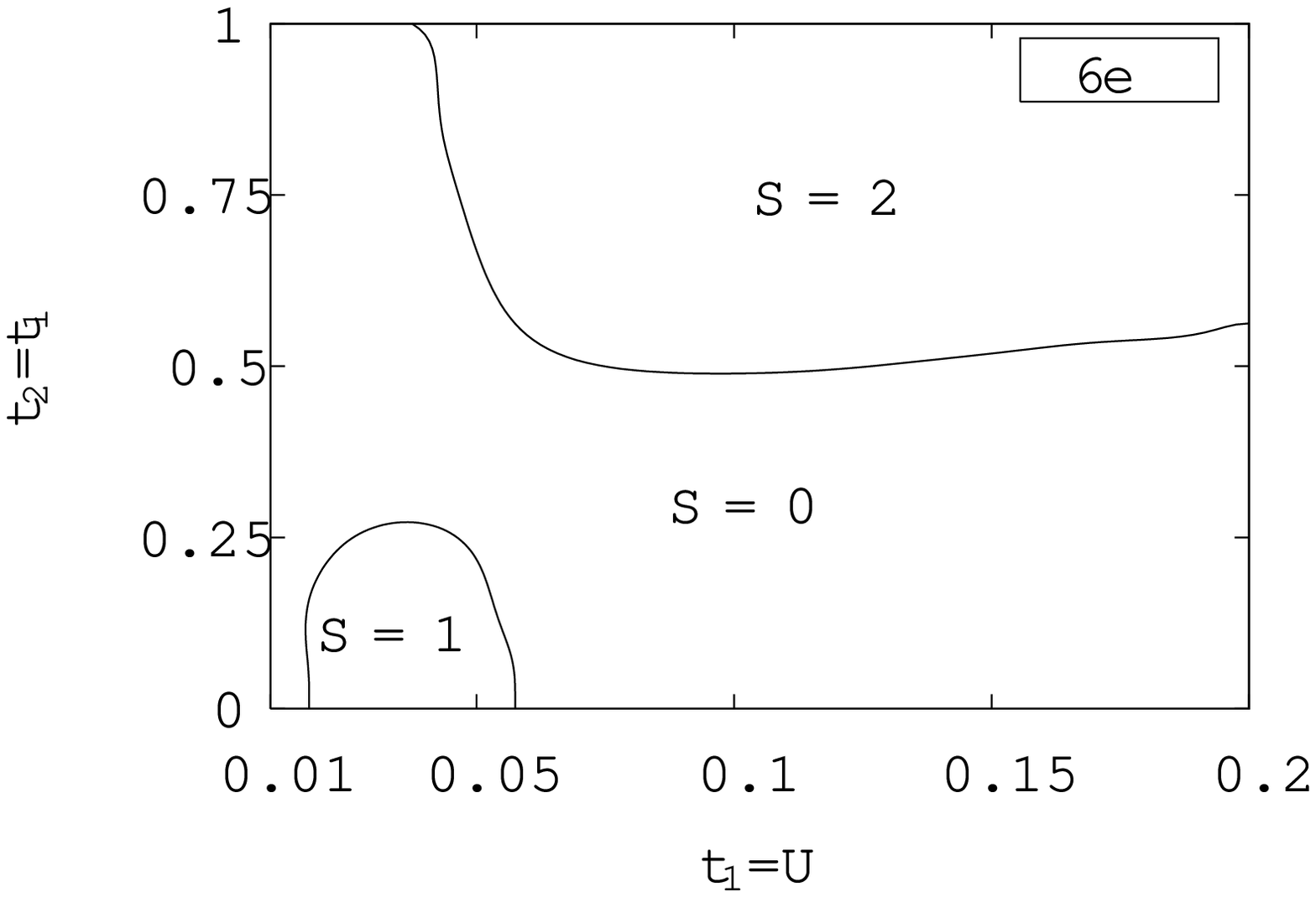}\vspace{0.1cm}} \\ \cline{2-4}
   & -1 & 1 &  \parbox{2.3in}{\vspace{0.1cm}\includegraphics[width=2.2in]{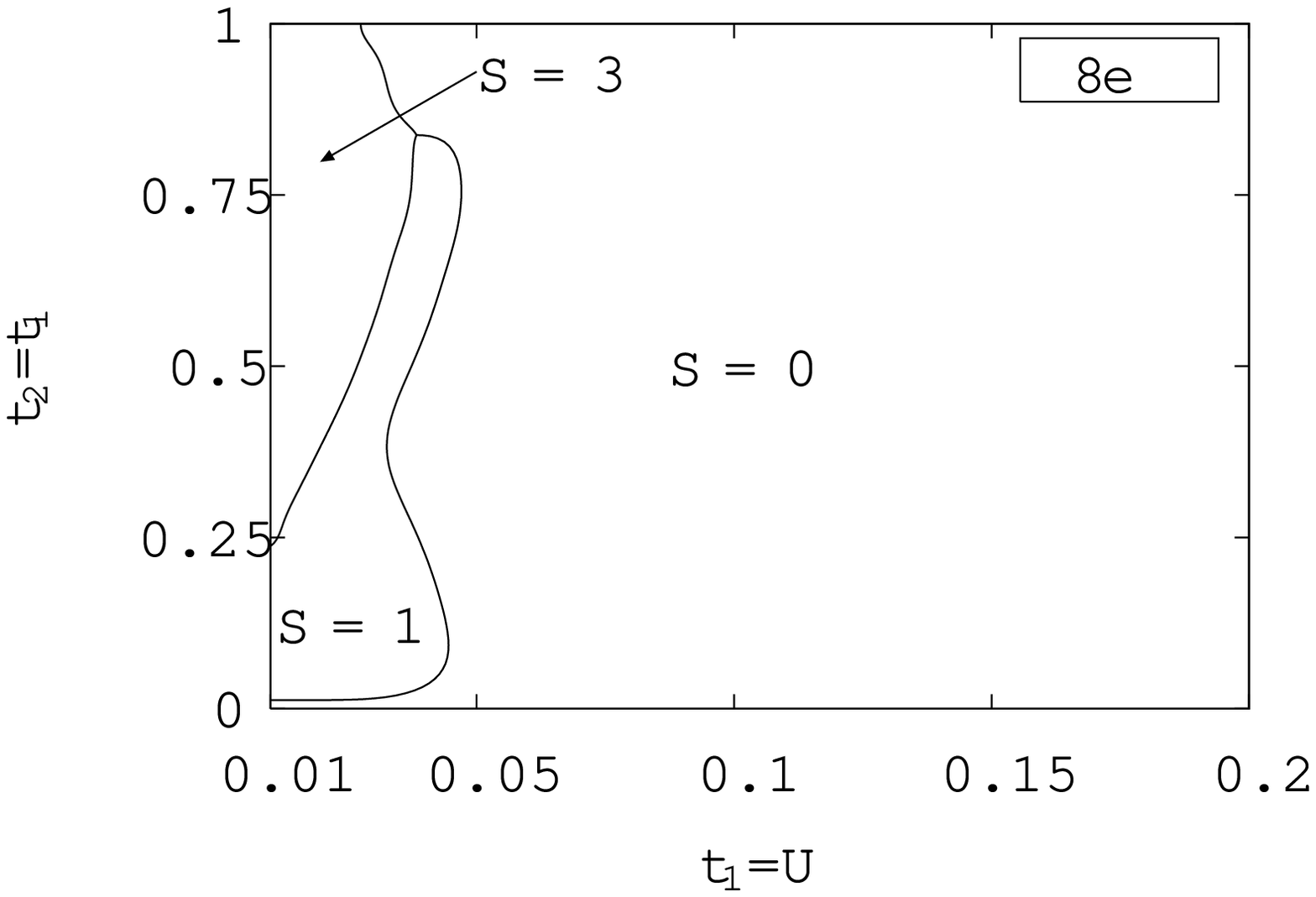}\vspace{0.1cm}} \\ \hline
\end{tabular}}
&
\parbox[t]{0.5\linewidth}{
\begin{tabular}{|c|c|c|c|} \hline
 \#$_{cl}$ & Q & $\tOuter/\tInner$ & Phase Diagram \\
\hline
18 & -1 & 5 & \parbox{2.3in}{\vspace{0.1cm}\includegraphics[width=2.2in]{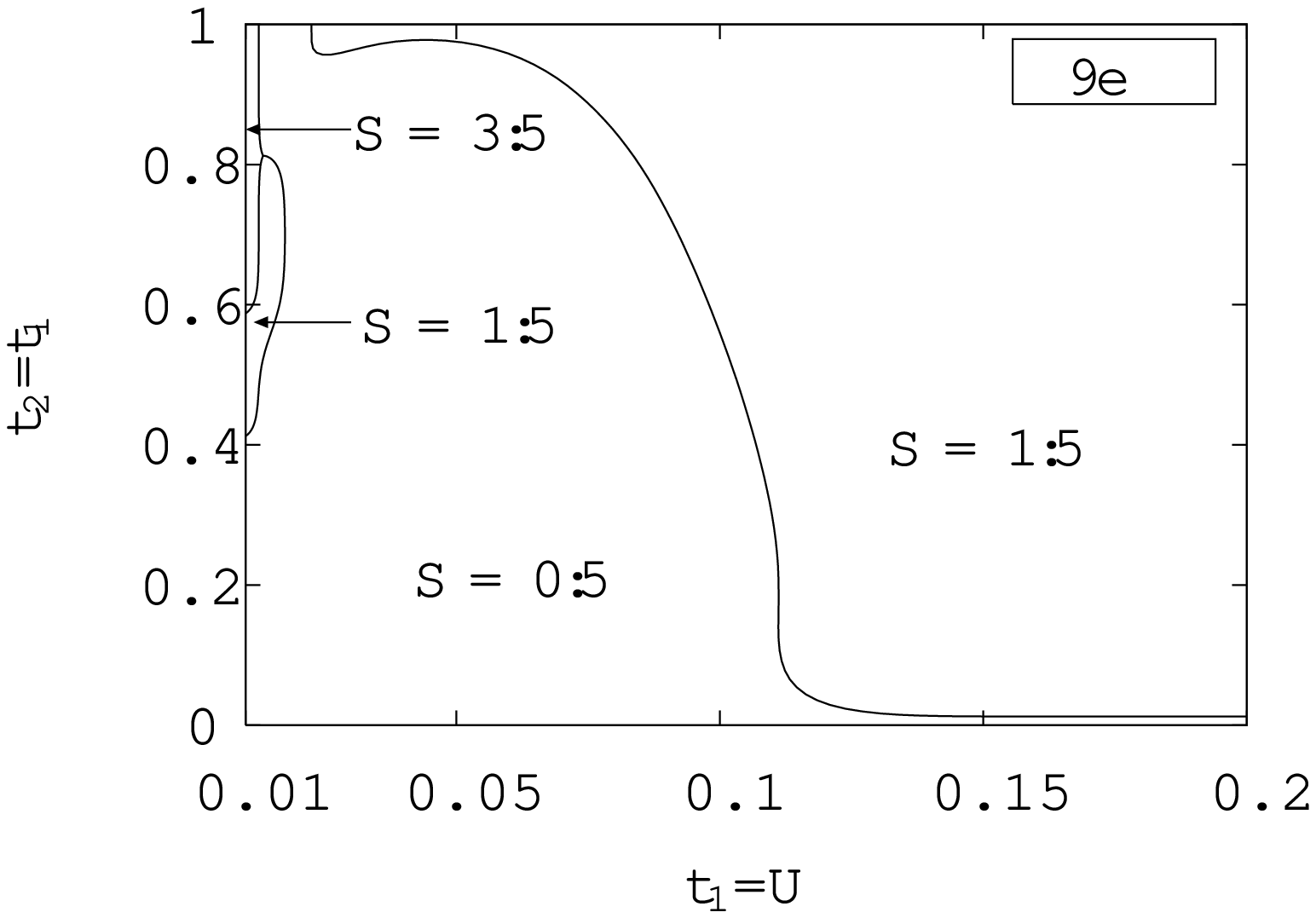}\vspace{0.1cm}} \\ \cline{3-4}
   &    & 10 & \parbox{2.3in}{\vspace{0.1cm}\includegraphics[width=2.2in]{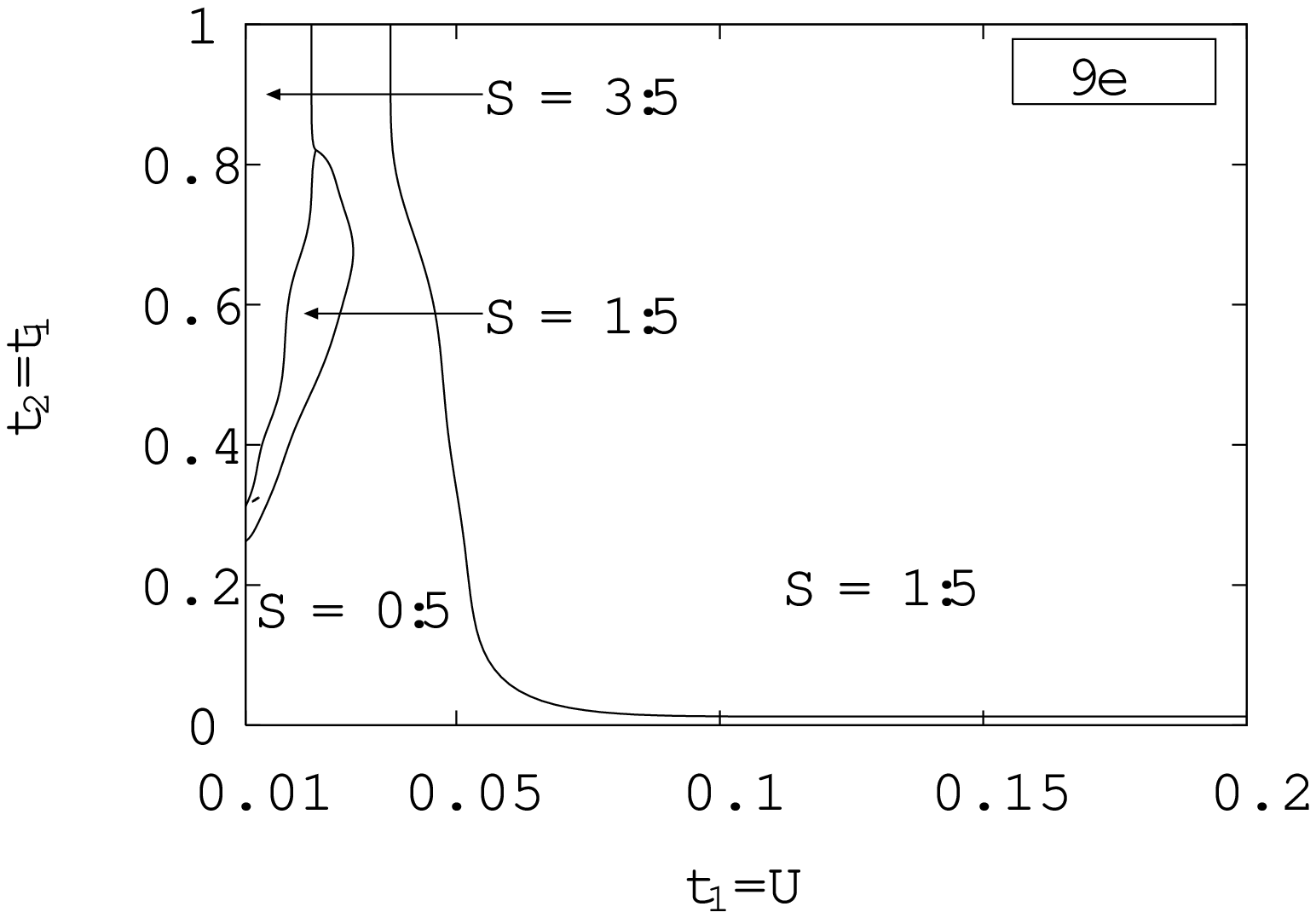}\vspace{0.1cm}} \\ \cline{2-4}
   & -2 & 5  & \parbox{2.3in}{\vspace{0.1cm}\includegraphics[width=2.2in]{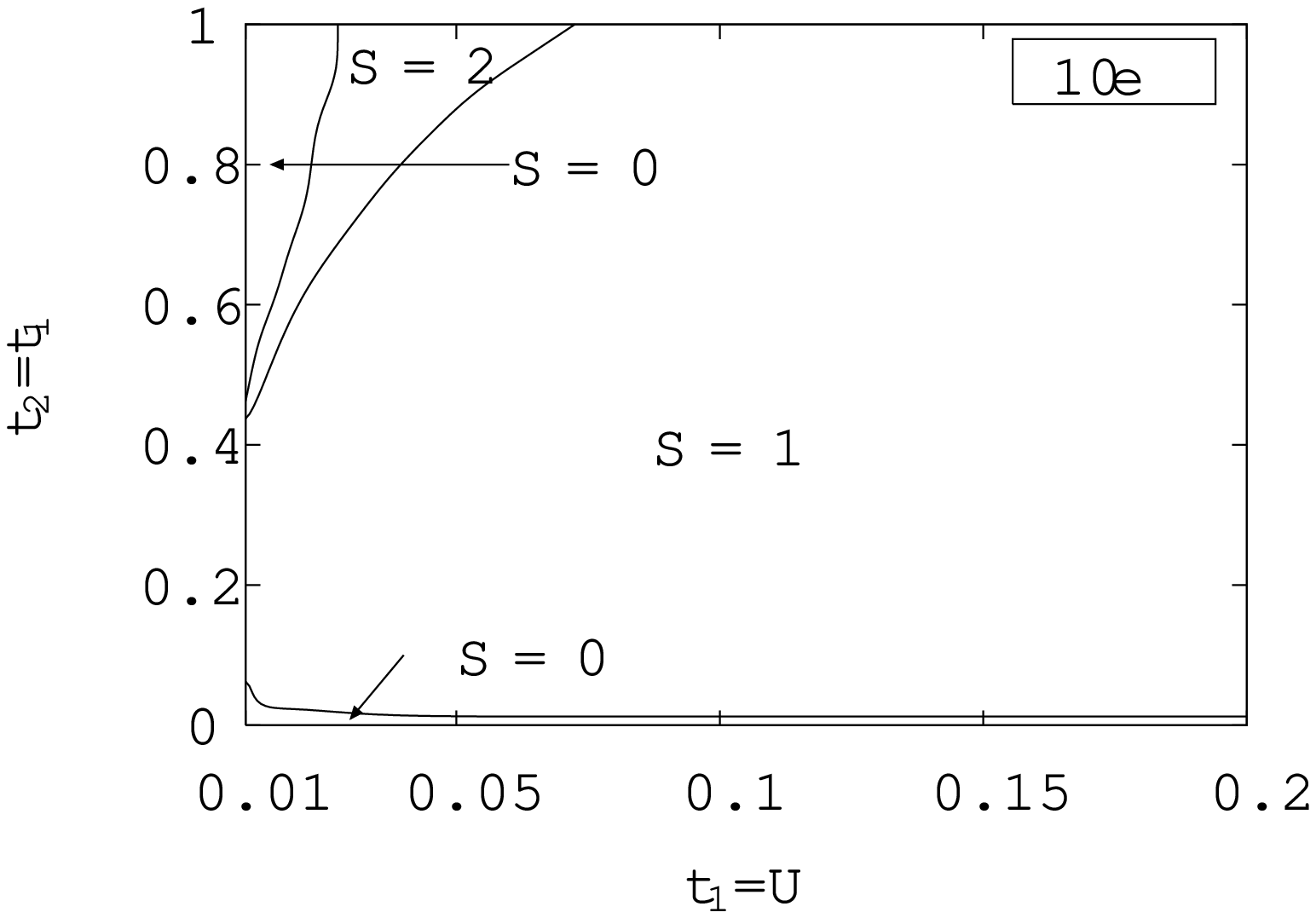}\vspace{0.1cm}} \\ \cline{3-4}
   &    & 10 & \parbox{2.3in}{\vspace{0.1cm}\includegraphics[width=2.2in]{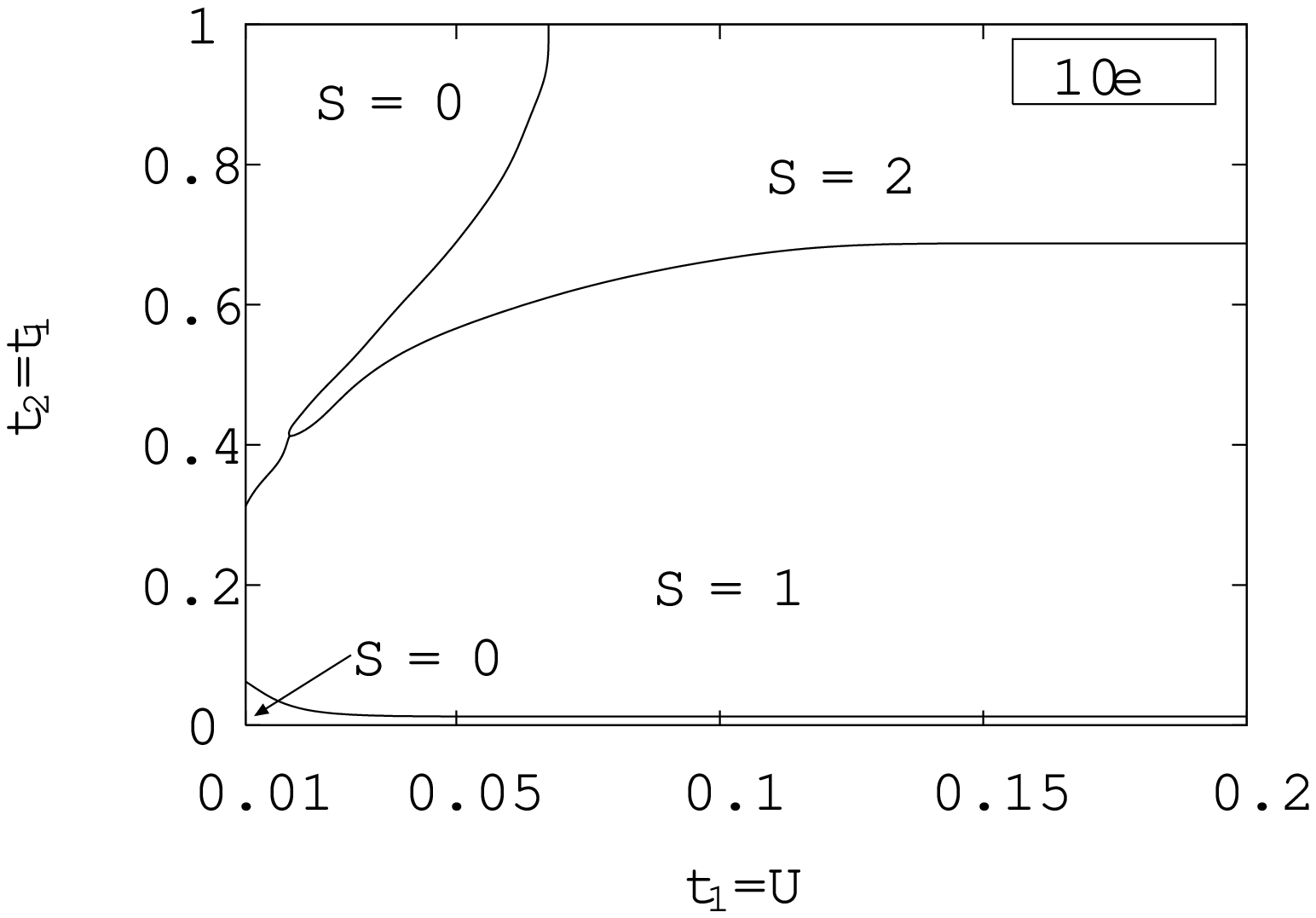}\vspace{0.1cm}} \\ \hline
   \multicolumn{4}{c}{\vspace{3.5cm}}
\end{tabular} }

\end{tabular}
\caption{Ground state phase diagrams that contain substantial regions of intermediate ground state spin (spin between the minimum and maximum attained in the explored parameter space). \label{figComplexDiagrams1}}
\end{figure}

\begin{figure}[b]
\begin{tabular}{cc}

\parbox[t]{0.5\linewidth}{
\begin{tabular}{|c|c|c|c|} \hline
 \#$_{cl}$ & Q & $\tOuter/\tInner$ & Phase Diagram \\
\hline
20 & -2 & 1 & \parbox{2.3in}{\vspace{0.1cm}\includegraphics[width=2.2in]{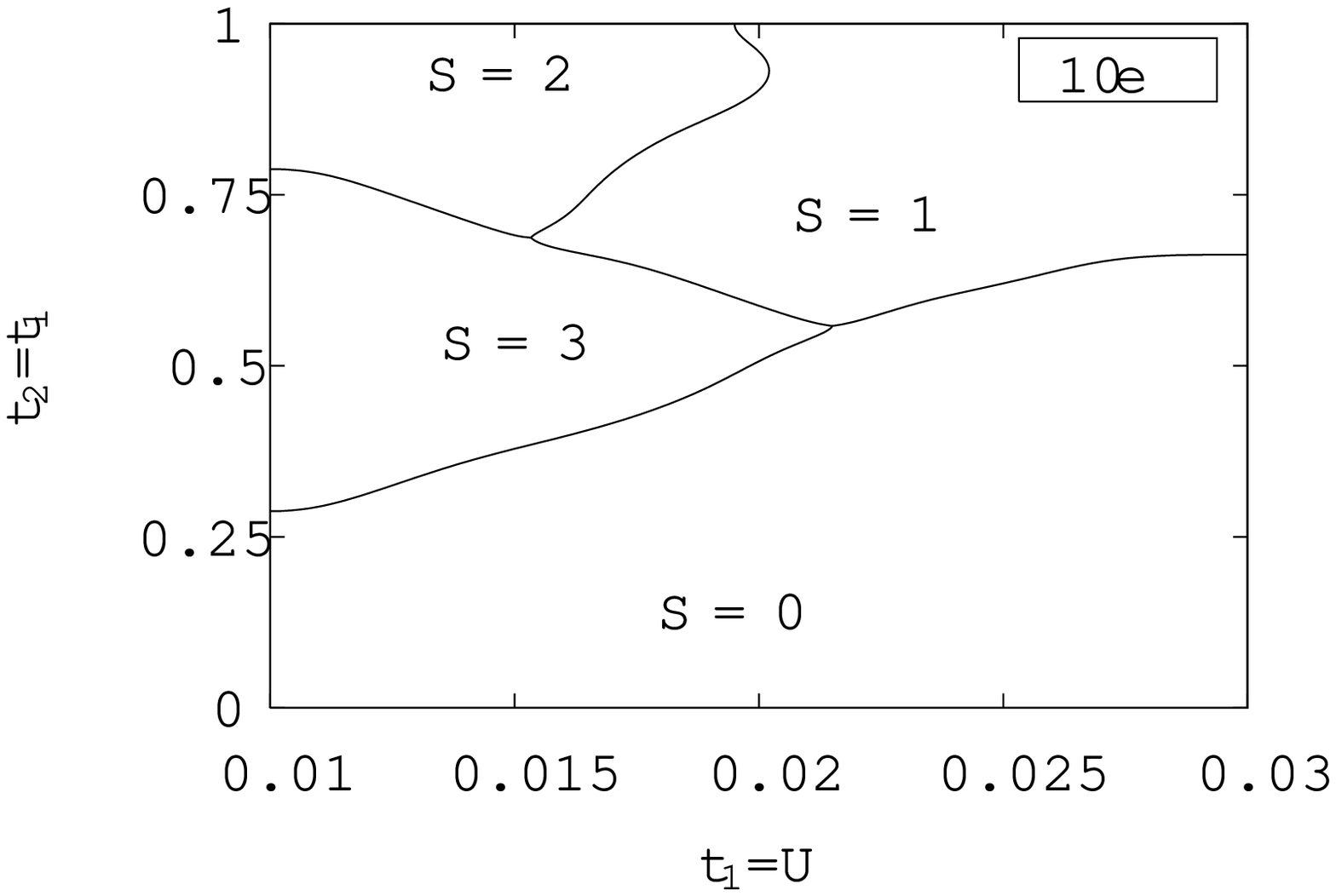}\vspace{0.1cm}} \\ \cline{3-4}
   &    & 5 & \parbox{2.3in}{\vspace{0.1cm}\includegraphics[width=2.2in]{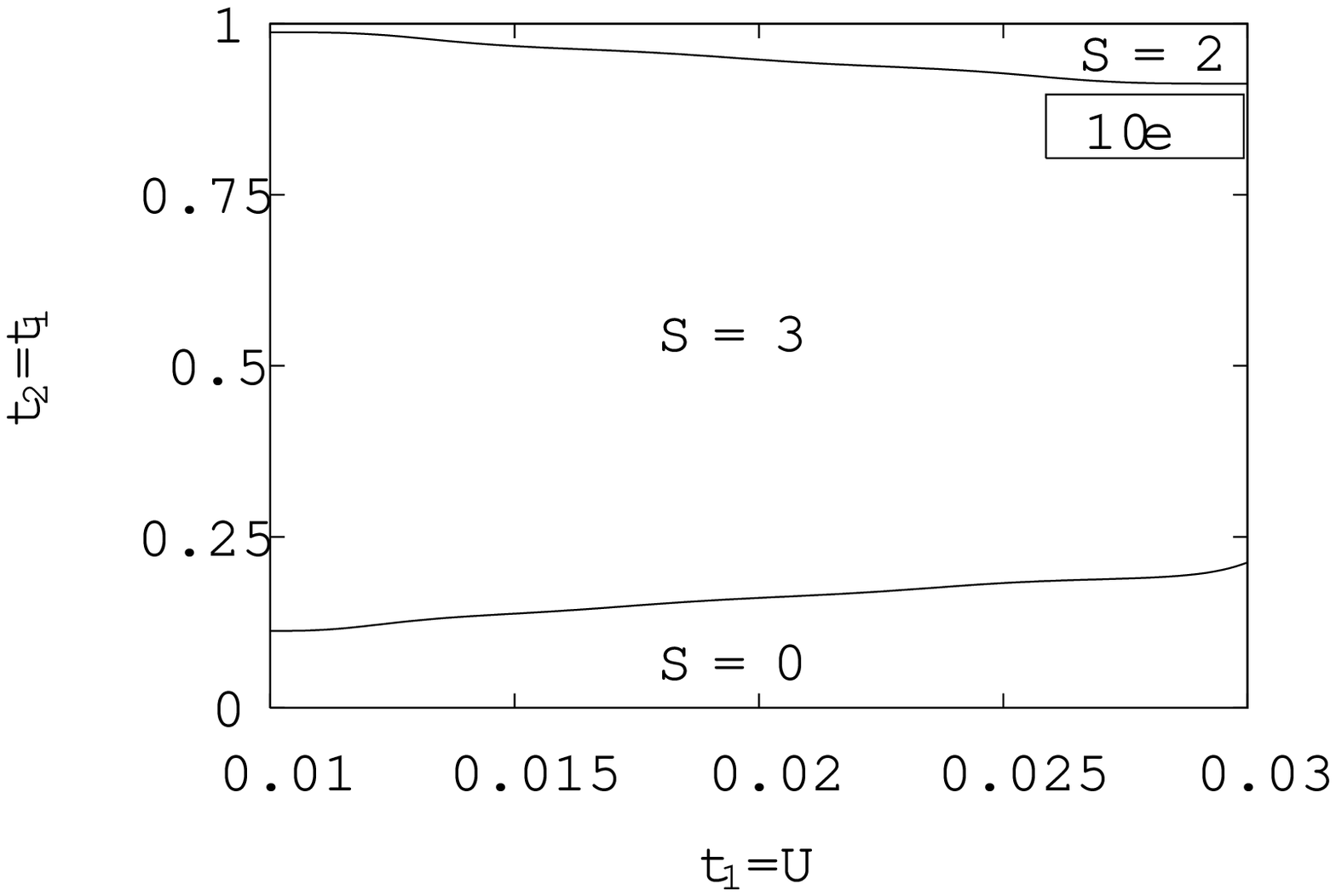}\vspace{0.1cm}} \\ \hline
21 & -1 & 1 & \parbox{2.3in}{\vspace{0.1cm}\includegraphics[width=2.2in]{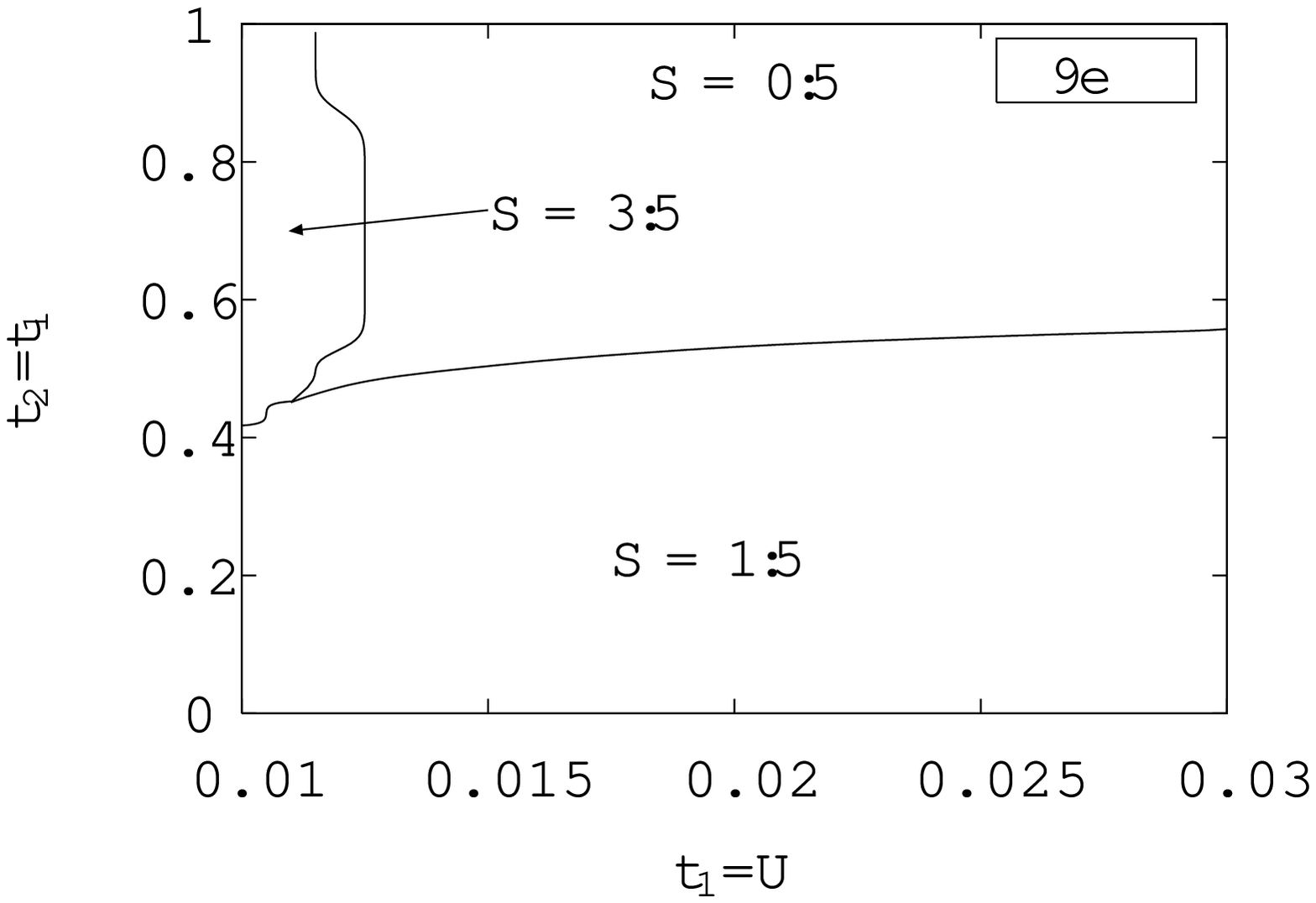}\vspace{0.1cm}} \\ \cline{2-4}
   & -2 & 5 & \parbox{2.3in}{\vspace{0.1cm}\includegraphics[width=2.2in]{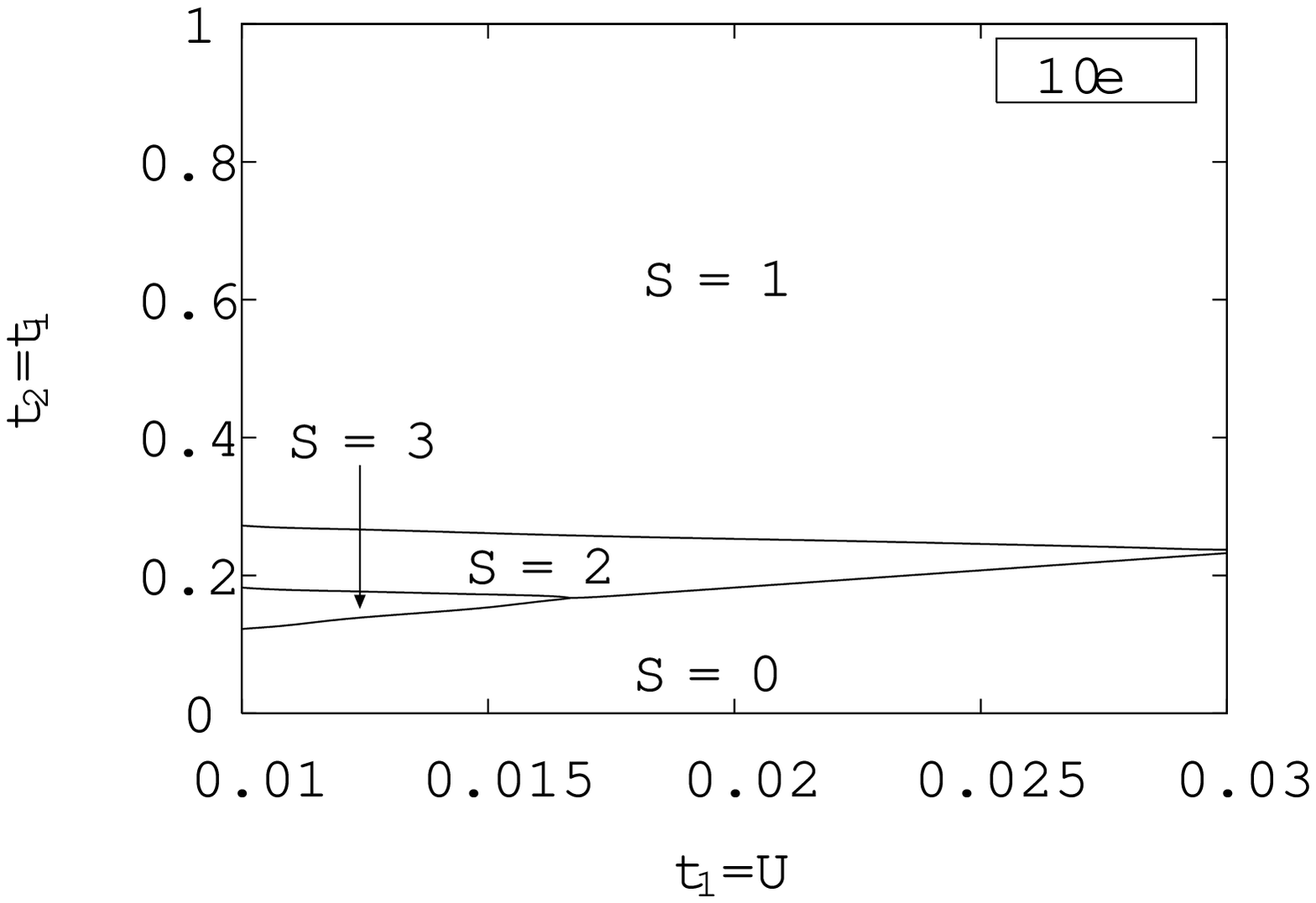}\vspace{0.1cm}} \\ \cline{3-4}
   &    & 10 & \parbox{2.3in}{\vspace{0.1cm}\includegraphics[width=2.2in]{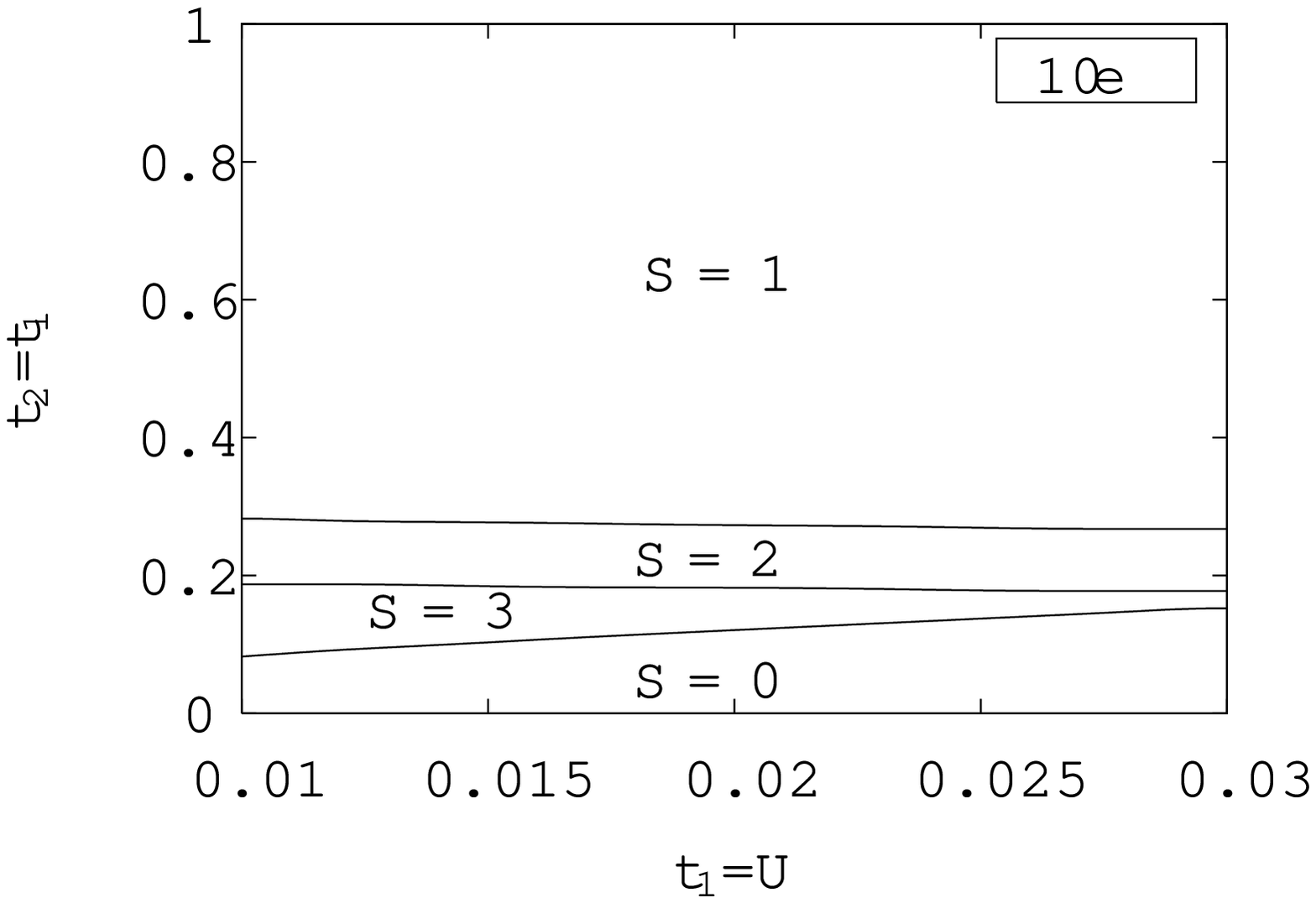}\vspace{0.1cm}} \\ \hline
\end{tabular}}

&

\parbox[t]{0.5\linewidth}{
\begin{tabular}{|c|c|c|c|} \hline
 \#$_{cl}$ & Q & $\tOuter/\tInner$ & Phase Diagram \\
\hline
22 & -1 & 1  & \parbox{2.3in}{\vspace{0.1cm}\includegraphics[width=2.2in]{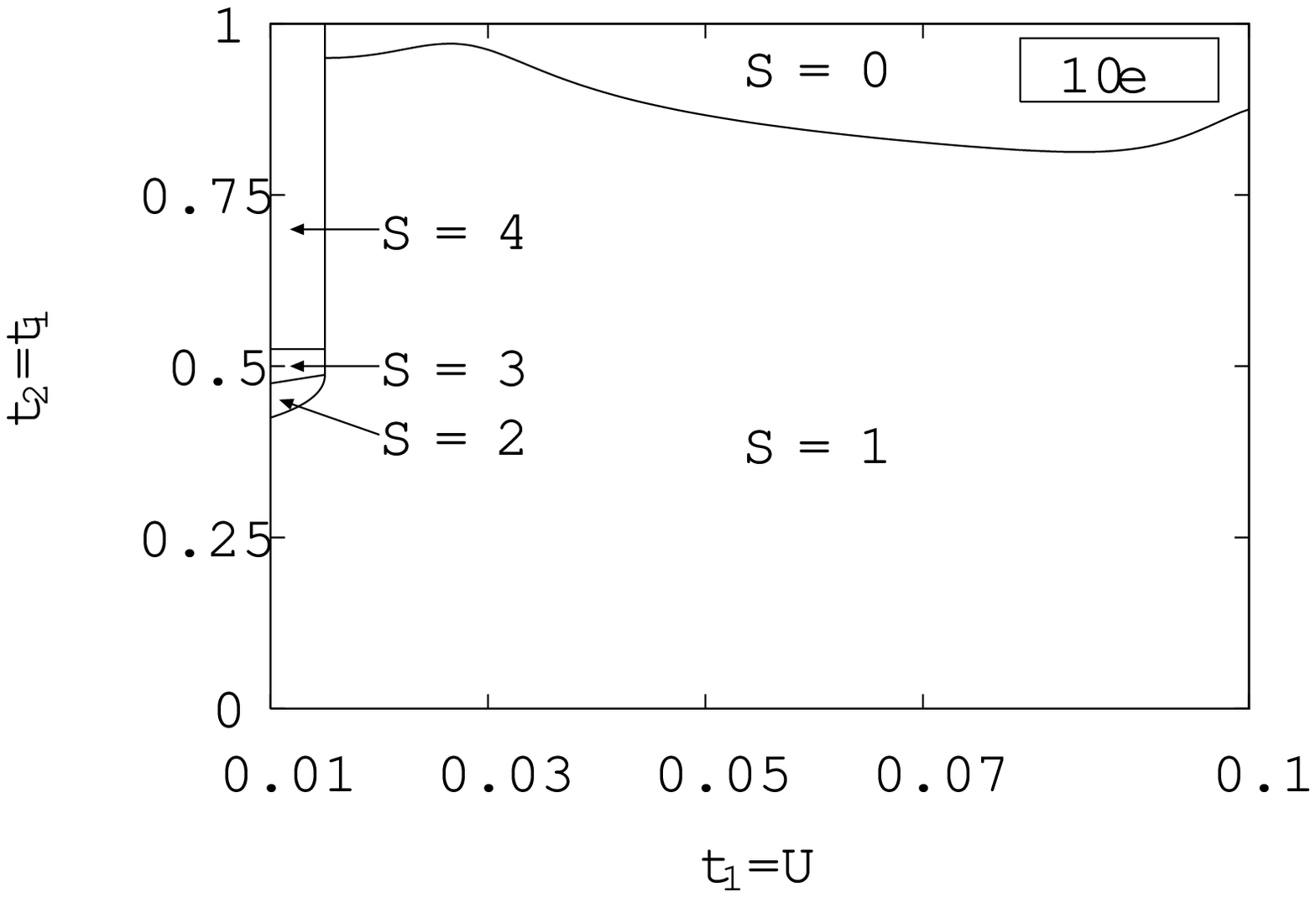}\vspace{0.1cm}} \\ \cline{3-4}
   &    & 5  & \parbox{2.3in}{\vspace{0.1cm}\includegraphics[width=2.2in]{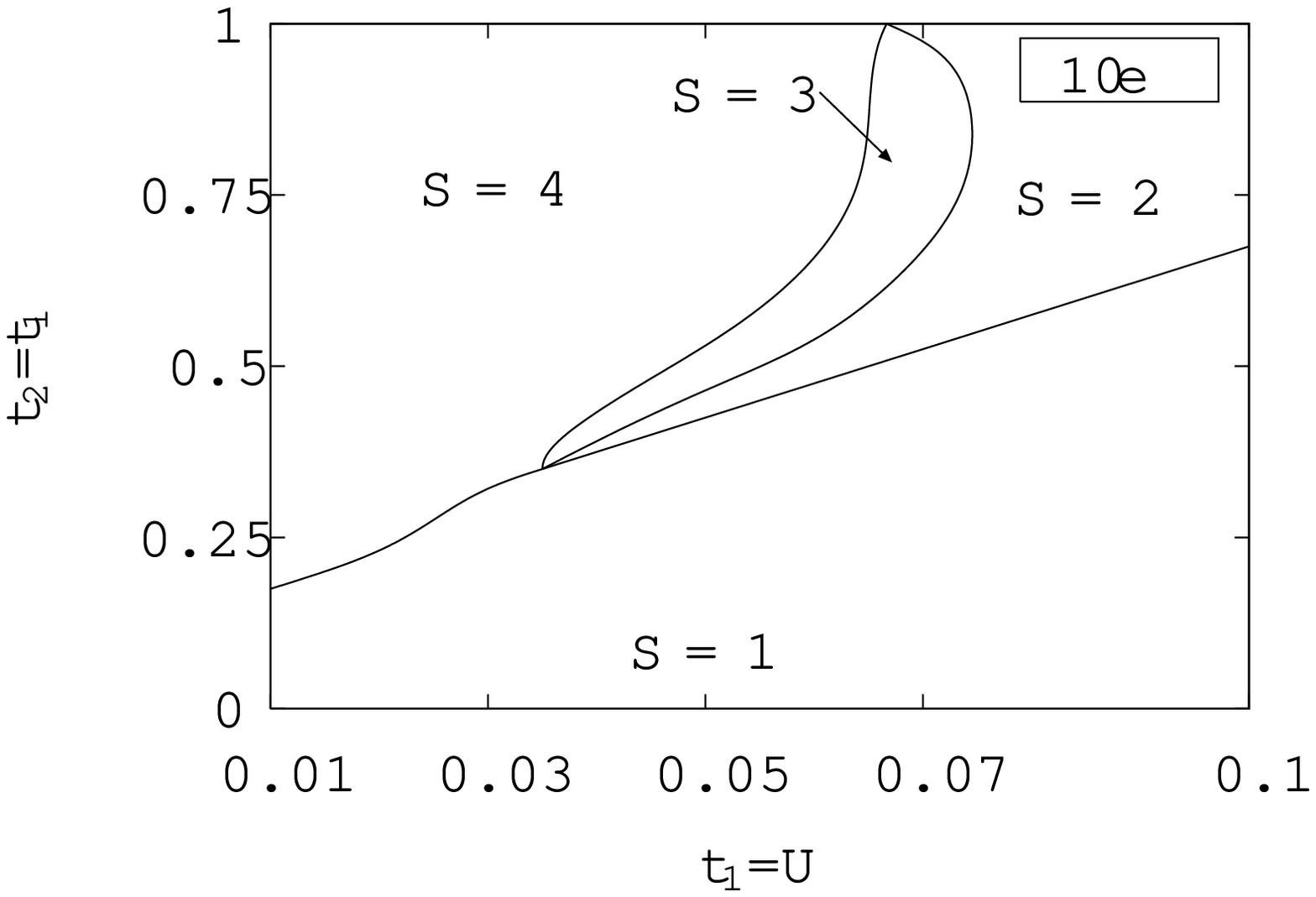}\vspace{0.1cm}} \\ \cline{3-4}
   &    & 10 & \parbox{2.3in}{\vspace{0.1cm}\includegraphics[width=2.2in]{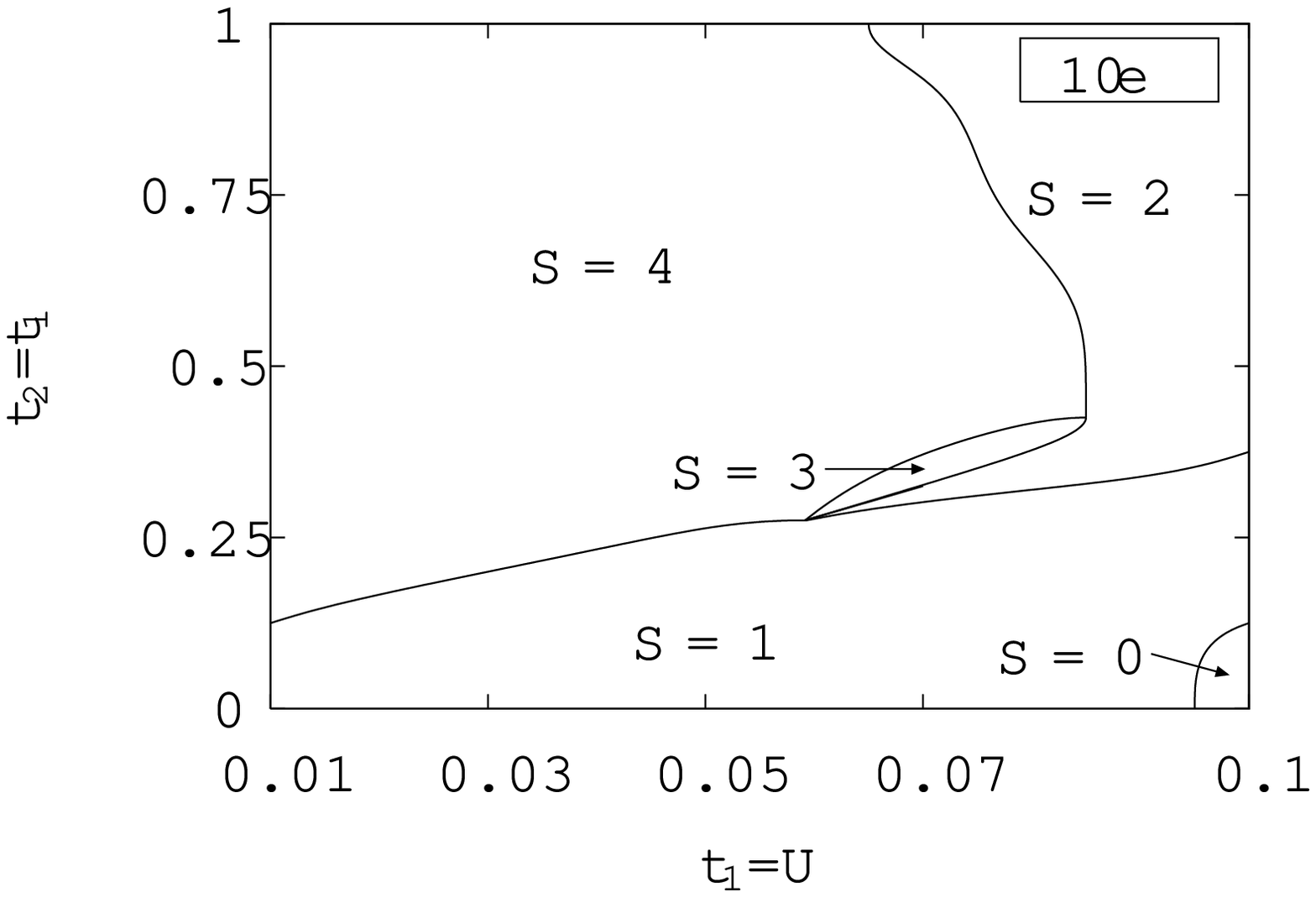}\vspace{0.1cm}} \\ \hline
\multicolumn{4}{c}{\vspace{8cm}}
\end{tabular}}
\end{tabular}
\caption{Ground state phase diagrams that contain substantial regions of intermediate ground state spin (spin between the minimum and maximum attained in the explored parameter space). \label{figComplexDiagrams2}}
\end{figure}

\begin{figure}[b]
\begin{tabular}{cc}
\parbox[t]{0.5\linewidth}{
\begin{tabular}{|c|c|c|c|} \hline
 \#$_{cl}$ & Q & $\tOuter/\tInner$ & Phase Diagram \\
\hline
23 & -1 & 1  & \parbox{2.3in}{\vspace{0.1cm}\includegraphics[width=2.2in]{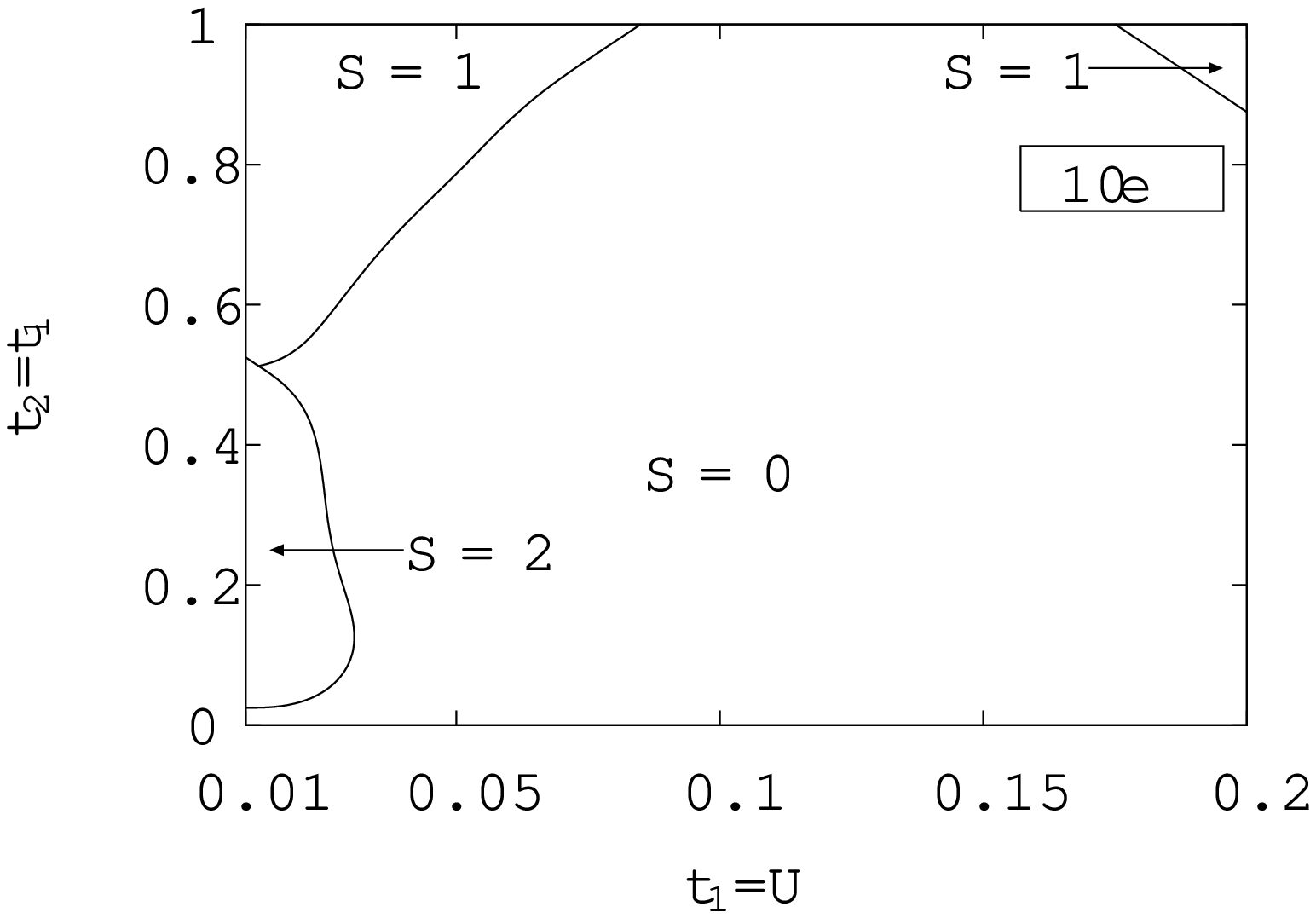}\vspace{0.1cm}} \\ \cline{3-4}
   &    & 5  & \parbox{2.3in}{\vspace{0.1cm}\includegraphics[width=2.2in]{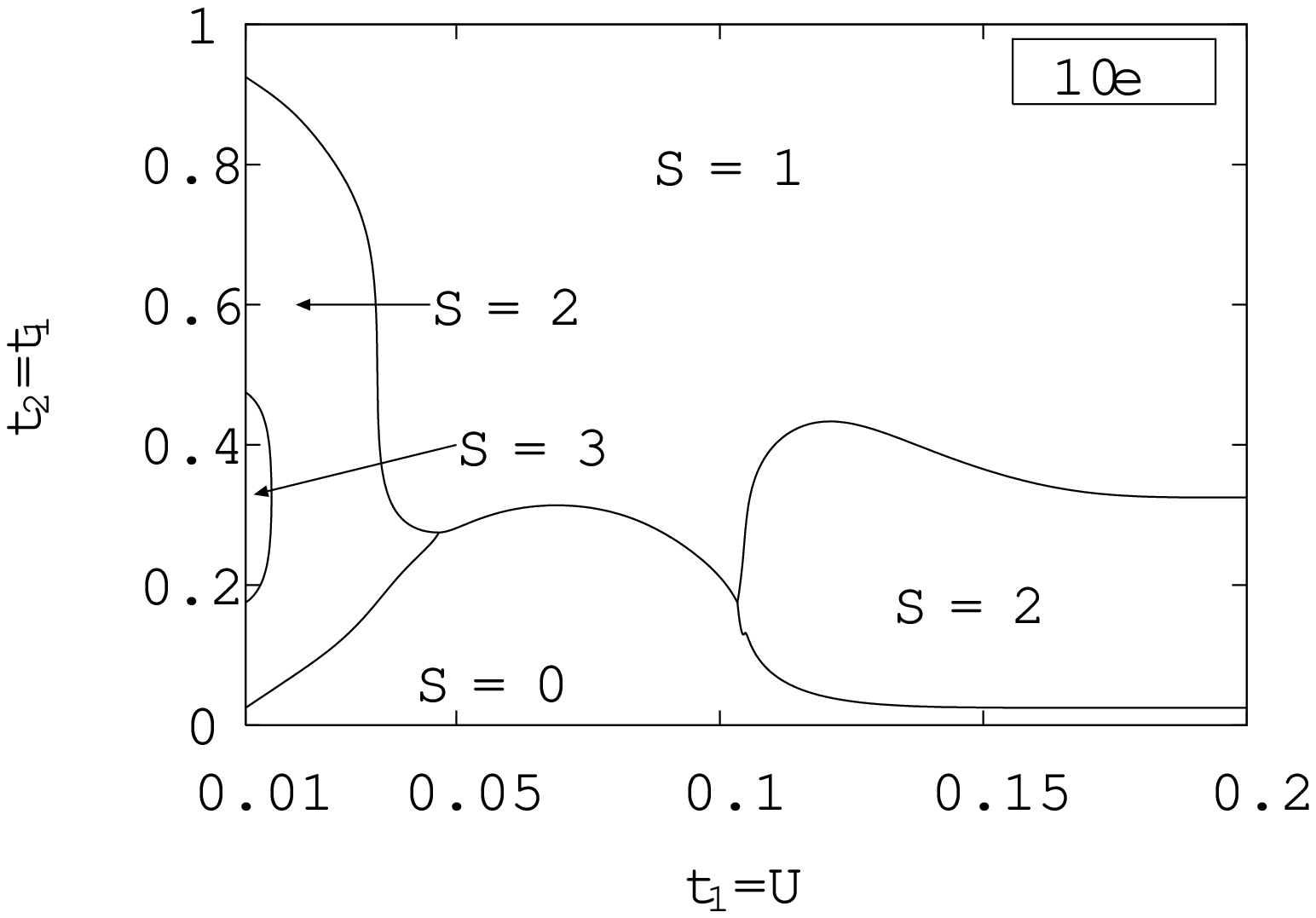}\vspace{0.1cm}} \\ \cline{3-4}
   &    & 10 & \parbox{2.3in}{\vspace{0.1cm}\includegraphics[width=2.2in]{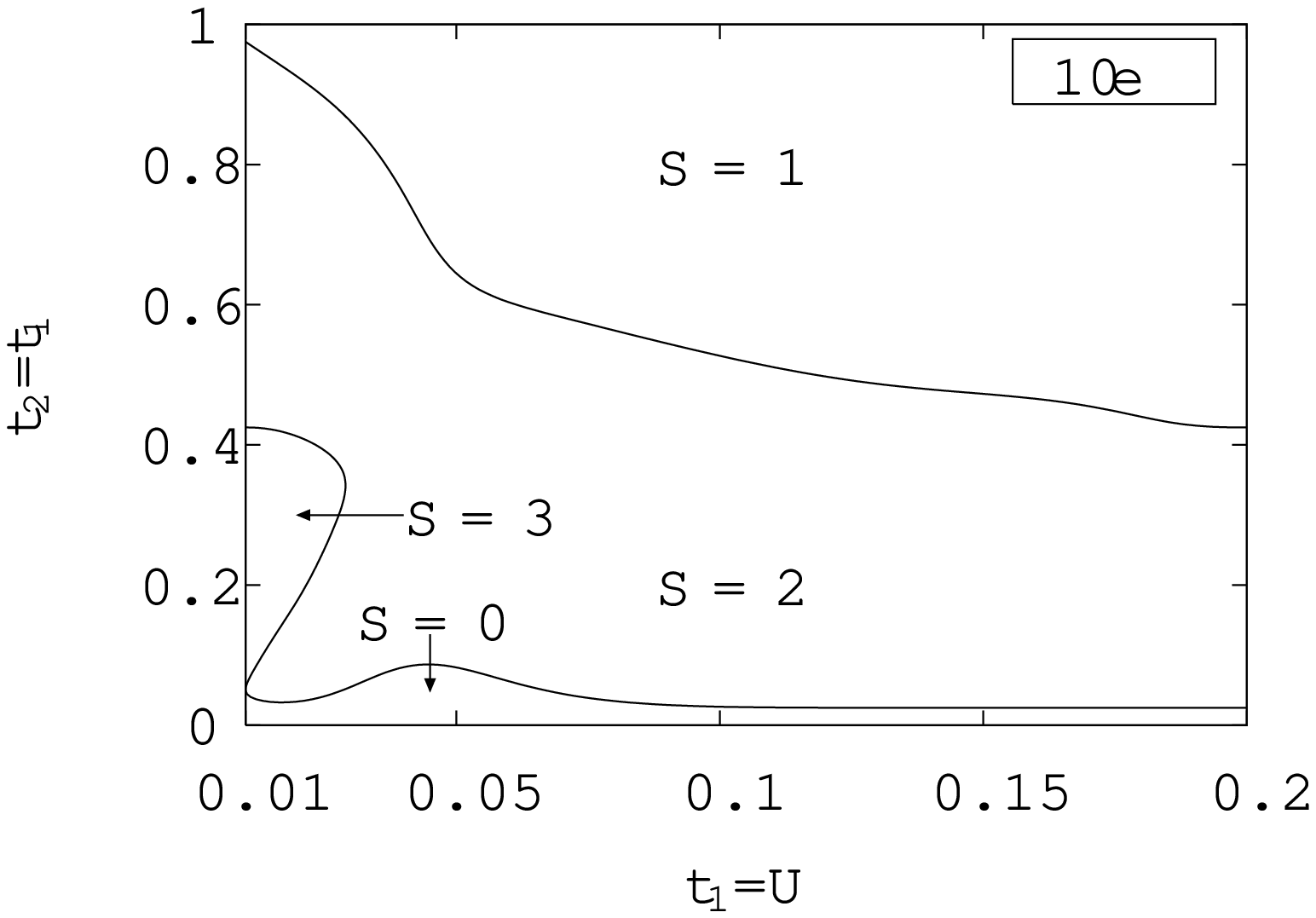}\vspace{0.1cm}} \\ \hline
\end{tabular}}
 
&

\parbox[t]{0.5\linewidth}{
\begin{tabular}{|c|c|c|c|} \hline
 \#$_{cl}$ & Q & $\tOuter/\tInner$ & Phase Diagram \\
\hline
23 & -2 & 1  & \parbox{2.3in}{\vspace{0.1cm}\includegraphics[width=2.2in]{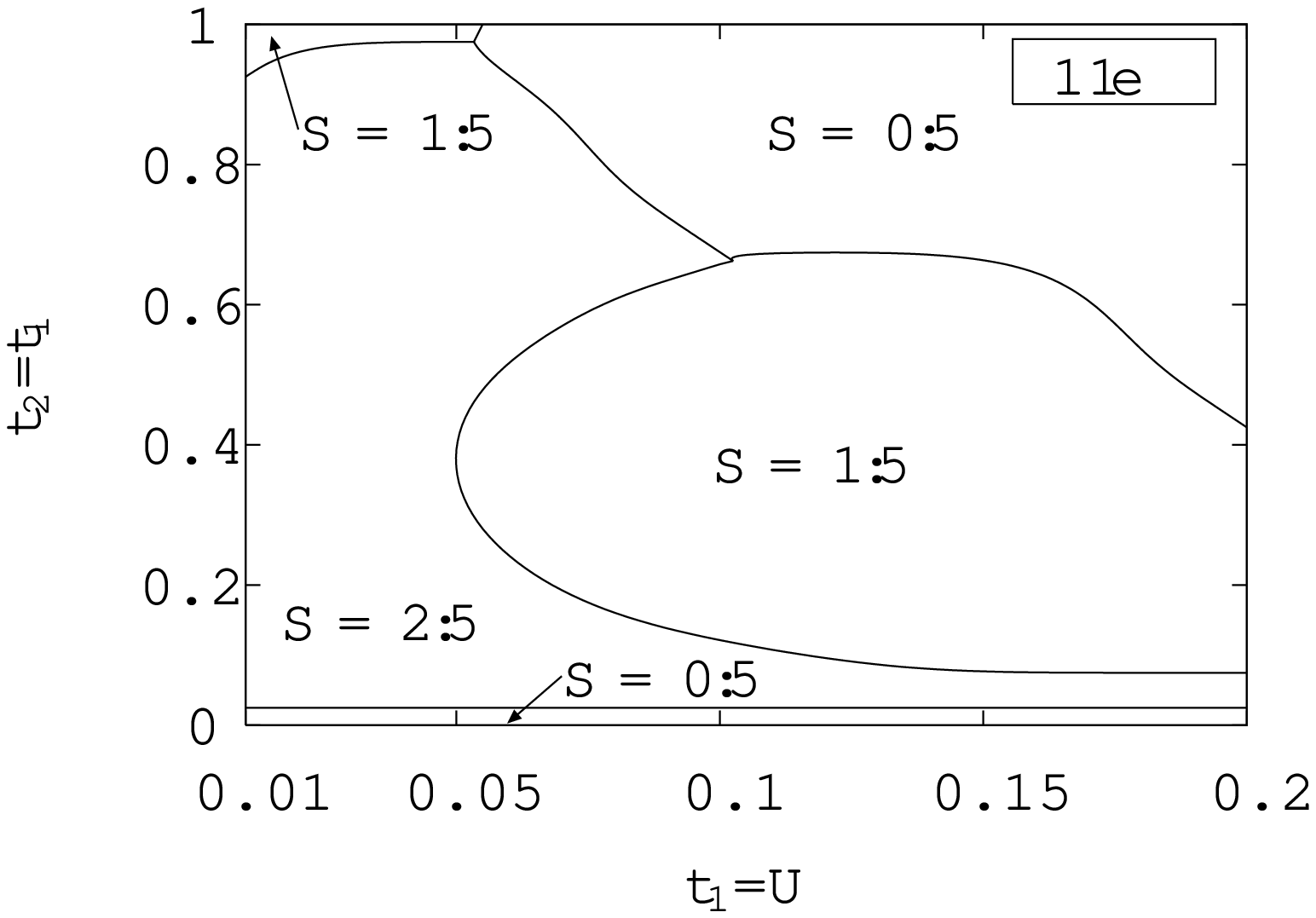}\vspace{0.1cm}} \\ \cline{3-4}
   &    & 5  & \parbox{2.3in}{\vspace{0.1cm}\includegraphics[width=2.2in]{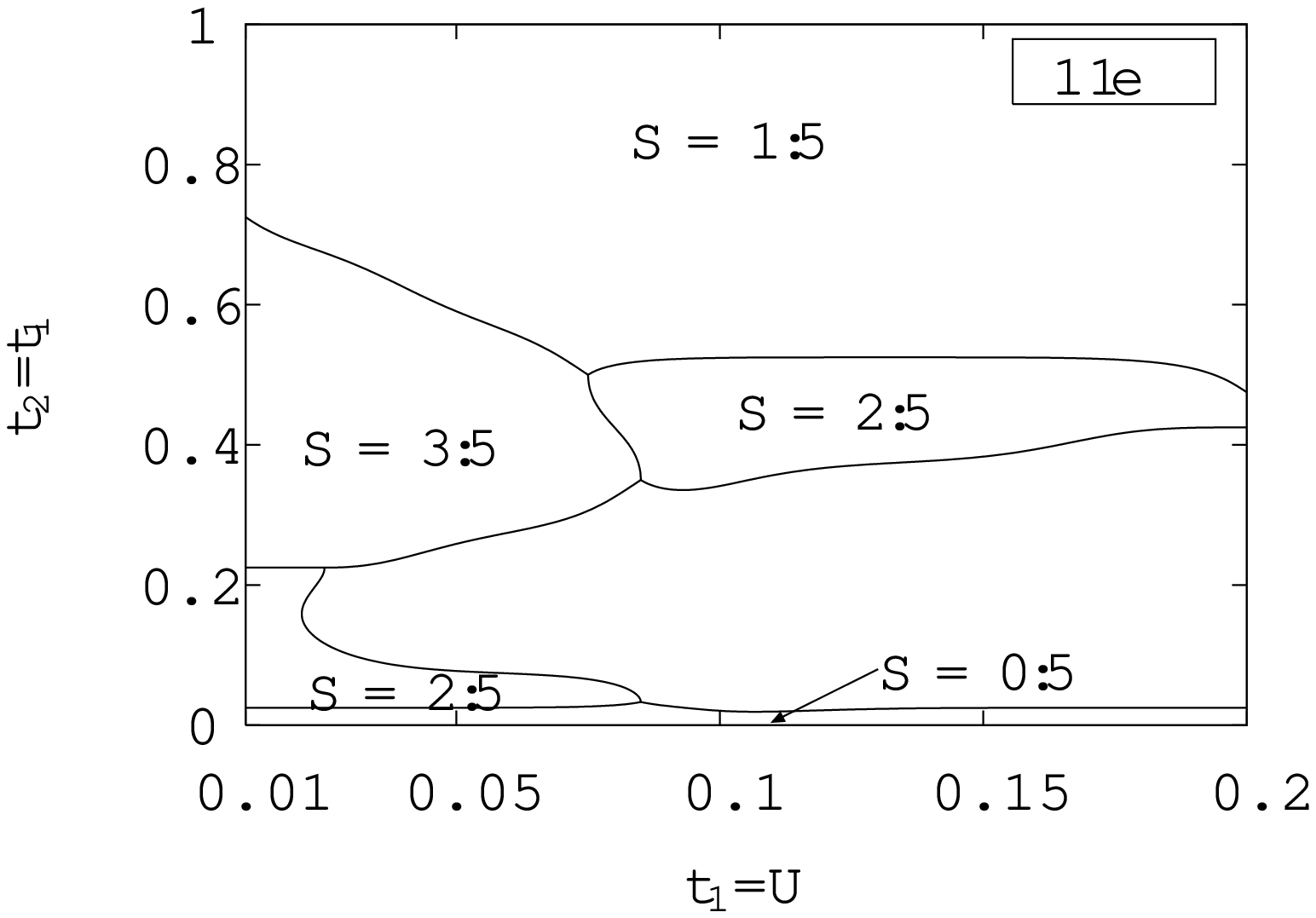}\vspace{0.1cm}} \\ \cline{3-4}
   &    & 10 & \parbox{2.3in}{\vspace{0.1cm}\includegraphics[width=2.2in]{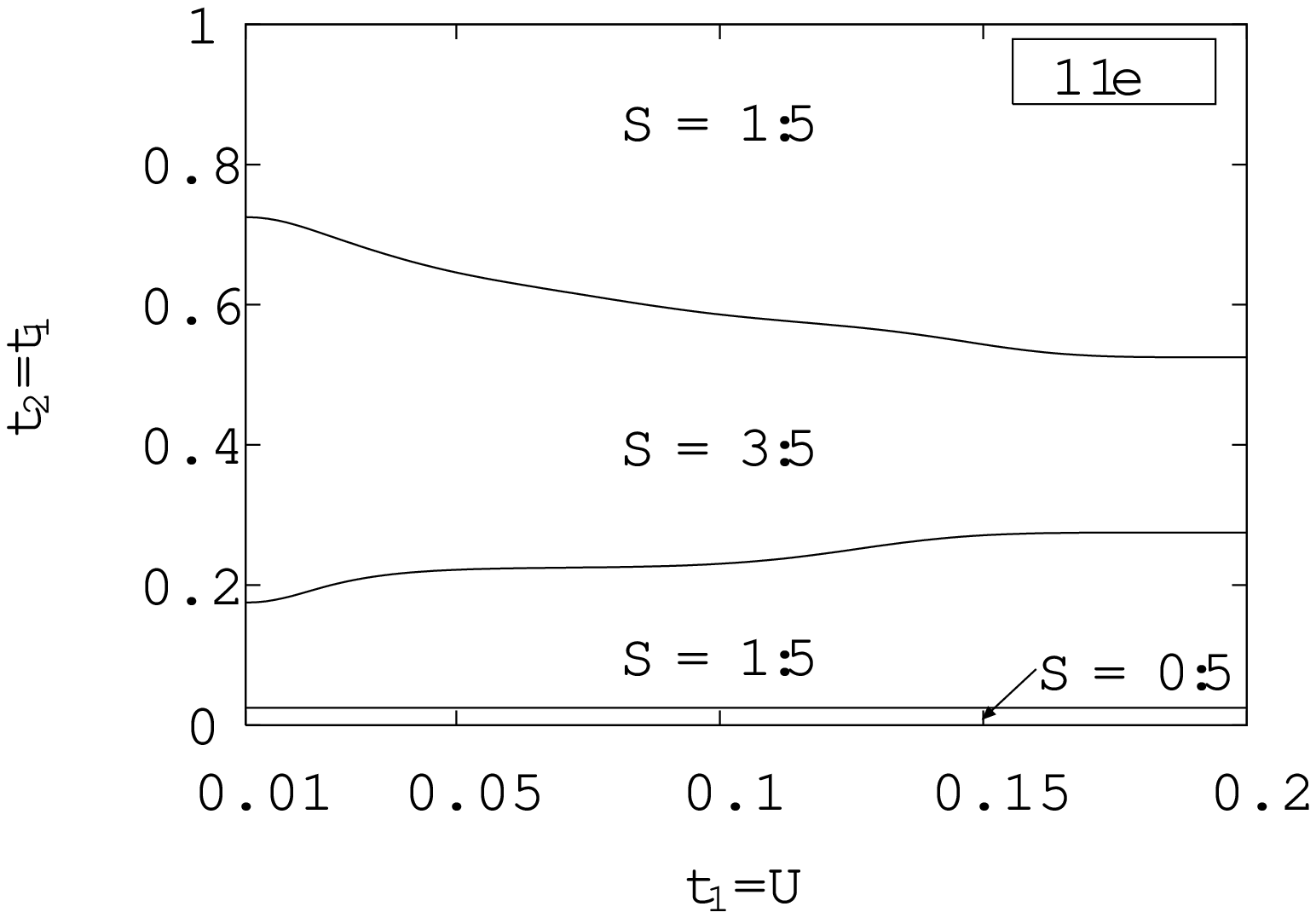}\vspace{0.1cm}} \\ \hline
\end{tabular}}
\end{tabular}
\caption{Ground state phase diagrams that contain substantial regions of intermediate ground state spin (spin between the minimum and maximum attained in the explored parameter space). \label{figComplexDiagrams3}}
\end{figure}

\end{widetext}

\clearpage

\section{Random Cluster Spin Data\label{appRandomClusterData}}

\subsection{Spin distribution data}

Spin distributions, given by the percentage of clusters with each possible spin, are given for two-dimensional clusters with open and periodic boundary conditions in Table \ref{tabRandClusters2D}.  In this table we only consider models 1 and 2 (see text, section \ref{secFixedDensityClusters}) for determining $\tOuter_{ij}$ -- a comparison of models 2 and 3 is given separately.  To facilitate comparison between the two types of boundary conditions, the difference between the two cases (left and right sides of Table \ref{tabRandClusters2D}) is shown in Table \ref{tabRandClusters2D_diff}.  The analogous results for 3-dimensional clusters are provided in Ref.~\onlinecite{NielsenThesis}.  Also included there is spin distribution data comparing models 2 and 3 for 3D clusters (analagous to the 2D results of Fig.~\ref{tabRandClusters2D_cband}).

\vspace{0.9in} 
\subsection{Average spin and percentage magnetic clusters of fixed density clusters}
This section presents complete results for the average spin and the percentage of magnetic clusters for fixed cluster size as a function of doping (zero doping = half-filled).  Results for 2D and 3D clusters with open and periodic boundary conditions are shown in the figures below, as indicated in their titles.

In particular, Fig.~\ref{afigAvgSpin2D} shows the average spin of 2D clusters with open and periodic boundary conditions, respectively.  To remove an even-odd effect, Fig.~\ref{afigAvgSpin2Dr} shows the same average spin but \emph{relative to $S_{min}$} (i.e. 0.5 is subtracted from cases of odd electron number).  Figure \ref{afigPcMag2D} shows the percentage of magnetic clusters (defined as those with greater than minimal ground state spin) as a function of doping for the different cluster sizes.  Corresponding results for three-dimensional clusters with open and periodic boundary conditions are given in Ref.~\onlinecite{NielsenThesis}.

\begin{widetext}

\begin{table*}[b]
\begin{center}
\begin{tabular}{|c|c|c||c|c|c|c||c|c|c|c||c|c|c|c||} \hline 
 & \multicolumn{2}{c||}{Dim \& b.c.} & \multicolumn{12}{c||}{2D, periodic b.c - open b.c}\\ \cline{2-15}
$N_s$ & \multicolumn{2}{c||}{\rule[-3mm]{0mm}{8mm}$\rho$} & \multicolumn{4}{c||}{$\frac{1}{1600} \approx 0.005 \rho_c^{2D}$} & \multicolumn{4}{c||}{$\frac{1}{160} \approx 0.05 \rho_c^{2D}$} & \multicolumn{4}{c||}{$\frac{3}{160} \approx 0.15 \rho_c^{2D}$}\\ \cline{2-15}
 & \multicolumn{2}{c||}{spin} & \textbf{0.5} & \textbf{1.5} & \textbf{2.5} & \textbf{3.5} & \textbf{0.5} & \textbf{1.5} & \textbf{2.5} & \textbf{3.5} & \textbf{0.5} & \textbf{1.5} & \textbf{2.5} & \textbf{3.5}\\ \hline\hline 
\raisebox{-0.75cm}[0pt][0pt]{4} & \raisebox{-0.3cm}[0pt][0pt]{$ \tilde{t} > t$} & 1h & 0 & -0 & 0 & 0 & -1 & 1 & 0 & 0 & 0 & -0 & 0 & 0\\ \cline{3-15} 
  &  & 1e & -14 & 14 & 0 & 0 & -4 & 4 & 0 & 0 & 5 & -5 & 0 & 0\\ \cline{2-15} 
  & \raisebox{-0.3cm}[0pt][0pt]{$ \tilde{t} = t$} & 1h & 0 & -0 & 0 & 0 & -1 & 1 & 0 & 0 & 0 & -0 & 0 & 0\\ \cline{3-15} 
  &  & 1e & -0 & 0 & 0 & 0 & -16 & 16 & 0 & 0 & -12 & 12 & 0 & 0\\ \hline 
\raisebox{-0.75cm}[0pt][0pt]{6} & \raisebox{-0.3cm}[0pt][0pt]{$ \tilde{t} > t$} & 1h & -1 & 1 & 0 & 0 & -1 & 1 & 0 & 0 & 3 & -3 & 0 & 0\\ \cline{3-15} 
  &  & 1e & -13 & 6 & 7 & 0 & -12 & -5 & 17 & 0 & 3 & -3 & -0 & 0\\ \cline{2-15} 
  & \raisebox{-0.3cm}[0pt][0pt]{$ \tilde{t} = t$} & 1h & -0 & 0 & -0 & 0 & -0 & 0 & 0 & 0 & -0 & 0 & 0 & 0\\ \cline{3-15} 
  &  & 1e & -0 & 0 & -0 & 0 & -4 & 3 & 1 & 0 & -5 & 4 & 1 & 0\\ \hline 
 & \multicolumn{2}{c||}{spin} & \textbf{0} & \textbf{1} & \textbf{2} & \textbf{3} & \textbf{0} & \textbf{1} & \textbf{2} & \textbf{3} & \textbf{0} & \textbf{1} & \textbf{2} & \textbf{3}\\ \hline\hline 
\raisebox{-0.75cm}[0pt][0pt]{5} & \raisebox{-0.3cm}[0pt][0pt]{$ \tilde{t} > t$} & 1h & -0 & 0 & -0 & 0 & 3 & -3 & 0 & 0 & 10 & -10 & 0 & 0\\ \cline{3-15} 
  &  & 1e & -3 & -5 & 8 & 0 & -4 & -8 & 12 & 0 & 3 & 0 & -3 & 0\\ \cline{2-15} 
  & \raisebox{-0.3cm}[0pt][0pt]{$ \tilde{t} = t$} & 1h & 5 & -5 & -0 & 0 & 2 & -2 & 0 & 0 & 1 & -1 & 0 & 0\\ \cline{3-15} 
  &  & 1e & -3 & 4 & -1 & 0 & -6 & 1 & 6 & 0 & -8 & 5 & 4 & 0\\ \hline 
\raisebox{-0.75cm}[0pt][0pt]{7} & \raisebox{-0.3cm}[0pt][0pt]{$ \tilde{t} > t$} & 1h & 3 & -2 & -1 & 0 & -12 & 11 & 1 & 0 & 19 & -17 & -1 & 0\\ \cline{3-15} 
  &  & 1e & -7 & -8 & 10 & 5 & -6 & -13 & -1 & 20 & -2 & 1 & 2 & -1\\ \cline{2-15} 
  & \raisebox{-0.3cm}[0pt][0pt]{$ \tilde{t} = t$} & 1h & 3 & -4 & 0 & 0 & 3 & -3 & 0 & 0 & 1 & -1 & 0 & 0\\ \cline{3-15} 
  &  & 1e & 5 & -5 & 0 & 0 & 0 & 2 & -2 & 0 & -3 & 2 & 1 & 0\\ \hline 
\end{tabular}

\caption{\emph{Difference} between distribution of ground state spin values for 2D random clusters with open and periodic boundary conditions.  Values are obtained by subtracting the sets of data in Table \ref{tabRandClusters2D} below.  Thus, entries show increase or decrease in percentage when switching from open to periodic boundary conditions. Estimated error $\pm0.7\%$.\label{tabRandClusters2D_diff}}
\end{center}
\end{table*}

\begin{turnpage}

\begin{table*}[b]
\begin{center}
\begin{tabular}{cc}
\begin{tabular}{|c|c|c||c|c|c|c||c|c|c|c||c|c|c|c||} \hline 
 & \multicolumn{2}{c||}{Dim \& b.c.} & \multicolumn{12}{c||}{2D, open b.c}\\ \cline{2-15}
$N_s$ & \multicolumn{2}{c||}{\rule[-3mm]{0mm}{8mm}$\rho$} & \multicolumn{4}{c||}{$\frac{1}{1600} \approx 0.005 \rho_c^{2D}$} & \multicolumn{4}{c||}{$\frac{1}{160} \approx 0.05 \rho_c^{2D}$} & \multicolumn{4}{c||}{$\frac{3}{160} \approx 0.15 \rho_c^{2D}$}\\ \cline{2-15}
 & \multicolumn{2}{c||}{spin} & \textbf{0.5} & \textbf{1.5} & \textbf{2.5} & \textbf{3.5} & \textbf{0.5} & \textbf{1.5} & \textbf{2.5} & \textbf{3.5} & \textbf{0.5} & \textbf{1.5} & \textbf{2.5} & \textbf{3.5}\\ \hline\hline 
\raisebox{-0.75cm}[0pt][0pt]{4} & \raisebox{-0.3cm}[0pt][0pt]{$ \tilde{t} > t$} & 1h & 89 & 11 & 0 & 0 & 100 & 0 & 0 & 0 & 100 & 0 & 0 & 0\\ \cline{3-15} 
  &  & 1e & 69 & 31 & 0 & 0 & 20 & 80 & 0 & 0 & 37 & 63 & 0 & 0\\ \cline{2-15} 
  & \raisebox{-0.3cm}[0pt][0pt]{$ \tilde{t} = t$} & 1h & 90 & 10 & 0 & 0 & 100 & 0 & 0 & 0 & 100 & 0 & 0 & 0\\ \cline{3-15} 
  &  & 1e & 87 & 13 & 0 & 0 & 90 & 10 & 0 & 0 & 93 & 7 & 0 & 0\\ \hline 
\raisebox{-0.75cm}[0pt][0pt]{6} & \raisebox{-0.3cm}[0pt][0pt]{$ \tilde{t} > t$} & 1h & 86 & 14 & 0 & 0 & 93 & 7 & 0 & 0 & 95 & 5 & 0 & 0\\ \cline{3-15} 
  &  & 1e & 60 & 32 & 8 & 0 & 16 & 43 & 41 & 0 & 17 & 59 & 24 & 0\\ \cline{2-15} 
  & \raisebox{-0.3cm}[0pt][0pt]{$ \tilde{t} = t$} & 1h & 89 & 11 & 0 & 0 & 99 & 1 & 0 & 0 & 100 & 0 & 0 & 0\\ \cline{3-15} 
  &  & 1e & 79 & 19 & 1 & 0 & 89 & 10 & 0 & 0 & 93 & 7 & 0 & 0\\ \hline 
 & \multicolumn{2}{c||}{spin} & \textbf{0} & \textbf{1} & \textbf{2} & \textbf{3} & \textbf{0} & \textbf{1} & \textbf{2} & \textbf{3} & \textbf{0} & \textbf{1} & \textbf{2} & \textbf{3}\\ \hline\hline 
\raisebox{-0.75cm}[0pt][0pt]{5} & \raisebox{-0.3cm}[0pt][0pt]{$ \tilde{t} > t$} & 1h & 48 & 51 & 0 & 0 & 68 & 32 & 0 & 0 & 70 & 30 & 0 & 0\\ \cline{3-15} 
  &  & 1e & 26 & 52 & 21 & 0 & 6 & 34 & 60 & 0 & 6 & 52 & 41 & 0\\ \cline{2-15} 
  & \raisebox{-0.3cm}[0pt][0pt]{$ \tilde{t} = t$} & 1h & 37 & 63 & 0 & 0 & 82 & 18 & 0 & 0 & 88 & 12 & 0 & 0\\ \cline{3-15} 
  &  & 1e & 46 & 43 & 11 & 0 & 58 & 39 & 2 & 0 & 64 & 35 & 1 & 0\\ \hline 
\raisebox{-0.75cm}[0pt][0pt]{7} & \raisebox{-0.3cm}[0pt][0pt]{$ \tilde{t} > t$} & 1h & 38 & 58 & 4 & 0 & 66 & 33 & 1 & 0 & 54 & 45 & 1 & 0\\ \cline{3-15} 
  &  & 1e & 29 & 54 & 15 & 2 & 7 & 26 & 41 & 26 & 6 & 30 & 51 & 13\\ \cline{2-15} 
  & \raisebox{-0.3cm}[0pt][0pt]{$ \tilde{t} = t$} & 1h & 49 & 50 & 1 & 0 & 76 & 24 & 0 & 0 & 87 & 13 & 0 & 0\\ \cline{3-15} 
  &  & 1e & 37 & 60 & 3 & 0 & 59 & 37 & 4 & 0 & 63 & 36 & 1 & 0\\ \hline 
\end{tabular}

\begin{tabular}{|c|c|c||c|c|c|c||c|c|c|c||c|c|c|c||} \hline 
 & \multicolumn{2}{c||}{Dim \& b.c.} & \multicolumn{12}{c||}{2D, periodic b.c}\\ \cline{2-15}
$N_s$ & \multicolumn{2}{c||}{\rule[-3mm]{0mm}{8mm}$\rho$} & \multicolumn{4}{c||}{$\frac{1}{1600} \approx 0.005 \rho_c^{2D}$} & \multicolumn{4}{c||}{$\frac{1}{160} \approx 0.05 \rho_c^{2D}$} & \multicolumn{4}{c||}{$\frac{3}{160} \approx 0.15 \rho_c^{2D}$}\\ \cline{2-15}
 & \multicolumn{2}{c||}{spin} & \textbf{0.5} & \textbf{1.5} & \textbf{2.5} & \textbf{3.5} & \textbf{0.5} & \textbf{1.5} & \textbf{2.5} & \textbf{3.5} & \textbf{0.5} & \textbf{1.5} & \textbf{2.5} & \textbf{3.5}\\ \hline\hline 
\raisebox{-0.75cm}[0pt][0pt]{4} & \raisebox{-0.3cm}[0pt][0pt]{$ \tilde{t} > t$} & 1h & 89 & 11 & 0 & 0 & 99 & 1 & 0 & 0 & 100 & 0 & 0 & 0\\ \cline{3-15} 
  &  & 1e & 55 & 45 & 0 & 0 & 16 & 84 & 0 & 0 & 43 & 57 & 0 & 0\\ \cline{2-15} 
  & \raisebox{-0.3cm}[0pt][0pt]{$ \tilde{t} = t$} & 1h & 90 & 10 & 0 & 0 & 99 & 1 & 0 & 0 & 100 & 0 & 0 & 0\\ \cline{3-15} 
  &  & 1e & 86 & 14 & 0 & 0 & 74 & 26 & 0 & 0 & 81 & 19 & 0 & 0\\ \hline 
\raisebox{-0.75cm}[0pt][0pt]{6} & \raisebox{-0.3cm}[0pt][0pt]{$ \tilde{t} > t$} & 1h & 85 & 15 & 0 & 0 & 93 & 7 & 0 & 0 & 98 & 2 & 0 & 0\\ \cline{3-15} 
  &  & 1e & 46 & 38 & 15 & 0 & 5 & 38 & 57 & 0 & 20 & 56 & 24 & 0\\ \cline{2-15} 
  & \raisebox{-0.3cm}[0pt][0pt]{$ \tilde{t} = t$} & 1h & 88 & 12 & 0 & 0 & 99 & 1 & 0 & 0 & 99 & 1 & 0 & 0\\ \cline{3-15} 
  &  & 1e & 79 & 20 & 1 & 0 & 85 & 13 & 2 & 0 & 88 & 11 & 1 & 0\\ \hline 
 & \multicolumn{2}{c||}{spin} & \textbf{0} & \textbf{1} & \textbf{2} & \textbf{3} & \textbf{0} & \textbf{1} & \textbf{2} & \textbf{3} & \textbf{0} & \textbf{1} & \textbf{2} & \textbf{3}\\ \hline\hline 
\raisebox{-0.75cm}[0pt][0pt]{5} & \raisebox{-0.3cm}[0pt][0pt]{$ \tilde{t} > t$} & 1h & 48 & 52 & 0 & 0 & 71 & 29 & 0 & 0 & 80 & 20 & 0 & 0\\ \cline{3-15} 
  &  & 1e & 23 & 48 & 29 & 0 & 1 & 27 & 72 & 0 & 9 & 53 & 39 & 0\\ \cline{2-15} 
  & \raisebox{-0.3cm}[0pt][0pt]{$ \tilde{t} = t$} & 1h & 42 & 58 & 0 & 0 & 84 & 16 & 0 & 0 & 89 & 11 & 0 & 0\\ \cline{3-15} 
  &  & 1e & 43 & 47 & 10 & 0 & 52 & 40 & 8 & 0 & 56 & 39 & 5 & 0\\ \hline 
\raisebox{-0.75cm}[0pt][0pt]{7} & \raisebox{-0.3cm}[0pt][0pt]{$ \tilde{t} > t$} & 1h & 41 & 56 & 3 & 0 & 54 & 44 & 2 & 0 & 73 & 27 & 0 & 0\\ \cline{3-15} 
  &  & 1e & 23 & 46 & 25 & 7 & 1 & 13 & 40 & 46 & 4 & 31 & 53 & 12\\ \cline{2-15} 
  & \raisebox{-0.3cm}[0pt][0pt]{$ \tilde{t} = t$} & 1h & 52 & 47 & 1 & 0 & 79 & 21 & 0 & 0 & 88 & 12 & 0 & 0\\ \cline{3-15} 
  &  & 1e & 41 & 55 & 3 & 0 & 59 & 39 & 2 & 0 & 60 & 38 & 2 & 0\\ \hline 
\end{tabular}

\end{tabular}
\caption{ Distribution of ground state spin values for 2D random clusters with \emph{open boundary conditions} (left) and \emph{periodic boundary conditions} (right). Table entries give the percentage of clusters with the ground state spin specified in the column header.  Results are the ensemble average of many clusters with fixed size $\Nsites$, density $\rho$, and doping = one electron (1e) or hole (1h).  $\tOuter > \tInner$ indicates that $\tOuter$ is set by our band calculation (to be compared with the case $\tOuter = \tInner$). Estimated error $\pm0.5\%$.\label{tabRandClusters2D}}
\end{center}
\end{table*}

\begin{figure}
\begin{center}
\begin{tabular}{cc}
\parbox{0.5\linewidth}{
\begin{tabular}{|c|c|} \hline 
$\rho$ & \textbf{2D \ : \ Average Spin \ : \ open b.c.}\\ \hline
$\frac{1}{1600}$ & \parbox{4in}{ 
\includegraphics[width=2in, angle=270]{figs/finalAvgS4-5-6-7_0.010_2D.ps}} \\ \hline 
$\frac{1}{160}$ & \parbox{4in}{
\includegraphics[width=2in, angle=270]{figs/finalAvgS4-5-6-7_0.100_2D.ps}} \\ \hline
$\frac{3}{160}$ & \parbox{4in}{ 
\includegraphics[width=2in, angle=270]{figs/finalAvgS4-5-6-7_0.300_2D.ps}} \\ \hline
\end{tabular}}
 &
\parbox{0.5\linewidth}{
\begin{tabular}{|c|c|} \hline
$\rho$ & \textbf{2D \ : \ Average Spin \ : \ periodic b.c.}\\ \hline
$\frac{1}{1600}$ & \parbox{4in}{
\includegraphics[width=2in, angle=270]{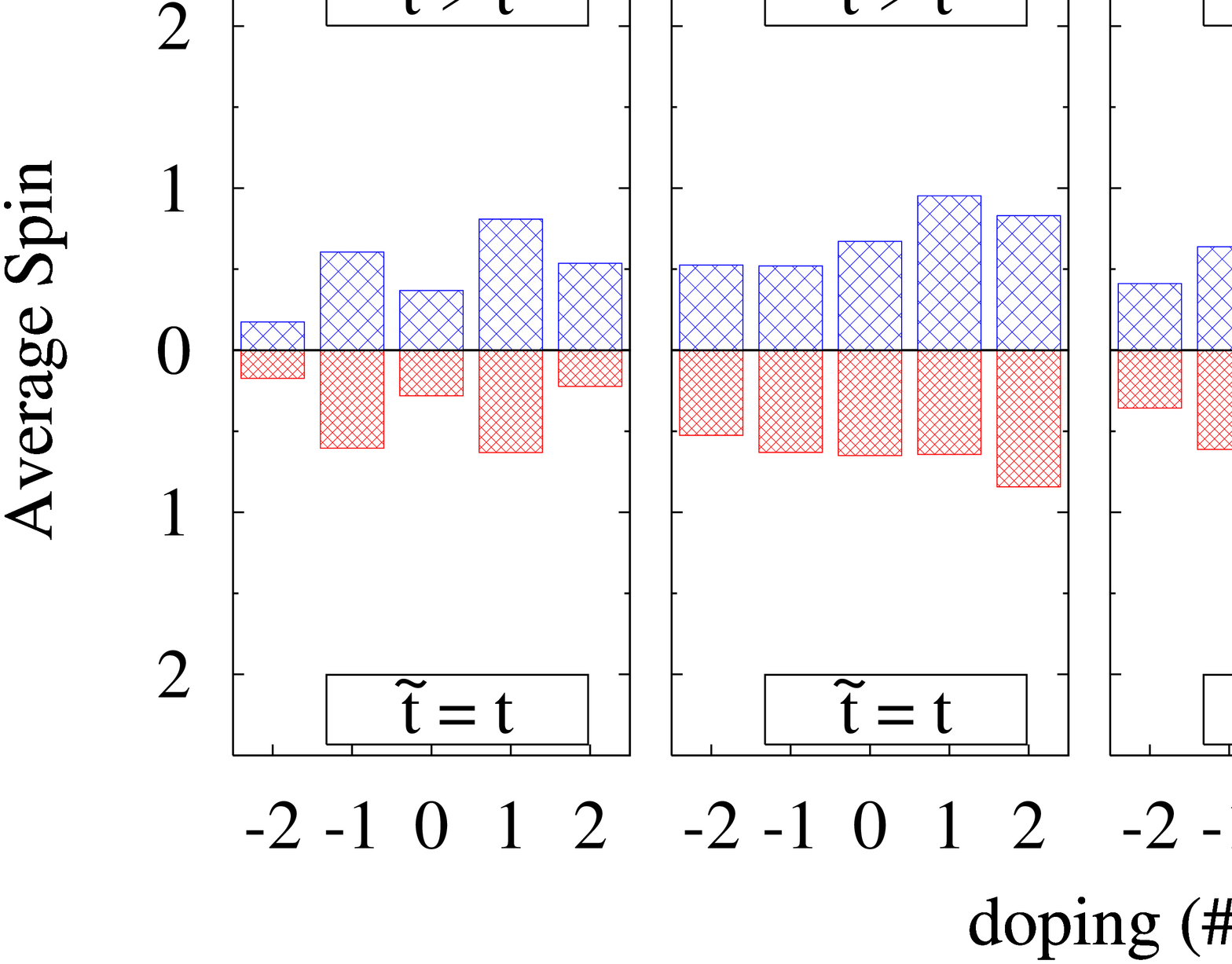}} \\ \hline
$\frac{1}{160}$ & \parbox{4in}{
\includegraphics[width=2in, angle=270]{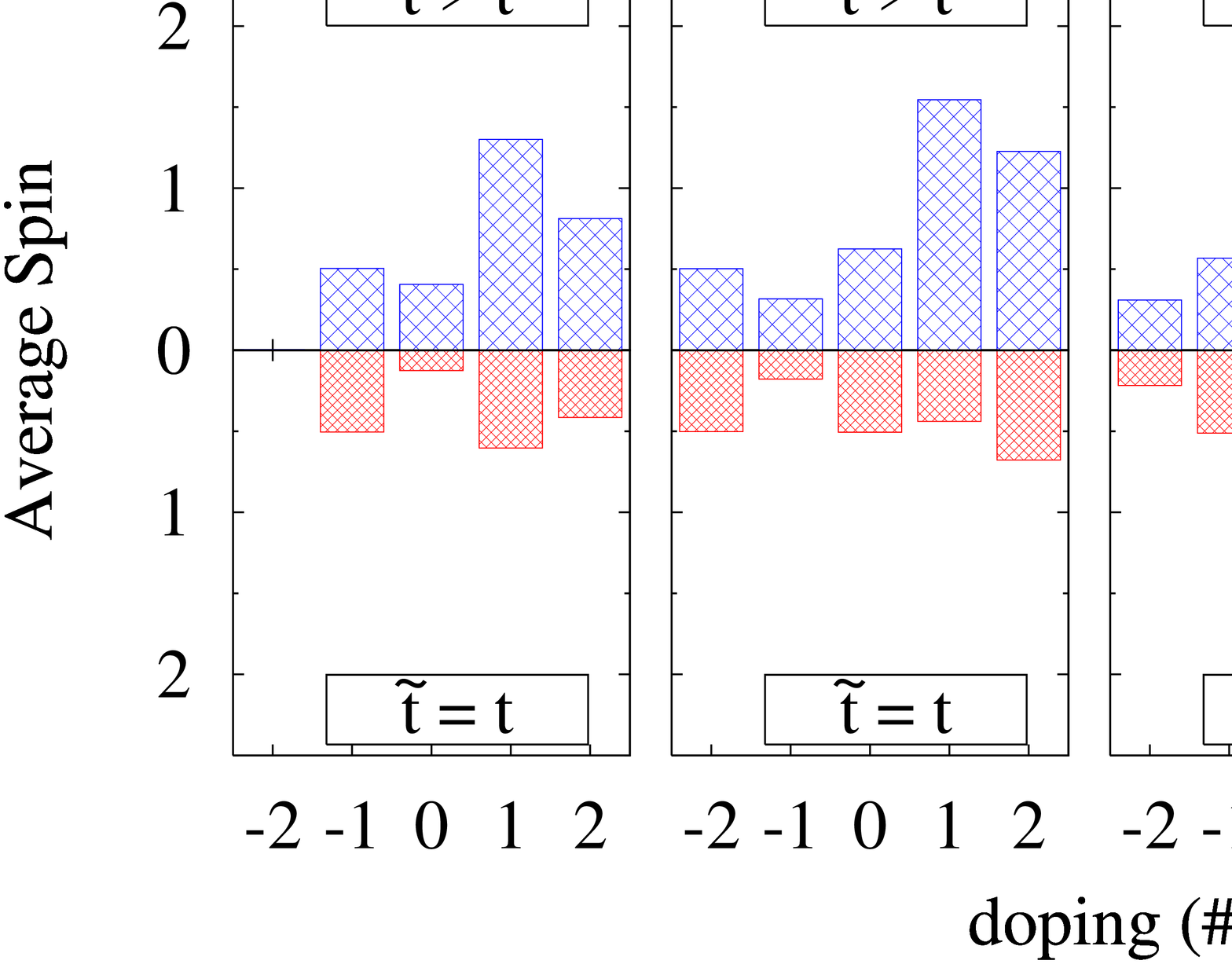}} \\ \hline
$\frac{3}{160}$ & \parbox{4in}{
\includegraphics[width=2in, angle=270]{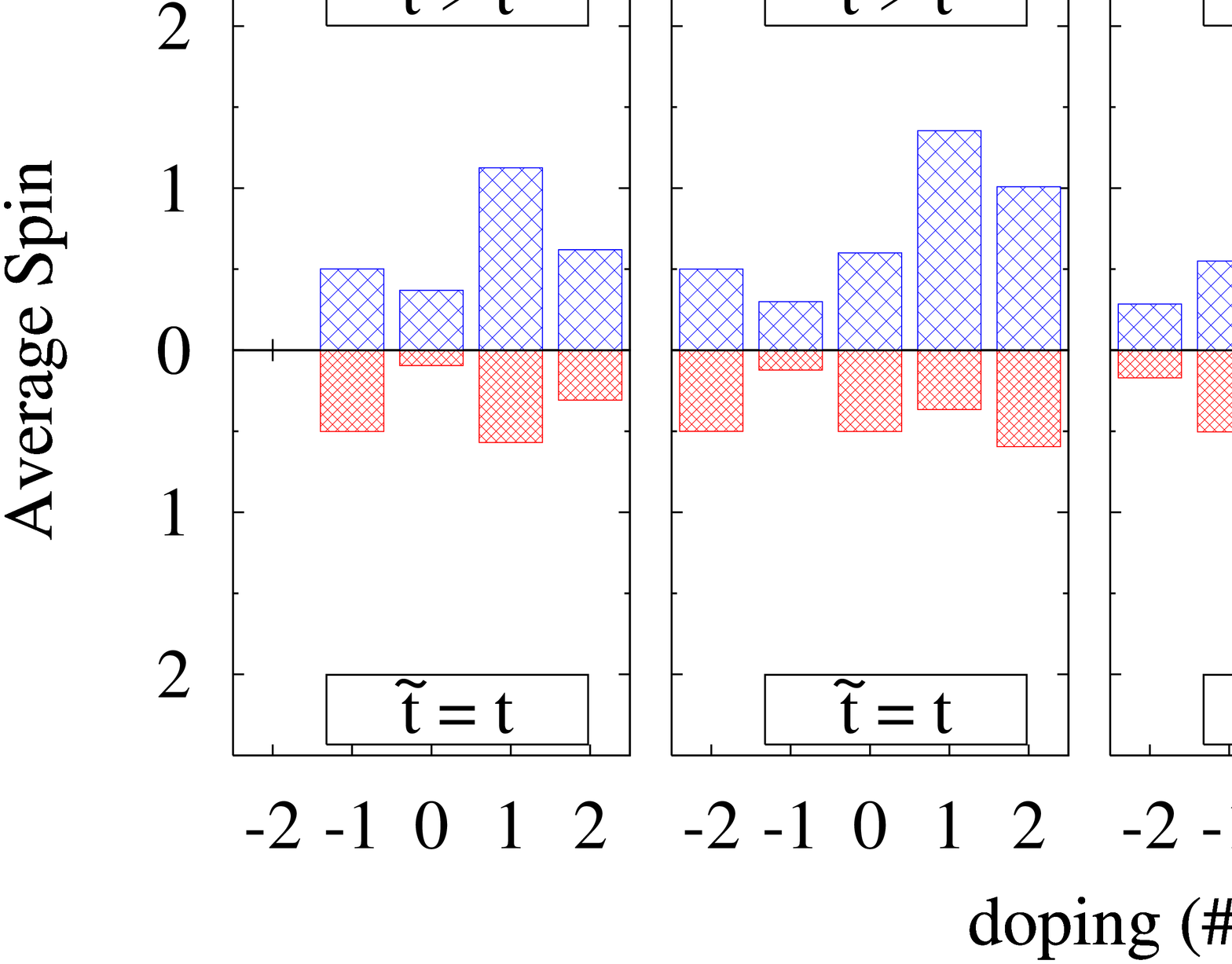}} \\ \hline
\end{tabular}}
\end{tabular}
\caption{Ground state average spin of 2D random clusters with fixed size and density, as a function of electron-doping (negative = hole-doping).  Data for systems with \emph{open and periodic boundary conditions} is shown in the left and right table respectively.  The lower half of plots are the result of setting $\tOuter_{ij}=\tInner_{ij}$, determined by the bandwidth of the lower Hubbard band.  The upper half use $\tOuter_{ij}$ determined by the bandwidth of the upper Hubbard ($D^-$) band. \label{afigAvgSpin2D}}
\end{center}
\end{figure}


\begin{figure}
\begin{center}
\begin{tabular}{cc}
\parbox{0.5\linewidth}{
\begin{tabular}{|c|c|} \hline
$\rho$ & \textbf{2D \ : \ Average Spin - $\mathbf{S_{min}}$\ : \ open b.c.}\\ \hline
$\frac{1}{1600}$ & \parbox{4in}{
\includegraphics[width=2in, angle=270]{figs/finalAvgS4-5-6-7_0.010_2Dr.ps}} \\ \hline
$\frac{1}{160}$ & \parbox{4in}{
\includegraphics[width=2in, angle=270]{figs/finalAvgS4-5-6-7_0.100_2Dr.ps}} \\ \hline
$\frac{3}{160}$ & \parbox{4in}{
\includegraphics[width=2in, angle=270]{figs/finalAvgS4-5-6-7_0.300_2Dr.ps}} \\ \hline
\end{tabular}}
&
\parbox{0.5\linewidth}{
\begin{tabular}{|c|c|} \hline
$\rho$ & \textbf{2D \ : \ Average Spin - $\mathbf{S_{min}}$\ : \ periodic b.c.}\\ \hline
$\frac{1}{1600}$ & \parbox{4in}{
\includegraphics[width=2in, angle=270]{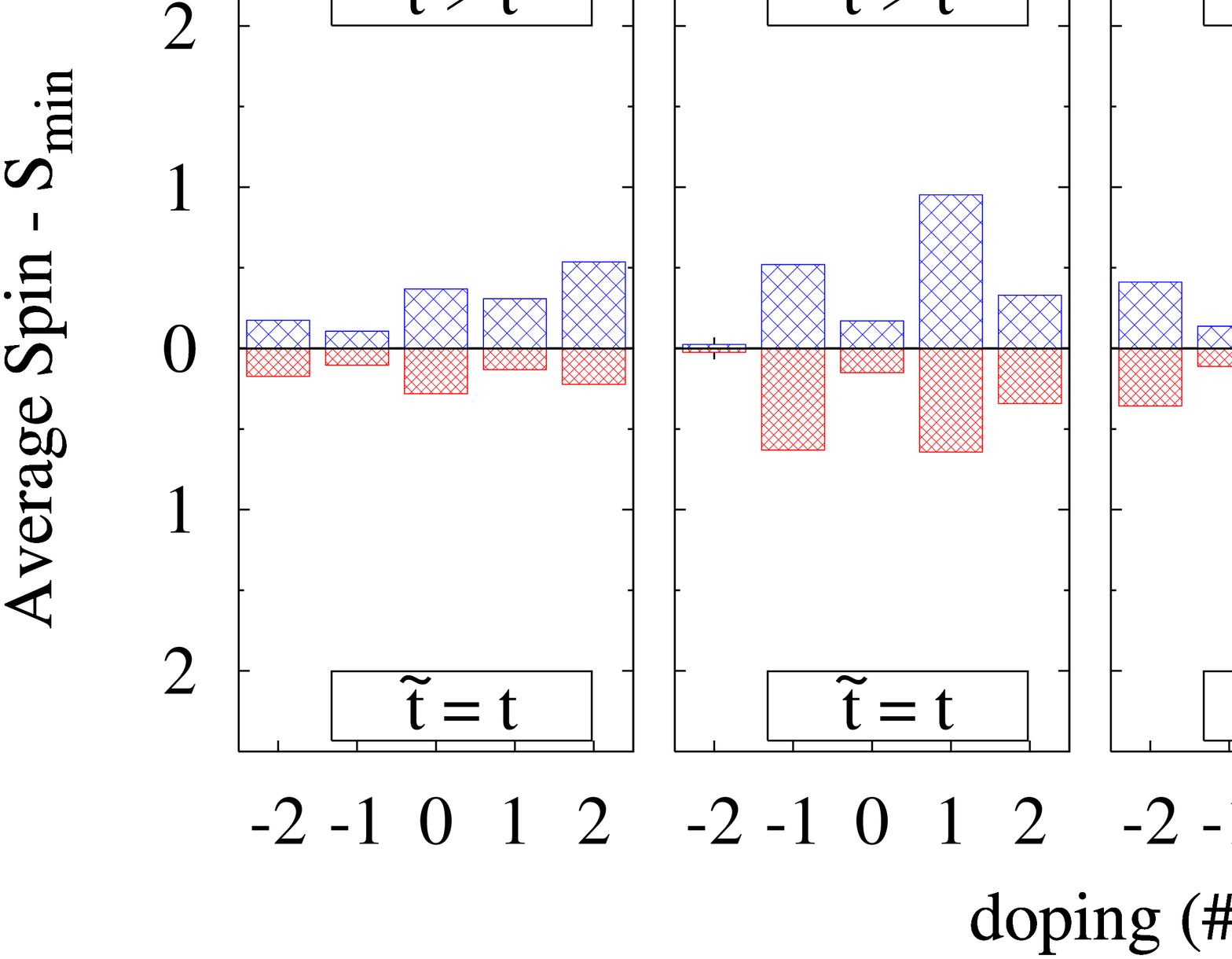}} \\ \hline
$\frac{1}{160}$ & \parbox{4in}{
\includegraphics[width=2in, angle=270]{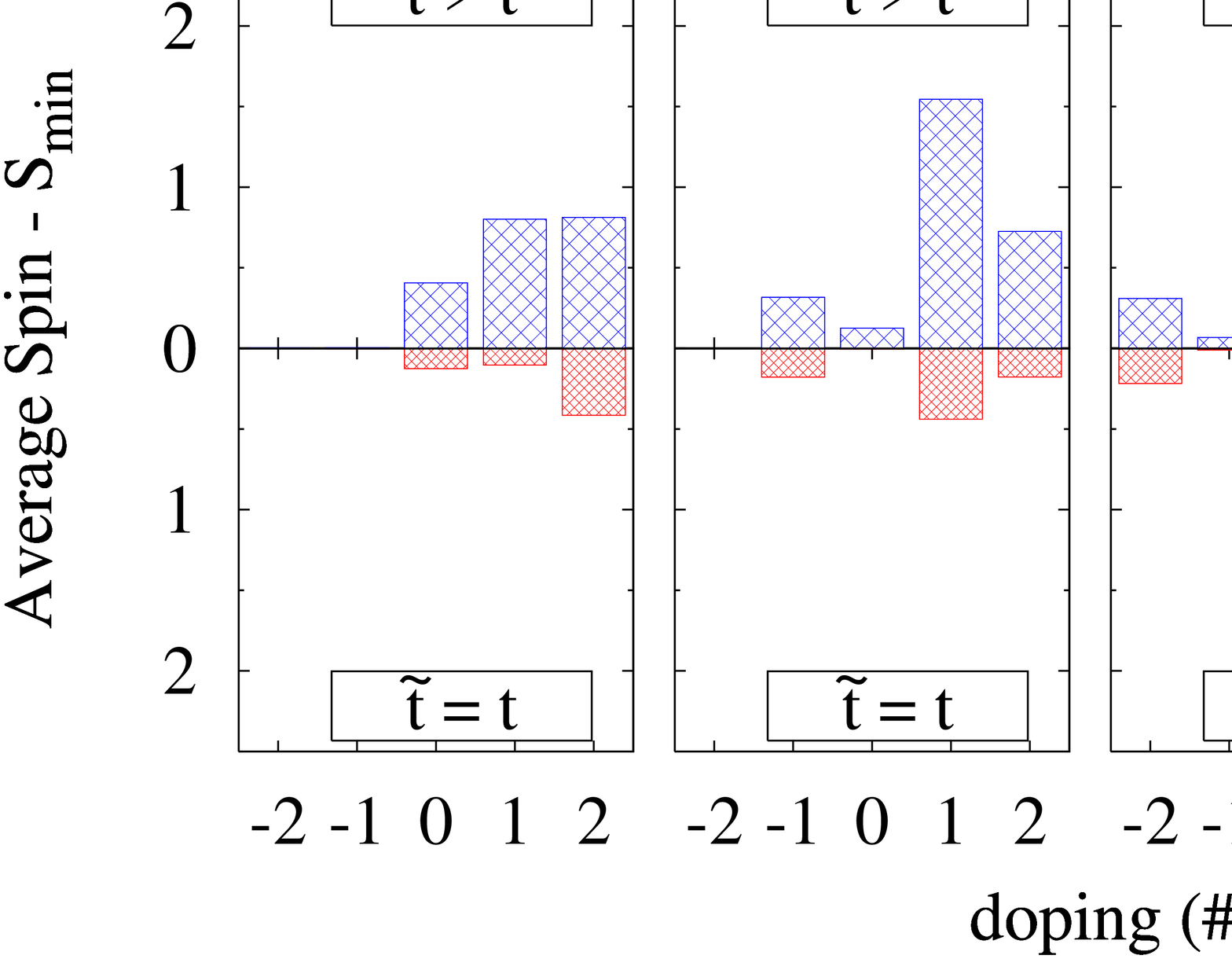}} \\ \hline
$\frac{3}{160}$ & \parbox{4in}{
\includegraphics[width=2in, angle=270]{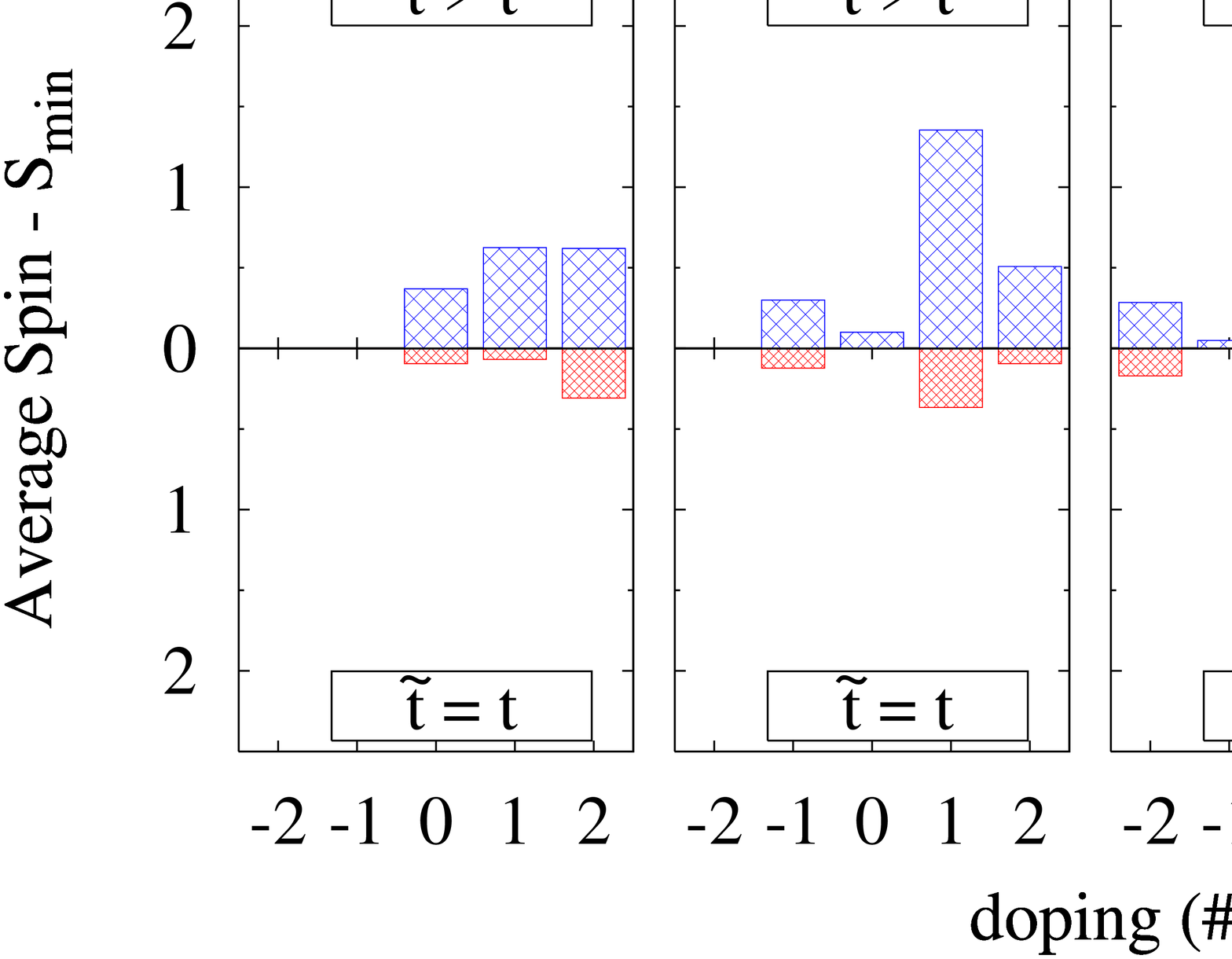}} \\ \hline
\end{tabular}}
\end{tabular}
\caption{Ground state average spin \emph{relative to minimum spin} of 2D random clusters with fixed size and density, as a function of electron-doping (negative = hole-doping). Data for systems with \emph{open and periodic boundary conditions} is shown in the left and right table respectively. The lower half of plots are the result of setting $\tOuter_{ij}=\tInner_{ij}$, determined by the bandwidth of the lower Hubbard band.  The upper half use $\tOuter_{ij}$ determined by the bandwidth of the upper Hubbard ($D^-$) band. \label{afigAvgSpin2Dr}}
\end{center}
\end{figure}


\begin{figure}
\begin{center}
\begin{tabular}{cc}
\parbox{0.5\linewidth}{
\begin{tabular}{|c|c|} \hline
$\rho$ & \textbf{2D \ : \ \% magnetic clusters \ : \ open b.c.}\\ \hline
$\frac{1}{1600}$ & \parbox{4in}{
\includegraphics[width=2in, angle=270]{figs/finalPcMag4-5-6-7_0.010_2D.ps}} \\ \hline
$\frac{1}{160}$ & \parbox{4in}{
\includegraphics[width=2in, angle=270]{figs/finalPcMag4-5-6-7_0.100_2D.ps}} \\ \hline
$\frac{3}{160}$ & \parbox{4in}{
\includegraphics[width=2in, angle=270]{figs/finalPcMag4-5-6-7_0.300_2D.ps}} \\ \hline
\end{tabular}}
&
\parbox{0.5\linewidth}{
\begin{tabular}{|c|c|} \hline
$\rho$ & \textbf{2D \ : \ \% magnetic clusters \ : \ periodic b.c.}\\ \hline
$\frac{1}{1600}$ & \parbox{4in}{
\includegraphics[width=2in, angle=270]{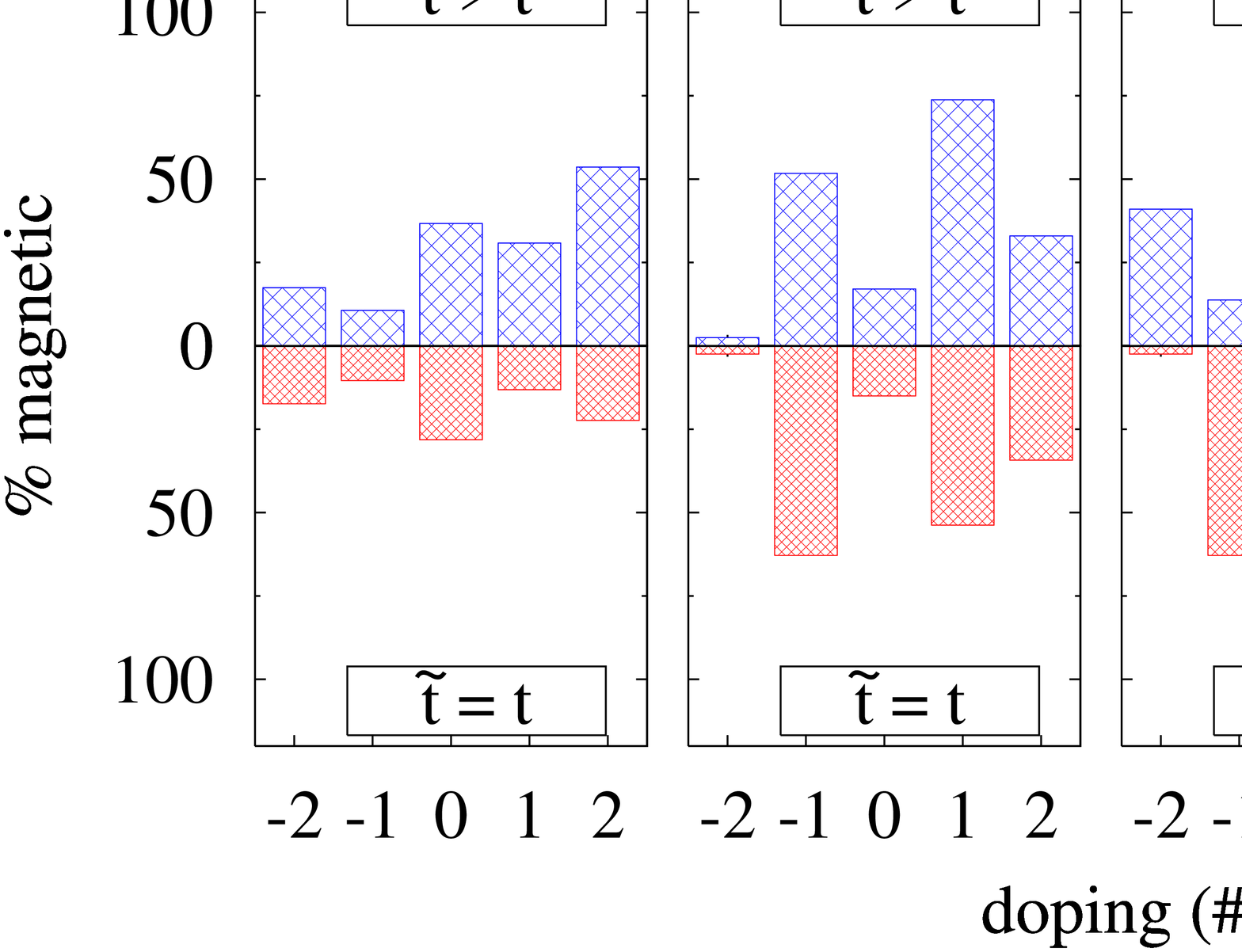}} \\ \hline
$\frac{1}{160}$ & \parbox{4in}{
\includegraphics[width=2in, angle=270]{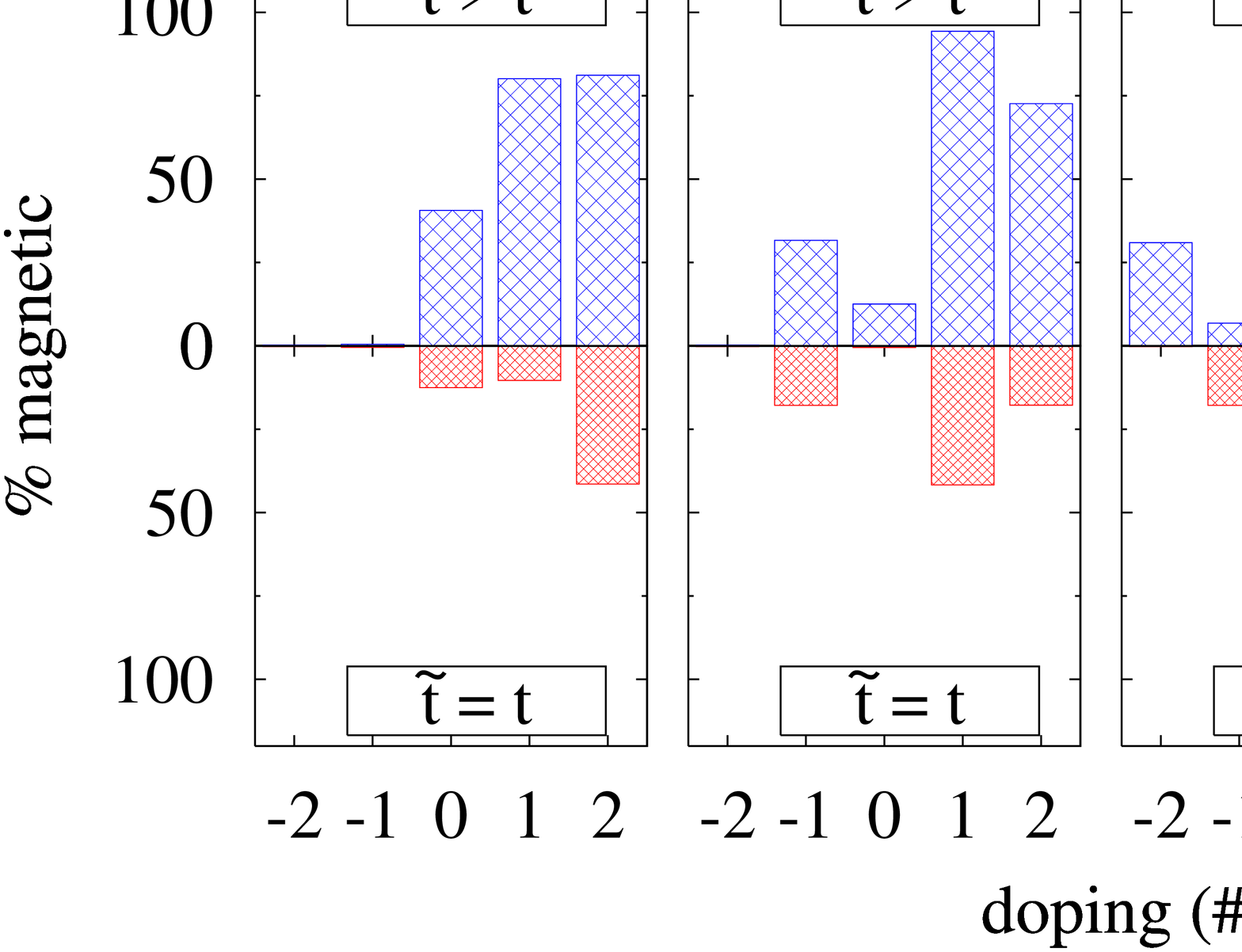}} \\ \hline
$\frac{3}{160}$ & \parbox{4in}{
\includegraphics[width=2in, angle=270]{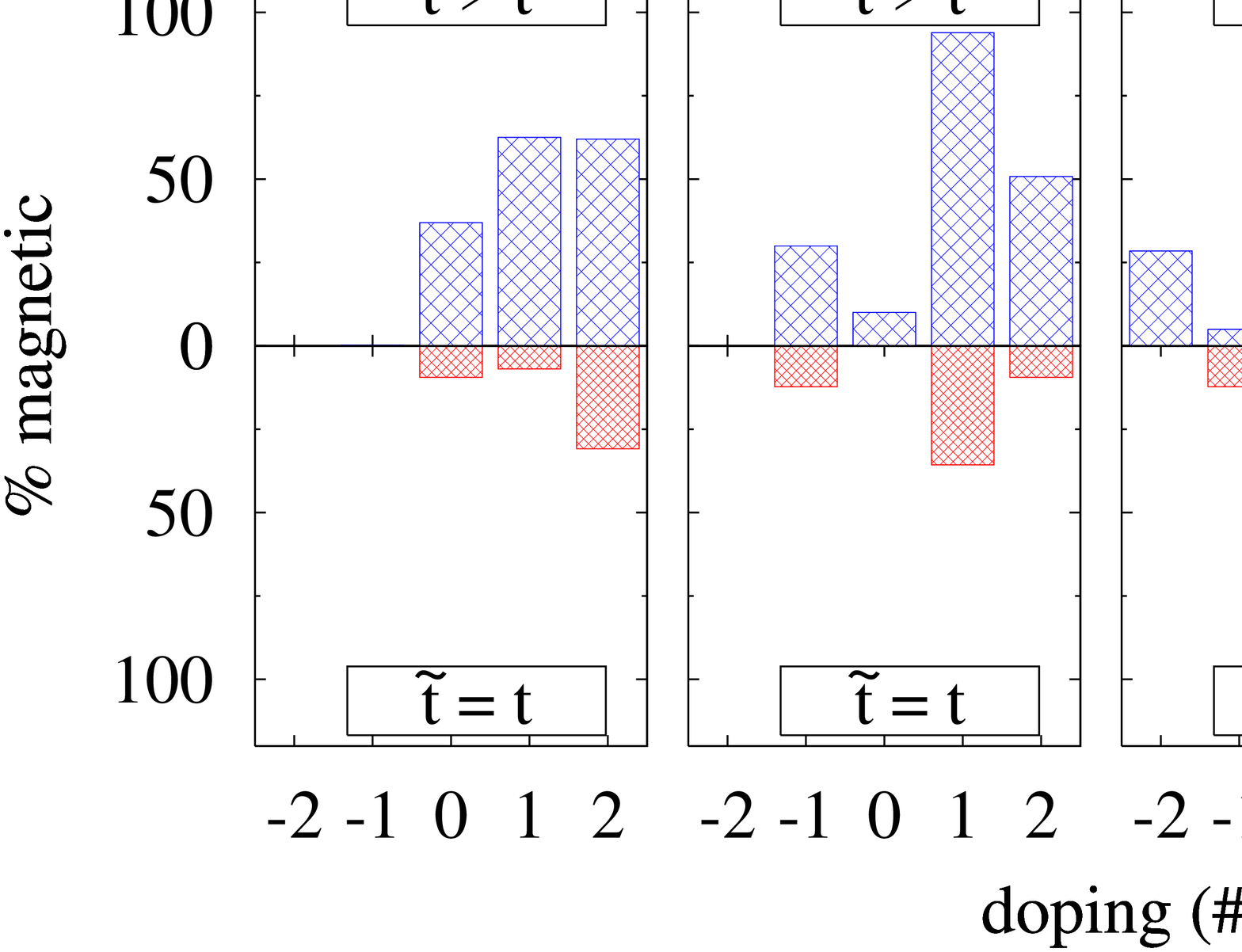}} \\ \hline
\end{tabular}}
\end{tabular}
\caption{Percentage of magnetic clusters (spin 1 or greater) in an ensemble of 2D random clusters with fixed size and density, as a function of electron-doping (negative = hole-doping). Data for systems with \emph{open and periodic boundary conditions} is shown in the left and right table respectively. The lower half of plots are the result of setting $\tOuter_{ij}=\tInner_{ij}$, determined by the bandwidth of the lower Hubbard band.  The upper half use $\tOuter_{ij}$ determined by the bandwidth of the upper Hubbard ($D^-$) band. \label{afigPcMag2D}}
\end{center}
\end{figure}


\end{turnpage}

\end{widetext}

\clearpage

\subsection{Average spin and percentage magnetic clusters from fixed density large systems}

We next consider large systems with a fixed number of sites $\Nsystem$ (10,000 to 1,000,000) and doping ($\Ntelec$ total electrons).  Each system is separately partitioned into clusters of size $\Nsites=2-7$, which are approximated as being independent, and then diagonalized.  The resulting data, averaged over many ($\sim 50$) large systems, gives the ensemble average distribution of clusters' ground state spin.  The same general trends appear here as for the clusters with fixed local density. For comparison we show the average spin and percentage of magnetic clusters in two dimensions in Figs.~\ref{afigAvgSpin2D_clFile} and \ref{afigPcMag2D_clFile} (for each cluster size separately).  We use the same plot format as for the clusters of fixed density, but show only the open boundary condition case.  Note that although the cluster size is fixed, there is substantial fluctuation in the local density of clusters;  only the average density of the \emph{entire} system is fixed.  Results for 3D clusters are given in Ref.~\onlinecite{NielsenThesis}.  

\begin{widetext}

\begin{figure*}[b]
\begin{center}
\begin{tabular}{|c|c|} \hline
$\bar{\rho}$ & \textbf{Large System \ : \ 2D \ : \ Average Spin \ : \ open b.c.}\\ \hline
$\frac{1}{1600}$ & \parbox{4in}{
\includegraphics[width=2in, angle=270]{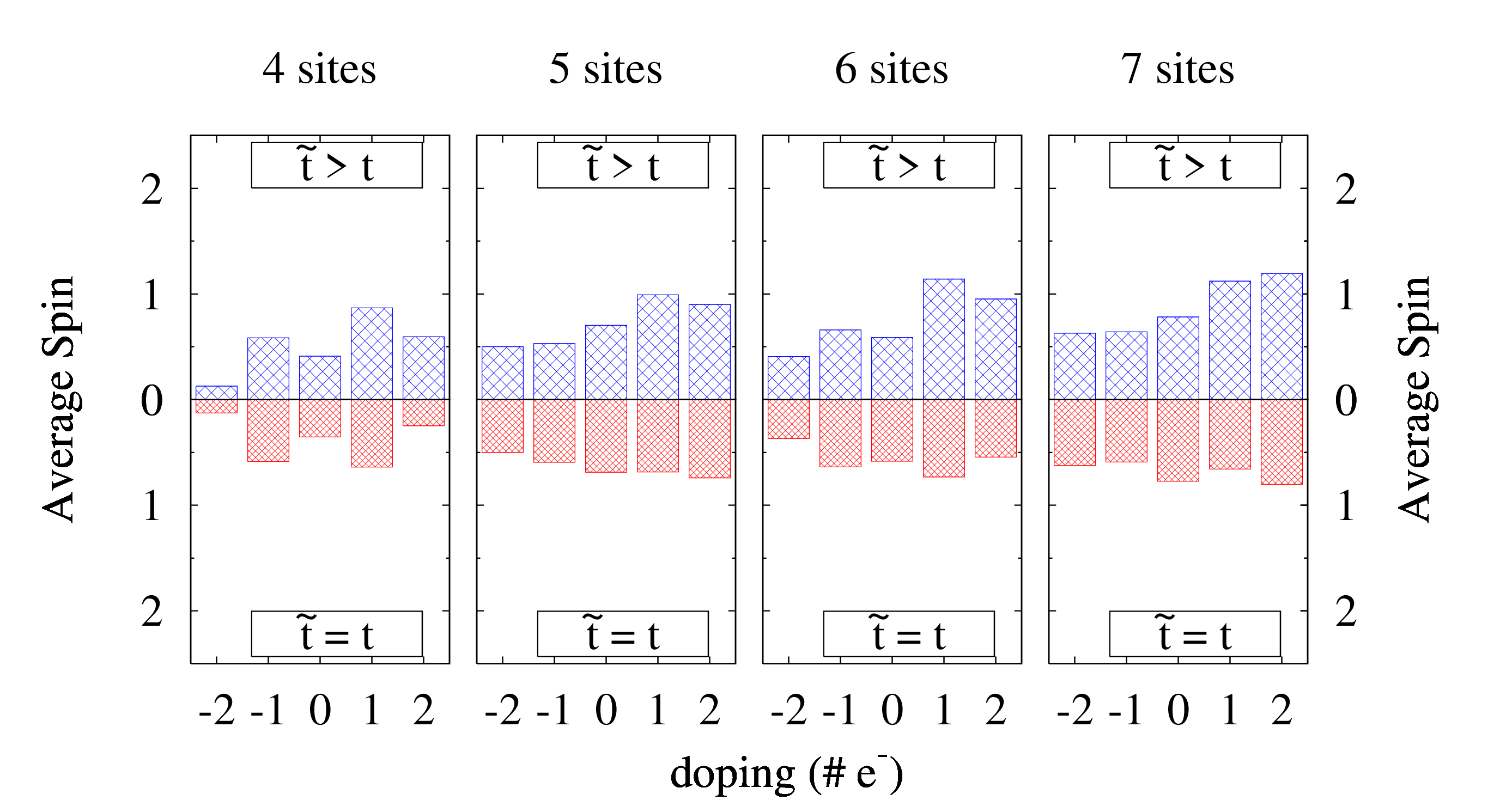}} \\ \hline
$\frac{1}{160}$ & \parbox{4in}{
\includegraphics[width=2in, angle=270]{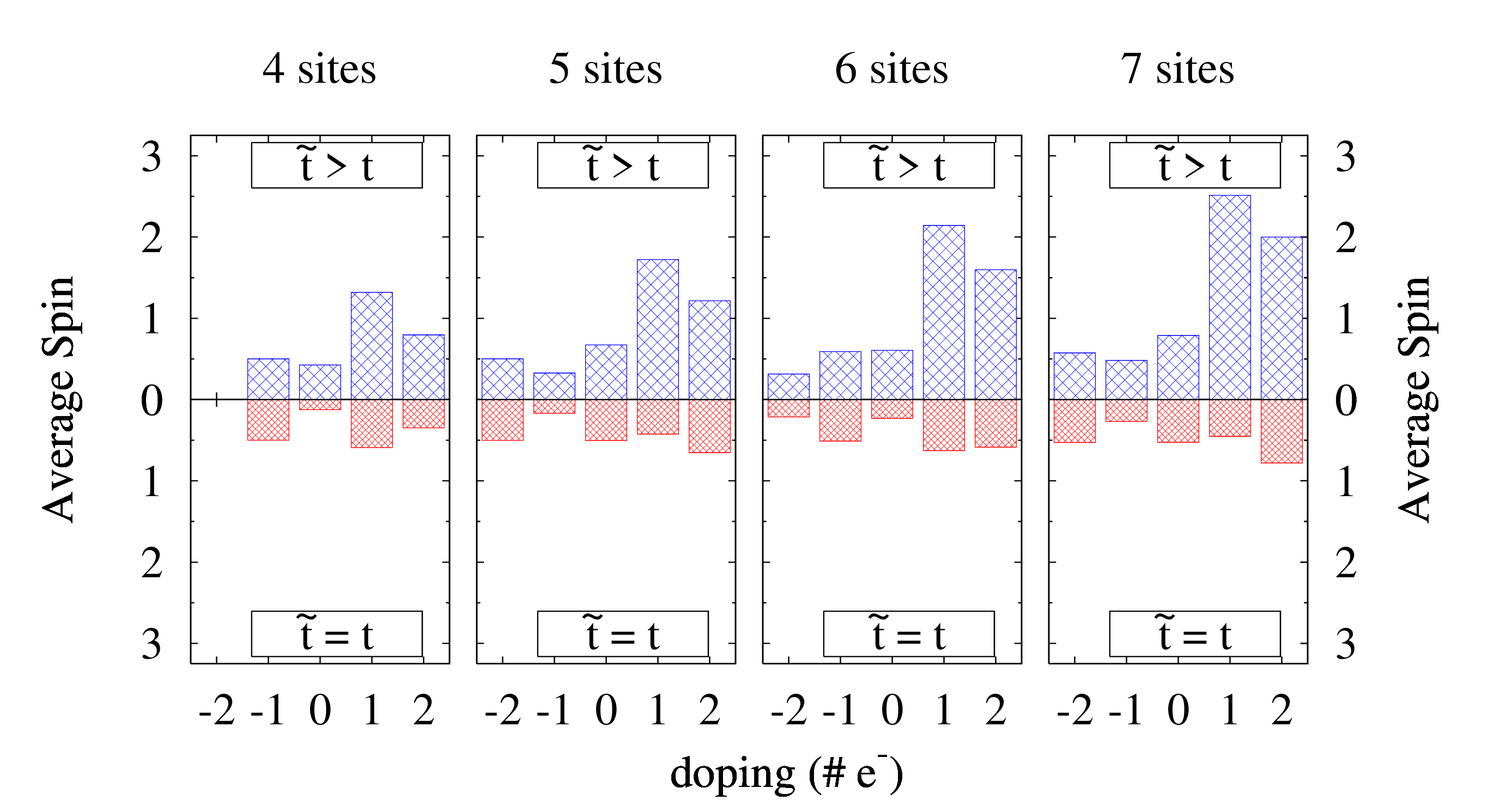}} \\ \hline
$\frac{3}{160}$ & \parbox{4in}{
\includegraphics[width=2in, angle=270]{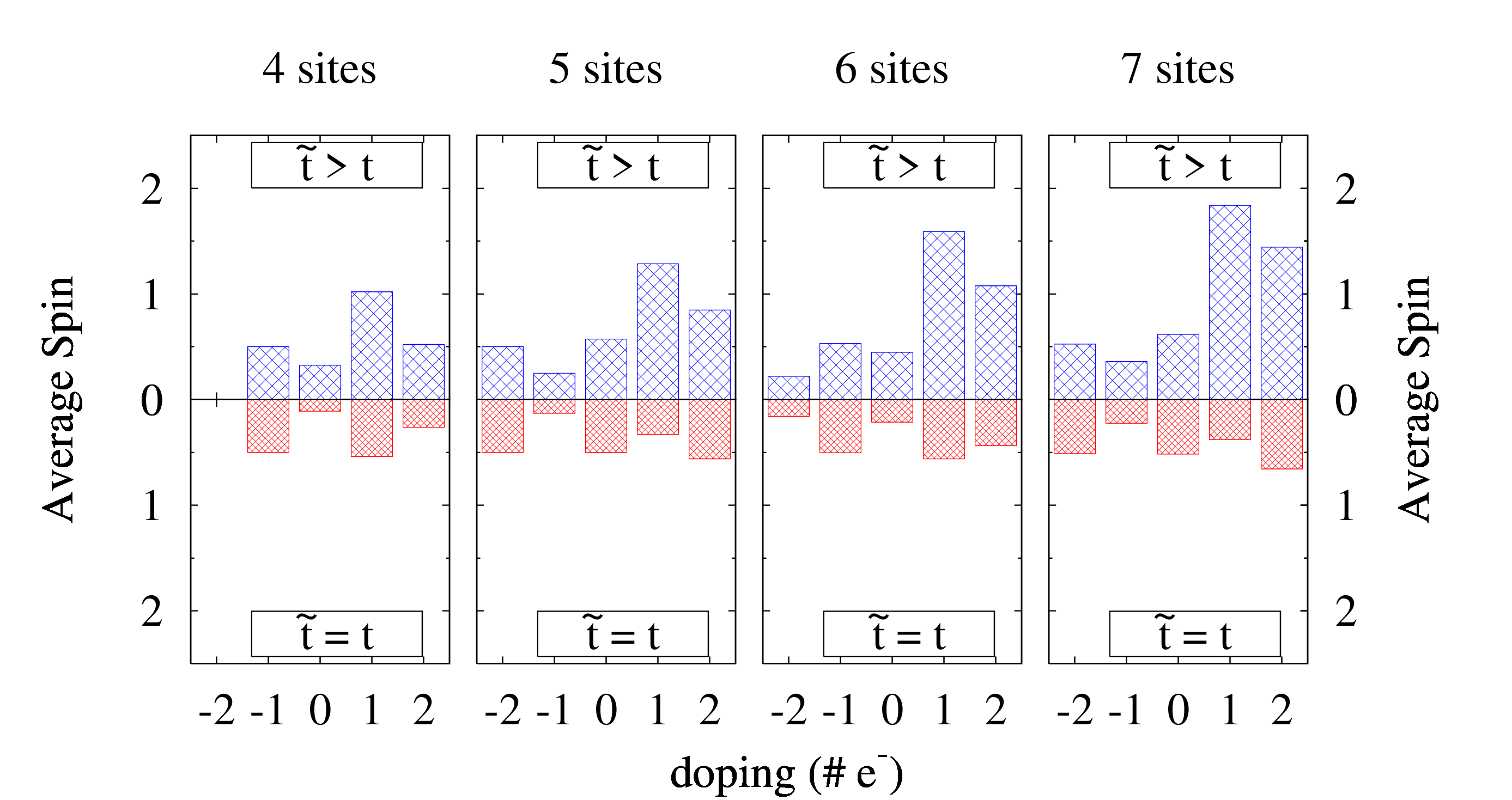}} \\ \hline
\end{tabular}
\caption{Ground state average spin of 2D random clusters (\emph{open b.c.}) obtained from large systems ($\Nsystem = 1\times 10^6$) with fixed average density $\bar{\rho}$, as a function of electron-doping (negative = hole-doping).  The lower half of plots are the result of setting $\tOuter_{ij}=\tInner_{ij}$, determined by the bandwidth of the lower Hubbard band.  The upper half use $\tOuter_{ij}$ determined by the bandwidth of the upper Hubbard ($D^-$) band. \label{afigAvgSpin2D_clFile}}
\end{center}
\end{figure*}

\begin{figure*}
\begin{center}
\begin{tabular}{|c|c|} \hline
$\bar{\rho}$ & \textbf{Large System \ : \ 2D \ : \ \% magnetic clusters \ : \ open b.c.}\\ \hline
$\frac{1}{1600}$ & \parbox{4in}{
\includegraphics[width=2in, angle=270]{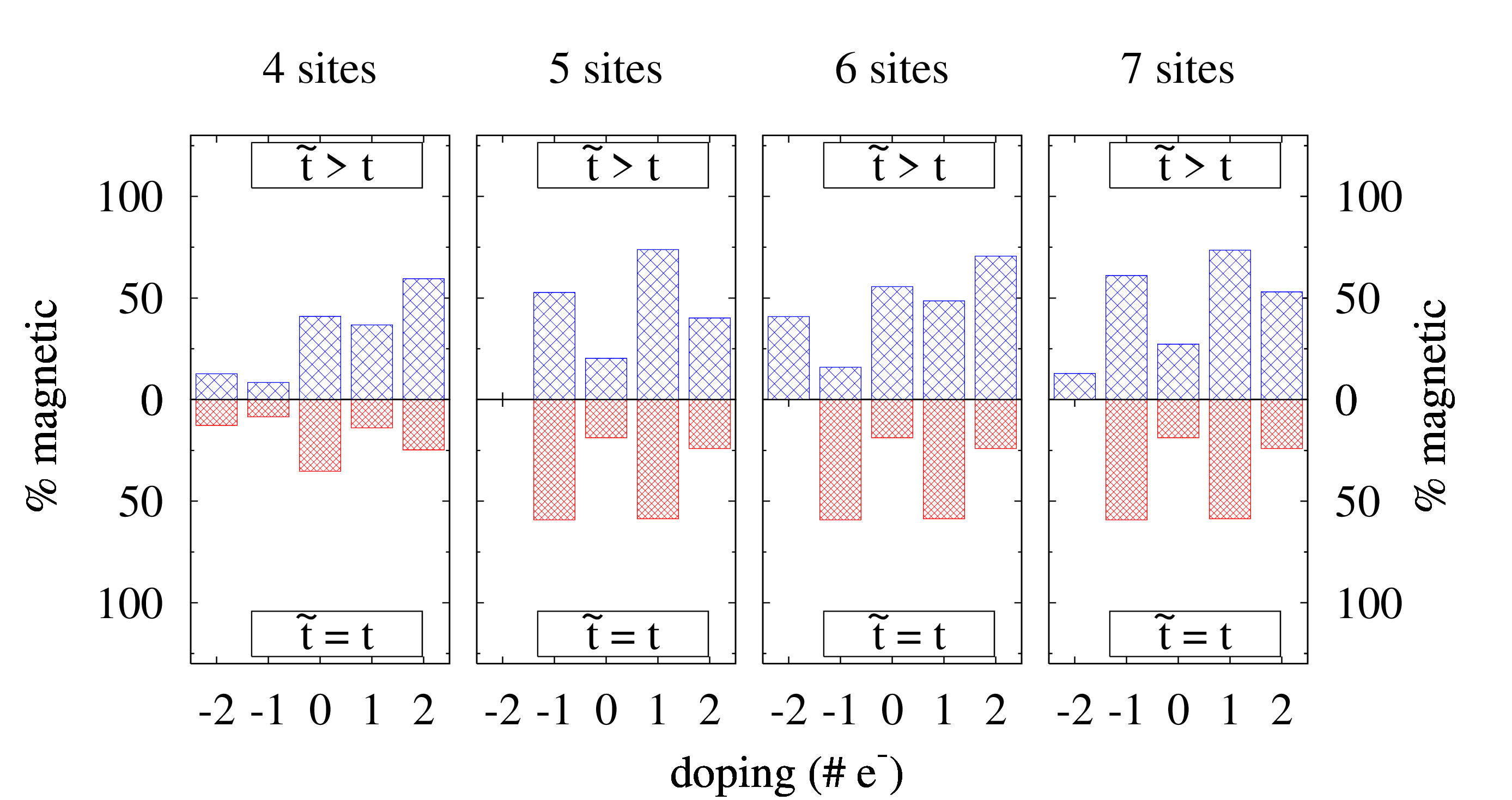}} \\ \hline
$\frac{1}{160}$ & \parbox{4in}{
\includegraphics[width=2in, angle=270]{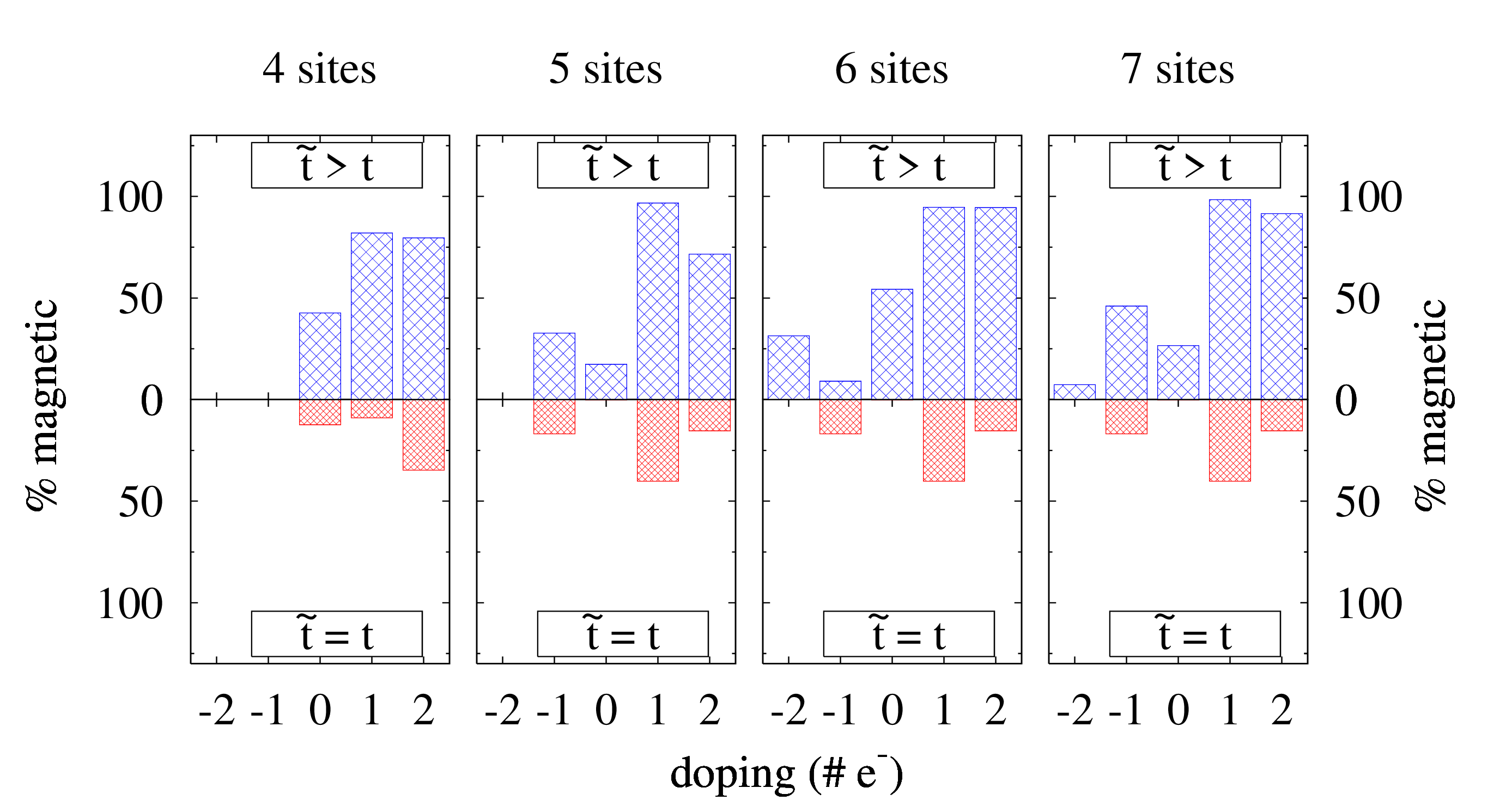}} \\ \hline
$\frac{3}{160}$ & \parbox{4in}{
\includegraphics[width=2in, angle=270]{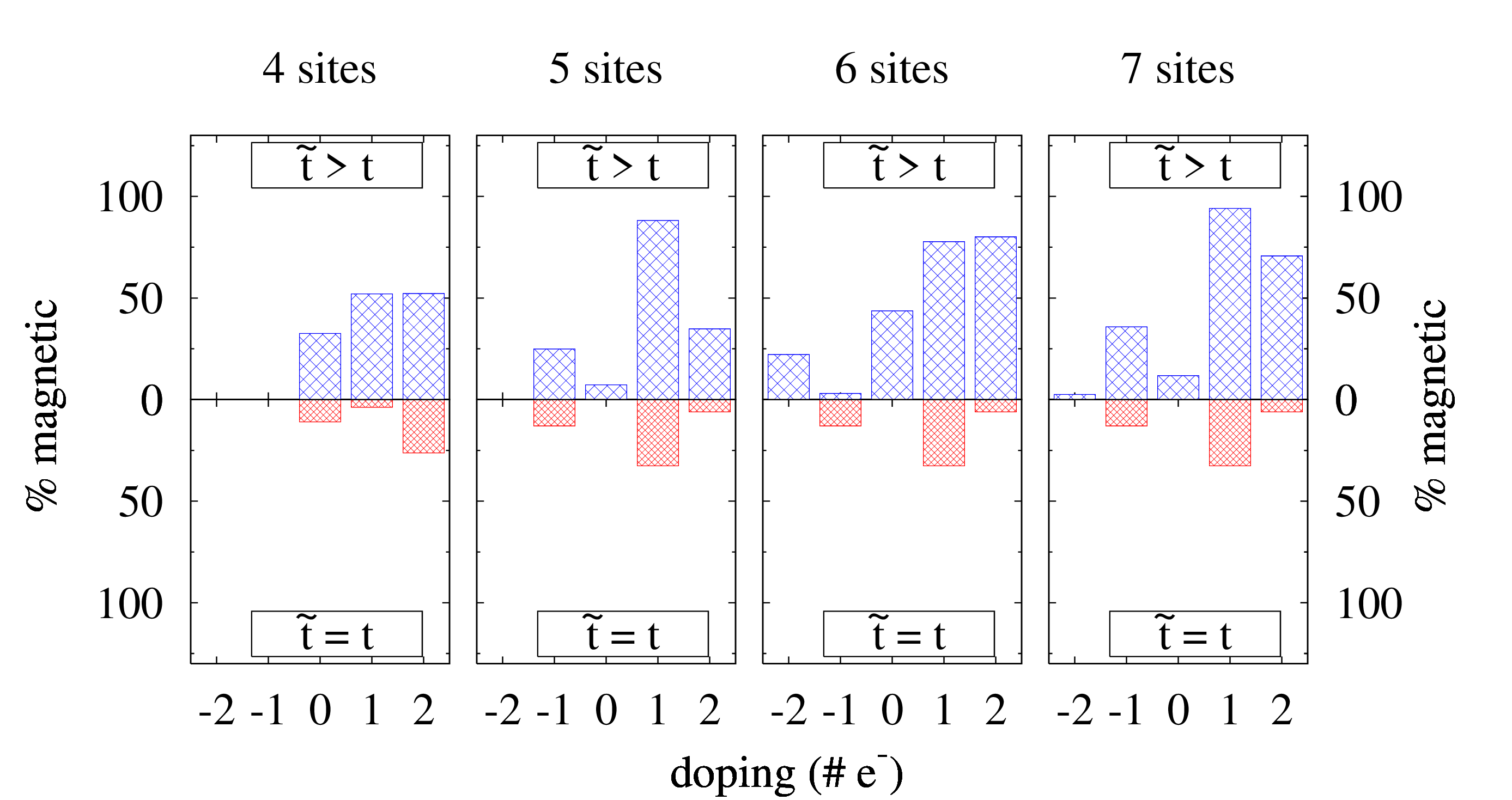}} \\ \hline
\end{tabular}
\caption{Percentage of magnetic clusters (spin 1 or greater) in an ensemble of 2D random clusters (\emph{open b.c.}) obtained from large systems ($\Nsystem = 1\times 10^6$) with fixed average density $\bar{\rho}$, plotted as a function of electron-doping (negative = hole-doping).  The lower half of plots are the result of setting $\tOuter_{ij}=\tInner_{ij}$, determined by the bandwidth of the lower Hubbard band.  The upper half use $\tOuter_{ij}$ determined by the bandwidth of the upper Hubbard ($D^-$) band. \label{afigPcMag2D_clFile}}
\end{center}
\end{figure*}

\end{widetext}

\clearpage

\bibliographystyle{unsrtnat}
\bibliography{hubbardPaperLarge}

\begin{thebibliography}{112}
\providecommand{\natexlab}[1]{#1}
\providecommand{\url}[1]{\texttt{#1}}
\expandafter\ifx\csname urlstyle\endcsname\relax
  \providecommand{\doi}[1]{doi: #1}\else
  \providecommand{\doi}{doi: \begingroup \urlstyle{rm}\Url}\fi

\bibitem[Hubbard(1963)]{Hubbard_1963}
J.~Hubbard.
\newblock \emph{Proc. Roy. Soc. A}, 276:\penalty0 238, 1963.

\bibitem[Gutzwiller(1963)]{Gutzwiller_1963}
Martin~C. Gutzwiller.
\newblock \emph{Phys. Rev. Lett.}, 10:\penalty0 159, 1963.

\bibitem[Kanamori(1963)]{Kanamori_1963}
J.~Kanamori.
\newblock \emph{Prog. Theor. Phys.}, 30:\penalty0 275, 1963.

\bibitem[Anderson(1963{\natexlab{a}})]{Anderson_HubModel_1963}
P.~W. Anderson.
\newblock \emph{Solid State Phys. (F. Seitz and D. Turnbull, eds.)},
  14:\penalty0 99, 1963{\natexlab{a}}.

\bibitem[Mott(1990)]{MottBook}
N.~F. Mott.
\newblock \emph{Metal-insulator Transitions 2nd Ed.}
\newblock Taylor and Francis London, 1990.
\newblock and references therein.

\bibitem[Anderson(1987)]{AndersonCuprates_1987}
P.~W. Anderson.
\newblock \emph{Science}, 235:\penalty0 1196, 1987.

\bibitem[Lee et~al.(2006)Lee, Nagaosa, and Wen]{LeeCupratesRMP_2006}
Patrick~A. Lee, Naoto Nagaosa, and Xiao-Gang Wen.
\newblock \emph{Rev. Mod. Phys.}, 78:\penalty0 17, 2006.

\bibitem[Izyumov(1997)]{IzyumovTJmodel_1997}
Yurii~A. Izyumov.
\newblock \emph{Physics-Uspekhi}, 40:\penalty0 445, 1997.

\bibitem[Macridin et~al.(2005)Macridin, Jarrell, Maier, and
  Sawatzky]{MacridinCuprates_2005}
A.~Macridin, M.~Jarrell, Th. Maier, and G.~A. Sawatzky.
\newblock \emph{Phys. Rev. B}, 71:\penalty0 134527, 2005.

\bibitem[Pyo et~al.(2005)Pyo, Ma, He, Xu, and Yang]{Pyo_2005}
Seungmoon Pyo, Liping Ma, Jun He, Qianfei Xu, and Yang Yang.
\newblock \emph{J. Appl. Phys.}, 98:\penalty0 054303, 2005.

\bibitem[Wu et~al.(2004)Wu, Ma, and Yang]{Wu_2004}
Jianhua Wu, Liping Ma, and Yang Yang.
\newblock \emph{Phys. Rev. B}, 69:\penalty0 115321, 2004.

\bibitem[Sing et~al.(2003)Sing, Schwingenschl\"{o}gl, Claessen, Blaha, Carmelo,
  Martelo, Sacramento, Dressel, and Jacobsen]{Sing_2003}
M.~Sing, U.~Schwingenschl\"{o}gl, R.~Claessen, P.~Blaha, J.~M.~P. Carmelo,
  L.~M. Martelo, P.~D. Sacramento, M.~Dressel, and C.~S. Jacobsen.
\newblock \emph{Phys. Rev. B}, 68:\penalty0 125111, 2003.

\bibitem[Weitering et~al.(1997)Weitering, Shi, Johnson, Chen, DiNardo, and
  Kempa]{Weitering_1997}
H.~H. Weitering, X.~Shi, P.~D. Johnson, J.~Chen, N.~J. DiNardo, and K.~Kempa.
\newblock \emph{Phys. Rev. Lett.}, 78:\penalty0 1331, 1997.

\bibitem[McWhan et~al.(1973)McWhan, Menth, Remeika, Brinkman, and
  Rice]{McWhan_1973}
D.~B. McWhan, A.~Menth, J.~P. Remeika, W.~F. Brinkman, and T.~M. Rice.
\newblock \emph{Phys. Rev. B}, 7:\penalty0 1920, 1973.

\bibitem[Carter et~al.(1991)Carter, Yang, Rosenbaum, Spalek, and
  Honig]{Carter_1991}
S.~A. Carter, J.~Yang, T.~F. Rosenbaum, J.~Spalek, and J.~M. Honig.
\newblock \emph{Phys. Rev. B}, 43:\penalty0 607, 1991.

\bibitem[Ogawa(1979)]{Ogawa_1979}
S.~Ogawa.
\newblock \emph{J. Appl. Phys.}, 50:\penalty0 2308, 1979.

\bibitem[Thio and Bennett(1994)]{ThioBennett_1994}
Tineke Thio and J.~W. Bennett.
\newblock \emph{Phys. Rev. B}, 50:\penalty0 10574, 1994.

\bibitem[Thio et~al.(1995)Thio, Bennett, and Thurston]{ThioBennett_1995}
Tineke Thio, J.~W. Bennett, and T.~R. Thurston.
\newblock \emph{Phys. Rev. B}, 52:\penalty0 3555, 1995.

\bibitem[da~Silva et~al.(1981)da~Silva, Kishore, and
  da~Cunha~Lima]{FerreiraSpecificHeat_1981}
A.~Ferreira da~Silva, R.~Kishore, and I.~C. da~Cunha~Lima.
\newblock \emph{Phys. Rev. B}, 23:\penalty0 4035, 1981.

\bibitem[Refolio et~al.(1996)Refolio, Sancho, and
  Rubio]{RefolioKSiInterface_1996}
M.~C. Refolio, J.~M.~L\'{o}pez Sancho, and J.~Rubio.
\newblock \emph{Phys. Rev. B}, 53:\penalty0 4791, 1996.

\bibitem[Rotani et~al.(1999)Rotani, Rossi, Manghi, and
  Molinari]{Massimo_hubInQDots_1999}
Massimo Rotani, Fausto Rossi, Franca Manghi, and Elisa Molinari.
\newblock \emph{Phys. Rev. B}, 59:\penalty0 10165, 1999.

\bibitem[Hanisch et~al.(1997)Hanisch, Uhrig, and
  M\"{u}ller-Hartmann]{Hanisch_diffLatt_1997}
Thoralf Hanisch, G\"{o}tz~S. Uhrig, and Erwin M\"{u}ller-Hartmann.
\newblock \emph{Phys. Rev. B}, 56:\penalty0 13960, 1997.

\bibitem[Wegner et~al.(1998)Wegner, Potthoff, and
  Nolting]{Wegner_diffLatt_1998}
T.~Wegner, M.~Potthoff, and W.~Nolting.
\newblock \emph{Phys. Rev. B}, 57:\penalty0 6211, 1998.

\bibitem[Penc et~al.(1996)Penc, Shiba, Mila, and Tsukagoshi]{Penc_multiBand}
Karlo Penc, Hiroyuki Shiba, Fr\`{e}d\'{e}ric Mila, and Takuya Tsukagoshi.
\newblock \emph{Phys. Rev. B}, 54:\penalty0 4056, 1996.

\bibitem[Fr\'{e}sard and Kotliar(1997)]{Fresard_degenBand}
Raymond Fr\'{e}sard and Gabriel Kotliar.
\newblock \emph{Phys. Rev. B}, 56:\penalty0 12909, 1997.

\bibitem[Kuei and Scalettar(1997)]{Kuei_degenBand}
J.~Kuei and R.~T. Scalettar.
\newblock \emph{Phys. Rev. B}, 55:\penalty0 14968, 1997.

\bibitem[Byczuk et~al.(2003)Byczuk, Ulmke, and Vollhardt]{Byczuk_alloyDisorder}
Krzysztof Byczuk, Martin Ulmke, and Dieter Vollhardt.
\newblock \emph{Phys. Rev. Lett.}, 90:\penalty0 196403, 2003.

\bibitem[Becca and Sorella(2001)]{Becca_largeU}
Federico Becca and Sandro Sorella.
\newblock \emph{Phys. Rev. Lett.}, 86:\penalty0 3396, 2001.

\bibitem[Obermeier et~al.(1997)Obermeier, Pruschke, and
  Keller]{Obermeier_largeU}
Thomas Obermeier, Thomas Pruschke, and Joachim Keller.
\newblock \emph{Phys. Rev. B}, 56:\penalty0 R8479, 1997.

\bibitem[Jaksch and Zoller(2005)]{JakschZoller_2005}
D.~Jaksch and P.~Zoller.
\newblock \emph{Ann. Phys.}, 315:\penalty0 52, 2005.

\bibitem[Thomas et~al.(1981)Thomas, Capizzi, DeRosa, Bhatt, and
  Rice]{Thomas_1981}
G.~A. Thomas, M.~Capizzi, F.~DeRosa, R.~N. Bhatt, and T.~M. Rice.
\newblock \emph{Phys. Rev. B}, 23:\penalty0 5472, 1981.

\bibitem[Andres et~al.(1981)Andres, Bhatt, Goalwin, Rice, and
  Walstedt]{AndresBhatt_1981}
K.~Andres, R.~N. Bhatt, P.~Goalwin, T.~M. Rice, and R.~E. Walstedt.
\newblock \emph{Phys. Rev. B}, 24:\penalty0 244, 1981.

\bibitem[Pekeris(1962)]{Pekeris_1962}
C.~L. Pekeris.
\newblock \emph{Phys. Rev.}, 126:\penalty0 1470, 1962.

\bibitem[Bethe and Salpeter(1977)]{BS_QMbook_1977}
Hans.~A. Bethe and Edwin~E. Salpeter.
\newblock \emph{Quantum Mechanics of 1 and 2 electron atoms}.
\newblock Springer, 1977.

\bibitem[Bhatt and Rice(1981)]{BhattRice_1981}
R.~N. Bhatt and T.~M. Rice.
\newblock \emph{Phys. Rev. B}, 23:\penalty0 1920, 1981.

\bibitem[Bhatt(1981)]{Bhatt_1981}
R.~N. Bhatt.
\newblock \emph{Phys. Rev. B}, 24:\penalty0 3630, 1981.

\bibitem[Anderson(1963{\natexlab{b}})]{AndersonEffHeisenberg_1963}
P.~W. Anderson.
\newblock \emph{Concepts in Solids}.
\newblock Benjamin, Reading, MA, 1963{\natexlab{b}}.

\bibitem[Manousakis(1991)]{ManousakisRMP_1991}
Efstratios Manousakis.
\newblock \emph{Rev. Mod. Phys.}, 63:\penalty0 1, 1991.

\bibitem[Chao et~al.(1977)Chao, Spalek, and Ol\'{e}s]{ChaoSpalekOles_1977}
K.~A. Chao, J.~Spalek, and A.~M. Ol\'{e}s.
\newblock \emph{J. PHys. C}, 10:\penalty0 271, 1977.

\bibitem[Chao et~al.(1978)Chao, Spalek, and Ol\'{e}s]{ChaoSpalekOles_1978}
K.~A. Chao, J.~Spalek, and A.~M. Ol\'{e}s.
\newblock \emph{Phys. Rev. B}, 18:\penalty0 3453, 1978.

\bibitem[Anderson(1959)]{AndersonExchange_1959}
P.~W. Anderson.
\newblock \emph{Phys. Rev.}, 115:\penalty0 2, 1959.

\bibitem[Fradkin(1991)]{FradkinBook_1991}
Eduardo Fradkin.
\newblock \emph{Field theories of condensed matter systems}.
\newblock Westview Press, 1991.

\bibitem[Georges et~al.(1996)Georges, Kotliar, Krauth, and
  Rozenberg]{GeorgesKotliar_1996}
Antoine Georges, Gabriel Kotliar, Werner Krauth, and Marcelo~J. Rozenberg.
\newblock \emph{Rev. Mod. Phys.}, 68:\penalty0 13, 1996.

\bibitem[Ulmke et~al.(1995)Ulmke, Jani\u{s}, and
  Vollhardt]{UlmkeJanisVollhart_1995}
M.~Ulmke, V.~Jani\u{s}, and D.~Vollhardt.
\newblock \emph{Phys. Rev. B}, 51:\penalty0 10411, 1995.

\bibitem[Chandra et~al.(1999)Chandra, Kollar, and
  Vollhardt]{ChandraKollarVollhart_1999}
N.~Chandra, M.~Kollar, and D.~Vollhardt.
\newblock \emph{Phys. Rev. B}, 59:\penalty0 10541, 1999.

\bibitem[Eckstein et~al.(2007)Eckstein, Kollar, Potthoff, and
  Vollhardt]{Eckstein_2007}
Martin Eckstein, Marcus Kollar, Michael Potthoff, and Dieter Vollhardt.
\newblock \emph{Phys. Rev. B}, 75:\penalty0 125103, 2007.

\bibitem[Brinkman and Rice(1970)]{BrinkmanRice_1970}
W.~F. Brinkman and T.~M. Rice.
\newblock \emph{Phys. Rev. B}, 2:\penalty0 1324, 1970.

\bibitem[Shraiman and Siggia(1988)]{ShraimonSiggia_1988}
Boris~I. Shraiman and Eric~D. Siggia.
\newblock \emph{Phys. Rev. Lett.}, 61:\penalty0 467, 1988.

\bibitem[Nagaoka(1966)]{Nagaoka_1966}
Yosuke Nagaoka.
\newblock \emph{Phys. Rev.}, 147:\penalty0 392, 1966.

\bibitem[Tasaki(2003)]{Tasaki_2003}
Hal Tasaki.
\newblock \emph{Commun. Math. Phys.}, 242:\penalty0 445, 2003.

\bibitem[Tian(1991)]{TianFewHoles_1991}
Guang-Shan Tian.
\newblock \emph{J. Phys. A}, 24:\penalty0 513, 1991.

\bibitem[Trugman(1990)]{Trugman_1990}
S.~A. Trugman.
\newblock \emph{Phys. Rev. B}, 42:\penalty0 6612, 1990.

\bibitem[Tian(1990)]{TianNagaokaProof_1990}
Guang-Shan Tian.
\newblock \emph{J. Phys. A}, 23:\penalty0 2231, 1990.

\bibitem[Tasaki(1998)]{TasakiNagaokaProof_1998}
Hal Tasaki.
\newblock \emph{Prog. Theor. Phys.}, 99:\penalty0 489, 1998.

\bibitem[Lieb(1989)]{LiebFerrimagnetism_1989}
Elliot~H. Lieb.
\newblock \emph{Phys. Rev. Lett.}, 62:\penalty0 1201, 1989.

\bibitem[Mielke(1991)]{MielkeFlatBands_1991}
A.~Mielke.
\newblock \emph{J. Phys. A}, 24:\penalty0 L73, 1991.

\bibitem[Tasaki(1992)]{TasakiFlatBands_1992}
Hal Tasaki.
\newblock \emph{Phys. Rev. Lett.}, 69:\penalty0 1608, 1992.

\bibitem[Denteneer(1996)]{Denteneer_1996}
P.~J.~H. Denteneer.
\newblock \emph{Phys. Rev. B}, 53:\penalty0 9764, 1996.

\bibitem[Long and Zotos(1993)]{LongZotos_1993}
M.~W. Long and X.~Zotos.
\newblock \emph{Phys. Rev. B}, 48:\penalty0 317, 1993.

\bibitem[Chiappe et~al.(1993)Chiappe, Louis, Gal\'{a}n, Guinea, and
  Verg\'{e}s]{Chiappe_1993}
G.~Chiappe, E.~Louis, J.~Gal\'{a}n, F.~Guinea, and J.~A. Verg\'{e}s.
\newblock \emph{Phys. Rev. B}, 48:\penalty0 16539, 1993.

\bibitem[Pastor et~al.(1994)Pastor, Hirsch, and
  M\"{u}hlschlegel]{PastorHirschMuhlschlegel_1994}
G.~M. Pastor, R.~Hirsch, and B.~M\"{u}hlschlegel.
\newblock \emph{Phys. Rev. Lett.}, 72:\penalty0 3879, 1994.

\bibitem[Barbieri et~al.(1990)Barbieri, Riera, and
  Young]{BarbieriRieraYoung_1990}
A.~Barbieri, J.~A. Riera, and A.~P. Young.
\newblock \emph{Phys. Rev. B}, 41:\penalty0 11697, 1990.

\bibitem[Pastor et~al.(1996)Pastor, Hirsch, and
  M\"{u}hlschlegel]{PastorHirschMuhlschlegel_1996}
G.~M. Pastor, R.~Hirsch, and B.~M\"{u}hlschlegel.
\newblock \emph{Phys. Rev. B}, 53:\penalty0 10382, 1996.

\bibitem[Strack and Vollhardt(1994)]{StrackVollhardt_1994}
Rainer Strack and Dieter Vollhardt.
\newblock \emph{Phys. Rev. Lett.}, 72:\penalty0 3425, 1994.

\bibitem[Kollar et~al.(1996)Kollar, Strack, and
  Vollhardt]{KollarStrackVollhardt_1996}
Marcus Kollar, Rainer Strack, and Dieter Vollhardt.
\newblock \emph{Phys. Rev. B}, 53:\penalty0 9225, 1996.

\bibitem[Wahle et~al.(1998)Wahle, Bl\"{u}mer, Schlipf, Held, and
  Vollhardt]{Wahle_1998}
J.~Wahle, N.~Bl\"{u}mer, J.~Schlipf, K.~Held, and D.~Vollhardt.
\newblock \emph{Phys. Rev. B}, 58:\penalty0 12749, 1998.

\bibitem[Park et~al.(2008)Park, Haule, Marianetti, and Kotliar]{Park_2008}
Hyowon Park, K.~Haule, C.~A. Marianetti, and G.~Kotliar.
\newblock \emph{Phys. Rev. B}, 77:\penalty0 035107, 2008.

\bibitem[Lieb and Mattis(1962)]{LiebMattis_1962}
Elliot Lieb and Daniel Mattis.
\newblock \emph{Phys. Rev.}, 125:\penalty0 164, 1962.

\bibitem[Daul and Noack(1998)]{DaulNoack_1998}
S.~Daul and R.~M. Noack.
\newblock \emph{Phys. Rev. B}, 58:\penalty0 2635, 1998.

\bibitem[Vollhardt et~al.(1999)Vollhardt, Bl\"{u}mer, Held, Kollar, Schlipf,
  Ulmke, and Wahle]{Vollhardt_1999}
D.~Vollhardt, N.~Bl\"{u}mer, K.~Held, M.~Kollar, J.~Schlipf, M.~Ulmke, and
  J.~Wahle.
\newblock \emph{Advances in Solid State Phys.}, 38:\penalty0 383, 1999.

\bibitem[Barbieri and Young(1991)]{BarbieriYoung_1991}
A.~Barbieri and A.~P. Young.
\newblock \emph{J. Phys.: Condens. Matter}, 3:\penalty0 1801, 1991.

\bibitem[Haerter and Shastry(2005)]{HaerterShastryAFTriangle_2005}
Jan~O. Haerter and B.~Sriram Shastry.
\newblock \emph{Phys. Rev. Lett.}, 95:\penalty0 087202, 2005.

\bibitem[Eisenberg et~al.(2002)Eisenberg, Berkovits, Huse, and
  Altshuler]{EisenbergHuseAltshuler_2002}
E.~Eisenberg, R.~Berkovits, David~A. Huse, and B.~L. Altshuler.
\newblock \emph{Phys. Rev. B}, 65:\penalty0 134437, 2002.

\bibitem[Dagotto et~al.(1992)Dagotto, Moreo, Ortolani, Poilblanc, and
  Riera]{DagottoPhaseSep_1992}
E.~Dagotto, A.~Moreo, F.~Ortolani, D.~Poilblanc, and J.~Riera.
\newblock \emph{Phys. Rev. B}, 45:\penalty0 10741, 1992.

\bibitem[Hirsch(1985)]{HirschMFT_1985}
J.~E. Hirsch.
\newblock \emph{Phys. Rev. B}, 31:\penalty0 4403, 1985.

\bibitem[Sasaki and Kinoshita(1968)]{SasakiKinoshita_1968}
Wataru Sasaki and J.~Kinoshita.
\newblock \emph{Jour. Phys. Soc. of Japan}, 25:\penalty0 1622, 1968.

\bibitem[Sasaki(1976)]{Sasaki_1976}
Wataru Sasaki.
\newblock \emph{J. Phys. (Paris) Colloq.}, 37:\penalty0 C4--307, 1976.

\bibitem[Ue and Maekawa(1971)]{Ue_1971}
Hiromoto Ue and Shigeru Maekawa.
\newblock \emph{Phys. Rev. B}, 3:\penalty0 4232, 1971.

\bibitem[Quirt and Marko(1973)]{Quirt_1973}
J.~D. Quirt and J.~R. Marko.
\newblock \emph{Phys. Rev. B}, 7:\penalty0 3842, 1973.

\bibitem[Hirsch et~al.(1992)Hirsch, Holcomb, Bhatt, and
  Paalanen]{Hirsch_NoExpFerro_1992}
M.~J. Hirsch, D.~F. Holcomb, R.~N. Bhatt, and M.~A. Paalanen.
\newblock \emph{Phys. Rev. Lett.}, 68:\penalty0 1418, 1992.

\bibitem[Kamimura(1978)]{Kamimura_1978}
H.~Kamimura.
\newblock In L.~R. Friedman and D.~P. Tunstall, editors, \emph{Proc. 19th
  Scottish Universities Summer School in Physicst}, pages 327--68. SUSSP, 1978.

\bibitem[Kamimura(1985)]{Kamimura_1985}
H.~Kamimura.
\newblock In M.~Pollack and A.~L. Efros, editors, \emph{Electron-electron
  Interactions in Disordered Systems}, pages 555--617. North-Holland, 1985.

\bibitem[Schofield et~al.(2003)Schofield, Curson, Simmons, Ruess, Hallam,
  Oberbeck, and Clark]{Schofield_2003}
S.~R. Schofield, N.~J. Curson, M.~Y. Simmons, F.~J. Ruess, T.~Hallam,
  L.~Oberbeck, and R.~G. Clark.
\newblock \emph{Phys. Rev. Lett.}, 91:\penalty0 136104, 2003.

\bibitem[Bhatt and Lee(1981)]{BhattLee_JAP_1981}
R.~N. Bhatt and P.~A. Lee.
\newblock \emph{J. Appl. Phys.}, 52:\penalty0 1703, 1981.

\bibitem[Bhatt and Lee(1982)]{BhattLee_1982}
R.~N. Bhatt and P.~A. Lee.
\newblock \emph{Phys. Rev. Lett.}, 48:\penalty0 344, 1982.

\bibitem[Bhatt(1990)]{Bhatt_1990}
R.~N. Bhatt.
\newblock In \emph{Proc. 20th Int. Conf. on Physics of Semiconductors}, page
  2633. World Scientific, Singapore, 1990.

\bibitem[Bhatt et~al.(1988)Bhatt, Paalanen, and Sachdev]{Bhatt_1988}
R.~N. Bhatt, M.~A. Paalanen, and S.~Sachdev.
\newblock \emph{Jour. de Physique}, 49:\penalty0 C8--1179, 1988.

\bibitem[Fisher(1994)]{Fisher_1994}
Daniel~S. Fisher.
\newblock \emph{Phys. Rev. B}, 50:\penalty0 3799, 1994.

\bibitem[Holcomb(1986)]{Holcomb_SUSSP_1986}
D.~Holcomb.
\newblock Localization and interaction in disordered metals and doped
  semiconductors.
\newblock In D.~M. Finlayson, editor, \emph{Proc. 31st Scottish Universities
  Summer School in Physics}, page 313. SUSSP, 1986.

\bibitem[Paalanen et~al.(1988)Paalanen, Graebner, Bhatt, and
  Sachdev]{Paalanen_1988}
M.~A. Paalanen, J.~E. Graebner, R.~N. Bhatt, and S.~Sachdev.
\newblock \emph{Phys. Rev. Lett.}, 61:\penalty0 597, 1988.

\bibitem[Bhatt(1986)]{Bhatt_1986}
R.~N. Bhatt.
\newblock \emph{Physica Scripta}, T14:\penalty0 7, 1986.

\bibitem[Nielsen and Bhatt()]{NielsenBhattTransport}
Erik Nielsen and R.~N. Bhatt.
\newblock Mobility edges, energy gaps and metal-insulator transition in doped
  semiconductors.
\newblock In preparation.

\bibitem[Nielsen and Bhatt(2007)]{ErikNanoscaleFM_2007}
Erik Nielsen and R.~N. Bhatt.
\newblock \emph{Phys. Rev. B}, 76:\penalty0 161202, 2007.

\bibitem[Herring and Flicker(1964)]{HerringFlicker_1964}
Conyers Herring and Michael Flicker.
\newblock \emph{Phys. Rev.}, 134:\penalty0 A362, 1964.

\bibitem[Chernyshev et~al.(2004)Chernyshev, Galanakis, Phillips, Rozhkov, and
  Tremblay]{ChernyshevEffTheories_2004}
A.~L. Chernyshev, D.~Galanakis, P.~Phillips, A.~V. Rozhkov, and A.-M.~S.
  Tremblay.
\newblock \emph{Phys. Rev. B}, 70:\penalty0 235111, 2004.

\bibitem[Lin and Hirsch(1995)]{HirschOccDepHopping_1995}
H.~Q. Lin and J.~E. Hirsch.
\newblock \emph{Phys. Rev. B}, 52:\penalty0 16155, 1995.

\bibitem[Ohno(1998)]{Ohno_1998}
H.~Ohno.
\newblock \emph{Science}, 281:\penalty0 951, 1998.

\bibitem[Chiba et~al.(2003)Chiba, Takamura, Matsukura, and
  Ohno]{ChibaOhno_2003}
D.~Chiba, K.~Takamura, F.~Matsukura, and H.~Ohno.
\newblock \emph{Appl. Phys. Lett.}, 82:\penalty0 3020, 2003.

\bibitem[Berciu and Bhatt(2001)]{Berciu_DMS_2001}
Mona Berciu and R.~N. Bhatt.
\newblock \emph{Phys. Rev. Lett.}, 87:\penalty0 107203, 2001.

\bibitem[Kennett et~al.(2002)Kennett, Berciu, and Bhatt]{Kennett_2002}
Malcolm~P. Kennett, Mona Berciu, and R.~N. Bhatt.
\newblock \emph{Phys. Rev. B}, 66:\penalty0 045207, 2002.

\bibitem[Berciu and Bhatt(2004)]{Berciu_DMS_2004}
Mona Berciu and R.~N. Bhatt.
\newblock \emph{Phys. Rev. B}, 69:\penalty0 045202, 2004.

\bibitem[Calvetti et~al.(1994)Calvetti, Reichel, and Sorensen]{Calvetti_1994}
D.~Calvetti, L.~Reichel, and D.~C. Sorensen.
\newblock \emph{ETNA}, 2:\penalty0 1, 1994.

\bibitem[Riera and Young(1989)]{RieraYoung_HubSq16_1989}
J.~A. Riera and A.~P. Young.
\newblock \emph{Phys. Rev. B}, 40:\penalty0 5285, 1989.

\bibitem[Gros(1996)]{Gros_1996}
Claudius Gros.
\newblock \emph{Phys. Rev. B}, 53:\penalty0 6865, 1996.

\bibitem[Peres et~al.(2004)Peres, Ara\'{u}jo, and Bozi]{Peres_2004}
N.~M.~R. Peres, M.~A.~N. Ara\'{u}jo, and Daniel Bozi.
\newblock \emph{Phys. Rev. B}, 70:\penalty0 195122, 2004.

\bibitem[Ghosh and Singh(2008)]{GhoshSingh_2008}
Saptarshi Ghosh and Avinash Singh.
\newblock \emph{Phys. Rev. B}, 77:\penalty0 094430, 2008.

\bibitem[Nielsen(2008)]{NielsenThesis}
Erik Nielsen.
\newblock \emph{Impurity Bands at Low Densities: Electronic States, Magnetism,
  and Transport}.
\newblock PhD thesis, Princeton University, 2008.

\bibitem[Efros and Shklovskii(1975)]{EfrosShklovskii_1975}
A.~L. Efros and B.~I. Shklovskii.
\newblock \emph{J. Phys. C}, 8:\penalty0 L49, 1975.

\bibitem[Efros(1976)]{Efros_1976}
A.~L. Efros.
\newblock \emph{J. Phys. C}, 9:\penalty0 2021, 1976.

\bibitem[Davies et~al.(1982)Davies, Lee, and Rice]{DaviesLeeRice_1982}
J.~H. Davies, P.~A. Lee, and T.~M. Rice.
\newblock \emph{Phys. Rev. Lett.}, 49:\penalty0 758, 1982.

\bibitem[Davies et~al.(1984)Davies, Lee, and Rice]{DaviesLeeRice_1984}
J.~H. Davies, P.~A. Lee, and T.~M. Rice.
\newblock \emph{Phys. Rev. B}, 29:\penalty0 4260, 1984.

\bibitem[Baranovskii et~al.(1979)Baranovskii, Efros, Gelmont, and
  Shklovskii]{Baranovskii_1979}
S.~D. Baranovskii, A.~L. Efros, B.~L. Gelmont, and B.~I. Shklovskii.
\newblock \emph{J. Phys. C}, 12:\penalty0 1023, 1979.

\end{thebibliography}

\end{document}